\documentclass[trackchanges]{aastex701}

\usepackage{caption}
\usepackage{supertabular}

\usepackage{newtxtext,newtxmath}

\usepackage[T1]{fontenc}

\DeclareRobustCommand{\VAN}[3]{#2}
\let\VANthebibliography\thebibliography
\def\thebibliography{\DeclareRobustCommand{\VAN}[3]{##3}\VANthebibliography}

\usepackage{graphicx}	
\usepackage{amsmath}	
\usepackage{amssymb}	
\newcommand{\angstrom}{\textup{\AA}}
\begin{document}

\title[Magnetic white dwarfs from DESI]{Magnetic white dwarfs from DESI}

\author[orcid=0000-0002-8617-9317]{L. L. Amorim}\email{E-mail: larissal.pesquisa@gmail.com}
\affiliation{Instituto de Física, Universidade Federal do Rio Grande do Sul, 91501-900 Porto-Alegre, RS, Brazil}

\author[orcid=0000-0002-7470-5703]{S. O. Kepler}\email{kepler@if.ufrgs.br}
\affiliation{Instituto de Física, Universidade Federal do Rio Grande do Sul, 91501-900 Porto-Alegre, RS, Brazil}

\author[orcid=0000-0002-0797-0507]{Alejandra D. Romero}\email{aleromero82@gmail.com}
\affiliation{Instituto de Física, Universidade Federal do Rio Grande do Sul, 91501-900 Porto-Alegre, RS, Brazil}

\begin{abstract}
A significant fraction of white dwarfs, the degenerate remnants of low- and intermediate-mass stars, host strong magnetic fields; yet, the origin and evolution of these magnetic fields remain poorly understood. Building a large, statistically robust sample of these magnetic white dwarfs (MWDs) is crucial for testing competing theories of field generation. We used the white dwarf candidates catalog from Gaia DR3 to select objects with spectra from the first data release of the Dark Energy Spectroscopic Instrument (DESI) survey. 
We identified candidate MWDs through visual inspection of their spectra, searching for the characteristic Zeeman splitting of absorption lines. After cross-matching with the literature, we present the discovery of 137 new MWDs.
Follow-up analysis and atmospheric modeling were used to determine magnetic field strengths ranging from approximately 1 to nearly 500~MG. Our findings demonstrate the exceptional capability of large-scale spectroscopic surveys, such as DESI, to uncover rare stellar populations and advance our understanding of compact-object astrophysics.
\end{abstract}


\section{Introduction}
If we consider single stellar evolution, stars with initial masses below $\sim 11 M_\odot$ evolve into white dwarfs (WD). Given the distribution of initial stellar masses in our galaxy, it is evident that this is the final evolutionary stage for 97\% of all stars \citep{2018Lauffer}. Recent studies have shown that up to 25\% of these stars have stable magnetic fields with intensities stronger than 1~MG \citep{2021Bagnulo}. These fields should last billions of years as the white dwarfs cool once the ohmic decay timescales are larger than the thermal evolution time \citep{1972Chanmugam}.

The first magnetic field detected in a WD was by \citet{1970Kemp} in 1970. Since then, the Sloan Digital Sky Survey (SDSS) has been an extensive survey that significantly augmented the catalog, with more than 800 objects identified \citep{2023Amorim}. 
Alternatively, a different approach is to use volume-limited samples, as in \citet{2021Bagnulo} and \citet{2025Moss}.

Despite having such a sample, we still do not have an answer to the question of the origin of these magnetic fields. It has been hypothesized that the fields measured in the compact phase are inherited from earlier stages of stellar evolution (fossil fields).  Alternatively, these fields could arise from binary evolution, either via dynamos during the Common-Envelope phase or via mergers. Finally, they could be generated, or at least enhanced, during the white dwarf phase due to crystallization or convection in their atmosphere. 

It has been shown that the fossil field alone is insufficient to explain the distribution of MWD observed \citep{2005WickFerrario}. \citet{2023Amorim} showed that although, after crystallization, magnetic fields are more intense, there are dozens of stars exhibiting magnetism before this stage. \citet{2021Bagnulo} called attention to the delay between when crystallization begins and the emergence of the magnetic field to the surface. \citet{2025Moss} reiterated that there may be two distinct populations of MWD. They stated that the young, high-field, and high-mass objects are likely remnants of mergers. In contrast, the old, less massive, low-field objects with magnetism likely result from a single evolutionary path, with the caveat that those with $M\leq 0.45M_\odot$ must have originated from binary interactions.

Volume-limited samples are superior in their completeness, especially for older, fainter stars \citep{2022Bagnulo}. Unfortunately, they are very time-intensive to observe; therefore, they are scarce. \citet{2025Moss}, a 100~pc sample, has only 163 MWD. More data is crucial to the sedimentation of the conclusions drawn so far. To increase the overall sample size, magnitude-limited samples are a frequently used technique.

DESI is an instrument that, although not specifically designed for the study of white dwarfs, is highly useful. In their goal to build a spectroscopic redshift map of the Universe, they ultimately observed the spectra of thousands of white dwarfs \citet{2023DESI_MWS}.

In this work, we used DESI DR1 spectroscopic data to find 137 new MWD. This extends the previous list of all magnetic WDs, which is dominated by SDSS spectra, and demonstrates DESI's capability to contribute to WD measurements.

\section{DESI data}
DESI is a highly multiplexed instrument mounted at the prime focus of the Mayall
4-meter telescope at Kitt Peak National Observatory (KPNO) in Arizona, USA. It has 
5,000 robotic fibers and a $3.2^o$ diameter field-of-view, enabling it to rapidly acquire optical spectrophotometry of tens of thousands of targets per night. 
We use the DESI Data Release 1 (DR1), which comprises all data acquired during the first 13 months of the DESI main survey \citet{2025DESIDR1}. They obtained the spectra of 4 million stars, even though that was not their primary scientific focus. 
The published spectra have a resolution of 0.8\,\AA \ and cover the wavelength range from 3600\,\AA \ to 9824\,\AA.

From DESI DR1, we use the Milky Way catalog and match the stars with the \citet{2021Gentile-Nicola-Gdr3} white dwarf candidates catalog, resulting in 46006 stars. The Milky Way Survey has an upper limit of magnitude r = 19. Regrettably, fewer than 2\% of these have a signal-to-noise ratio above 50. We applied an SNR cut of 10 to ensure that the Zeeman splitting is visible above the noise [e.g., \citet{2013Kepler}]. We then visually inspected all 16,847 remaining spectra for split lines.


\section{Magnetic Field Determination}
Due to their strong surface gravity, white dwarfs typically have a simple atmospheric composition, and the absorption lines they exhibit in their spectra are broad, allowing for visual identification but restricting B < 1 MG identification, except for SNR>20. At lower magnetic fields, where magnetism divides each H absorption line into three components, it is easy to identify MWDs above fractions of MG for $\simeq 1\angstrom$ spectral resolution with good SNR. As this is not the case for most available spectra, we apply 5-point smoothing. This compromises our detections below $\sim$1 MG. Additionally, this method is biased against larger fields, which exhibit quadratic magnetic effects. With this method, we found 137 new MWDs. After visual identification, we estimated their magnetic field intensities using either surface fitting or line positions alone, as described below.

Two distinct methods stand out in the literature for probing magnetic fields in white dwarfs. One of them is based solely on the separation of the line components and assumes a uniform field distribution over the stellar surface. This is clearly unrealistic, but in some cases, it is the best available option. The other approach considers an off-centered and inclined dipole distribution. We used YAWP, a code presented by \citet{2009Kulebi} that follows this second approach. Unfortunately, it has a limitation of $T_\mathrm{eff}\geq 8000$\,K models and
a detection limit around 1\,MG at this spectral resolution. For cooler stars, we used the first method, applying theoretical line-splitting data from \citet{2014Schimeczek}.

YAWP is a code designed to fit observed spectra with atmospheric models that depend on effective temperature and surface gravity. Although approximate treatments of magnetic field effects on spectra exist \citep{2023Hardy}, previous studies have shown that allowing surface gravity to vary does not significantly improve the determination of magnetic field intensity \citep{2023Amorim}. Therefore, in this work, we adopt models with a fixed value of $\log g = 8$.

The YAWP code allows us to adjust the effective temperature ($T_{eff}$), but it has too many free parameters and a degenerate problem, particularly for such low-resolution spectra. It is noticeable that the uncertainties of the measured $T_{eff}$ are large (around 500 K) and do not play a significant role in the determination of the magnetic field. Therefore, we opted to use Gaia \citet{2021Gentile-Nicola-Gdr3} measurements of temperature and treat it as a fixed parameter in the fitting process.

The code also allows for more complex field structures beyond dipoles (e.g., quadrupoles, octopoles). However, this quickly increases computational costs with little improvement in model predictions. Instead, we used inclined offset dipoles, yielding a complex field distribution over the stellar surface with fewer free parameters. We will adopt the dipole off-center along the z-axis, measured in units of stellar radius. With null inclination and offset, we recover a simple dipole field.  

\begin{figure}
\centering
	\includegraphics[width=0.7\columnwidth]{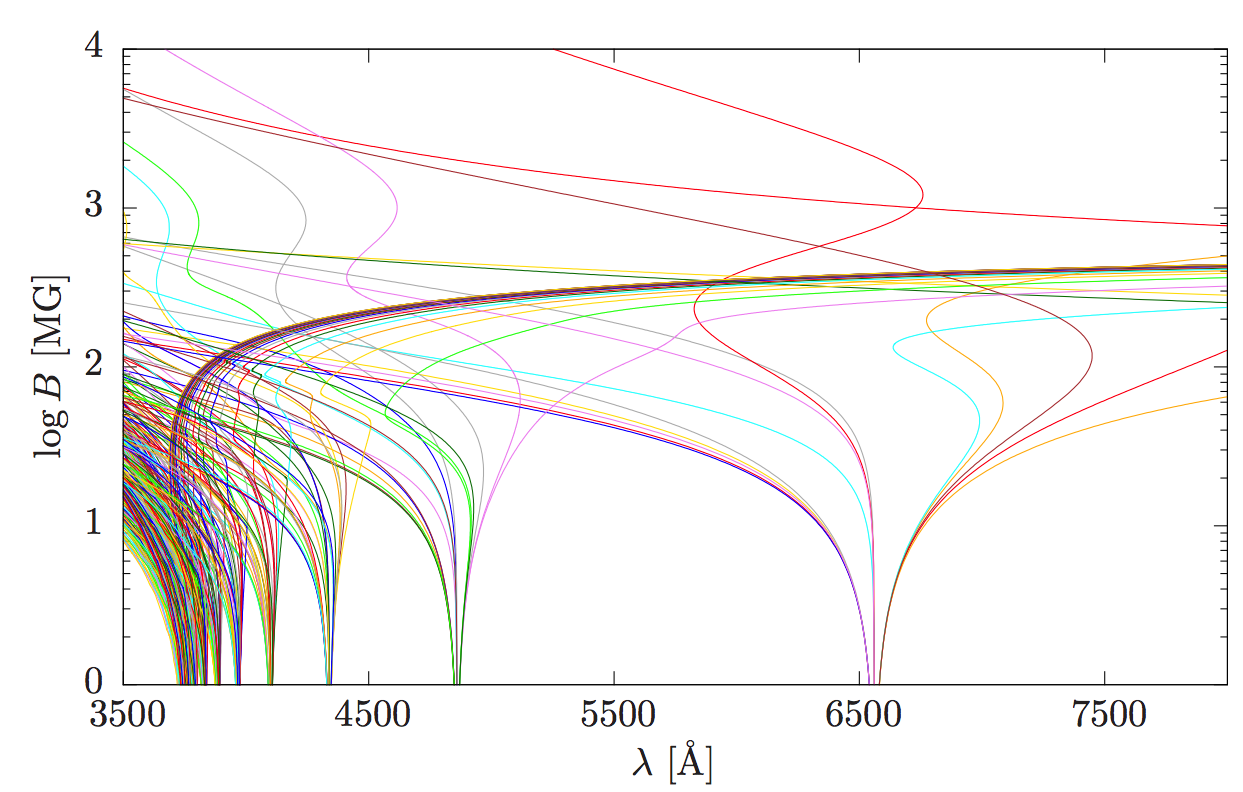}
    \caption{Magnetic field strength as a function of the wavelength of the first 325 transitions in the Balmer series, which emerge from the field-free Balmer transitions up to principal quantum numbers n = 11.
(A color version of this figure is available in the online journal.)}
    \label{fig:fields}
\end{figure}

Due to the faint characteristic of white dwarfs, which are small in size, many spectra are very noisy, especially in the red region. Additionally, we note that for low fields, the absorption line components remain below approximately 7000~$\angstrom$, as shown in Fig.~\ref{fig:fields}. Therefore, we cropped the spectra of stars with a field below 50~MG at this wavelength to proceed with fitting, thereby reducing noise and potential interference in our results. For stars with stronger magnetic fields, we used data up to 9000~$\angstrom$, the limit of the YAWP models.

\begin{figure}
    \centering
    \includegraphics[width=\linewidth]{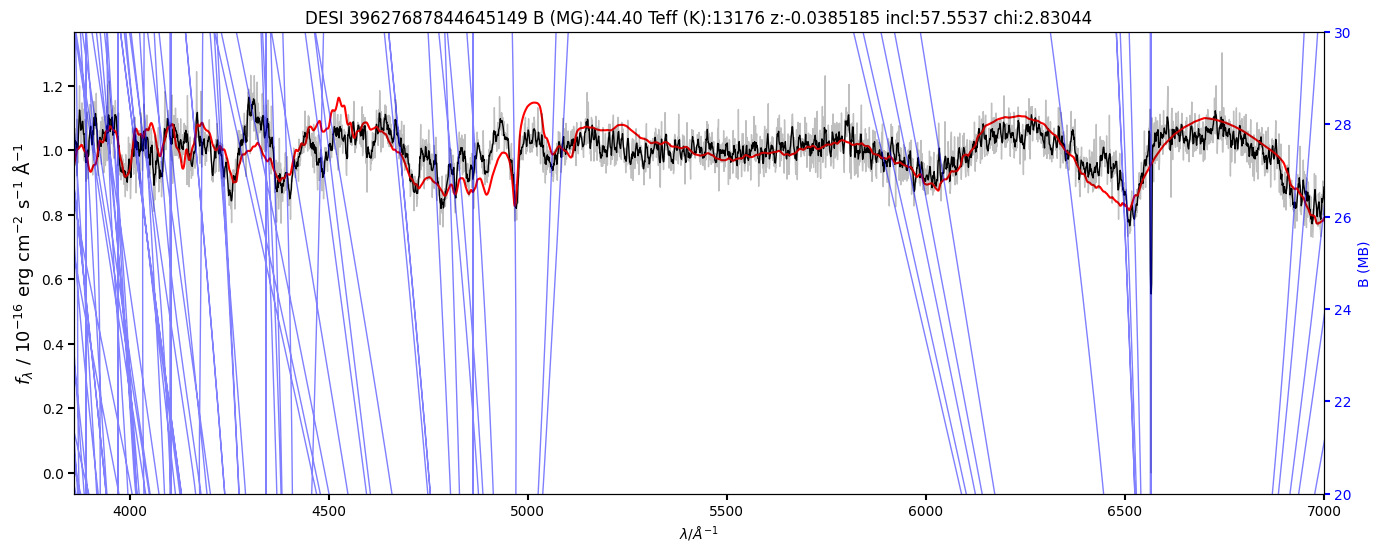}
    \caption{In black is the observed spectra for the star WD\,J225325.81-040822.38 with a running mean of 5 points. In gray is the original data. The best model to fit the data, identified with YAWP, is in red. Both are normalized. In blue is the position of the absorption lines for a certain magnetic field intensity, as in Fig.~\ref{fig:fields}. This is the most magnetic star in our sample for which we cropped the spectra at 7000~$\angstrom$, with a magnetic field intensity of 44~MG.}
    \label{fig:Balto}
\end{figure}

This was a reasonable decision, once our most magnetic WD below 50 MG was well fitted. We present this star as an example of our field determination with YAWP in Fig.~\ref{fig:Balto} (the spectra of all stars will be available digitally in \ref{app}). As noted previously, we also used the spaghetti method, which relies solely on line positions. It is important to note that these two methods are measuring different quantities. In  Fig.~\ref{fig:Balto}, we show in blue that if we used the spaghetti method, we would measure a magnetic field intensity of around 25~MG. This is the mean surface field, an average over the visible hemisphere of the local field modulus |B| derived from the mean line splitting. The YAWP value is almost twice as high because it measures the dipole polar field strength. The method applied to each individual star can be identified in Table~\ref{t:names} by the presence or absence of the inclination angle and the de-centering parameter among the listed quantities. For further information on our field measurement methods, see \citet{2023Amorim}, including the associated uncertainty estimates, which are approximately 10\%.

\begin{figure}
    \centering
    \includegraphics[width=\linewidth]{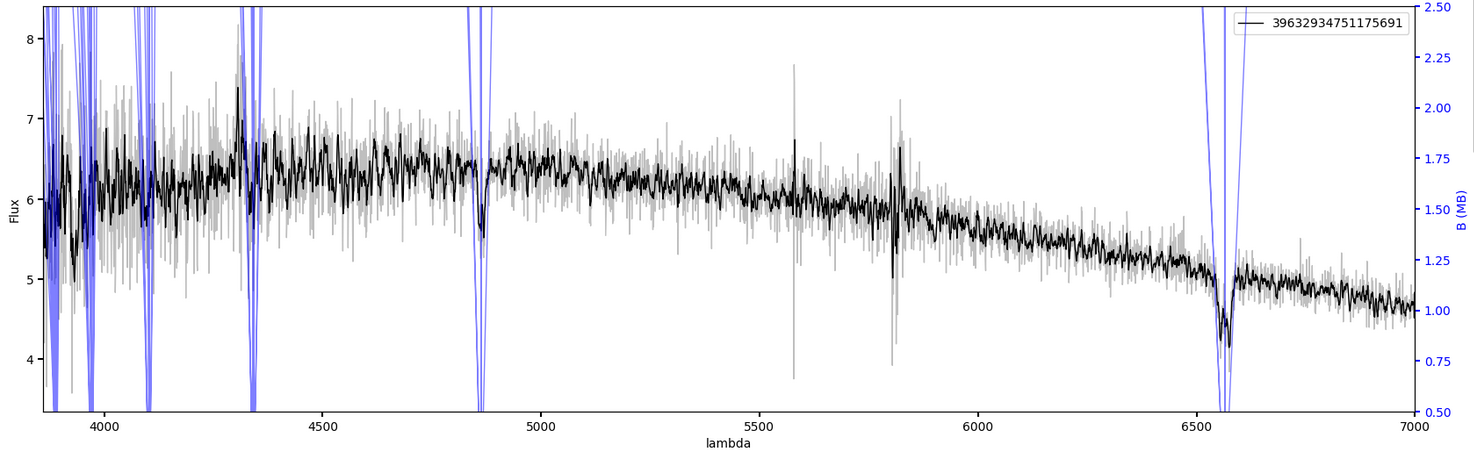}
    \caption{In black is the observed normalized spectra for the star Gaia~DR3~881355910802027008 (DESI name in the label) with a running mean of 5 points. In gray is the original data.} In blue is the position of the absorption lines for a certain magnetic field intensity, as in Fig.~\ref{fig:fields}. In addition to having a $T_{eff} < 8000\ K$, this is our smallest estimated magnetic field intensity, at 0.75~MG.
    \label{fig:T8000}
\end{figure}

In Fig.~\ref{fig:T8000}, we present our star with the lowest estimated magnetic field intensity, below 1 MG. It is an example of the spaghetti method. We determine the field intensity by focusing on the deepest portions of the absorption lines, starting with the Balmer alpha line, and then checking whether the field thus determined also explains, within a 10\% error margin, the other lines.

\section{Results}
Our sample is presented in Table~\ref{t:names} of the Appendix~\ref{app}, along with the stellar properties from the literature and the magnetic field properties derived from our YAWP model/spaghetti determination. We compare the distributions of mass, temperature, and magnetic field intensity with those from \citet{2023Amorim} and \citet{2025Moss}, which are the largest magnitude-limited and volume-limited samples of MWDs in the literature, respectively. We note that all stars in this work have data from \citet{2021Gentile-Nicola-Gdr3}, but only 632 out of 808 from \citet{2023Amorim} have this information.

\begin{figure}[h]
    \centering
    \includegraphics[width=0.49\linewidth]{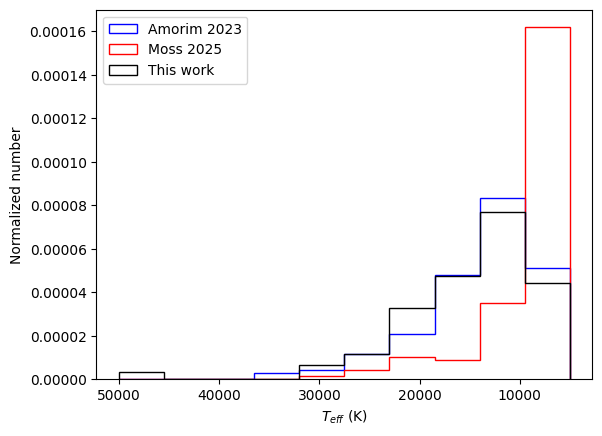}
    \includegraphics[width=0.46\linewidth]{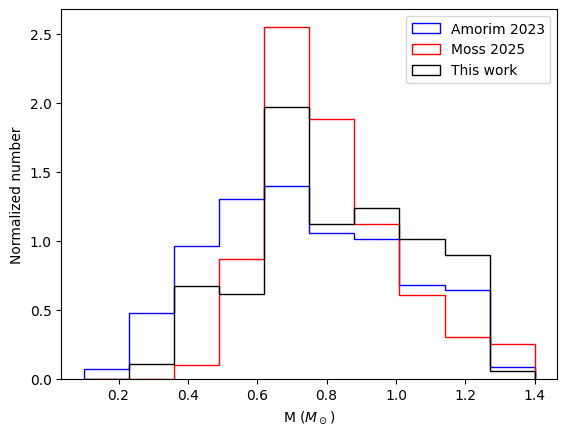}
    \caption{Distribution of effective temperature and Distribution of mass, respectively. In black are stars from this work, and in blue and red are the data from \citet{2023Amorim} and \citet{2025Moss}, respectively. By convention, the x-axis is inverted for the temperature, with a range between 5000~K and 50\,000~K. The histograms are normalized so that the total area under their curves equals 1. This helps to compare different-sized samples.}
    \label{fig:hist_T}
\end{figure}

It is easy to see in the left panel of Fig.~\ref{fig:hist_T} that the distribution of effective temperatures for the stars in this work is, overall, compatible with the other magnitude-limited samples from the literature. However, compared with a volume-limited sample, our sample exhibits bias. Cooler stars are fainter and, therefore, harder to spot in a magnitude-limited sample, even though we know that there are overall more WDs with effective temperatures below 10000 K than above it. Additionally, in cooler stars, the absorption lines become shallower, making them harder to distinguish from the noise.

The same behavior is not clearly repeated for the mass distribution, which can be seen in the right panel of the same figure. The magnitude-limited sample has a broad distribution around $0.73\ M_\odot$, while the volume-limited sample has a skewed distribution with a mean of $0.81\ M_\odot$. Our sample exhibits a rapidly increasing mass on the left side of the distribution, whereas the right side is broad. This is an intermediate behavior, although the highest peak of each distribution is the same.

\begin{figure}
    \centering
    \includegraphics[width=0.47\linewidth]{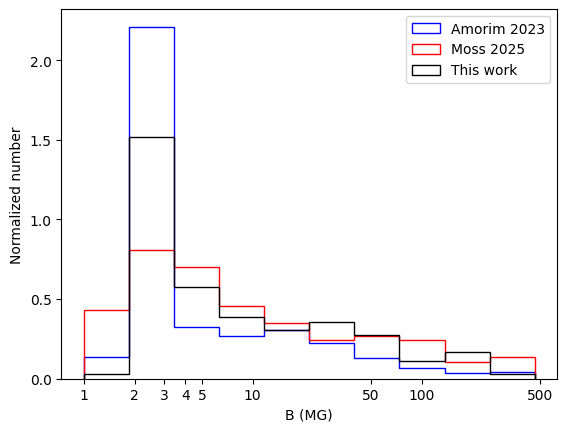}
    \includegraphics[width=0.47\linewidth]{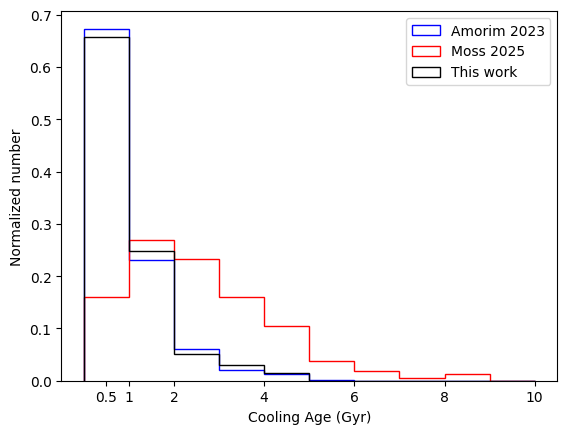}
    \caption{Distribution of magnetic field intensity and cooling age, respectively. In blue and red are stars from the literature, with magnetic field intensity range between 1 MG and 500 MG. In black are the data from this work. Note that the bins are equally spaced on both the logarithmic and linear scales for improved visualization.}
    \label{fig:hist_B}
\end{figure}

To better visualize the comparison of magnetic field intensity distribution, we limit the literature data to the interval of 1~MG to 500~MG. This is presented in the left panel of Fig.~\ref{fig:hist_B}. Although all distributions peak in the same bin, their formats differ. Again, our distribution is intermediate between the two literature cases. It does not have such a characteristic high peak around 2-3~MG when compared to the rest of the distribution as the other magnitude-limited sample does. However, the peak is still present, contrary to the volume-limited sample, which has a broader distribution.


In addition to the properties mentioned above, we decided to investigate the cooling age of these stars and see how they compare with data from the literature. To determine the cooling ages for the three samples analyzed here, we used the evolutionary sequences from \citet{2012MNRAS.420.1462R, 2013ApJ...779...58R} for white dwarf masses between 0.493 and 1.023 M$_{\odot}$, complemented at low and high stellar mass with computations from \citet{ 2013A&A...557A..19A} and \citet{2018MNRAS.480.1547L}, respectively. These computations account for the full evolution, from the zero-age main sequence through the central-burning and mass-loss stages to the white dwarf cooling sequence. The beginning of the white dwarf cooling sequence is defined as the point at which the effective temperature is higher during the post-AGB phase \citep[see, for example,][]{2015MNRAS.450.3708R}. It is from this point that the cooling age is computed. In the right panel of Fig.~\ref{fig:hist_B}, it is easy to see that the magnitude-limited samples have a bias against older stars. This is analogous to the temperature bias: older stars have had more time to cool and are therefore cooler. We draw attention to the fact that all new MWDs presented here are DAs, despite our search for any spectral type MWDs.

\section{Discussion}
\begin{figure}
    \centering
    \includegraphics[width=0.48\linewidth]{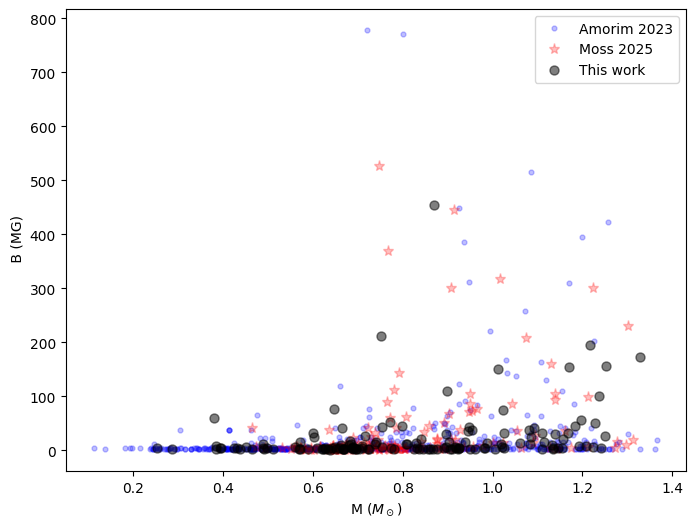}
    \includegraphics[width=0.48\linewidth]{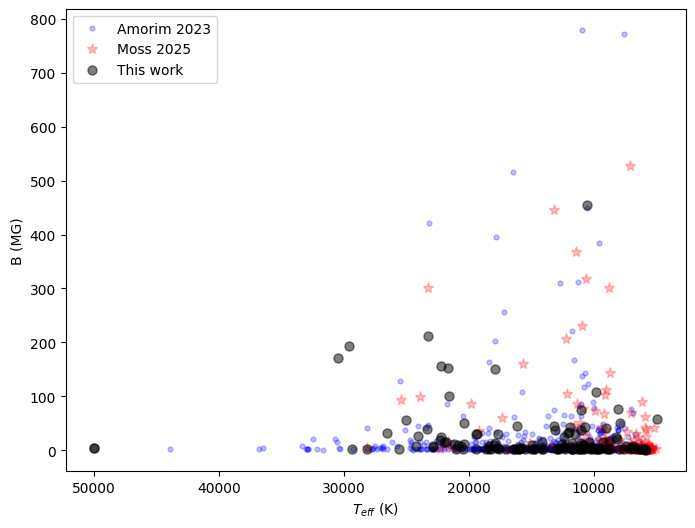}
    \caption{On the left, plot of the Magnetic field intensity versus Mass. On the right, plot of the Magnetic field intensity versus Effective temperature. In black are the literature data, limited to $500 MG$. In red are the stars from this work.}
    \label{fig:BxM}
\end{figure}

To investigate the relationship between stellar properties, we plotted magnetic field intensity versus stellar mass. As shown in Fig.~\ref{fig:BxM}, the only prominent feature is the absence of low-mass, highly magnetic stars. This presents a strong argument against the generation of these fields during binary interactions, as low-mass stars necessarily evolve in multiple systems.

Similarly, for the distribution of magnetic field intensity as a function of temperature, the only visible trend in Fig.~\ref{fig:BxM} is the absence of hot white dwarfs with high magnetic field intensity in all the data. This suggests a generation of fields during the cooling of the white dwarfs.

\begin{figure}
    \centering
    \includegraphics[width=0.75\linewidth]{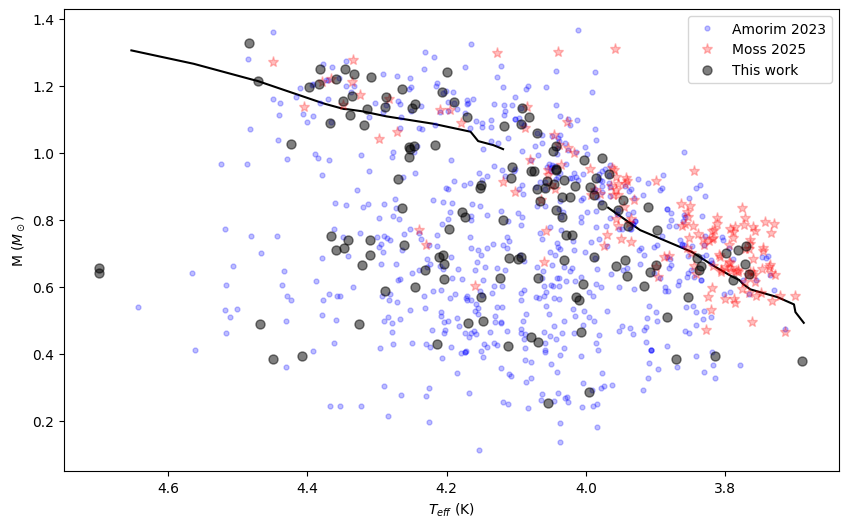}
    \caption{Plot of the $T_{eff}$ X Mass. The black line presents the beginning of the crystallization at the core of the white dwarfs. The colors represent the intensity of the magnetic field. As the stars cool down, they move to the right, so the x-axis can also be seen as proportional to the age of the white dwarf.  }
    \label{fig:cryst}
\end{figure}

A diagram that has become a classic in the study of MWDs is the $T_\mathrm{eff}$ vs. Mass. Here we reproduce it in Fig.~\ref{fig:cryst}. We observe the typical trend of more magnetic stars being located near the crystallization line. Additionally, we observe many magnetic stars that have not yet initiated this process. Additionally, we observe no delay in the emergence of an internal field. Thus, the role of crystallization must be to enhance preexisting fields.

\section{Conclusions}

We have investigated the DESI DR1 survey for isolated MWDs and identified 137 new objects, all with effective temperatures and masses measured in the literature. This represents an augmentation of more than 10\% in the total sample. Regarding the general properties of the objects found, we observe a similar distribution in effective temperature and cooling age to that reported in the magnitude-limited sample from the literature, within the interval $5000 - 50000\ K$. The mass distribution is right-skewed with a mean $0.82 ~M_\odot$. 

Although we searched for MWDs across all spectral types and magnetic field intensities, we found only DAs, with $B_\mathrm{MAX} < 500\ MG$. Our findings are consistent with the literature regarding the absence of both highly magnetic, hot stars and highly magnetic low-mass stars. This indicates that the field are not always derived from binary evolution, since all stars below $0.3~M_\odot$ necessarily evolved in multiple systems. And the field increases as the white dwarf cools.

We note that our initial sample includes previously studied MWDs, and we reserve the comparison of DESI data with data from other telescopes for future work.

\bibliography{sample701}{}
\bibliographystyle{aasjournal}
\appendix
\label{app}
\startlongtable
\begin{deluxetable}{lllllllll}
    \tablecaption{Object Names and stellar properties\label{t:names}}
    \tablehead{
        \colhead{Gaia Source ID} & \colhead{DESI Name} & \colhead{$T_{eff}\ (K)$} & \colhead{Mass ($M_\odot$)} & \colhead{B (MG)} & \colhead{zoffset ($r_*$)}& \colhead{Incl (°)}& \colhead{$\tau$ (Gyrs)}& \colhead{G (mag)}
    }
    \startdata
881355910802027008	&	39632934751175691	&	5831	&	0.638	&	0.75	&	NA	&	NA	&	3.209	&	19.3	\\
3937106051052425856	&	39628198891229530	&	5896	&	0.722	&	1.5	&	NA	&	NA	&	4.297	&	19.6	\\
697894183729407616	&	39628516555230071	&	18234	&	0.725	&	1.92	&	0.11	&	65.35	&	0.159	&	19.7	\\
3229556024930689024	&	39627767788082401	&	6145	&	0.620	&	2.0	&	NA	&	NA	&	2.468	&	18.3	\\
1412818462543444992	&	39633271843195931	&	6284	&	0.702	&	2.0	&	NA	&	NA	&	3.208	&	19.5	\\
1101423953375120896	&	39633437438509791	&	9725	&	0.690	&	2.05	&	-0.26	&	74.46	&	0.851	&	18.7	\\
846164701125190272	&	39633365241954558	&	19433	&	0.589	&	2.20	&	0.17	&	0.89	&	0.077	&	18.7	\\
4541481724254126592	&	39628100685797752	&	9887	&	0.285	&	2.23	&	-0.28	&	65.45	&	0.376	&	18.8	\\
3588092168153411456	&	39627565039616142	&	8743	&	0.634	&	2.27	&	-0.34	&	70.65	&	0.973	&	17.9	\\
4432128386565433856	&	39627825136798655	&	17013	&	0.651	&	2.28	&	0.31	&	90.00	&	0.156	&	18.8	\\
910479671719076352	&	39633052858583357	&	18596	&	0.922	&	2.28	&	0.30	&	85.80	&	0.274	&	19.7	\\
1319045616935970816	&	39628492119216740	&	14783	&	0.491	&	2.28	&	0.30	&	70.42	&	0.153	&	17.9	\\
4231486854731892736	&	39627814059647373	&	15993	&	0.670	&	2.28	&	-0.30	&	83.55	&	0.201	&	18.6	\\
1049950797557300992	&	39633383772391197	&	16049	&	0.696	&	2.29	&	0.30	&	85.74	&	0.217	&	18.6	\\
6897423254937635840	&	39627579358972174	&	18397	&	0.837	&	2.30	&	-0.09	&	79.81	&	0.220	&	18.6	\\
1047607983092185344	&	39633383818528274	&	20992	&	0.666	&	2.30	&	-0.17	&	75.14	&	0.071	&	18.7	\\
1739636512805502976	&	39627958876378070	&	10170	&	0.465	&	2.49	&	-0.37	&	9.66	&	0.494	&	18.0	\\
3761926017359305984	&	39627552796442946	&	5919	&	0.669	&	2.5	&	NA	&	NA	&	3.451	&	19.2	\\
3641304266032854912	&	39627631569670412	&	7128	&	0.569	&	2.5	&	NA	&	NA	&	1.420	&	17.8	\\
1089362757495335680	&	39633434850627832	&	7656	&	0.510	&	2.5	&	NA	&	NA	&	1.044	&	18.4	\\
854100945990370688	&	39633345524531710	&	12760	&	0.960	&	2.50	&	-0.36	&	47.59	&	0.882	&	18.6	\\
927017700990423680	&	39633173964916177	&	10326	&	0.571	&	2.51	&	-0.37	&	12.01	&	0.551	&	19.4	\\
4448354012799681280	&	39628094725691907	&	9734	&	0.875	&	2.52	&	-0.36	&	71.69	&	1.548	&	18.7	\\
2505076268513436928	&	39627742936826028	&	11027	&	0.949	&	2.53	&	-0.38	&	50.38	&	1.277	&	18.3	\\
1681361258365865472	&	2305843021737890912	&	11123	&	0.909	&	2.54	&	-0.38	&	51.62	&	1.080	&	17.7	\\
898706444806421632	&	39633023951440377	&	10822	&	0.868	&	2.55	&	-0.36	&	78.28	&	1.029	&	18.0	\\
3954757198407694848	&	39628290180253631	&	14141	&	0.906	&	2.56	&	0.35	&	89.99	&	0.569	&	18.1	\\
3170947523978412928	&	39627363528476751	&	8841	&	0.860	&	2.56	&	-0.36	&	65.90	&	1.998	&	17.6	\\
4582807659058030592	&	39628402042346485	&	15055	&	0.825	&	2.57	&	0.35	&	89.25	&	0.392	&	18.5	\\
846376120890633856	&	39633377854229136	&	11710	&	0.437	&	2.57	&	-0.38	&	61.92	&	0.348	&	18.7	\\
5109540386655680768	&	39627415554627258	&	11732	&	0.627	&	2.57	&	-0.39	&	52.51	&	0.439	&	18.2	\\
1300118314538891008	&	39628374527705634	&	7943	&	0.667	&	2.58	&	-0.36	&	11.34	&	1.348	&	18.2	\\
2717954192235991424	&	39628030997431350	&	14807	&	1.109	&	2.58	&	-0.36	&	63.66	&	0.859	&	18.4	\\
2372015982660258304	&	39627432323451204	&	11167	&	1.006	&	2.58	&	-0.38	&	53.78	&	1.421	&	19.3	\\
601996738060168576	&	39628074991487053	&	8691	&	0.781	&	2.59	&	-0.50	&	20.12	&	1.591	&	19.1	\\
3187858837446546560	&	39627623134925056	&	20422	&	0.695	&	2.73	&	-0.39	&	38.04	&	0.088	&	17.5	\\
1081503242221071744	&	39633341804186929	&	9475	&	0.846	&	2.75	&	-0.37	&	82.08	&	1.485	&	18.6	\\
898170849500461184	&	39633028732949110	&	15967	&	0.623	&	2.76	&	0.25	&	1.14	&	0.177	&	18.8	\\
3914885024134562560	&	39628010252404869	&	28111	&	0.384	&	2.77	&	-0.39	&	82.25	&	0.002	&	18.2	\\
2573823610997487744	&	39628025788108100	&	12337	&	1.134	&	2.83	&	-0.50	&	32.78	&	1.199	&	19.9	\\
2831008417345284096	&	39628235771743300	&	11622	&	0.857	&	2.85	&	-0.50	&	32.60	&	0.839	&	19.2	\\
1796123437350527360	&	39628364679481717	&	14039	&	0.498	&	2.98	&	0.43	&	18.42	&	0.187	&	19.2	\\
3205638623366144256	&	39627731490573214	&	12905	&	0.687	&	2.98	&	-0.41	&	56.11	&	0.403	&	17.8	\\
1000654988295741056	&	39633348317941347	&	6055	&	0.710	&	3.0	&	NA	&	NA	&	3.752	&	18.4	\\
2156682093558677888	&	39633396917340766	&	6532	&	0.394	&	3.0	&	NA	&	NA	&	1.886	&	18.9	\\
896833461108376704	&	39632974928414719	&	25596	&	0.395	&	3.00	&	0.44	&	88.77	&	0.005	&	18.9	\\
3743180001236999936	&	39628129114787126	&	10738	&	0.919	&	3.00	&	0.40	&	88.85	&	1.240	&	18.2	\\
2124982073818813696	&	39633275781644606	&	8987	&	0.831	&	3.06	&	-0.38	&	64.57	&	1.680	&	18.7	\\
2869251836941053696	&	39628462188660734	&	11770	&	0.892	&	3.06	&	-0.50	&	43.78	&	0.882	&	18.7	\\
2706493948180257152	&	39627941201579083	&	29365	&	0.488	&	3.07	&	-0.50	&	45.02	&	0.011	&	18.5	\\
2722111961097470464	&	39627995035471841	&	11345	&	0.253	&	3.08	&	-0.50	&	48.36	&	0.211	&	18.1	\\
4443197646862649472	&	39627993752018839	&	12944	&	0.425	&	3.23	&	0.47	&	6.00	&	0.255	&	20.0	\\
4439469409091233536	&	39627945660121415	&	9658	&	0.925	&	3.31	&	-0.50	&	39.61	&	1.770	&	18.8	\\
3072456708334858112	&	39627738604113642	&	17613	&	1.146	&	3.31	&	-0.50	&	21.14	&	0.562	&	18.2	\\
850121366732852608	&	39633304009310388	&	16262	&	0.691	&	3.33	&	0.47	&	4.27	&	0.206	&	18.0	\\
2713837689421703552	&	39628001217872467	&	15478	&	1.154	&	3.36	&	0.49	&	5.48	&	0.762	&	18.4	\\
2820819105651650176	&	39628201382643871	&	16471	&	1.025	&	3.38	&	0.50	&	3.59	&	0.555	&	18.9	\\
3146644949750999168	&	39628015423980059	&	16371	&	0.430	&	3.42	&	0.50	&	9.54	&	0.110	&	18.4	\\
315548051782005504	&	39628499304058947	&	15725	&	0.774	&	3.47	&	0.48	&	4.77	&	0.299	&	17.3	\\
1189508853532763776	&	39628070788794009	&	13303	&	0.627	&	3.48	&	0.49	&	5.52	&	0.314	&	18.2	\\
1468401111790054272	&	39628512260262820	&	10089	&	0.610	&	3.51	&	-0.50	&	20.07	&	0.634	&	20.0	\\
1003378611740398336	&	39633392446213860	&	10059	&	0.980	&	3.56	&	0.49	&	1.82	&	1.912	&	18.9	\\
3073014263809462528	&	39627732518180674	&	9468	&	0.984	&	3.81	&	-0.50	&	6.20	&	2.266	&	17.7	\\
2260095457118429440	&	39633490207051422	&	6923	&	0.687	&	4.0	&	NA	&	NA	&	2.179	&	17.7	\\
666234953295747072	&	39628339551409865	&	50000	&	0.642	&	4.00	&	-0.20	&	33.47	&	0.002	&	16.6	\\
1742220566993734016	&	39628030645114777	&	50000	&	0.658	&	4.00	&	-0.20	&	33.47	&	0.002	&	16.8	\\
2248972488255487616	&	39633468363113902	&	17752	&	1.135	&	4.48	&	-0.24	&	62.00	&	0.549	&	19.1	\\
2873304808599755520	&	39628488570836041	&	12830	&	0.926	&	4.56	&	-0.49	&	22.60	&	0.784	&	18.9	\\
2150473598074058112	&	39633315824668928	&	15839	&	1.241	&	4.66	&	-0.49	&	31.55	&	0.841	&	19.2	\\
819648260233854464	&	39633215819875500	&	11369	&	0.917	&	4.80	&	-0.50	&	36.90	&	1.034	&	18.9	\\
1486484367214724992	&	39633015705438333	&	6876	&	0.650	&	5.0	&	NA	&	NA	&	1.898	&	18.0	\\
1463829960916504064	&	39628443947633168	&	10464	&	0.754	&	5.01	&	-0.49	&	0.19	&	0.859	&	19.0	\\
1737045234480847488	&	39627964819703778	&	17641	&	1.020	&	5.07	&	-0.25	&	58.99	&	0.425	&	19.0	\\
4392632584570197376	&	39627915721182771	&	11988	&	0.450	&	5.33	&	-0.49	&	17.31	&	0.312	&	16.8	\\
3252609416507834496	&	39627737467455642	&	14192	&	0.896	&	5.35	&	-0.47	&	47.25	&	0.554	&	18.0	\\
3775549619263136512	&	39627618525386108	&	10219	&	0.560	&	5.35	&	-0.48	&	48.20	&	0.560	&	18.2	\\
2424632939808824064	&	39627502762590888	&	12397	&	0.689	&	5.90	&	-0.46	&	30.38	&	0.456	&	17.9	\\
4557629495620225792	&	39628358216057626	&	22860	&	1.223	&	6.29	&	-0.50	&	6.29	&	0.385	&	17.9	\\
3201873728048770048	&	39627701341917076	&	22848	&	0.710	&	7.75	&	-0.36	&	59.94	&	0.056	&	18.1	\\
313432522691279616	&	39628514755871286	&	18374	&	1.192	&	7.91	&	-0.39	&	60.34	&	0.576	&	19.0	\\
4392416049497430656	&	39627915742151049	&	24205	&	1.207	&	8.14	&	-0.39	&	0.17	&	0.314	&	19.2	\\
1179907093365481088	&	39628123544749626	&	7424	&	0.384	&	8.28	&	-0.43	&	48.73	&	1.407	&	19.8	\\
4552411449655668736	&	39628257334665173	&	20634	&	1.133	&	8.83	&	-0.06	&	65.09	&	0.370	&	17.3	\\
2460110885797604992	&	39627473591209517	&	6829	&	0.663	&	9.0	&	NA	&	NA	&	2.042	&	17.0	\\
1693083357988702336	&	39633535320982677	&	10741	&	0.681	&	9.52	&	-0.35	&	58.66	&	0.643	&	18.6	\\
1069174006022174336	&	39633467373259484	&	20863	&	1.083	&	10.32	&	-0.34	&	64.03	&	0.321	&	19.6	\\
987973083742743040	&	39633331515556118	&	10826	&	0.808	&	10.33	&	-0.34	&	24.93	&	0.889	&	19.8	\\
3786801922478222208	&	39627678906582382	&	17961	&	1.018	&	10.49	&	-0.35	&	76.19	&	0.397	&	18.6	\\
2568520803855523840	&	39627959992060857	&	14917	&	0.810	&	10.86	&	-0.09	&	43.10	&	0.379	&	18.1	\\
311301462998935424	&	39628473056101438	&	20403	&	0.739	&	10.90	&	-0.36	&	28.17	&	0.107	&	17.8	\\
4447531719836391168	&	39628047355222534	&	12530	&	0.684	&	11.16	&	-0.23	&	19.20	&	0.435	&	18.0	\\
2617163713666198528	&	39627537785031725	&	21205	&	0.489	&	11.42	&	-0.35	&	78.03	&	0.037	&	18.3	\\
4441843735731284096	&	39627957760690933	&	11771	&	1.061	&	12.94	&	-0.32	&	47.49	&	1.253	&	18.3	\\
1760816985111417856	&	39628183477164770	&	11055	&	0.829	&	13.23	&	-0.29	&	61.53	&	0.900	&	18.6	\\
89110744507050624	&	39628299026042654	&	17938	&	0.988	&	13.80	&	-0.30	&	53.95	&	0.374	&	18.5	\\
1750240404807147776	&	39628000693591095	&	10370	&	0.903	&	14.86	&	-0.27	&	39.94	&	1.324	&	17.9	\\
1494203797835698560	&	39633162355081945	&	9275	&	0.939	&	15.13	&	-0.24	&	53.81	&	2.089	&	19.0	\\
104716353558063232	&	39628338007903614	&	14142	&	0.570	&	15.21	&	-0.28	&	51.02	&	0.231	&	18.9	\\
74659175988656128	&	39628108818547184	&	21821	&	1.114	&	15.39	&	-0.23	&	25.47	&	0.304	&	19.5	\\
2505660658943681024	&	39627742894886499	&	21945	&	0.741	&	15.80	&	0.31	&	83.42	&	0.081	&	18.7	\\
1334260177805941376	&	39628528458664214	&	8787	&	0.682	&	16.74	&	-0.31	&	27.53	&	1.085	&	18.0	\\
1075472803195499008	&	39633509182081591	&	22339	&	1.157	&	19.81	&	-0.15	&	67.22	&	0.319	&	18.5	\\
0073137589334850560	&	39628091118584785	&	8151	&	0.840	&	20.98	&	-0.40	&	24.49	&	2.398	&	19.0	\\
922956452930142336	&	39633121695500646	&	12400	&	1.086	&	24.16	&	-0.20	&	28.82	&	1.274	&	19.3	\\
1153877014571472384	&	39627848910112258	&	8244	&	0.603	&	24.23	&	-0.11	&	48.59	&	1.044	&	20.0	\\
0585494545996081024	&	39627907798144402	&	22248	&	0.716	&	25.34	&	-0.20	&	15.80	&	0.064	&	18.7	\\
2466389333415191808	&	39627598447249395	&	24084	&	1.251	&	26.66	&	-0.14	&	21.23	&	0.375	&	18.2	\\
2794162270870033920	&	39628224631668806	&	11467	&	0.896	&	29.62	&	-0.09	&	49.67	&	0.968	&	19.1	\\
1388763622029050368	&	39633068222317776	&	19401	&	1.139	&	30.43	&	-0.14	&	51.67	&	0.441	&	18.3	\\
2700929702084328064	&	39627971014692238	&	17620	&	0.600	&	31.00	&	-0.05	&	61.12	&	0.115	&	16.5	\\
4441321776947248256	&	39627969810927847	&	19363	&	1.168	&	31.10	&	-0.14	&	75.98	&	0.478	&	18.9	\\
2439020186897492992	&	39627585885309547	&	26547	&	1.026	&	31.60	&	0.02	&	25.31	&	0.111	&	19.1	\\
1081238569156345984	&	39633331624610769	&	12054	&	1.109	&	32.43	&	-0.10	&	64.32	&	1.320	&	18.1	\\
2577531885760653568	&	39627959807510205	&	11994	&	0.946	&	34.94	&	-0.11	&	54.67	&	1.002	&	19.1	\\
5154241542982144384	&	39627415311356333	&	13127	&	1.080	&	36.90	&	-0.10	&	49.43	&	1.082	&	18.2	\\
5114689949364424576	&	39627485876327104	&	11026	&	0.954	&	37.12	&	-0.03	&	0.13	&	1.309	&	18.6	\\
2810559429509847680	&	39628048911306342	&	23340	&	1.091	&	40.44	&	-0.06	&	70.79	&	0.234	&	18.8	\\
1305727541826537472	&	39628455368722618	&	9048	&	0.664	&	40.86	&	-0.50	&	24.18	&	0.969	&	18.8	\\
4605118017001976960	&	39632967206700319	&	10664	&	0.754	&	42.65	&	-0.04	&	65.45	&	0.818	&	18.5	\\
1484295827976469632	&	39633020461778726	&	11858	&	0.948	&	42.74	&	-0.03	&	77.10	&	1.048	&	18.3	\\
1545134482103819264	&	39633199923463535	&	16103	&	1.183	&	44.38	&	-0.23	&	38.23	&	0.742	&	18.9	\\
2648173549340448640	&	39627687844645149	&	13176	&	0.799	&	44.40	&	-0.04	&	57.55	&	0.521	&	17.6	\\
1318204460477280512	&	39628481679593814	&	20365	&	1.228	&	50.96	&	-0.50	&	49.46	&	0.504	&	17.1	\\
1371714251131163136	&	39632966707577054	&	7912	&	0.771	&	51.21	&	-0.45	&	22.23	&	2.017	&	18.3	\\
2161602408093396480	&	39633448876379894	&	24985	&	1.197	&	56.01	&	-0.49	&	79.40	&	0.278	&	18.3	\\
4226641105124840448	&	39627741649180668	&	4898	&	0.380	&	58.79	&	-0.46	&	10.45	&	4.061	&	18.2	\\
2550870790371314688	&	39627863443381580	&	11034	&	1.022	&	75.09	&	-0.15	&	21.39	&	1.487	&	18.9	\\
2691666724647619072	&	39627850545893762	&	8078	&	0.646	&	76.39	&	-0.46	&	10.55	&	1.212	&	18.5	\\
1769524911044920064	&	39628160421074880	&	21558	&	1.236	&	100.95	&	0.11	&	82.97	&	0.458	&	17.9	\\
2592249329934861824	&	39628161276708160	&	9854	&	0.898	&	108.59	&	-0.05	&	14.08	&	1.556	&	18.2	\\
4557397837965773312	&	39628330504297684	&	17893	&	1.012	&	150.13	&	0.31	&	88.63	&	0.397	&	17.9	\\
3848213662682651648	&	39627883789943773	&	21677	&	1.170	&	153.20	&	-0.18	&	21.37	&	0.364	&	18.1	\\
1418237478616604160	&	39633329636509286	&	22197	&	1.252	&	155.79	&	-0.21	&	67.21	&	0.454	&	19.1	\\
0712758790663253760	&	39628526701249225	&	30490	&	1.329	&	171.75	&	0.11	&	10.46	&	0.287	&	19.6	\\
3076058841571735040	&	39627744568410786	&	29604	&	1.216	&	194.17	&	0.06	&	21.44	&	0.185	&	18.5	\\
0591307423453771904	&	39628021639939880	&	23266	&	0.751	&	211.15	&	-0.07	&	13.04	&	0.063	&	19.4	\\
4575337950720459264	&	39628455528107188	&	10554	&	0.869	&	455.10	&	-0.03	&	18.37	&	1.109	&	18.8
\enddata
\end{deluxetable}
\captionof{figure}{DESI spectra of DAH in black. Best YAWP model in red. In blue are the positions of the absorption lines for each magnetic field strength.}
    \label{fig:minhasimagens}
    \begin{supertabular}{c}
\includegraphics[width=0.9\linewidth]{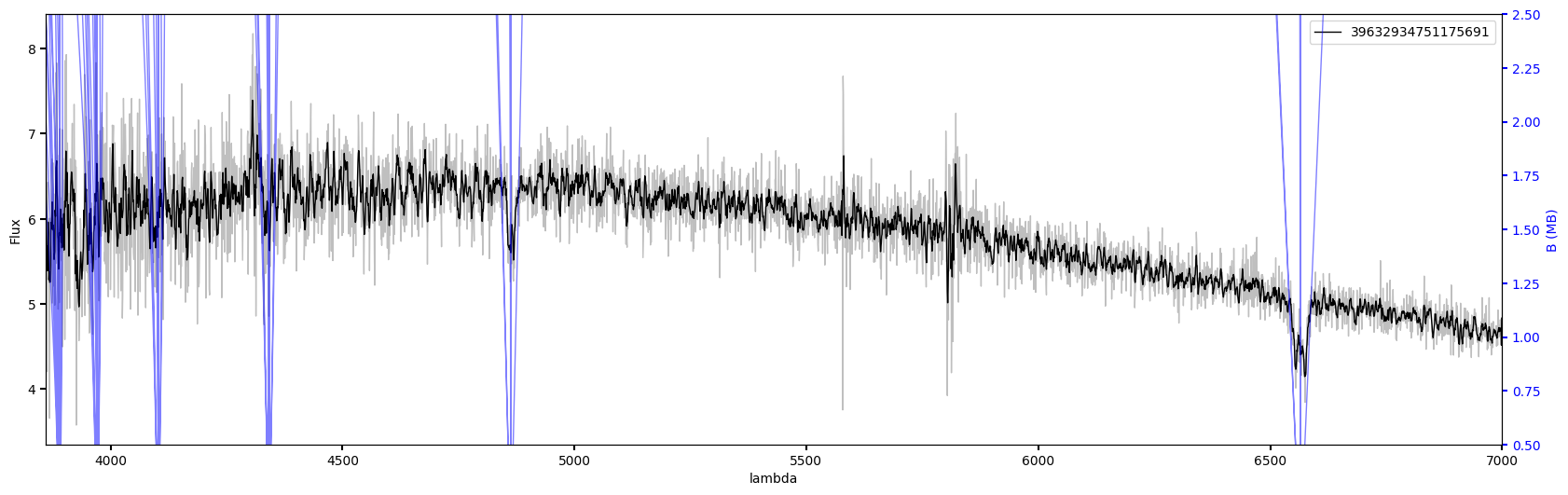}\\
\includegraphics[width=0.9\linewidth]{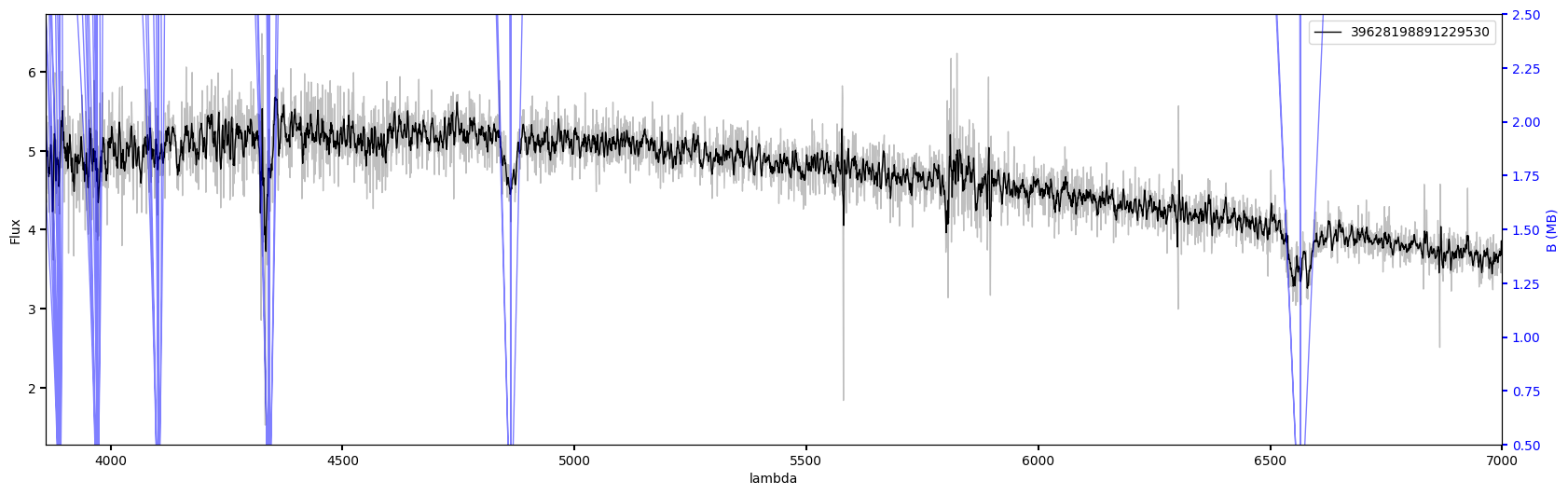}\\
\includegraphics[width=0.9\linewidth]{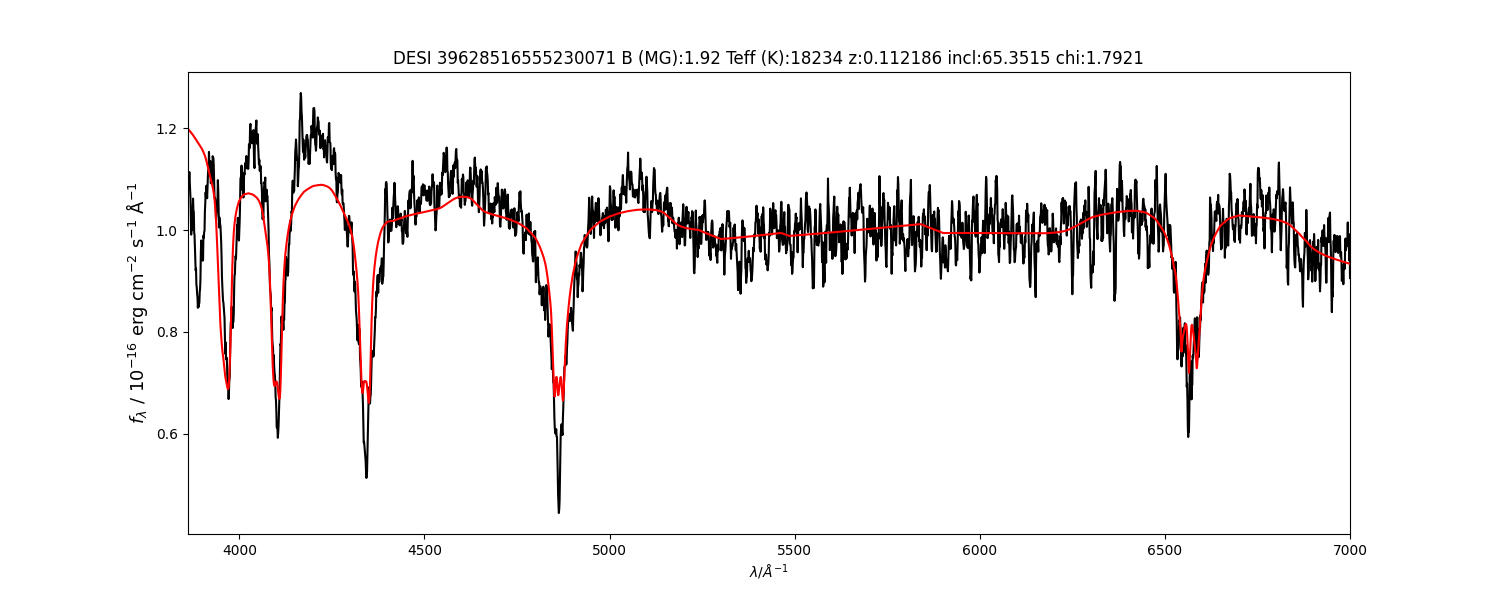}\\ 
\end{supertabular}
 \newpage \captionof{figure}{cont.}
\begin{supertabular}{c}
\includegraphics[width=0.9\linewidth]{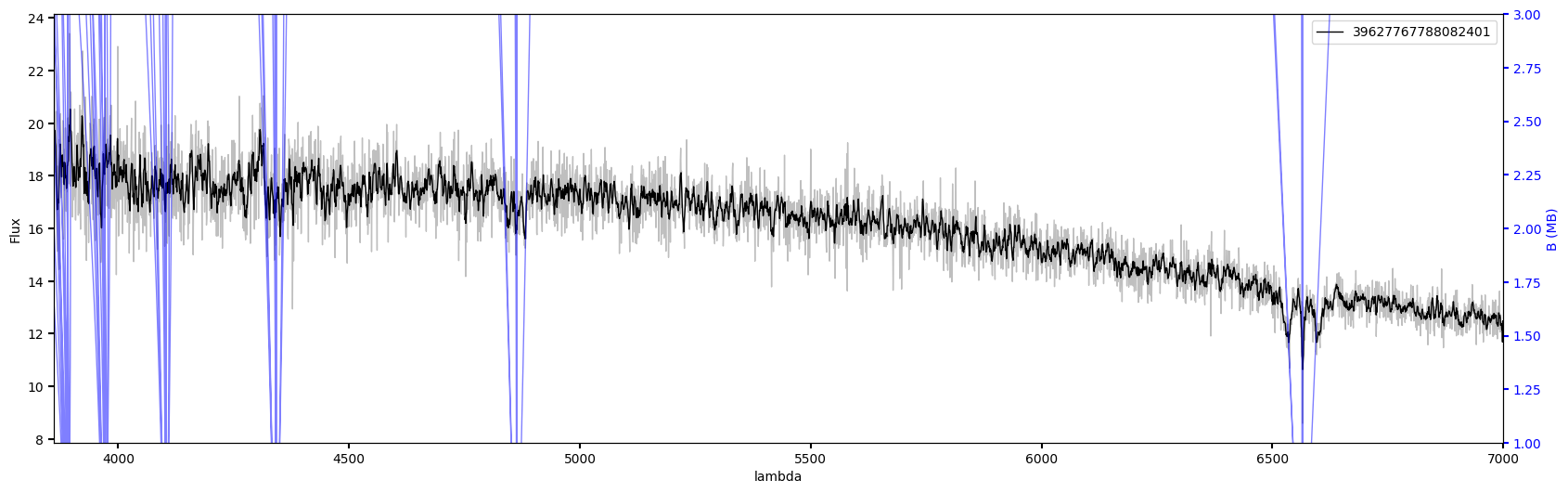}\\
\includegraphics[width=0.9\linewidth]{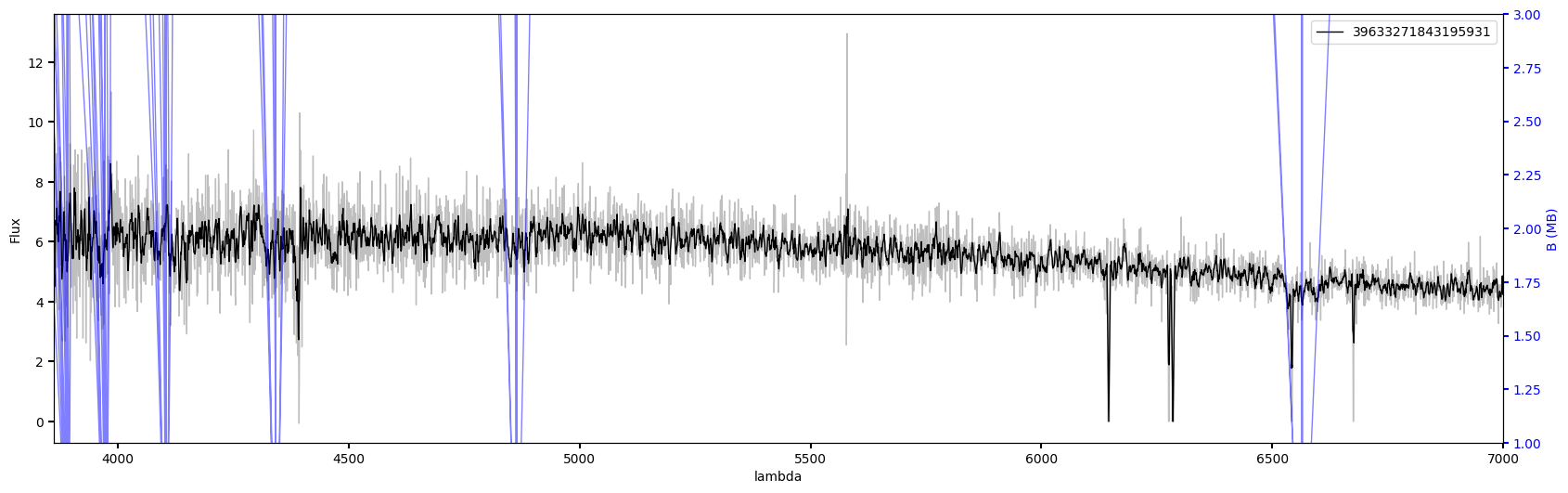}\\
\includegraphics[width=0.9\linewidth]{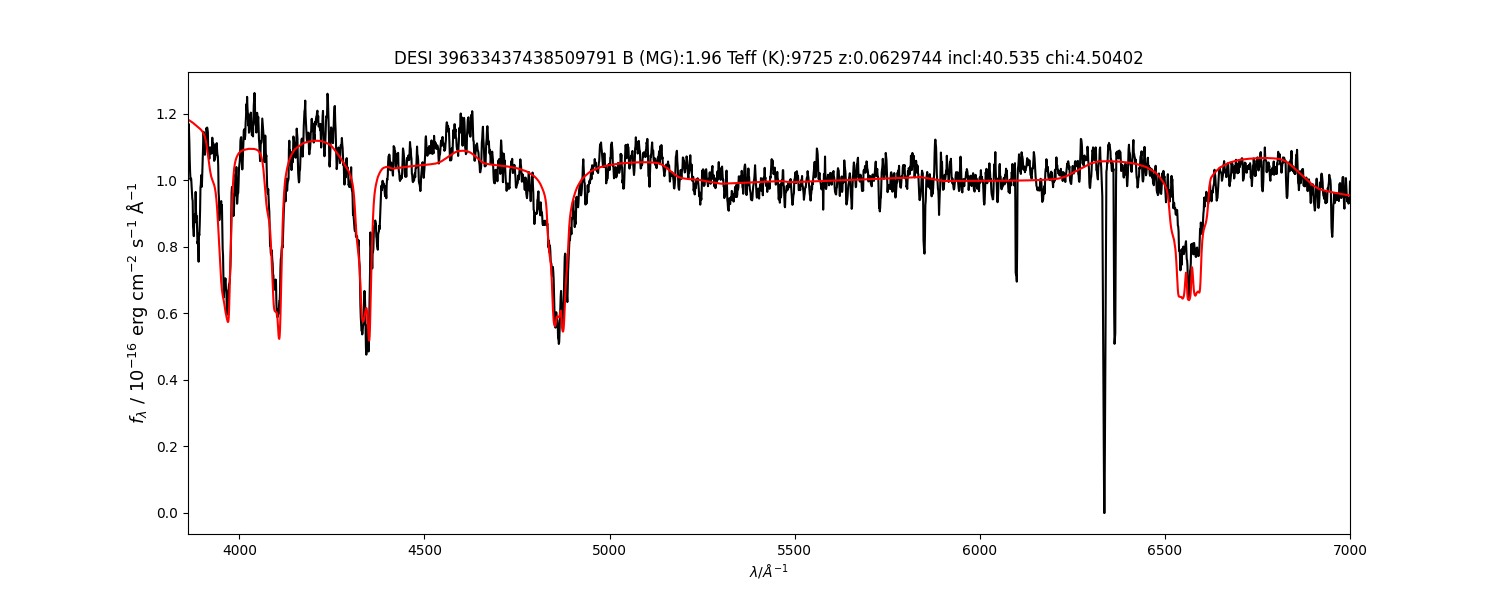}\\ 
\end{supertabular}
 \newpage \captionof{figure}{cont.}
\begin{supertabular}{c}
\includegraphics[width=0.9\linewidth]{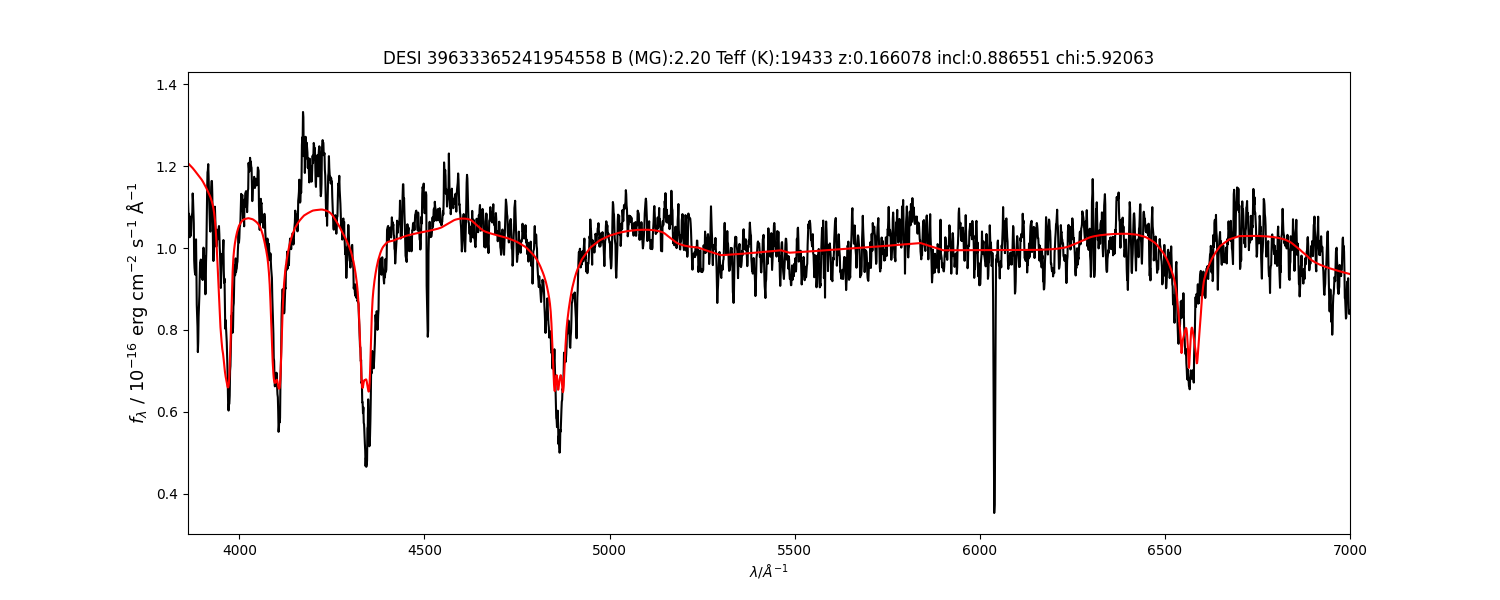}\\
\includegraphics[width=0.9\linewidth]{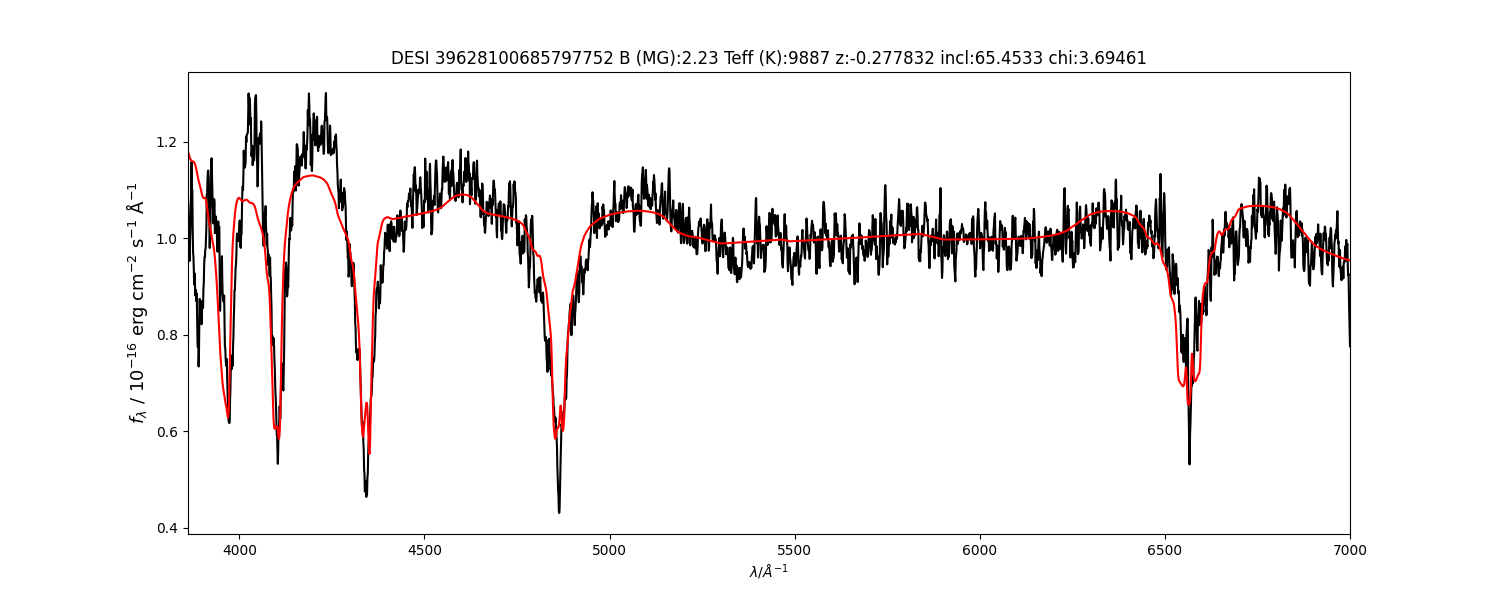}\\
\includegraphics[width=0.9\linewidth]{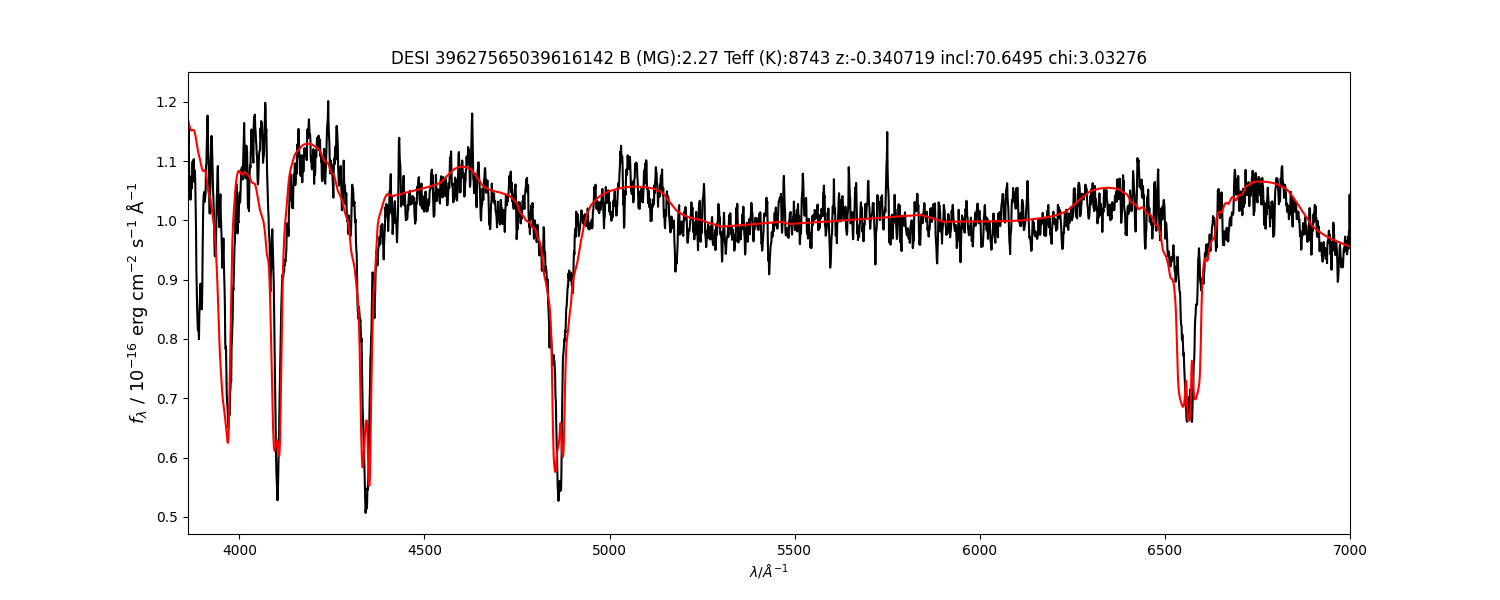}\\ 
\end{supertabular}
 \newpage \captionof{figure}{cont.}
\begin{supertabular}{c}
\includegraphics[width=0.9\linewidth]{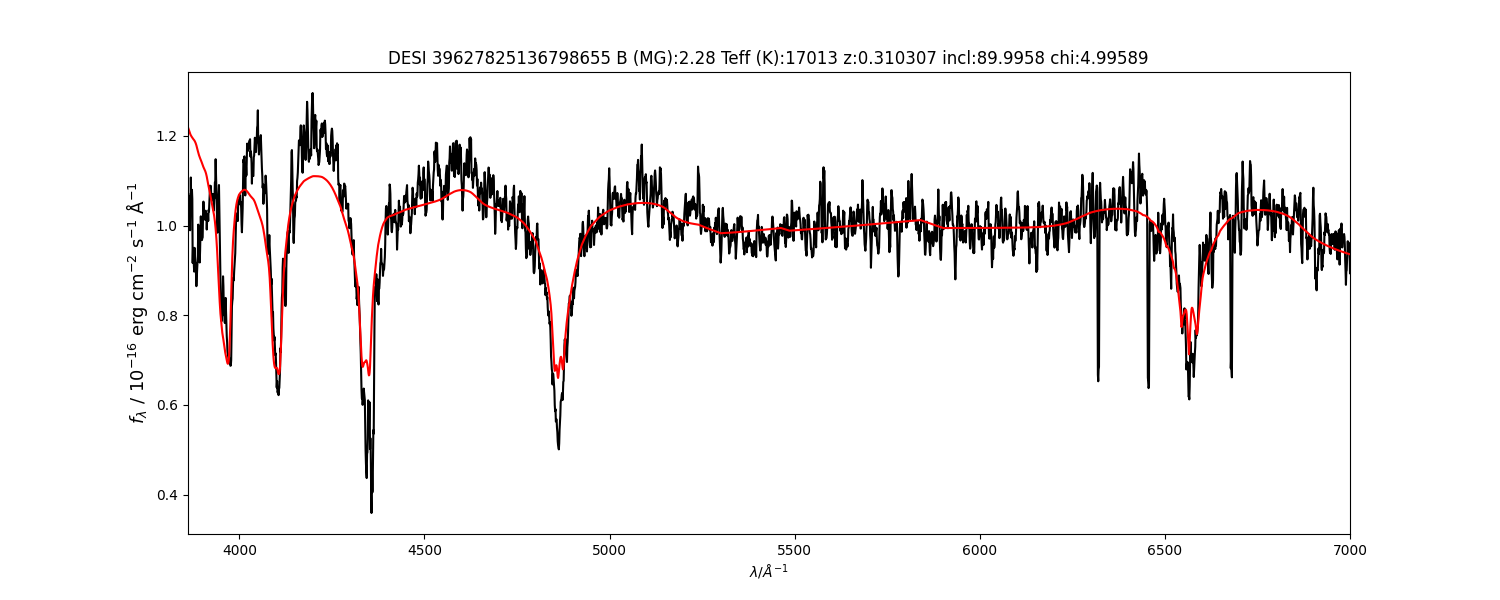}\\
\includegraphics[width=0.9\linewidth]{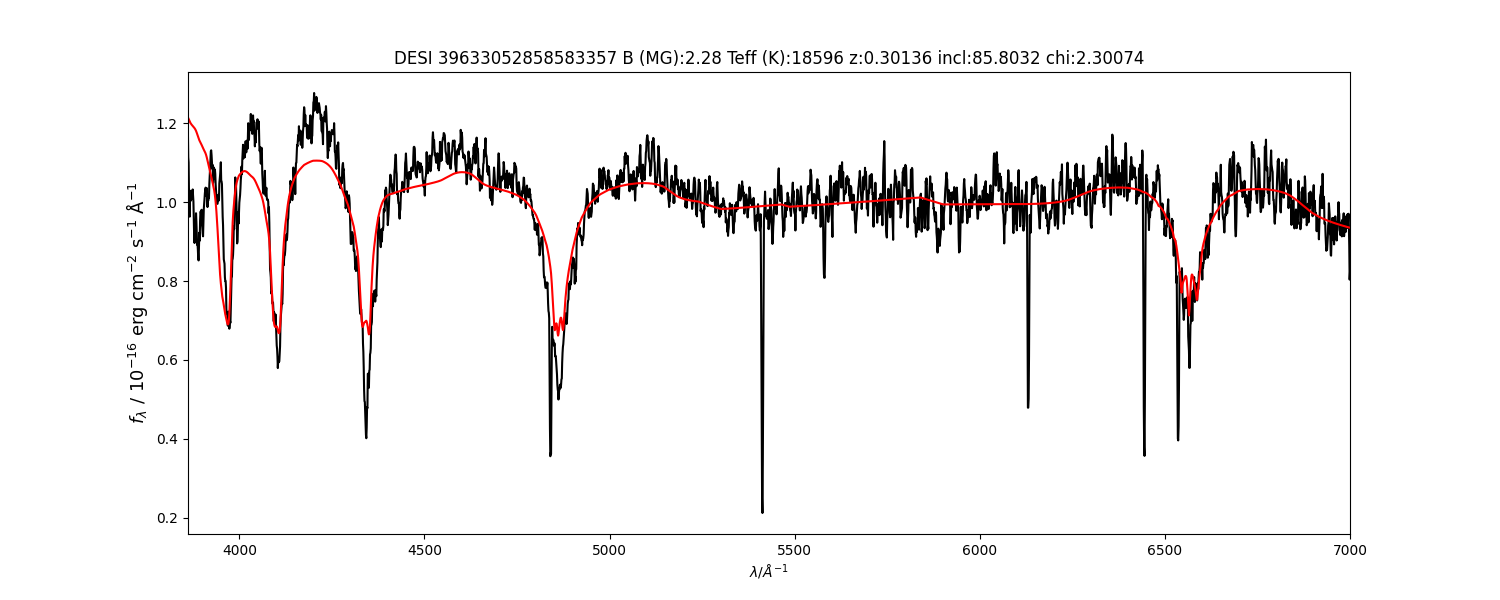}\\
\includegraphics[width=0.9\linewidth]{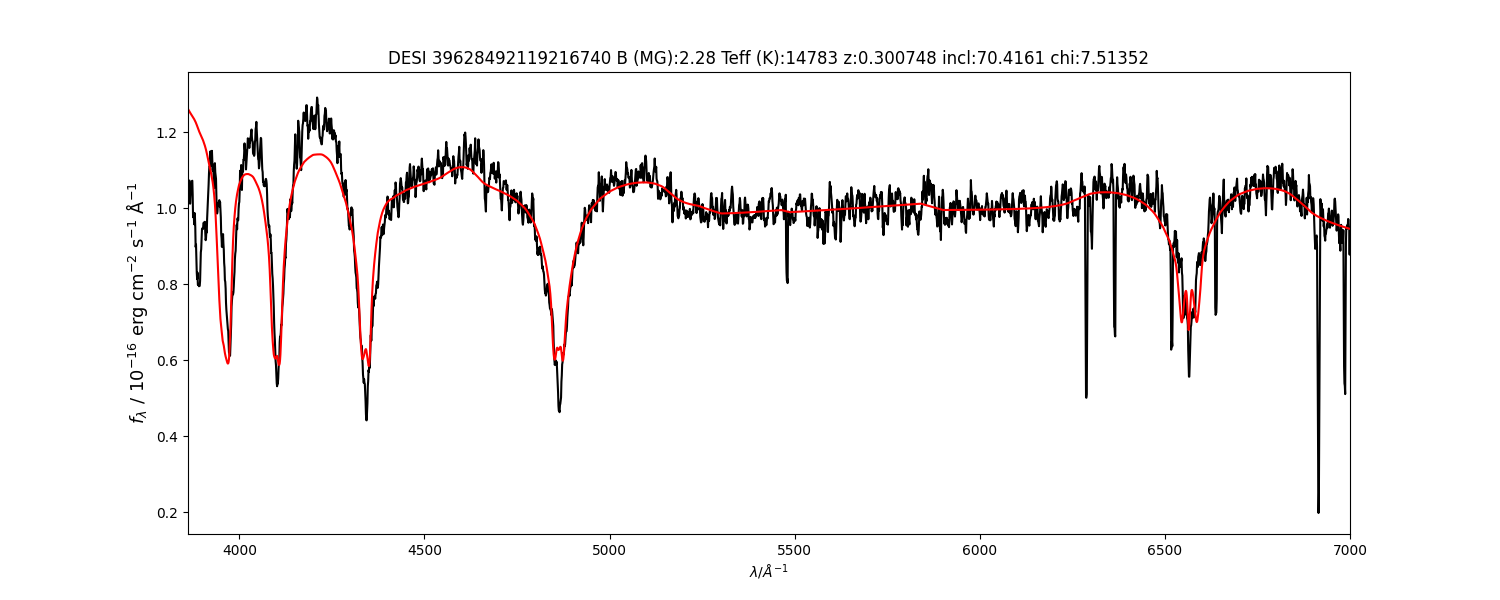}\\ 
\end{supertabular}
 \newpage \captionof{figure}{cont.}
\begin{supertabular}{c}
\includegraphics[width=0.9\linewidth]{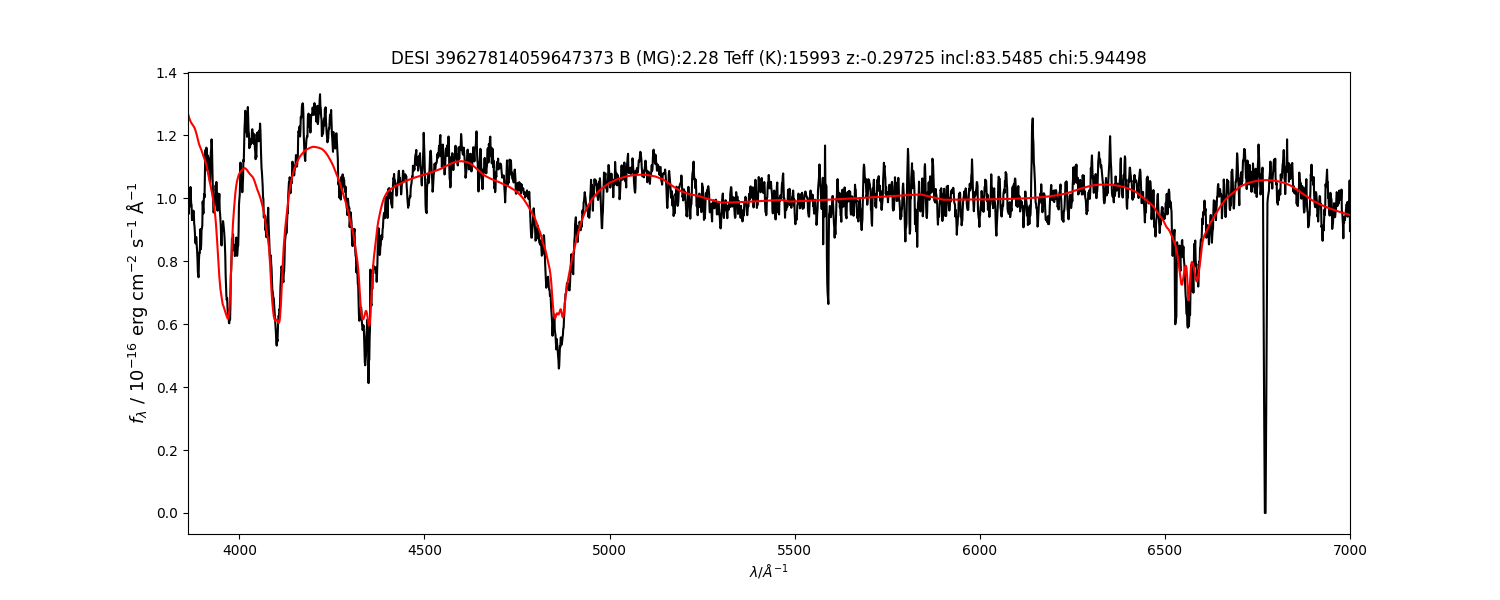}\\
\includegraphics[width=0.9\linewidth]{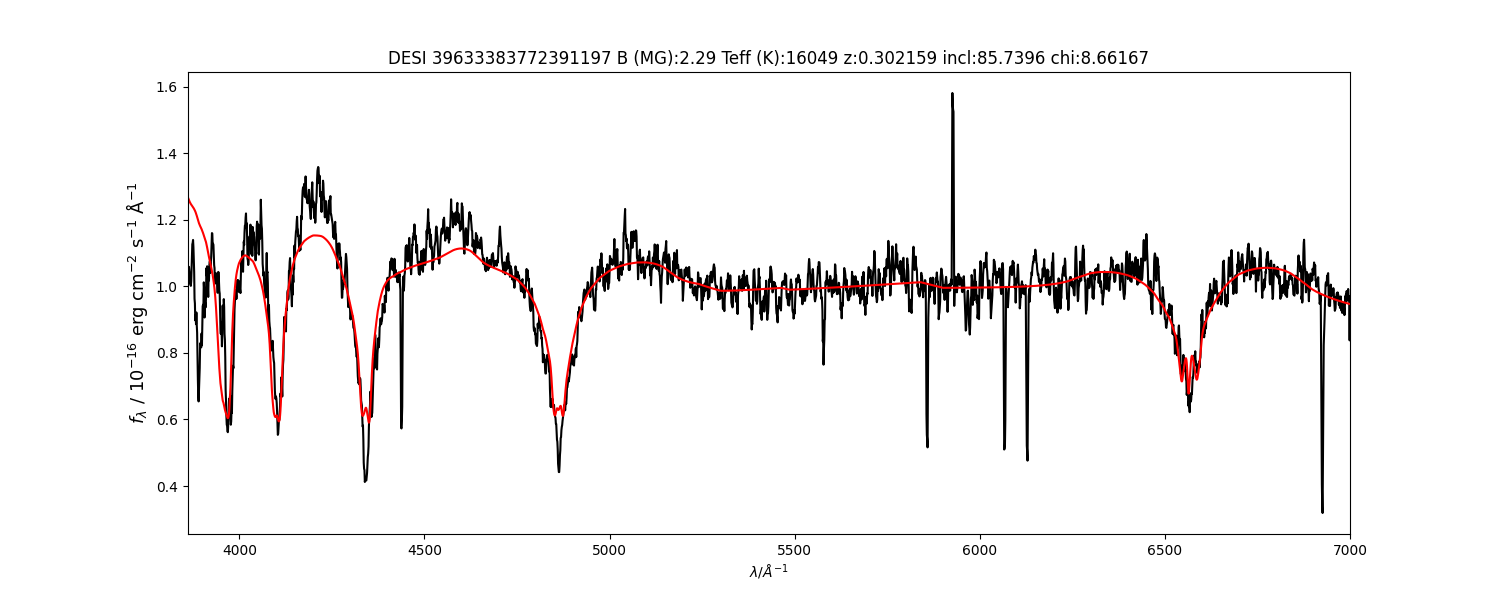}\\
\includegraphics[width=0.9\linewidth]{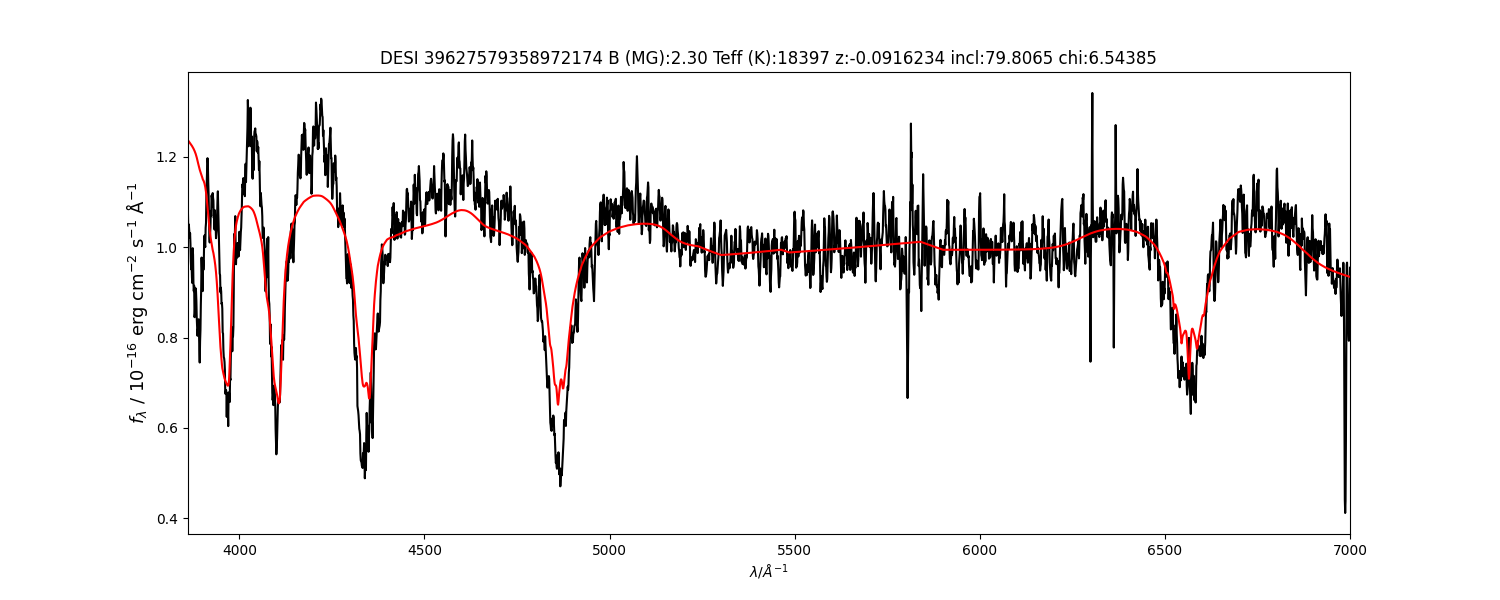}\\ 
\end{supertabular}
 \newpage \captionof{figure}{cont.}
\begin{supertabular}{c}
\includegraphics[width=0.9\linewidth]{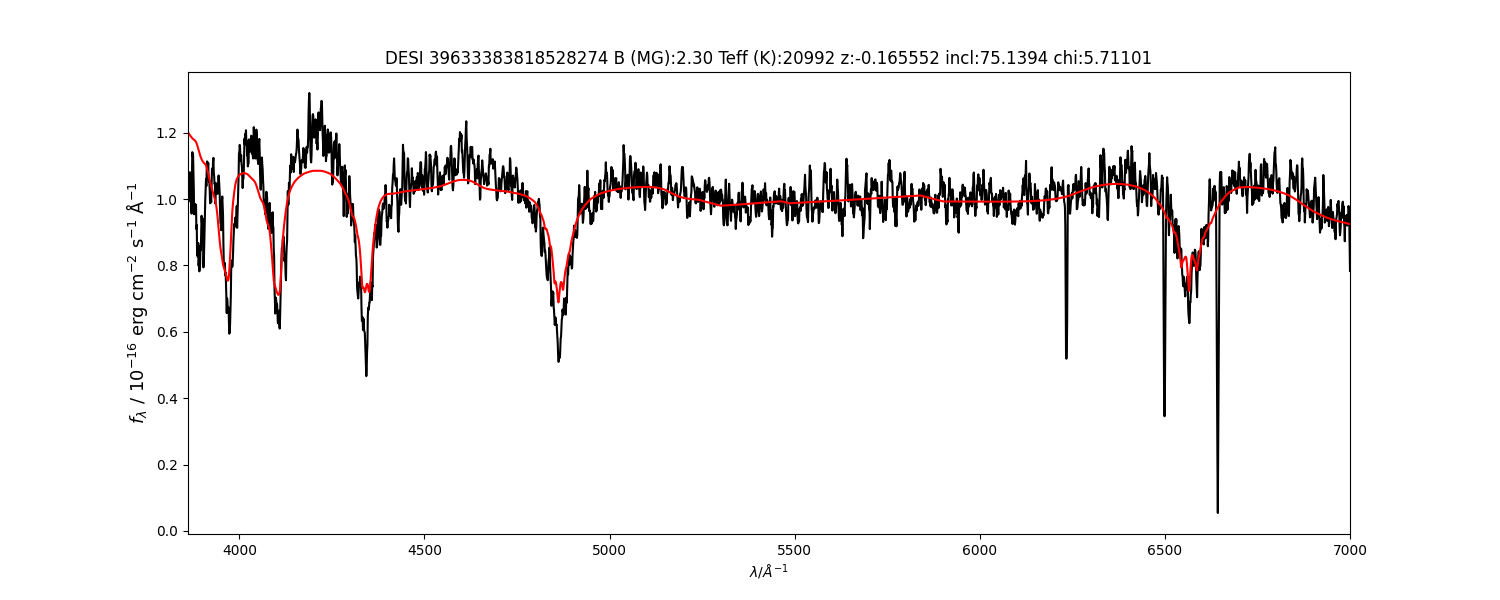}\\
\includegraphics[width=0.9\linewidth]{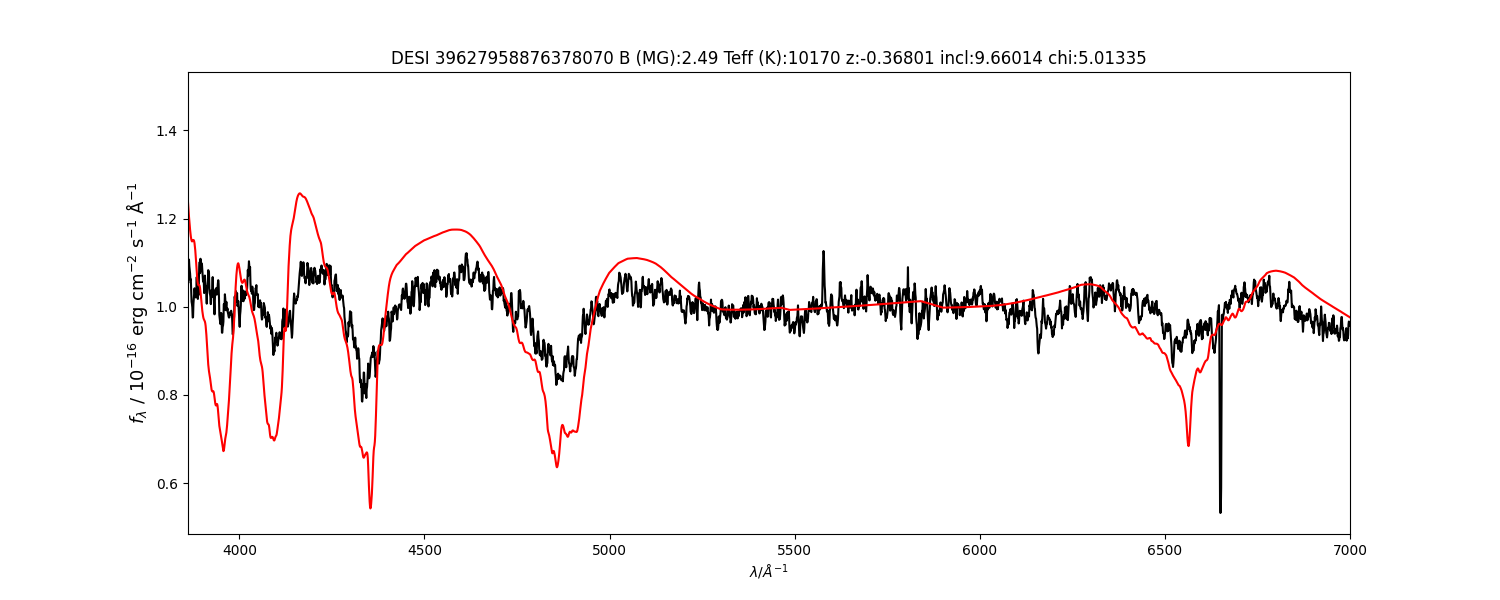}\\
\includegraphics[width=0.9\linewidth]{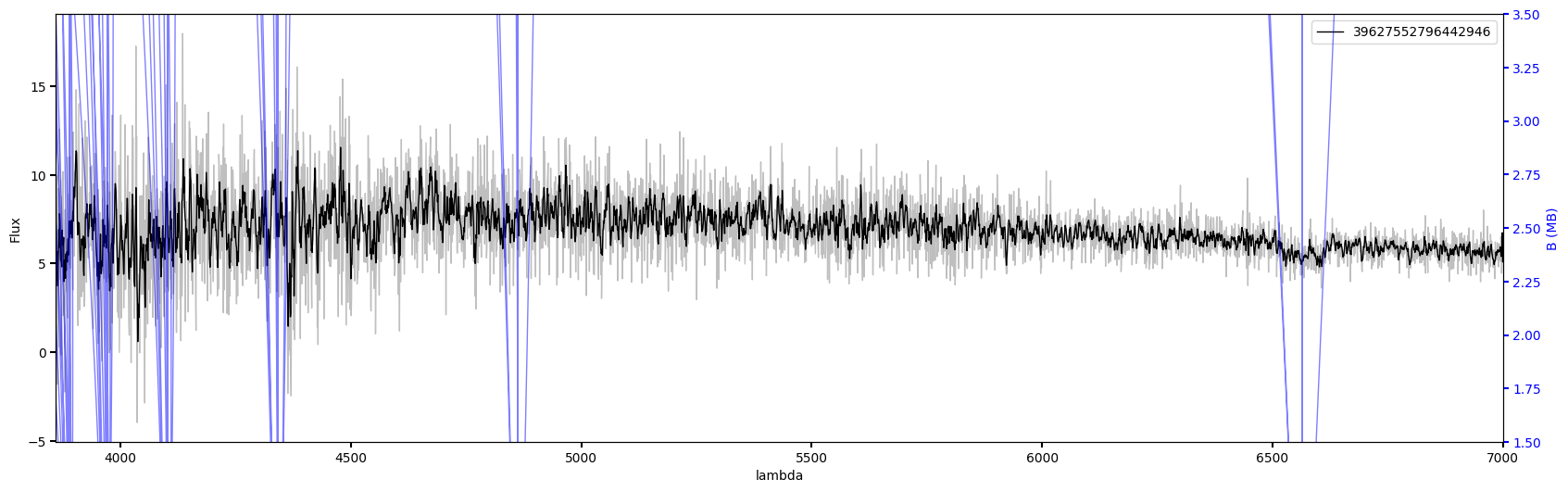}\\ 
\end{supertabular}
 \newpage \captionof{figure}{cont.}
\begin{supertabular}{c}
\includegraphics[width=0.9\linewidth]{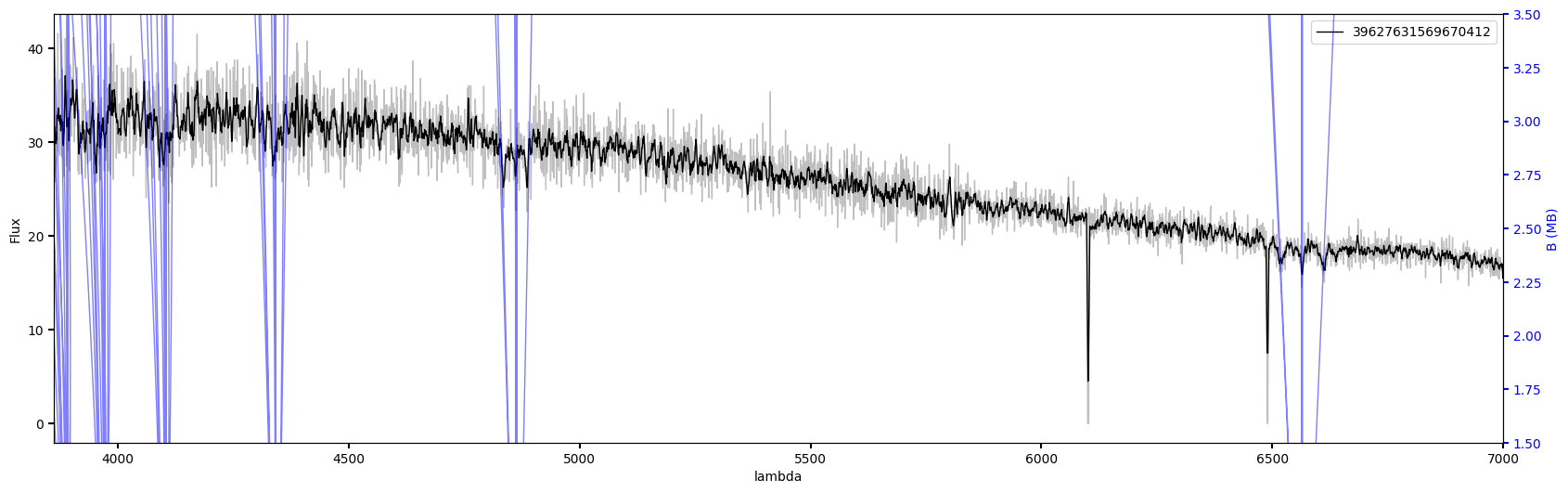}\\
\includegraphics[width=0.9\linewidth]{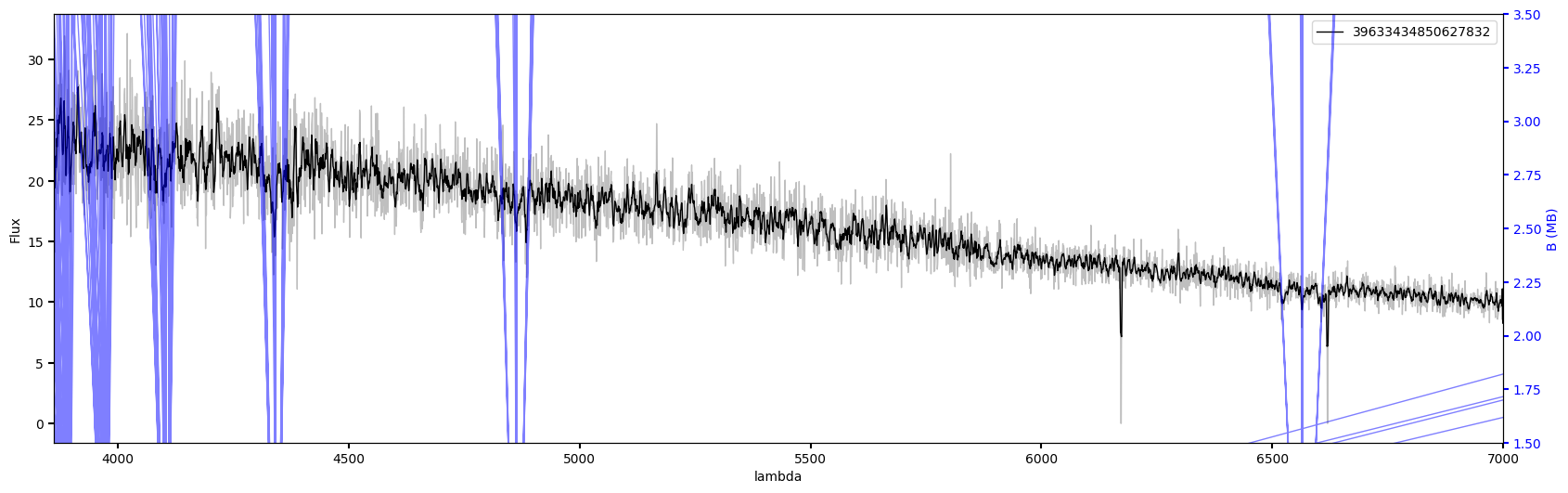}\\
\includegraphics[width=0.9\linewidth]{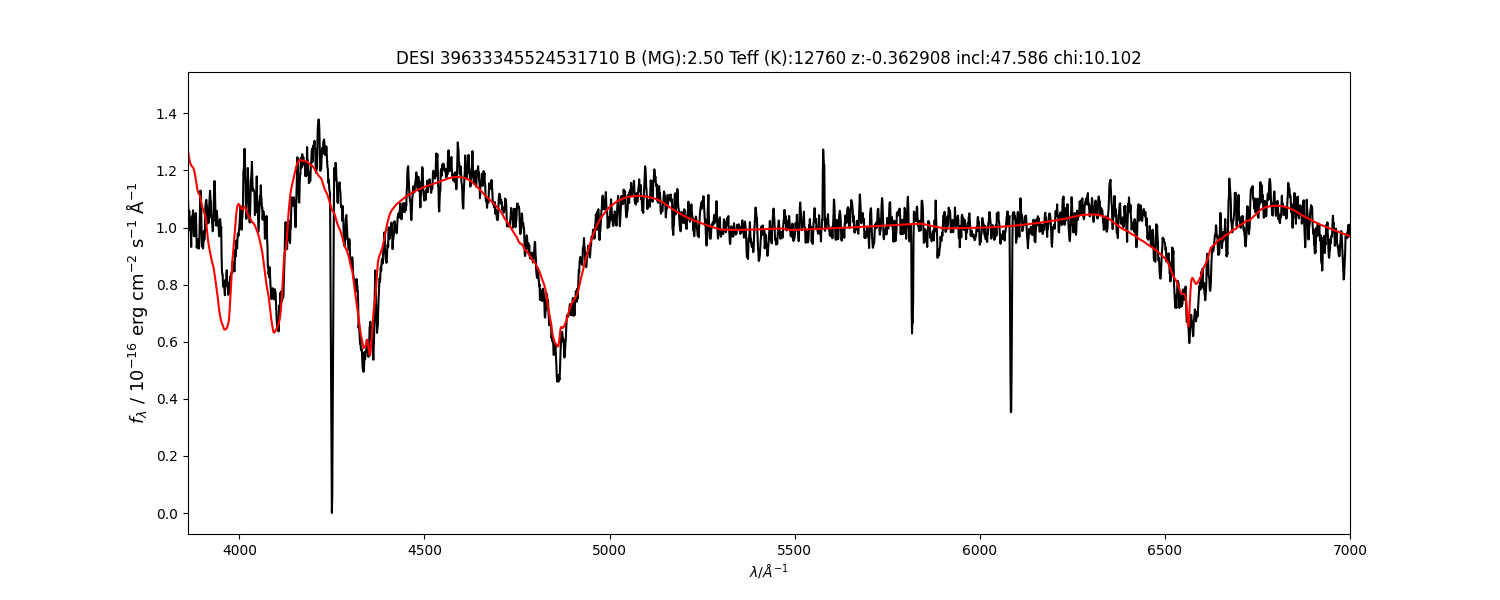}\\ 
\end{supertabular}
 \newpage \captionof{figure}{cont.}
\begin{supertabular}{c}
\includegraphics[width=0.9\linewidth]{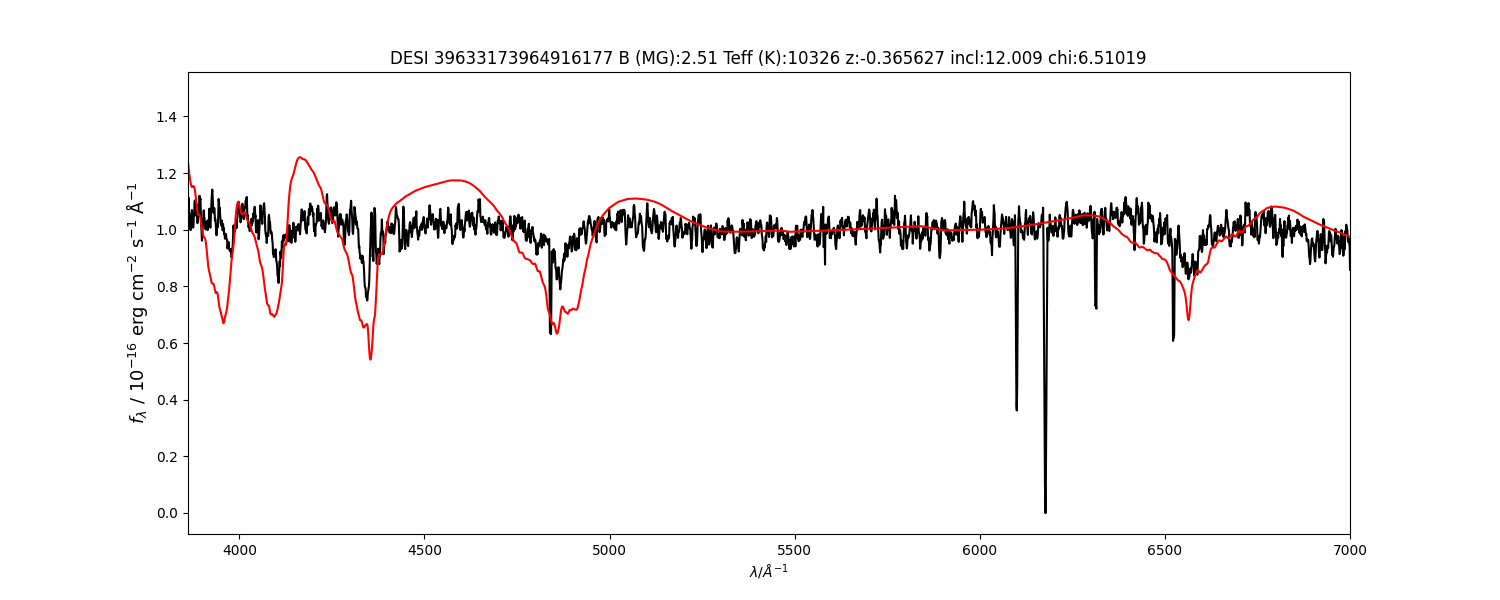}\\
\includegraphics[width=0.9\linewidth]{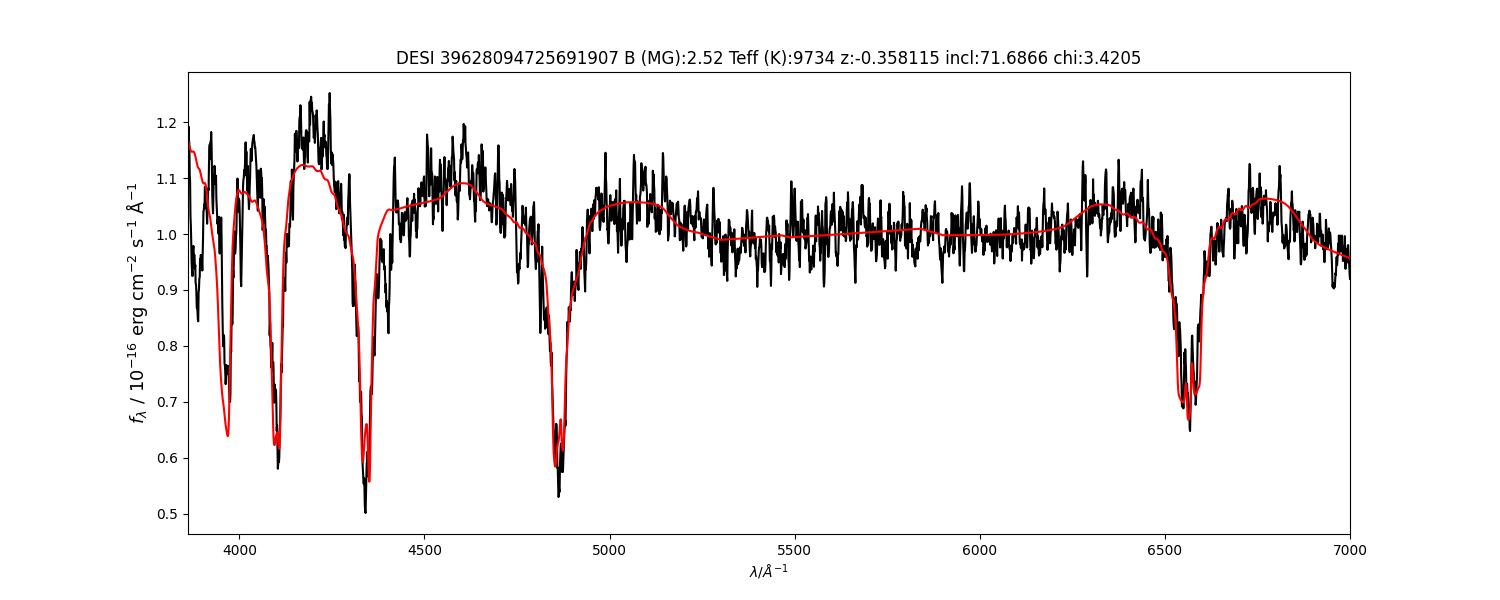}\\
\includegraphics[width=0.9\linewidth]{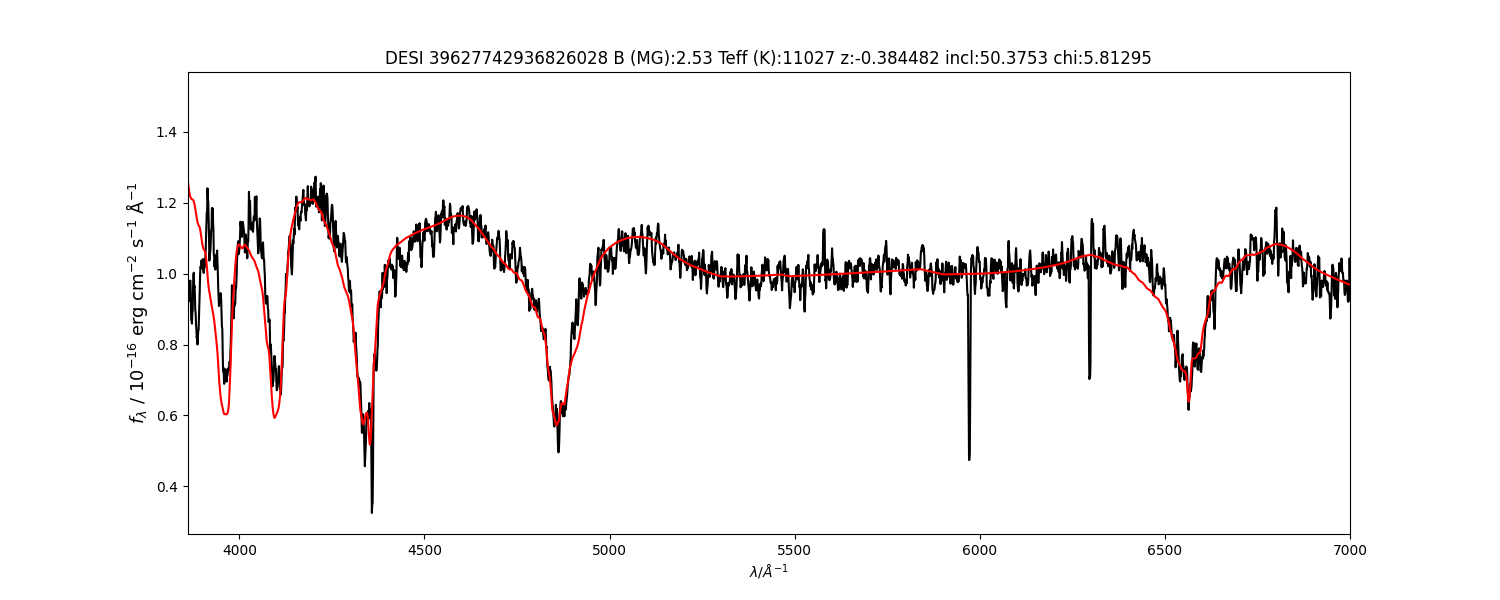}\\ 
\end{supertabular}
 \newpage \captionof{figure}{cont.}
\begin{supertabular}{c}
\includegraphics[width=0.9\linewidth]{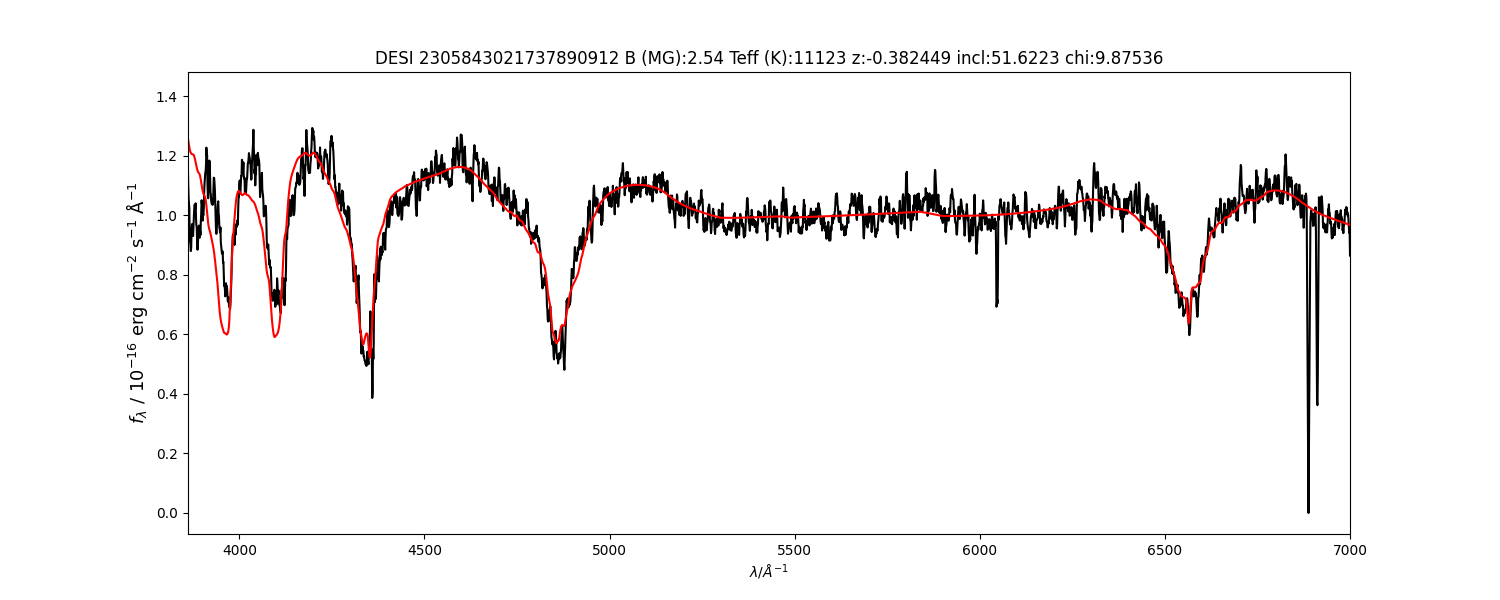}\\
\includegraphics[width=0.9\linewidth]{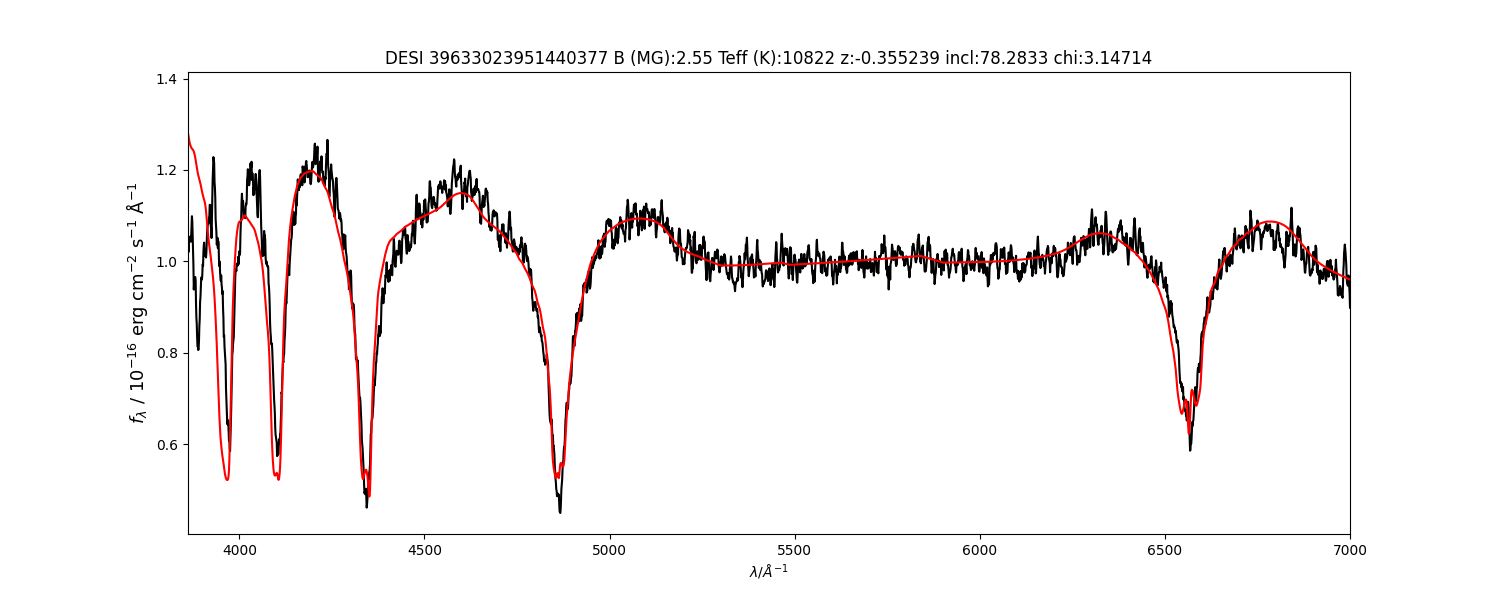}\\
\includegraphics[width=0.9\linewidth]{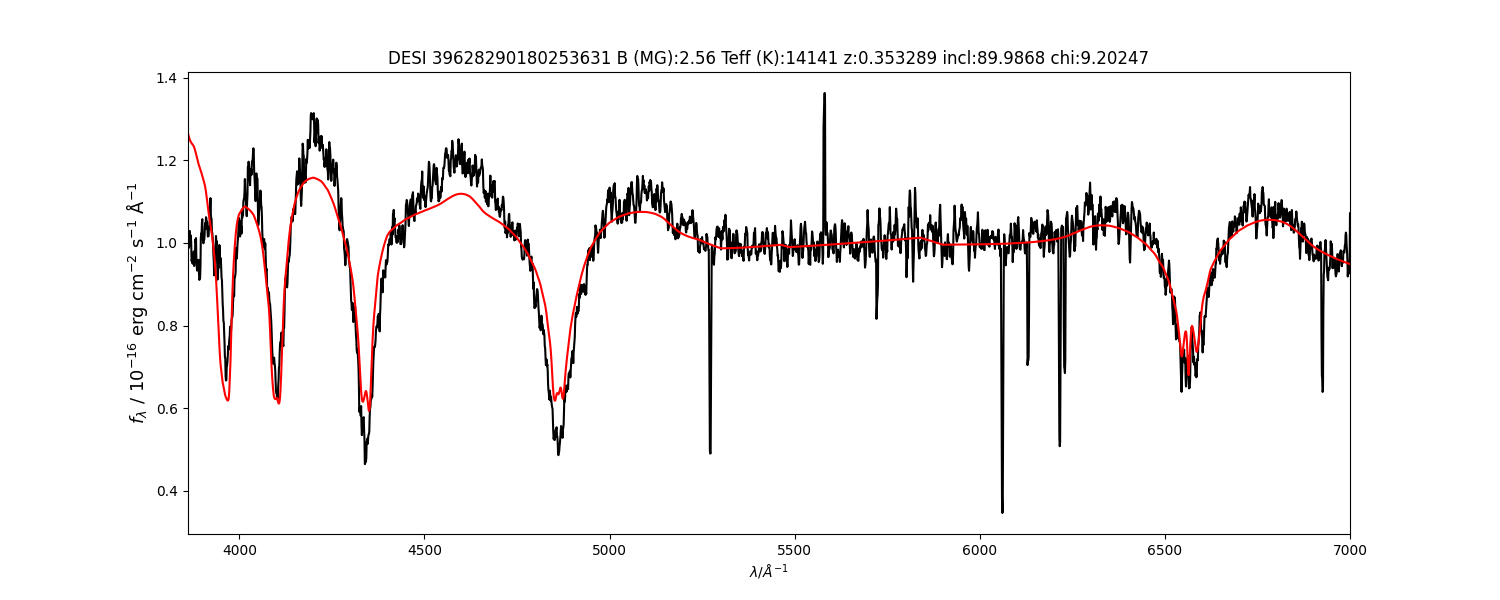}\\ 
\end{supertabular}
 \newpage \captionof{figure}{cont.}
\begin{supertabular}{c}
\includegraphics[width=0.9\linewidth]{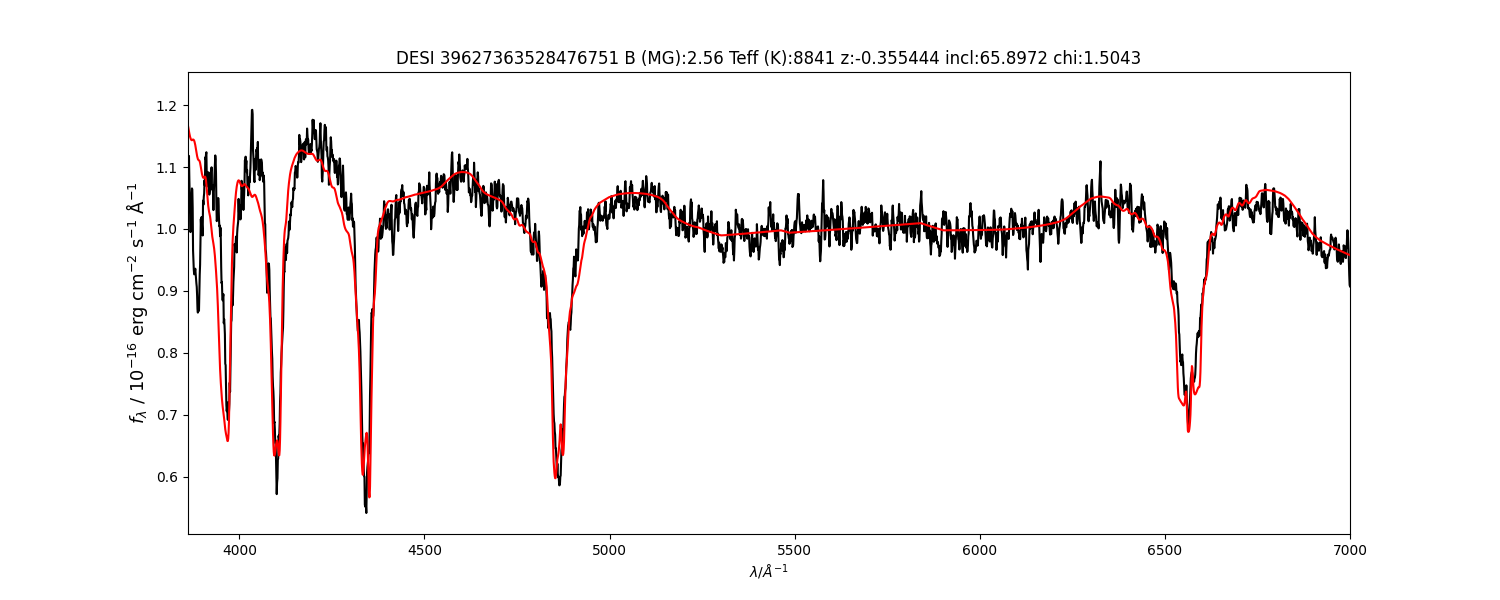}\\
\includegraphics[width=0.9\linewidth]{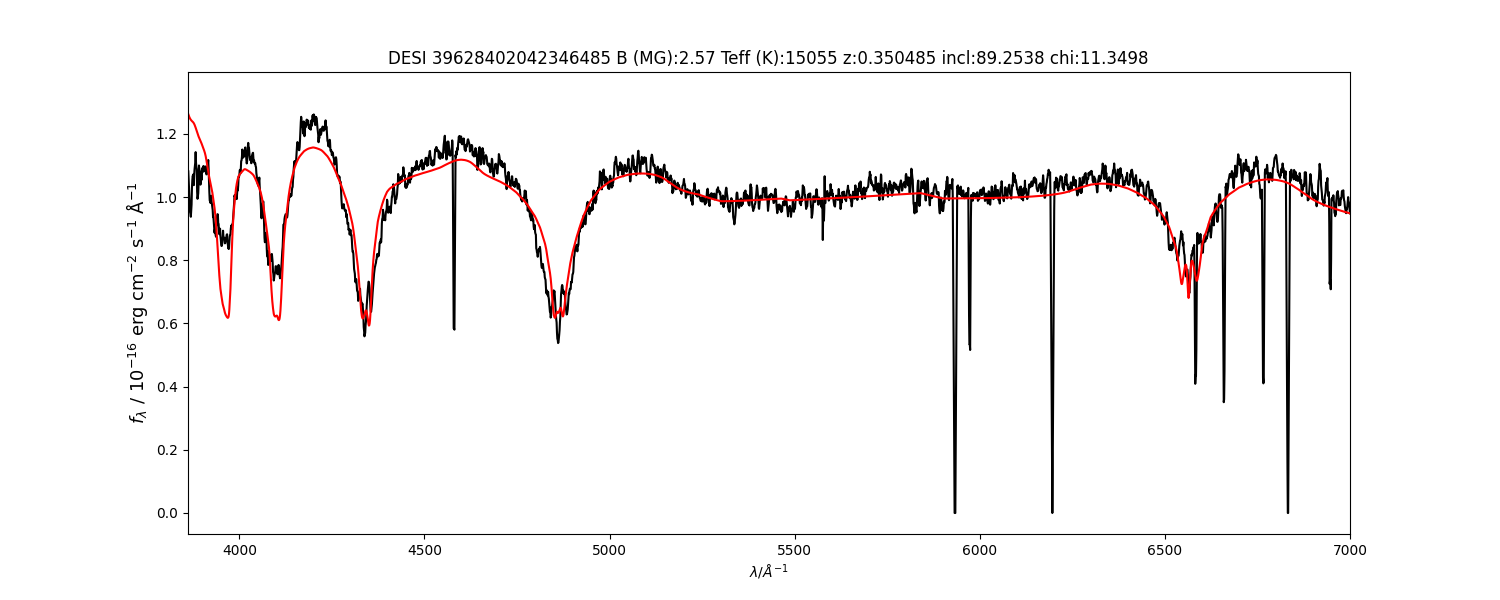}\\
\includegraphics[width=0.9\linewidth]{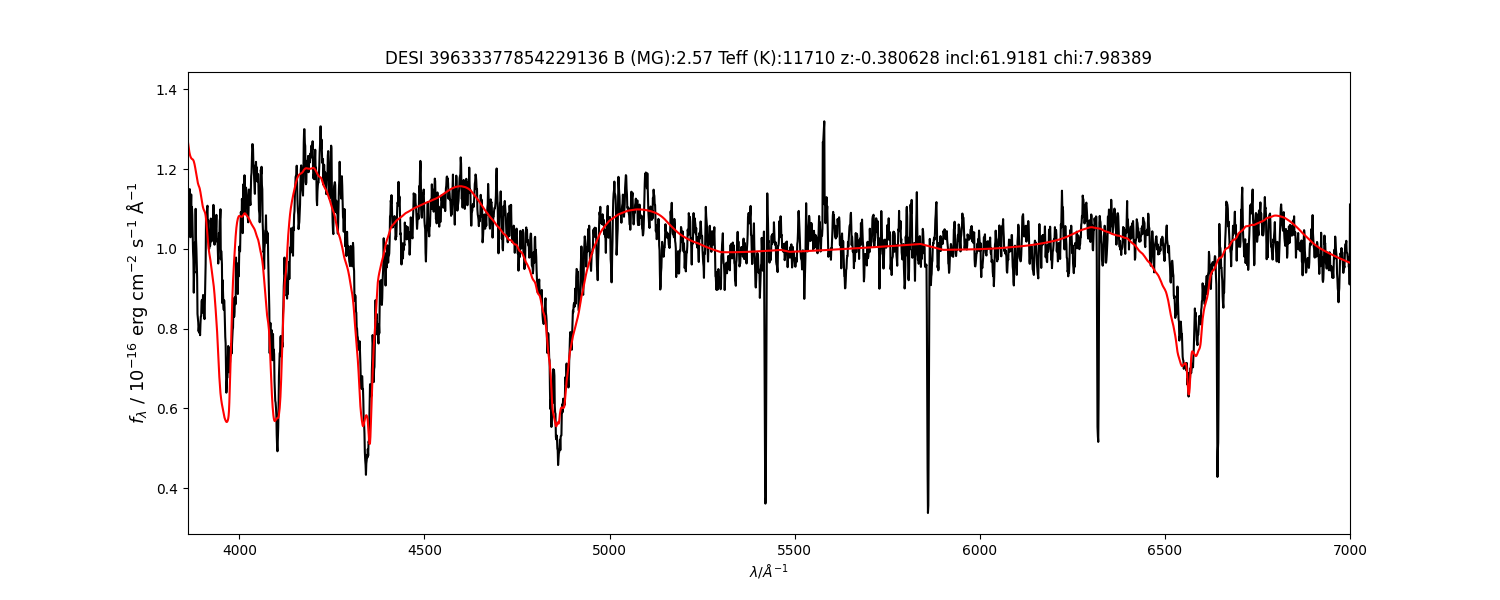}\\ 
\end{supertabular}
 \newpage \captionof{figure}{cont.}
\begin{supertabular}{c}
\includegraphics[width=0.9\linewidth]{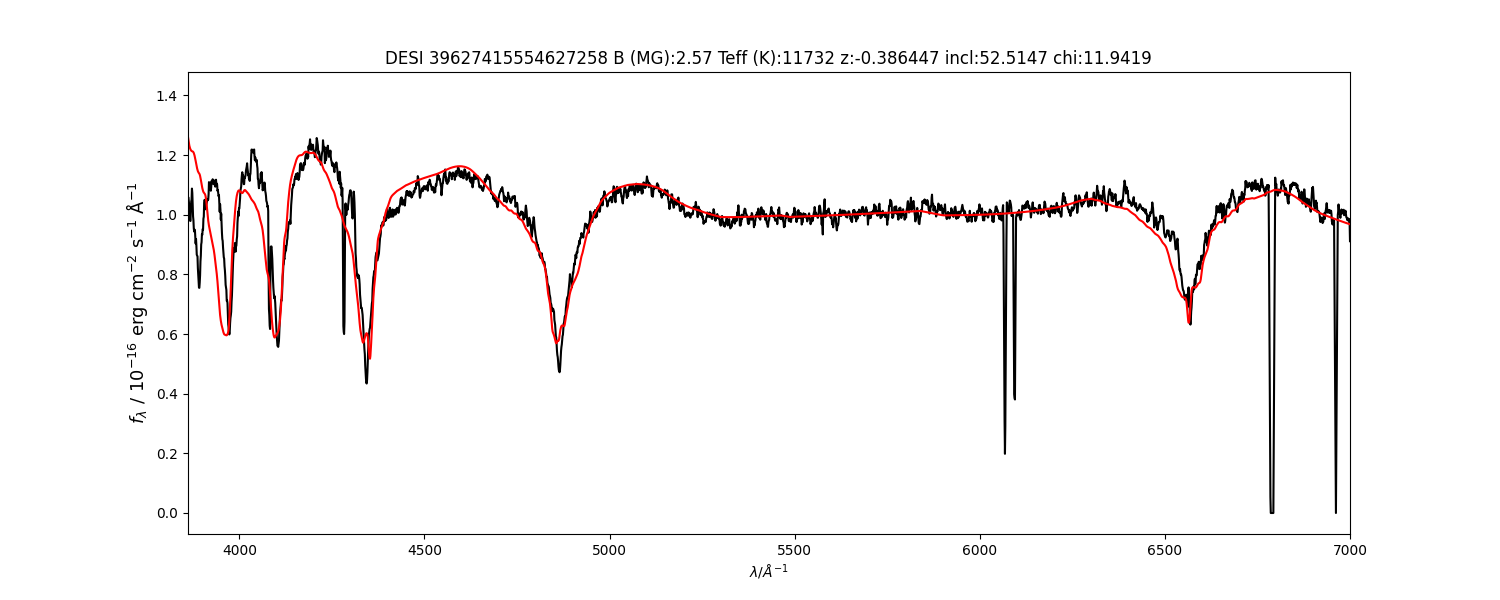}\\
\includegraphics[width=0.9\linewidth]{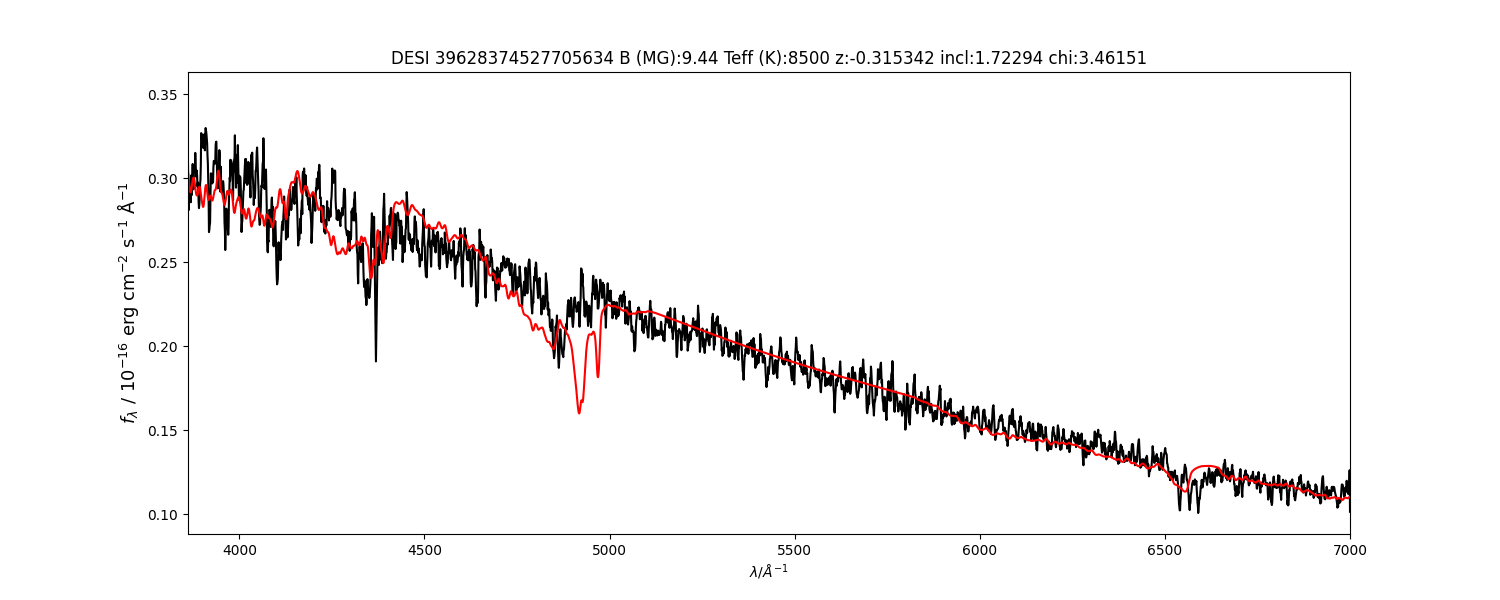}\\
\includegraphics[width=0.9\linewidth]{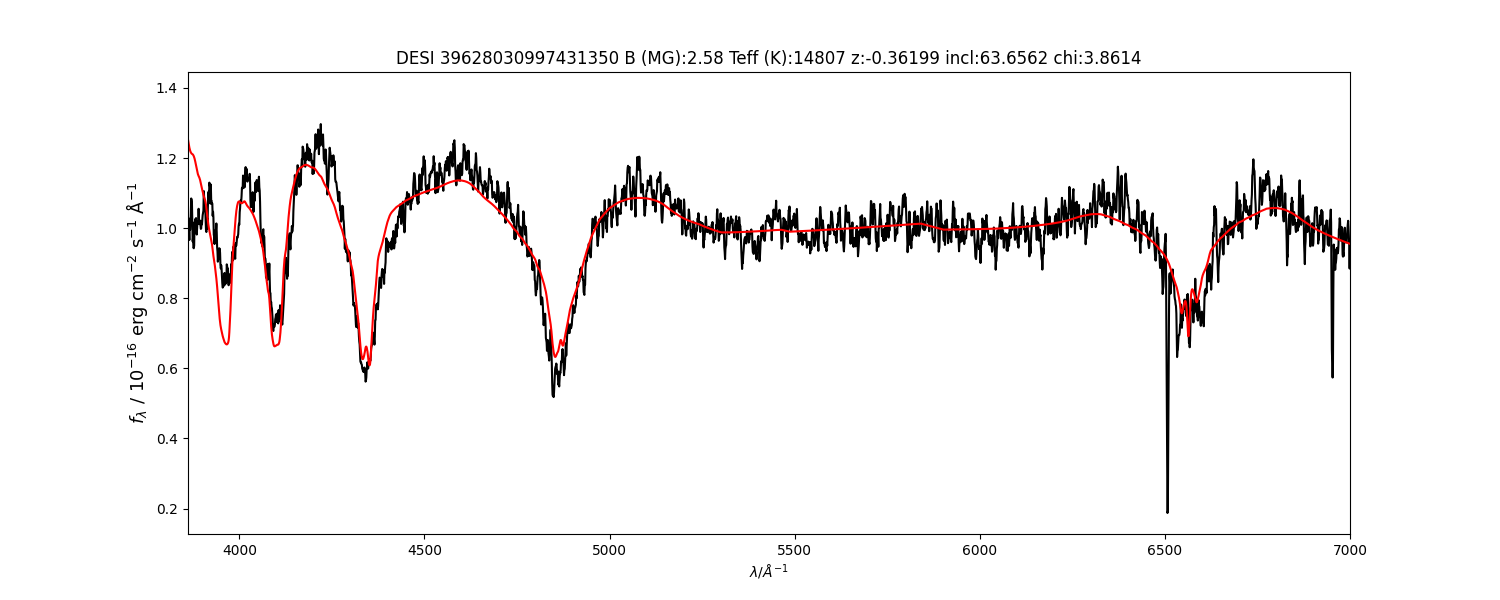}\\ 
\end{supertabular}
 \newpage \captionof{figure}{cont.}
\begin{supertabular}{c}
\includegraphics[width=0.9\linewidth]{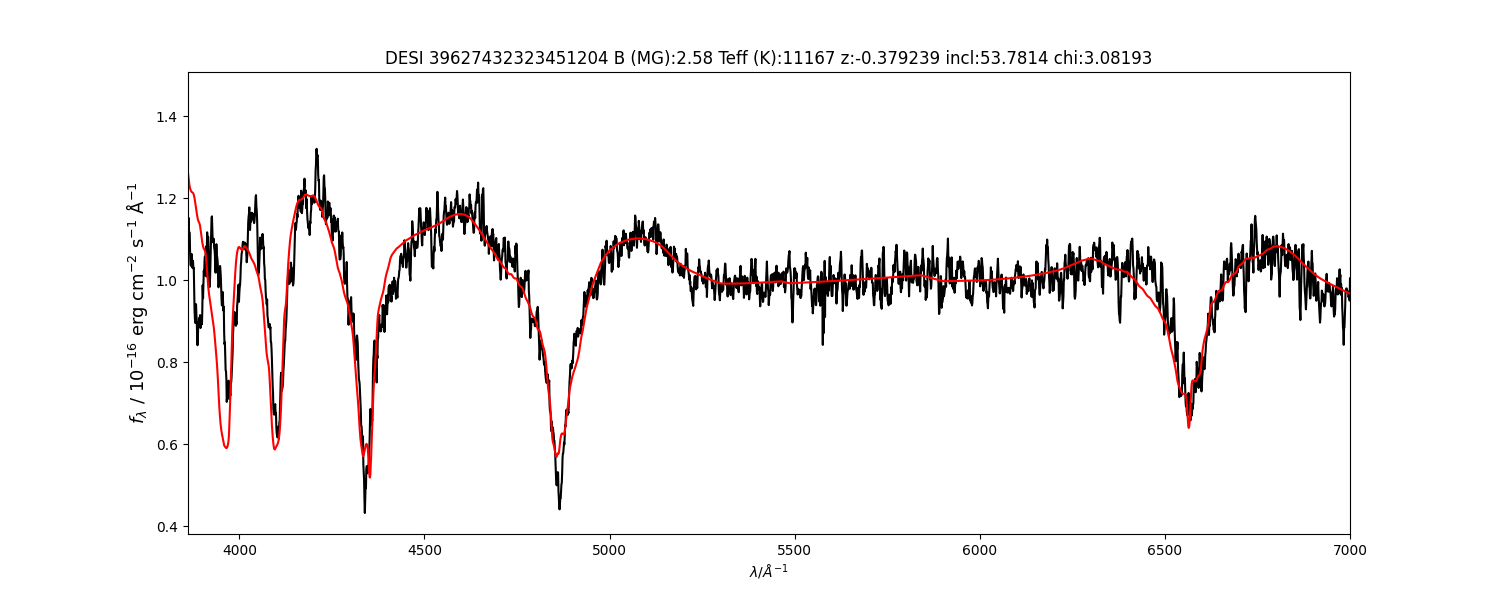}\\
\includegraphics[width=0.9\linewidth]{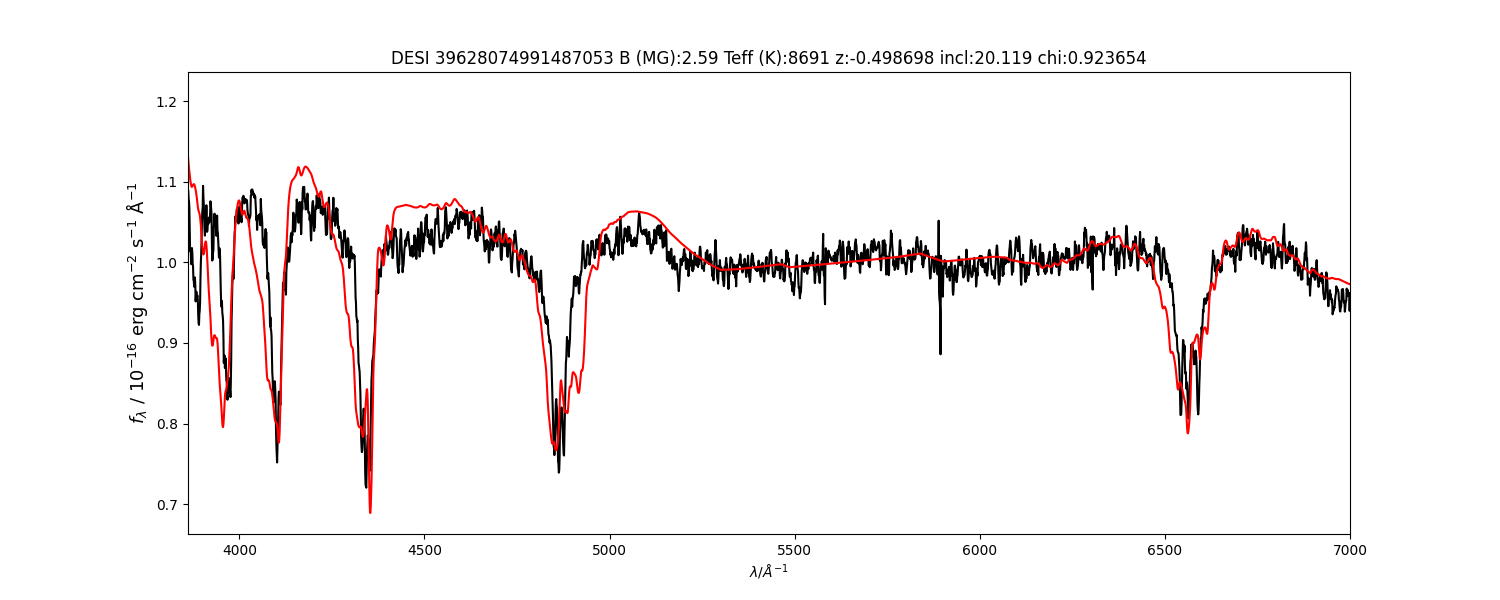}\\
\includegraphics[width=0.9\linewidth]{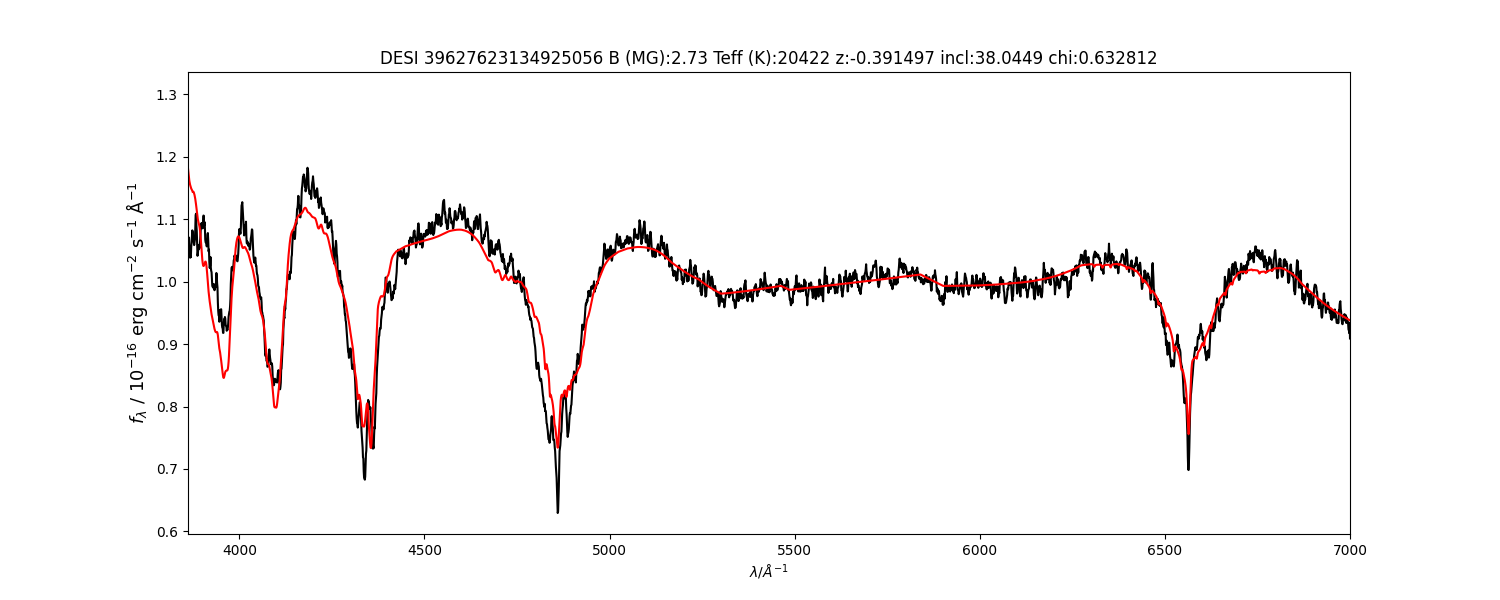}\\ 
\end{supertabular}
 \newpage \captionof{figure}{cont.}
\begin{supertabular}{c}
\includegraphics[width=0.9\linewidth]{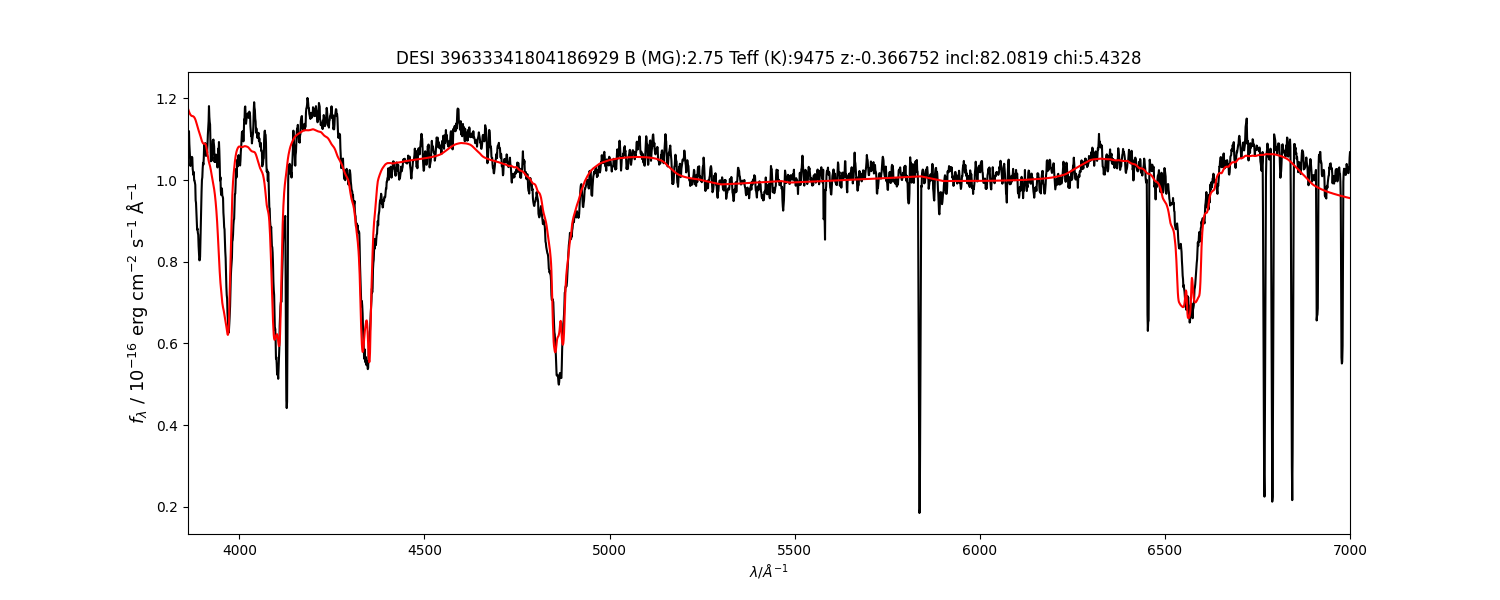}\\
\includegraphics[width=0.9\linewidth]{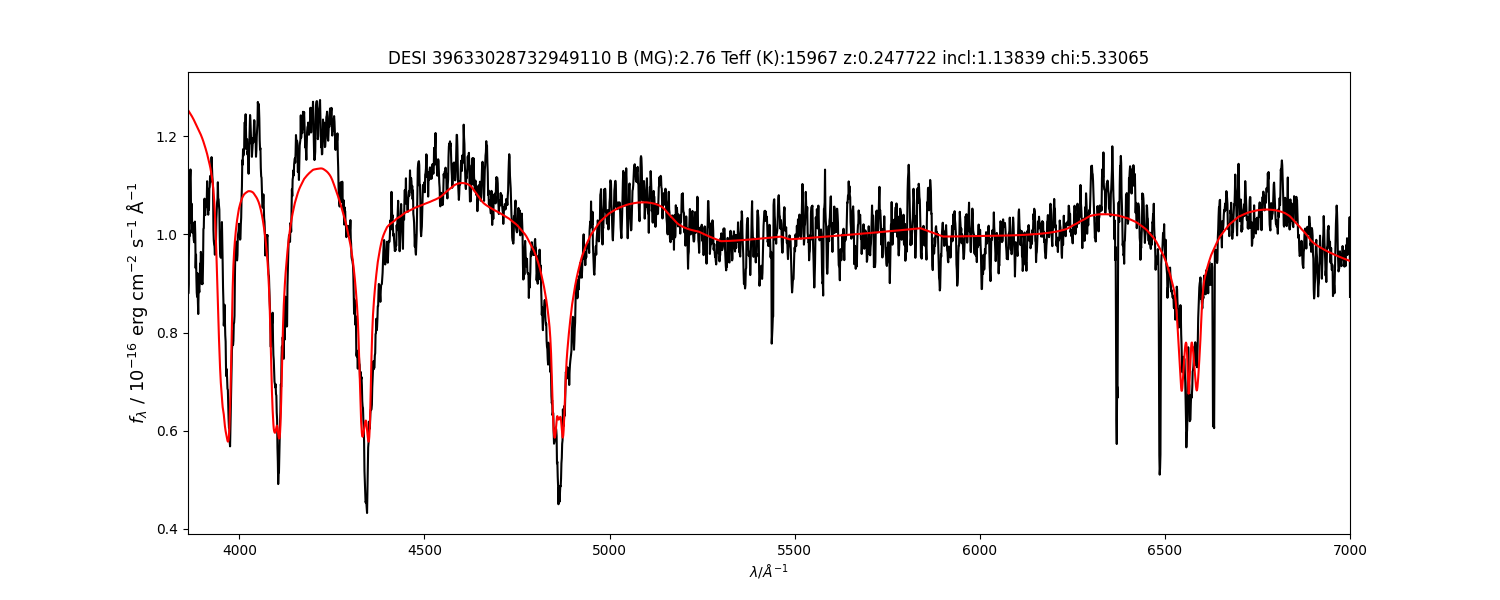}\\
\includegraphics[width=0.9\linewidth]{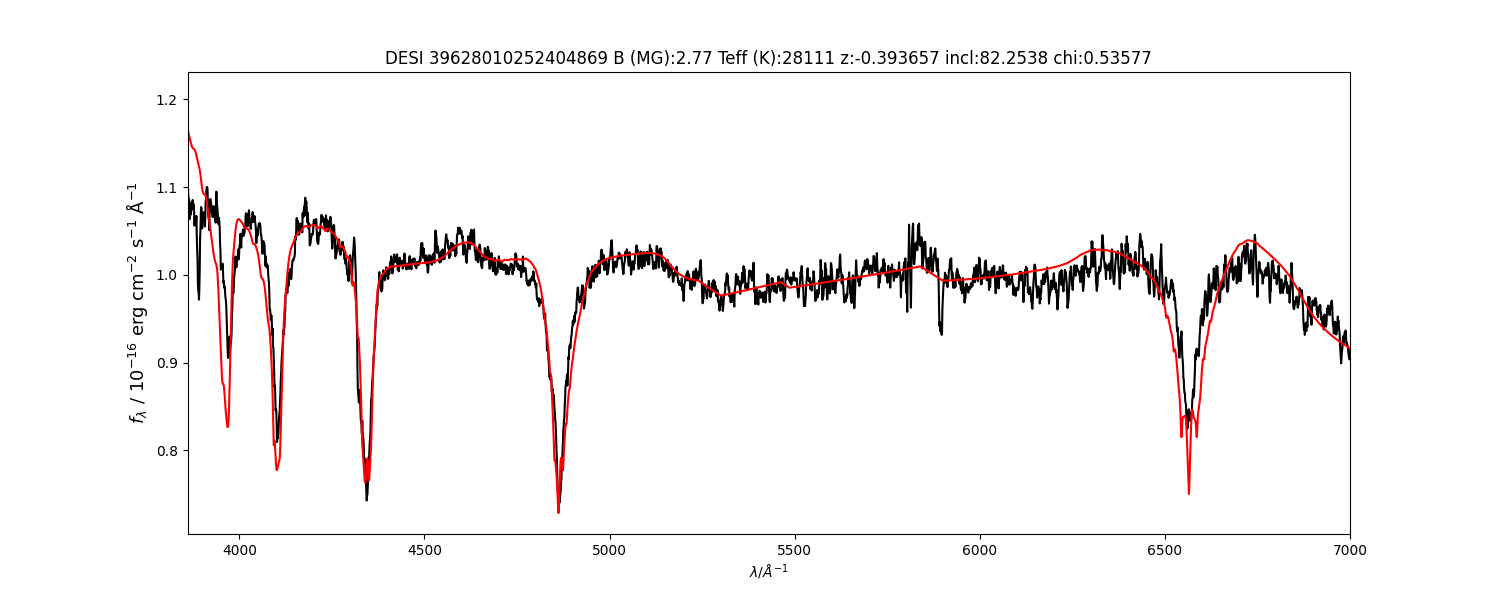}\\ 
\end{supertabular}
 \newpage \captionof{figure}{cont.}
\begin{supertabular}{c}
\includegraphics[width=0.9\linewidth]{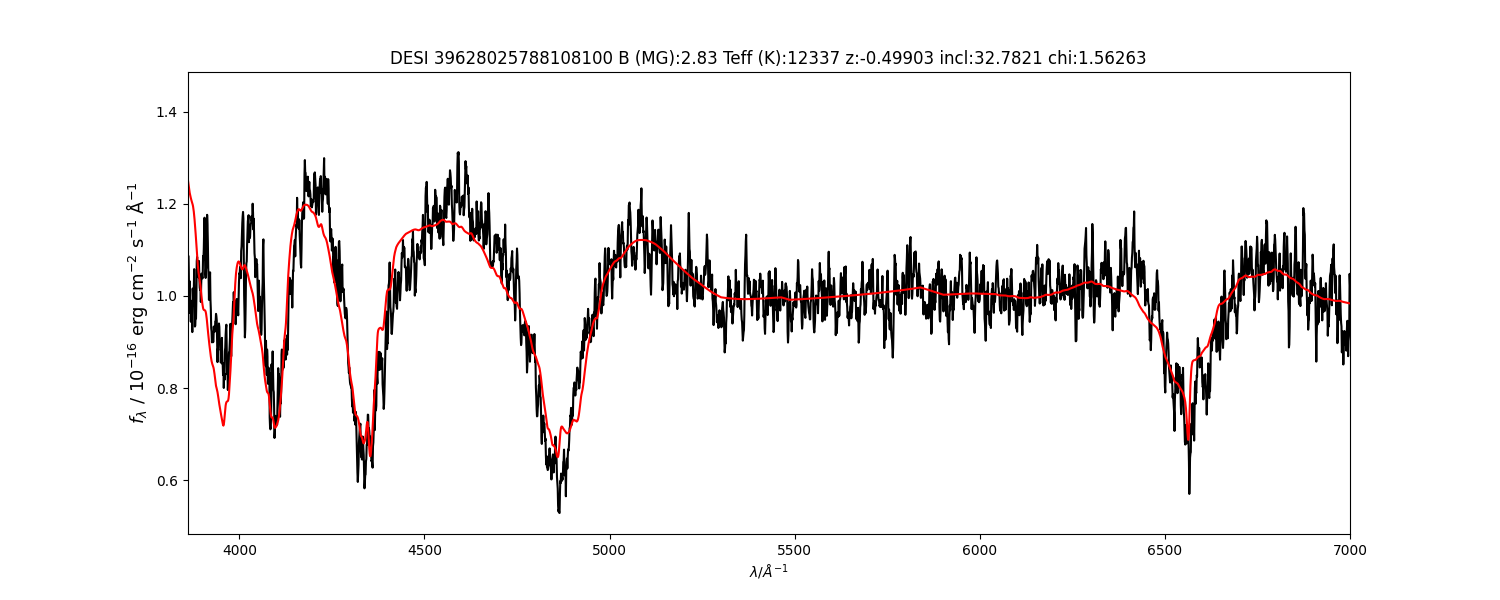}\\
\includegraphics[width=0.9\linewidth]{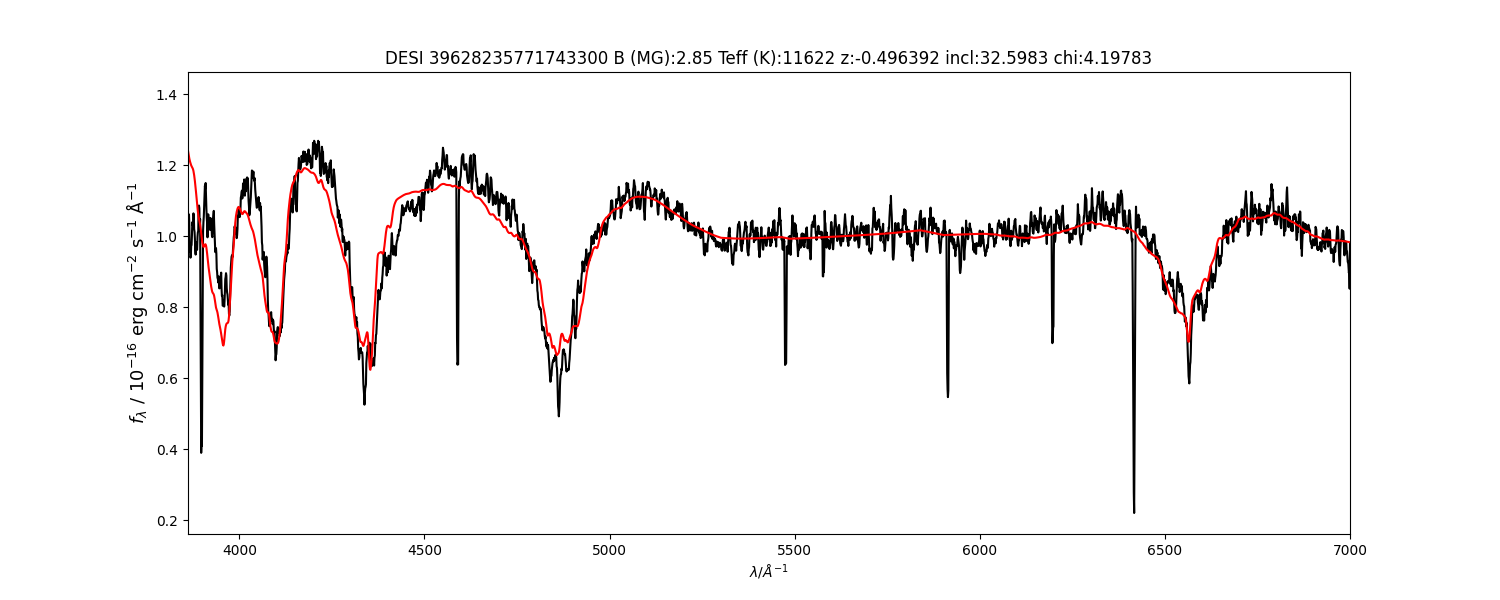}\\
\includegraphics[width=0.9\linewidth]{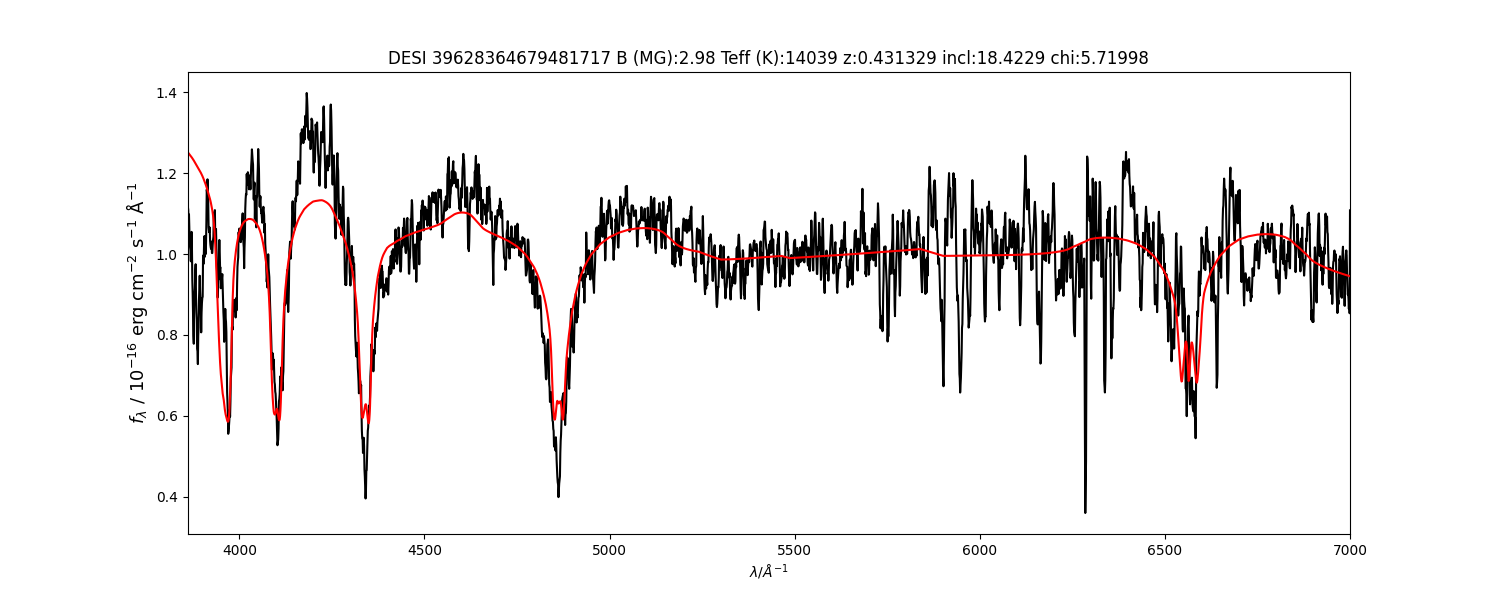}\\ 
\end{supertabular}
 \newpage \captionof{figure}{cont.}
\begin{supertabular}{c}
\includegraphics[width=0.9\linewidth]{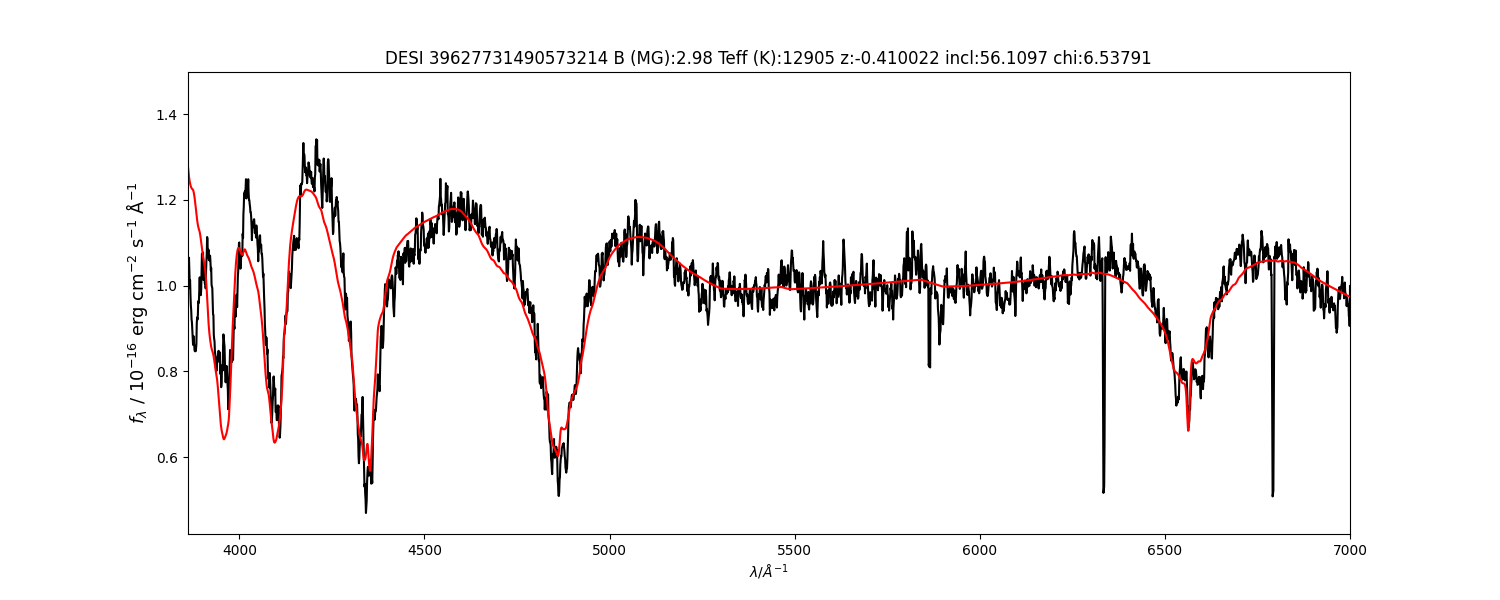}\\
\includegraphics[width=0.9\linewidth]{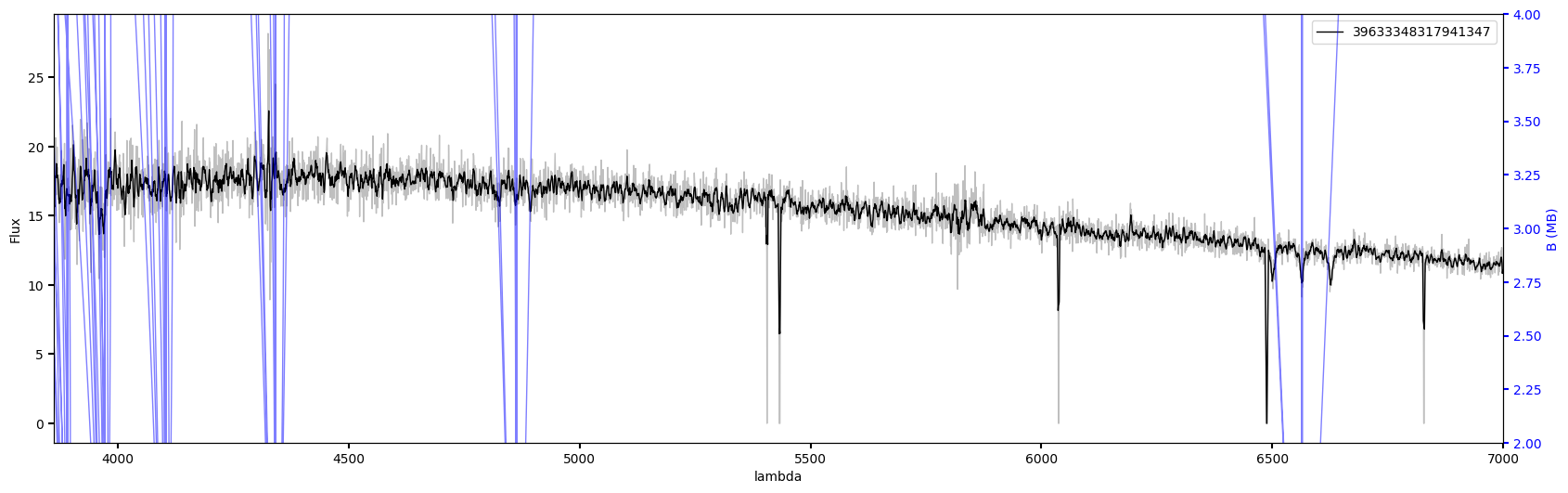}\\
\includegraphics[width=0.9\linewidth]{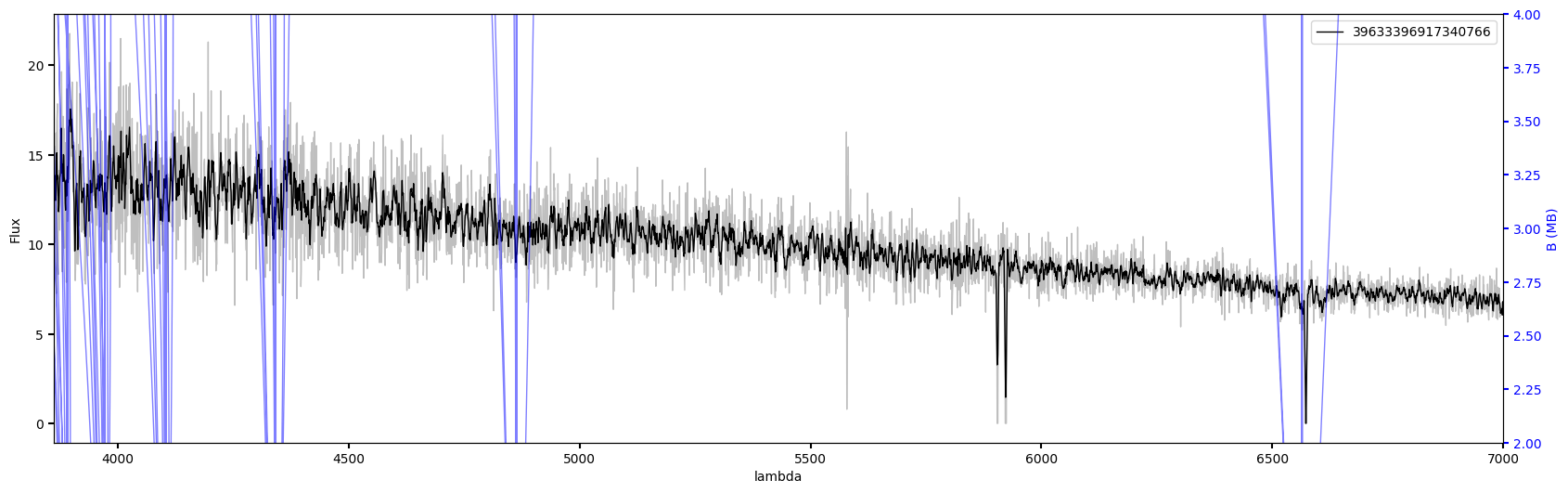}\\ 
\end{supertabular}
 \newpage \captionof{figure}{cont.}
\begin{supertabular}{c}
\includegraphics[width=0.9\linewidth]{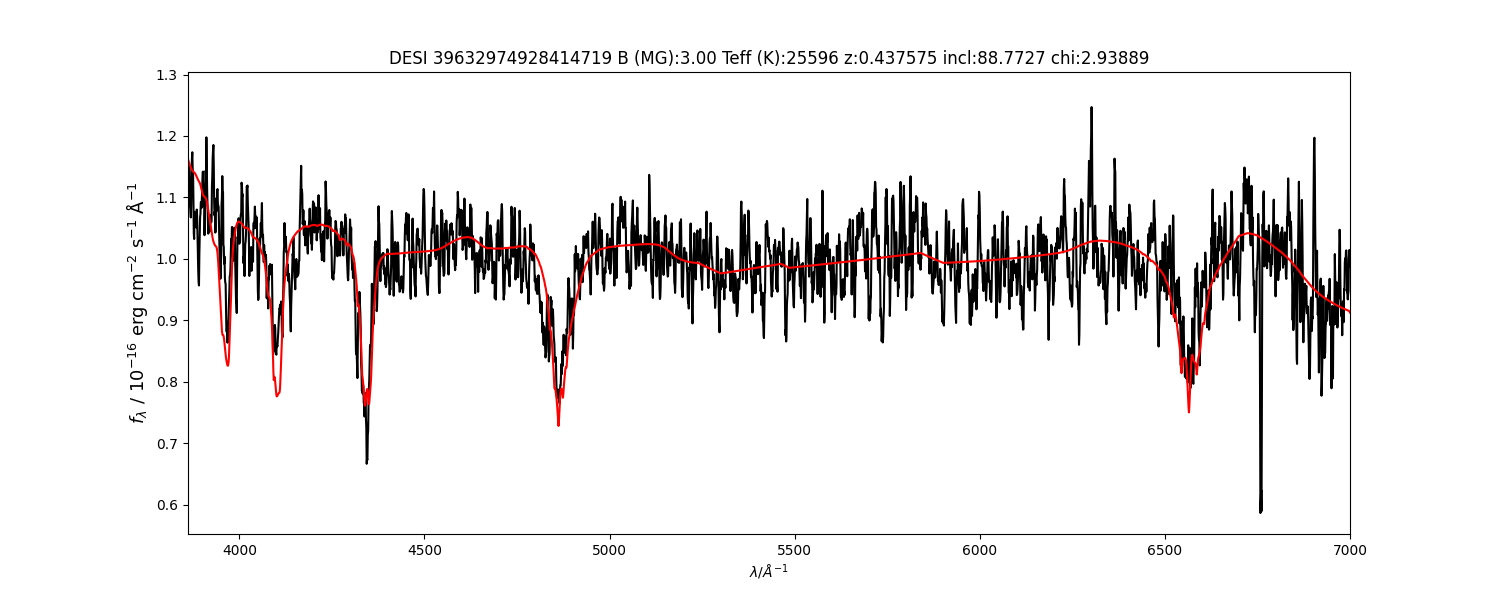}\\
\includegraphics[width=0.9\linewidth]{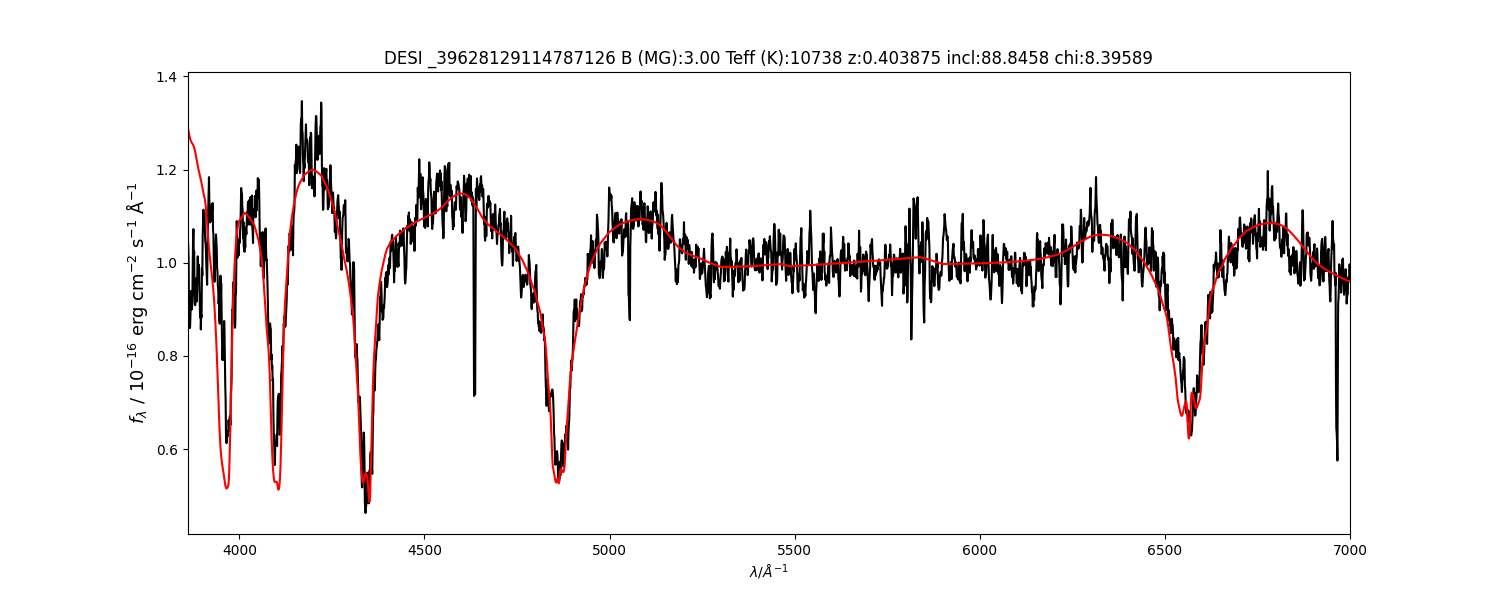}\\
\includegraphics[width=0.9\linewidth]{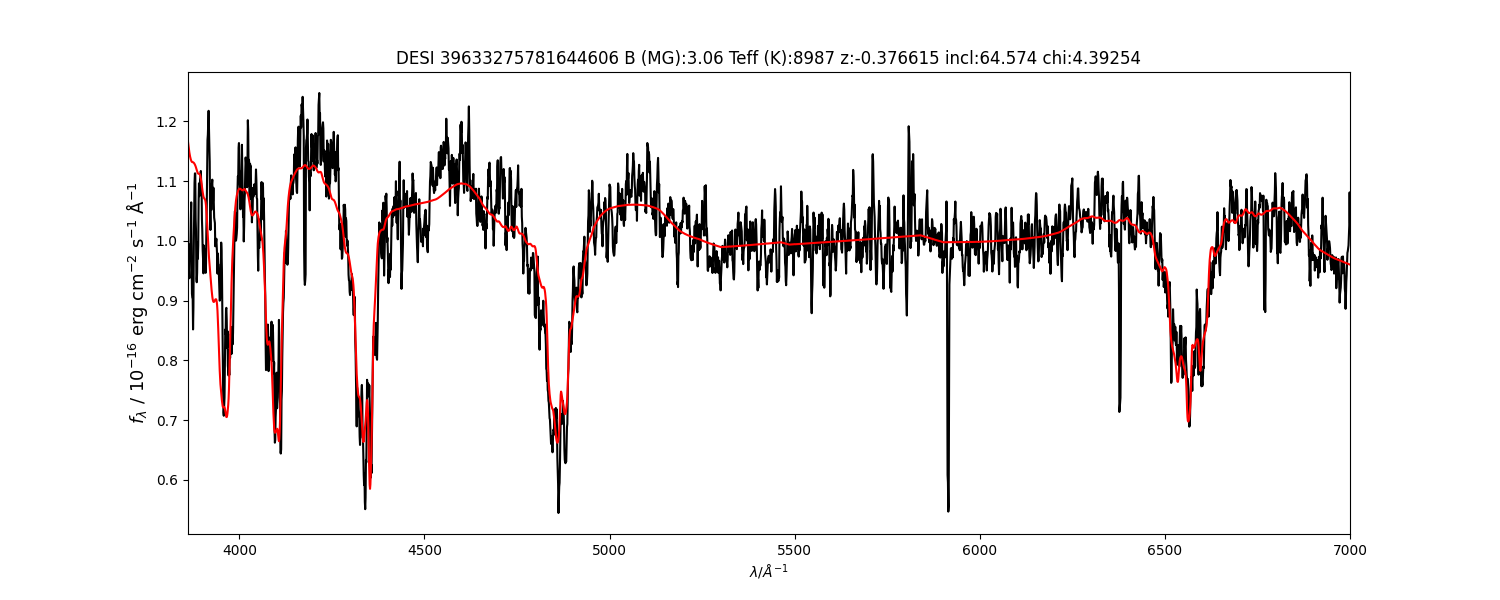}\\ 
\end{supertabular}
 \newpage \captionof{figure}{cont.}
\begin{supertabular}{c}
\includegraphics[width=0.9\linewidth]{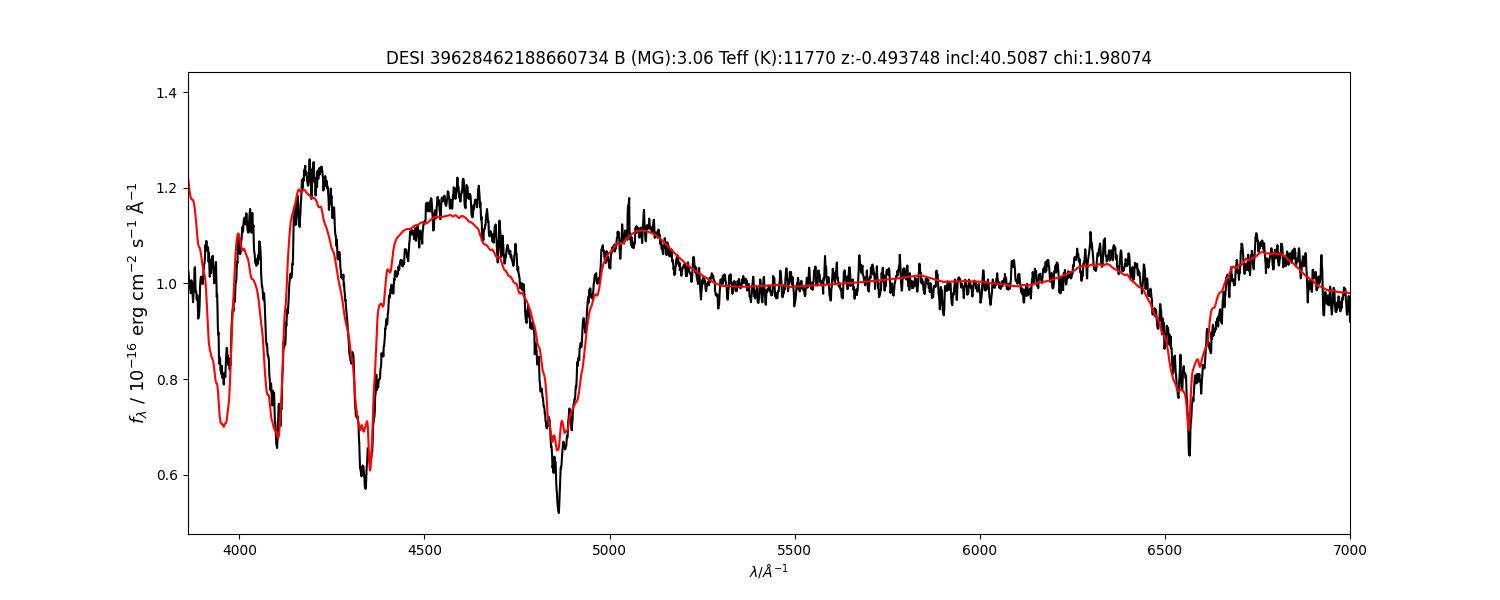}\\
\includegraphics[width=0.9\linewidth]{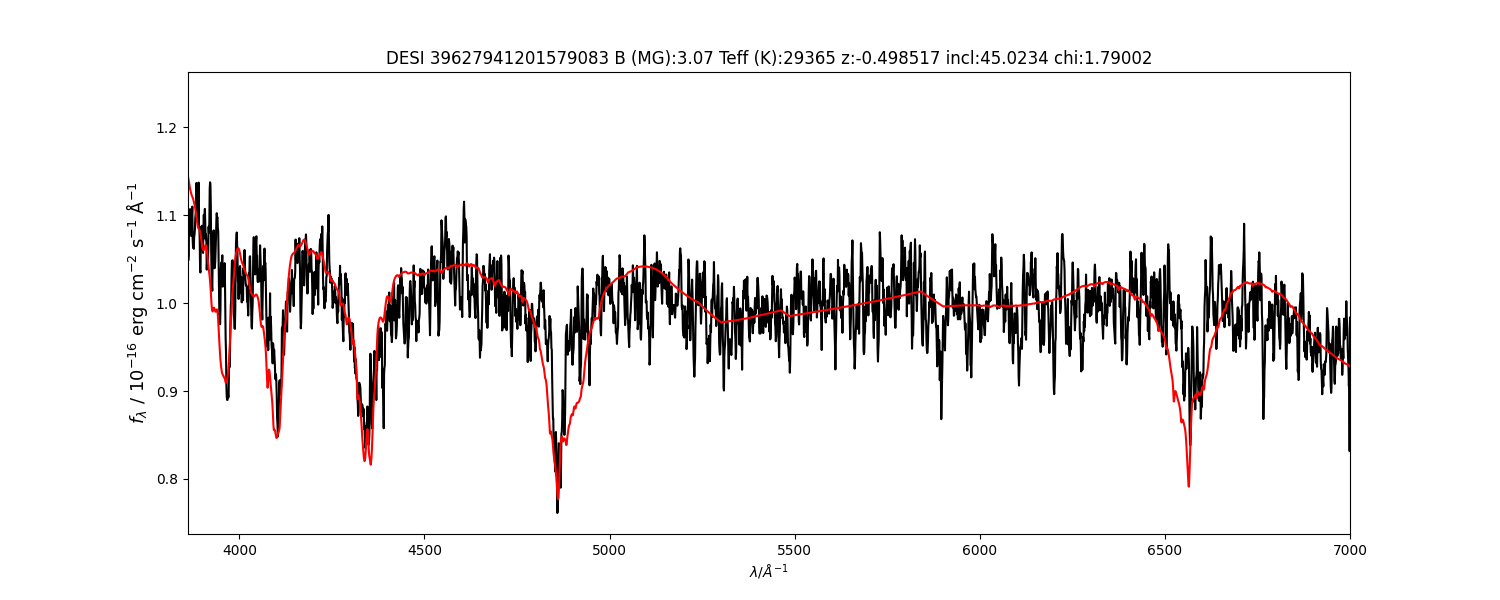}\\
\includegraphics[width=0.9\linewidth]{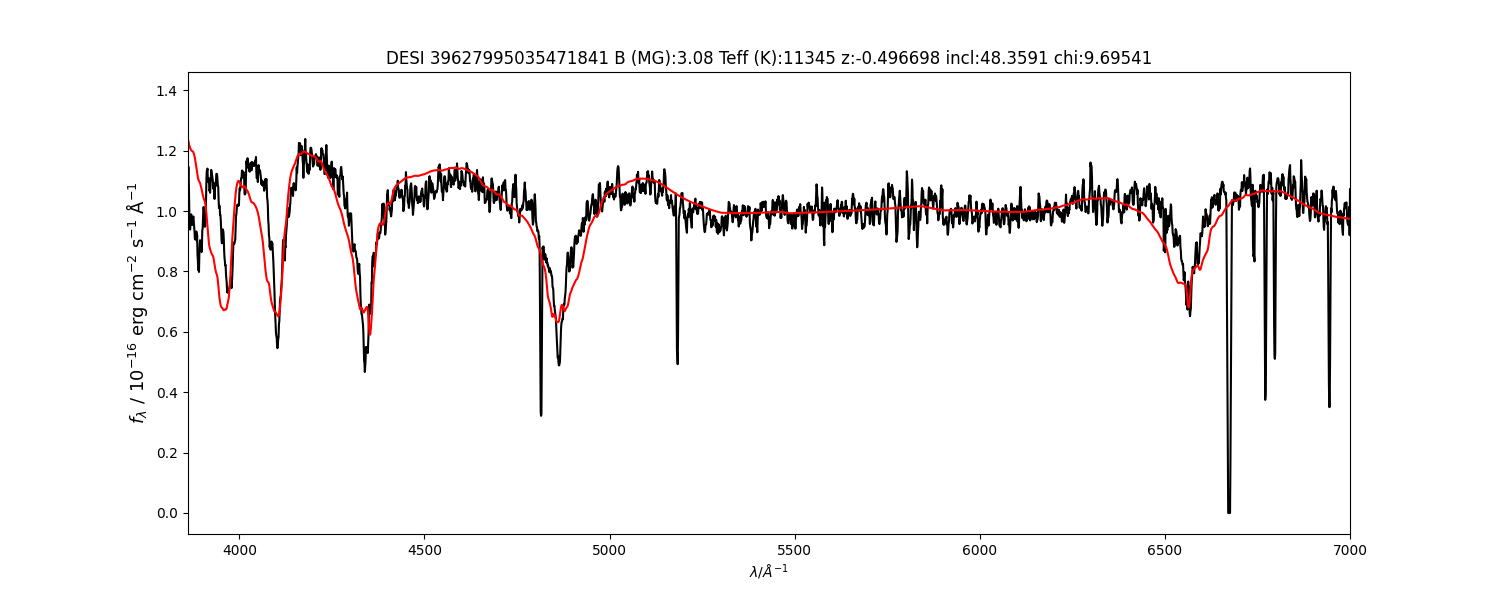}\\ 
\end{supertabular}
 \newpage \captionof{figure}{cont.}
\begin{supertabular}{c}
\includegraphics[width=0.9\linewidth]{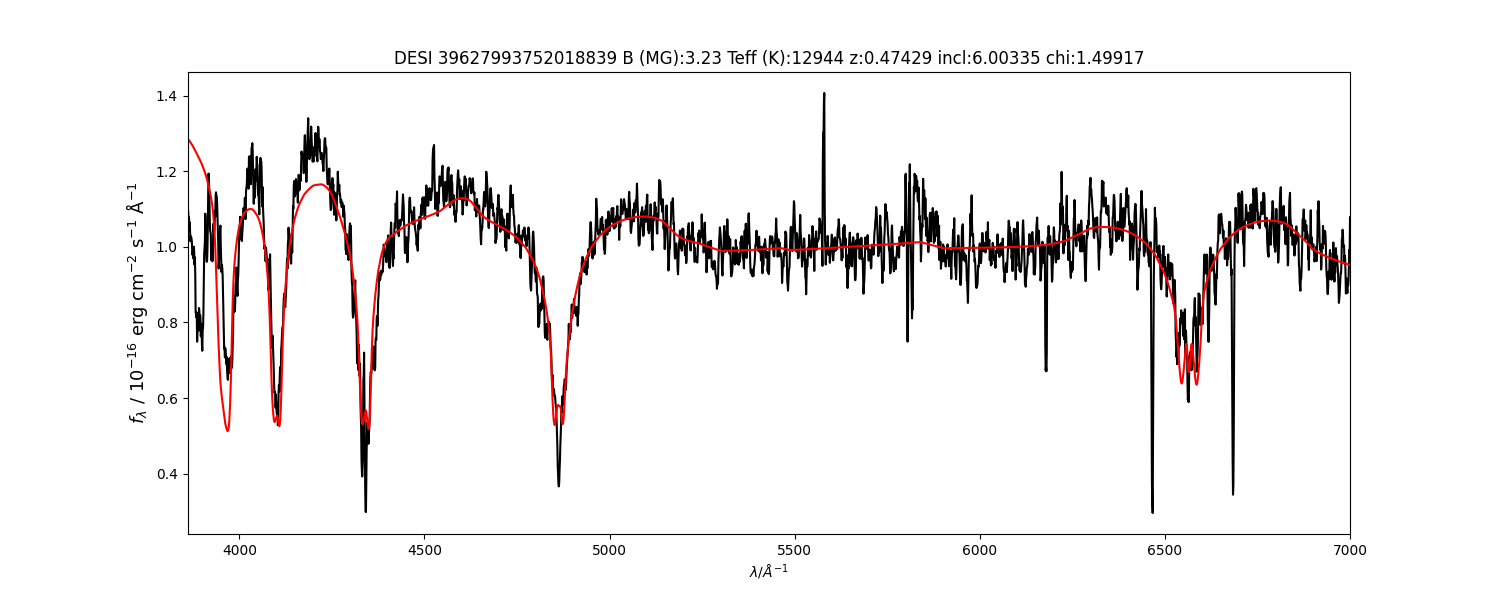}\\
\includegraphics[width=0.9\linewidth]{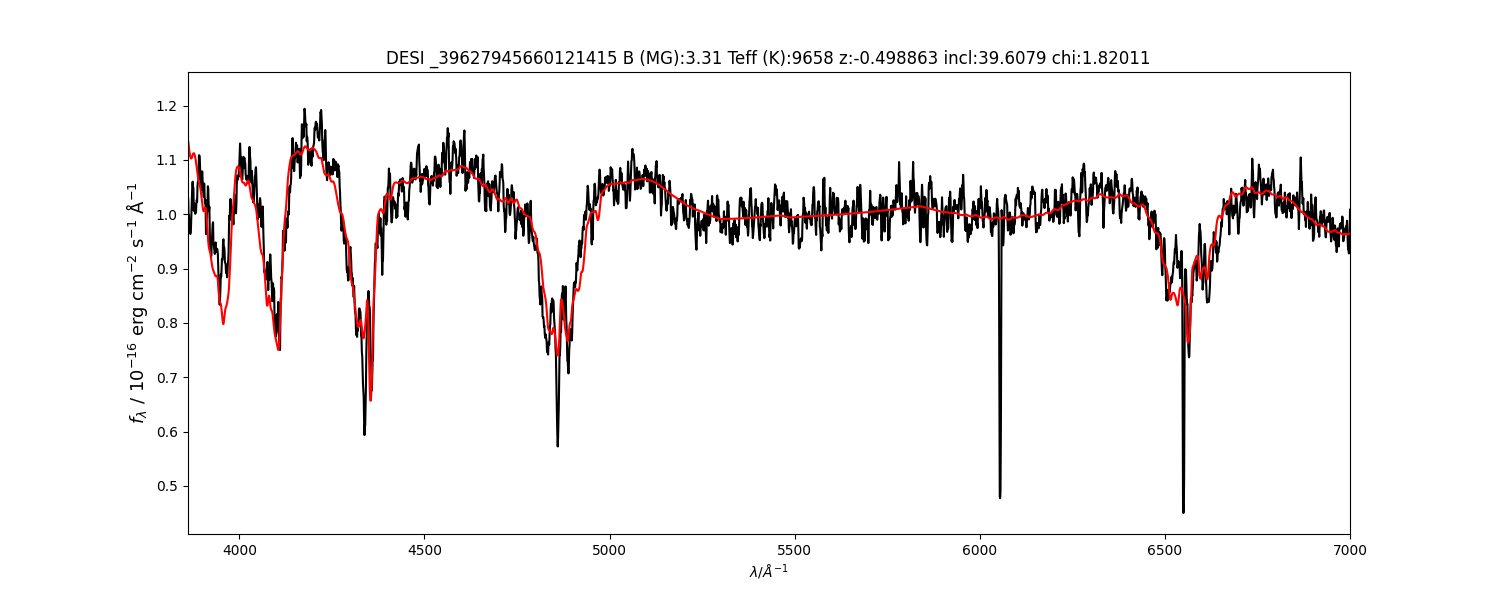}\\
\includegraphics[width=0.9\linewidth]{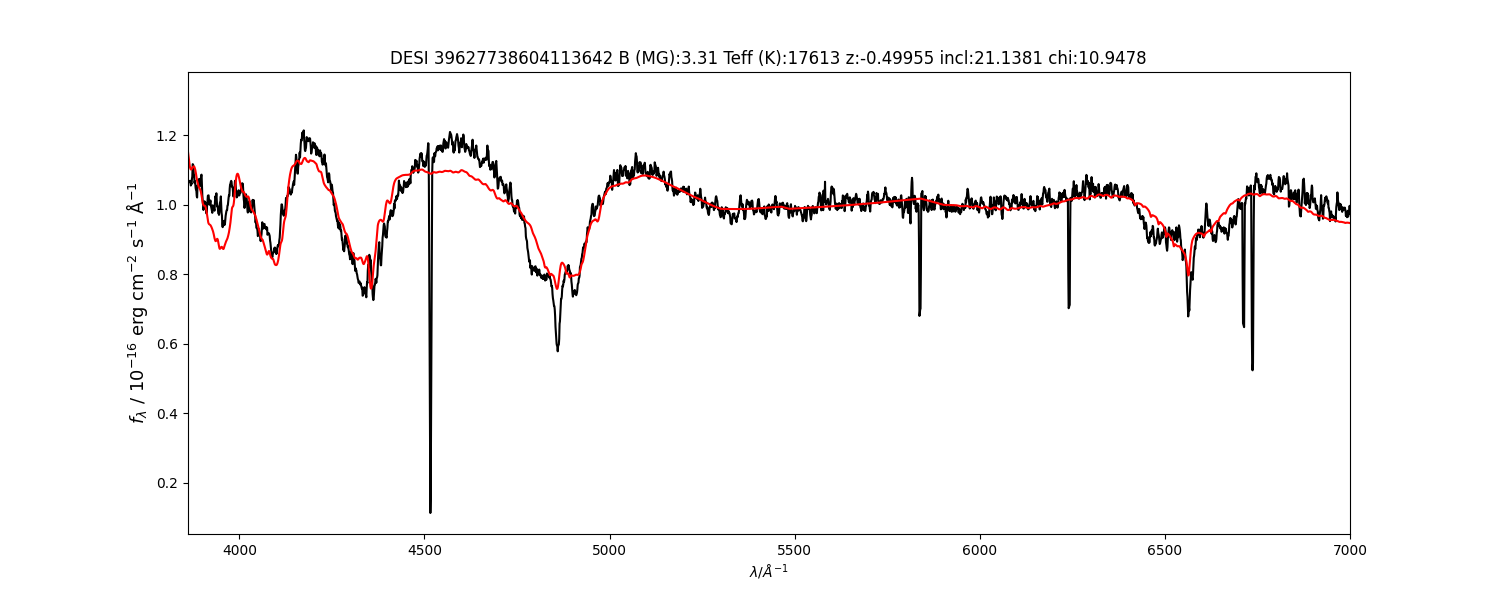}\\ 
\end{supertabular}
 \newpage \captionof{figure}{cont.}
\begin{supertabular}{c}
\includegraphics[width=0.9\linewidth]{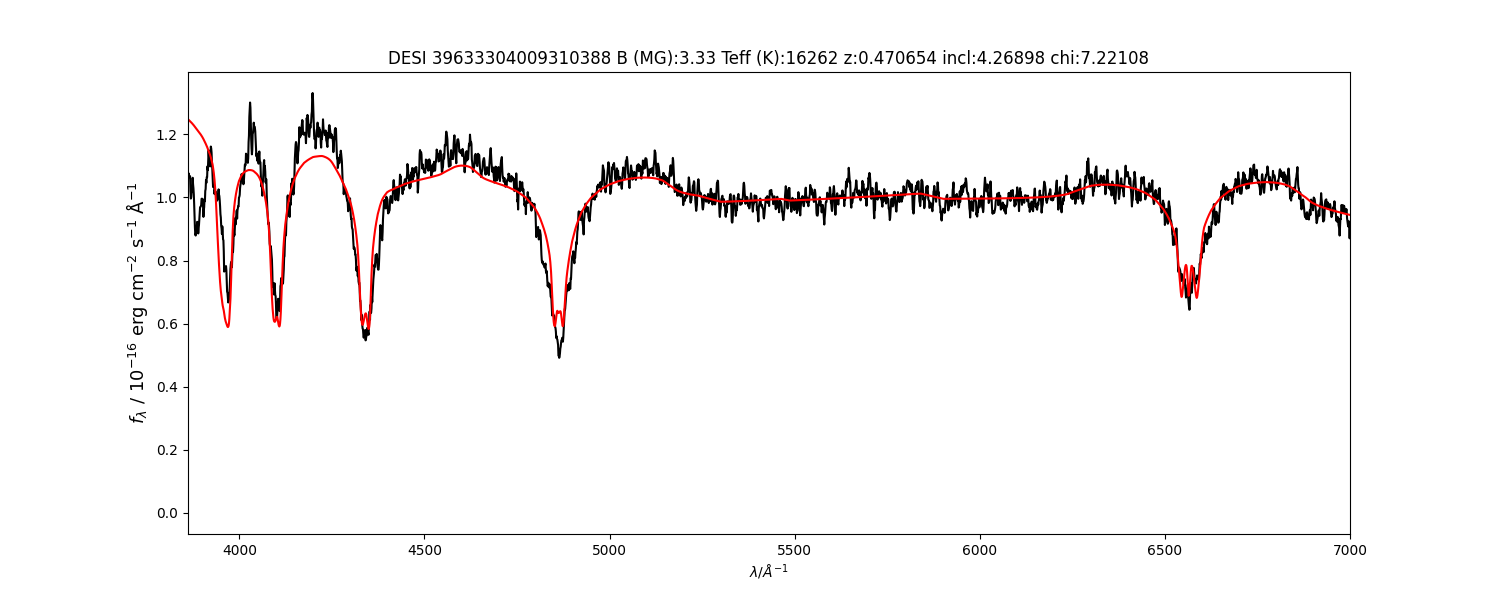}\\
\includegraphics[width=0.9\linewidth]{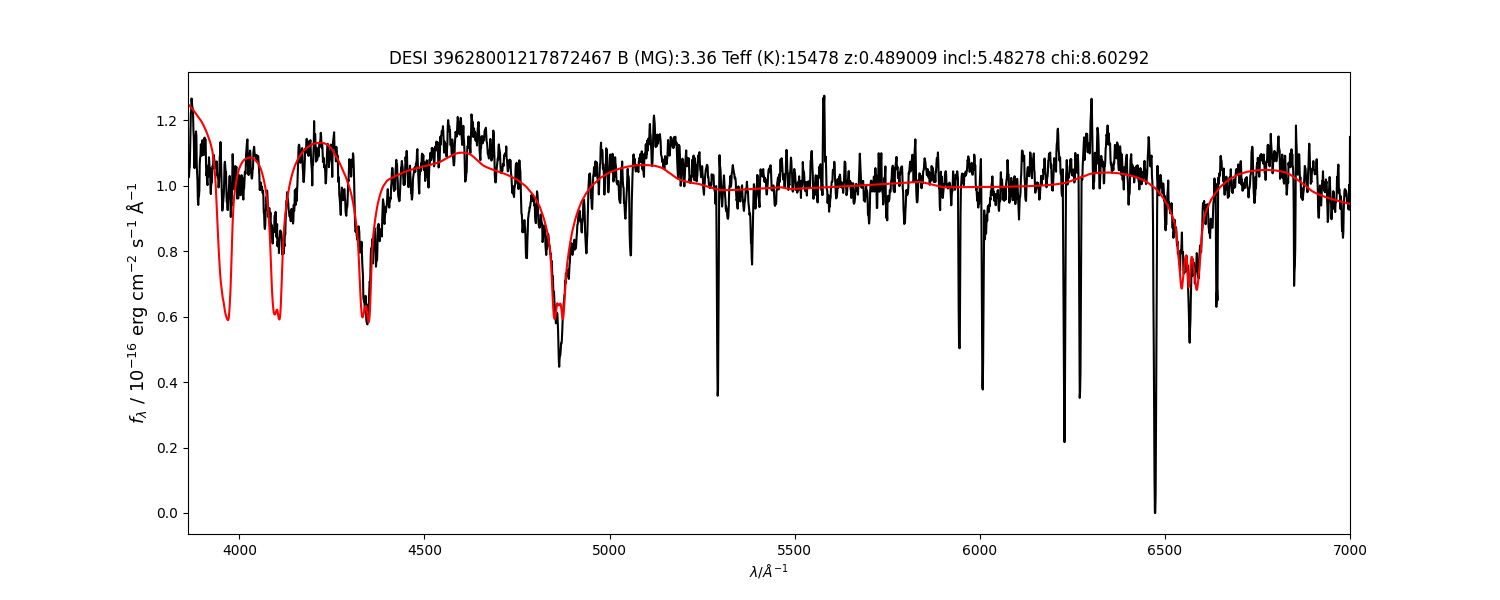}\\
\includegraphics[width=0.9\linewidth]{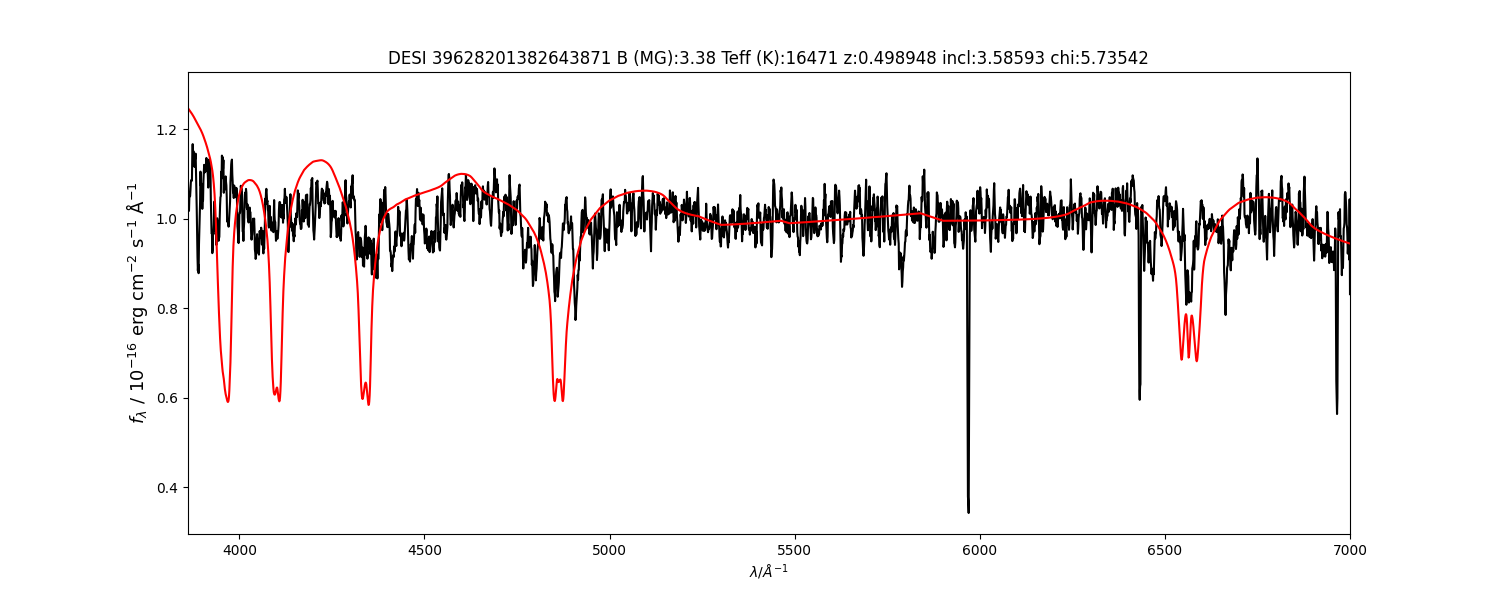}\\ 
\end{supertabular}
 \newpage \captionof{figure}{cont.}
\begin{supertabular}{c}
\includegraphics[width=0.9\linewidth]{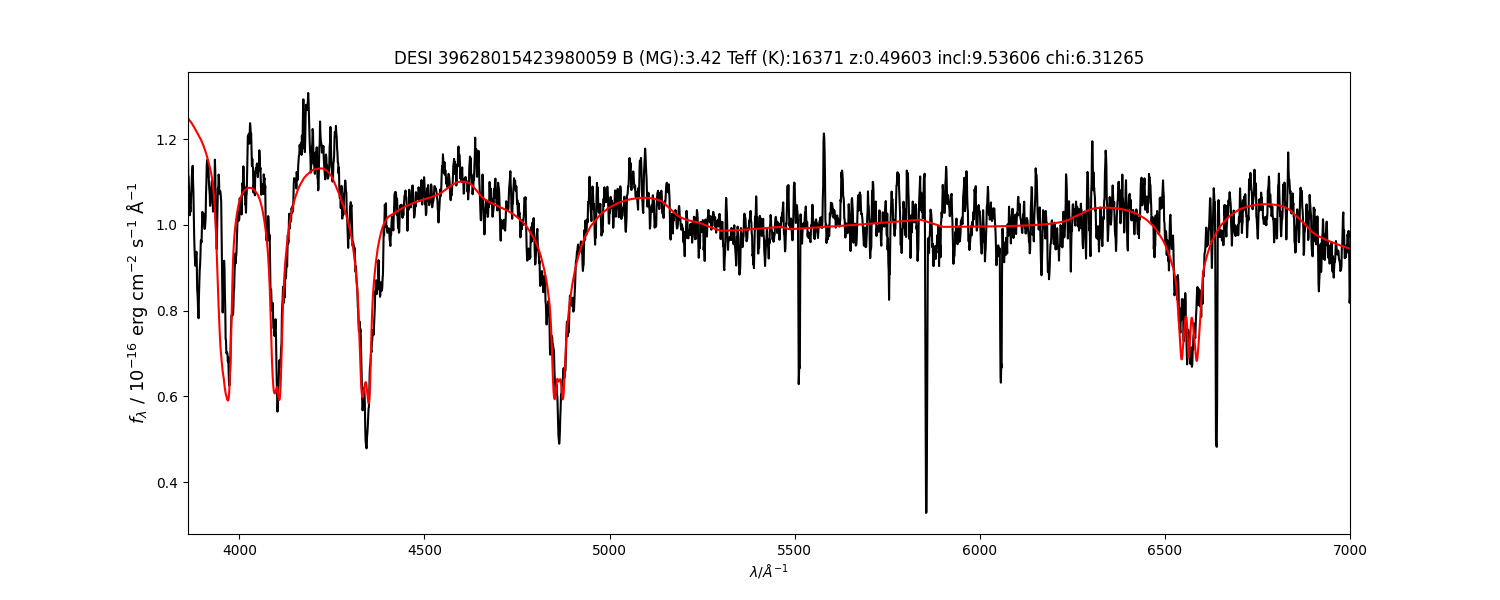}\\
\includegraphics[width=0.9\linewidth]{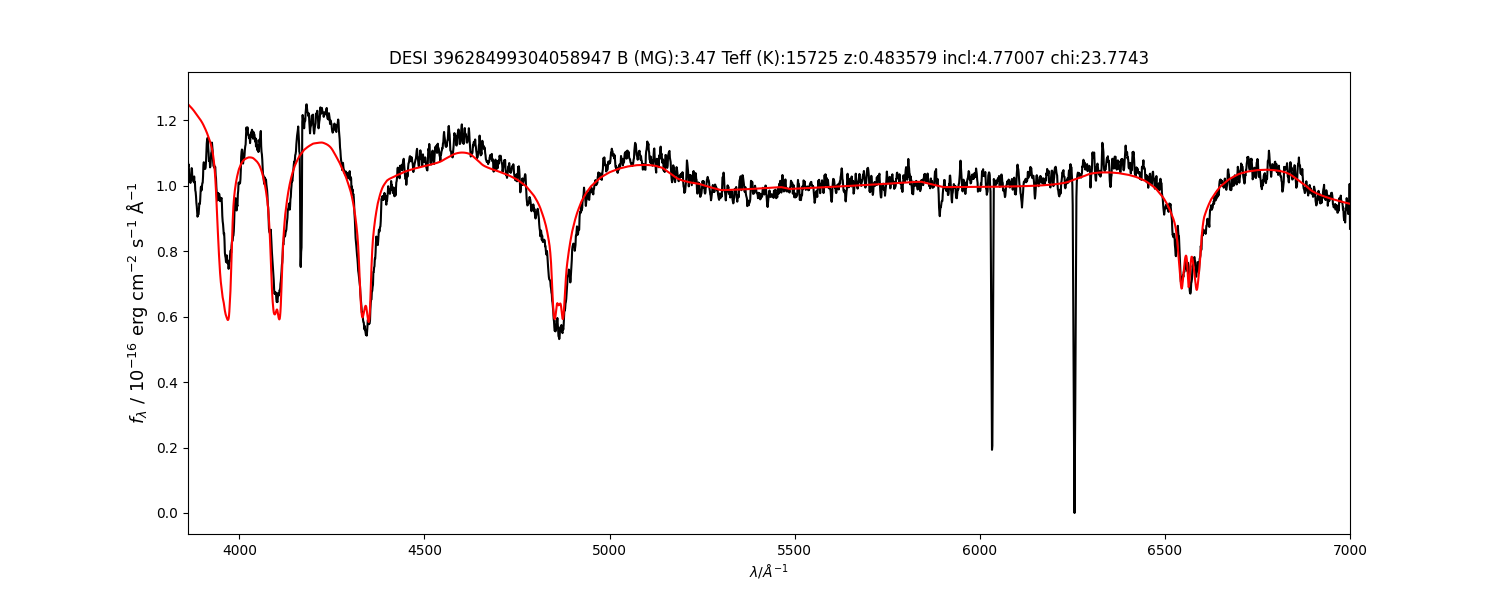}\\
\includegraphics[width=0.9\linewidth]{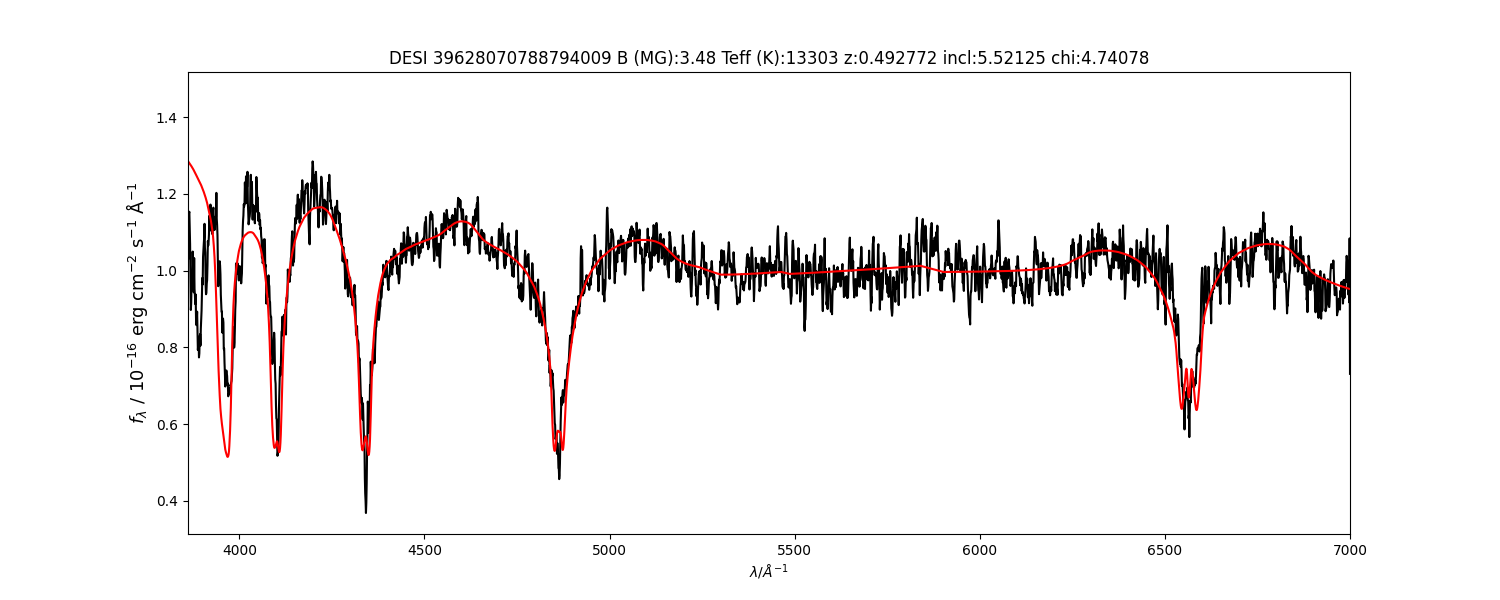}\\ 
\end{supertabular}
 \newpage \captionof{figure}{cont.}
\begin{supertabular}{c}
\includegraphics[width=0.9\linewidth]{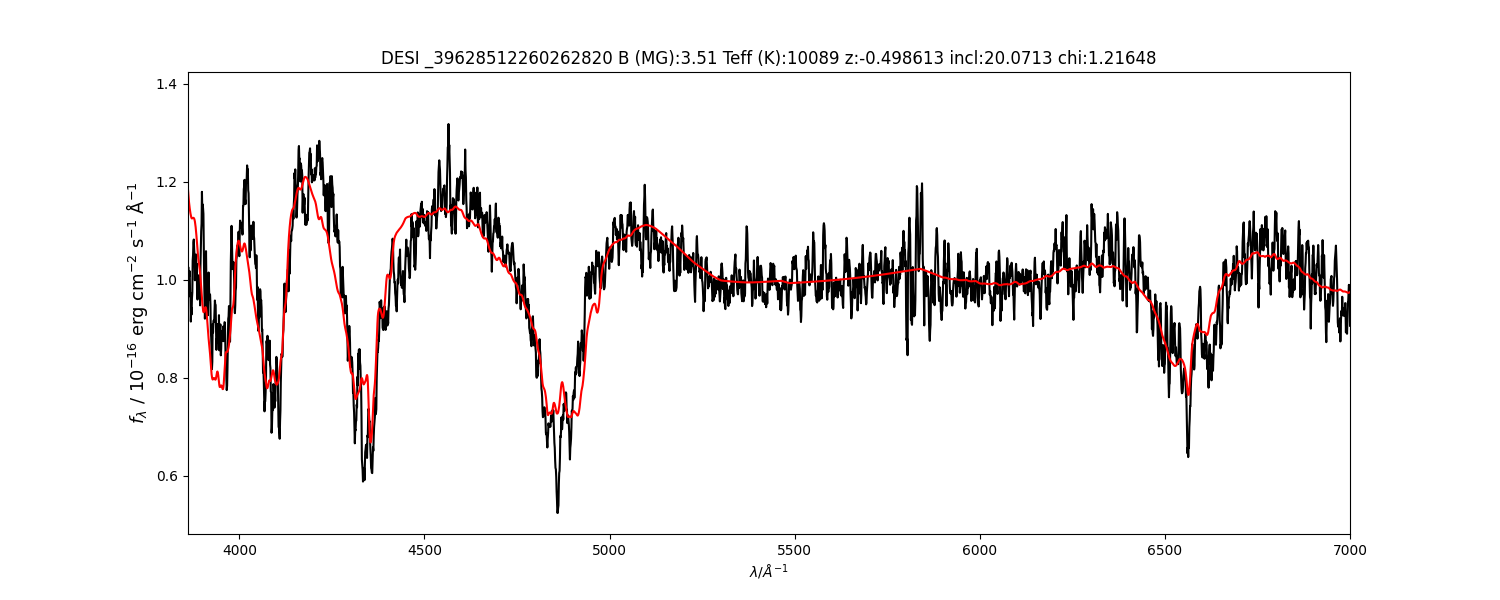}\\
\includegraphics[width=0.9\linewidth]{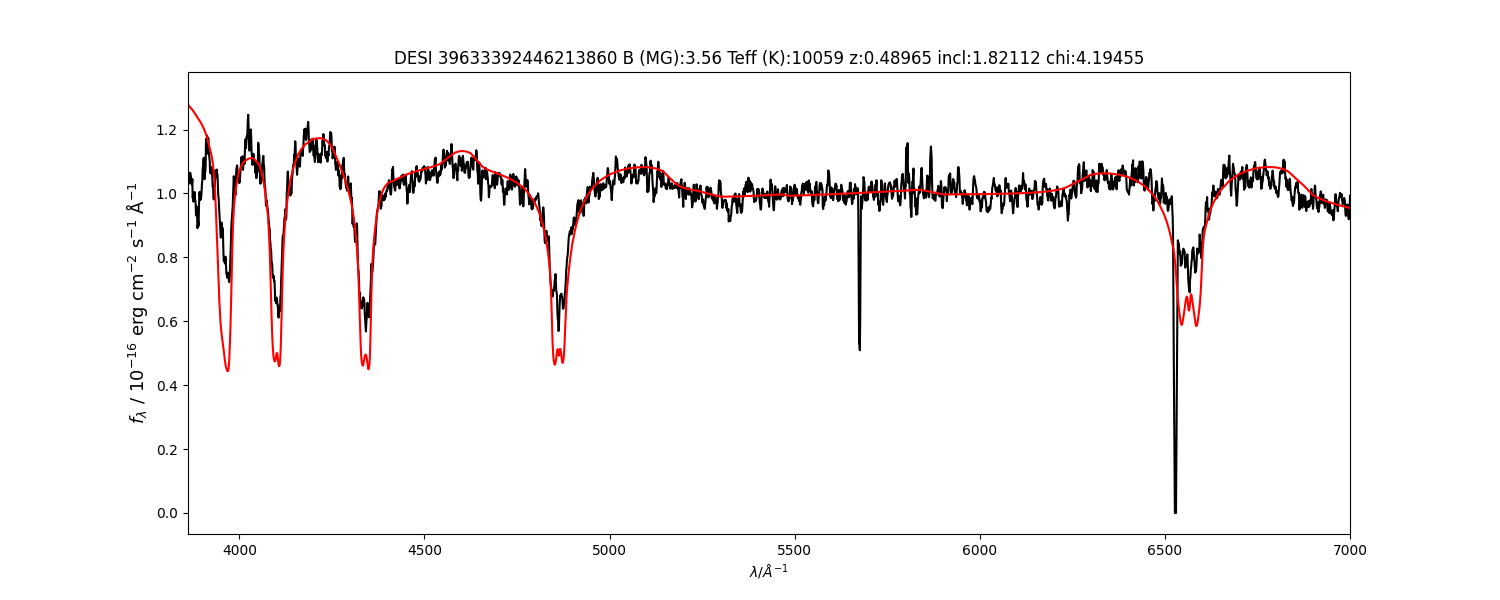}\\
\includegraphics[width=0.9\linewidth]{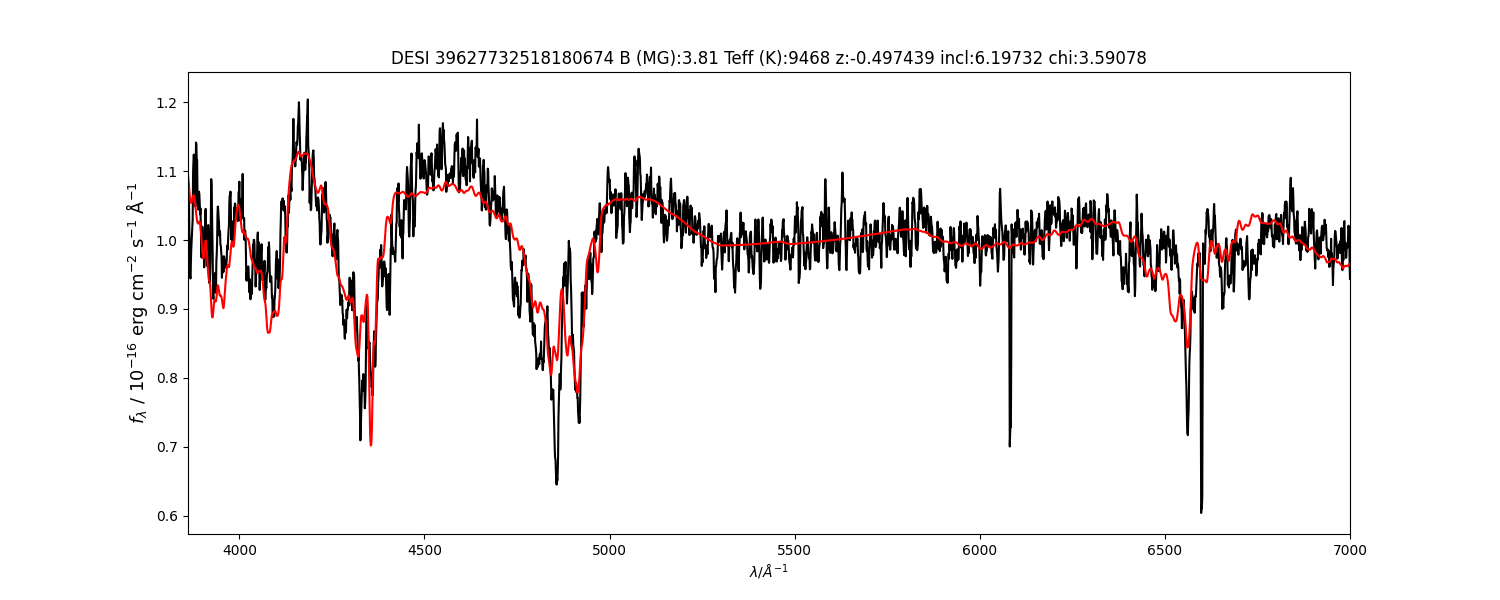}\\ 
\end{supertabular}
 \newpage \captionof{figure}{cont.}
\begin{supertabular}{c}
\includegraphics[width=0.9\linewidth]{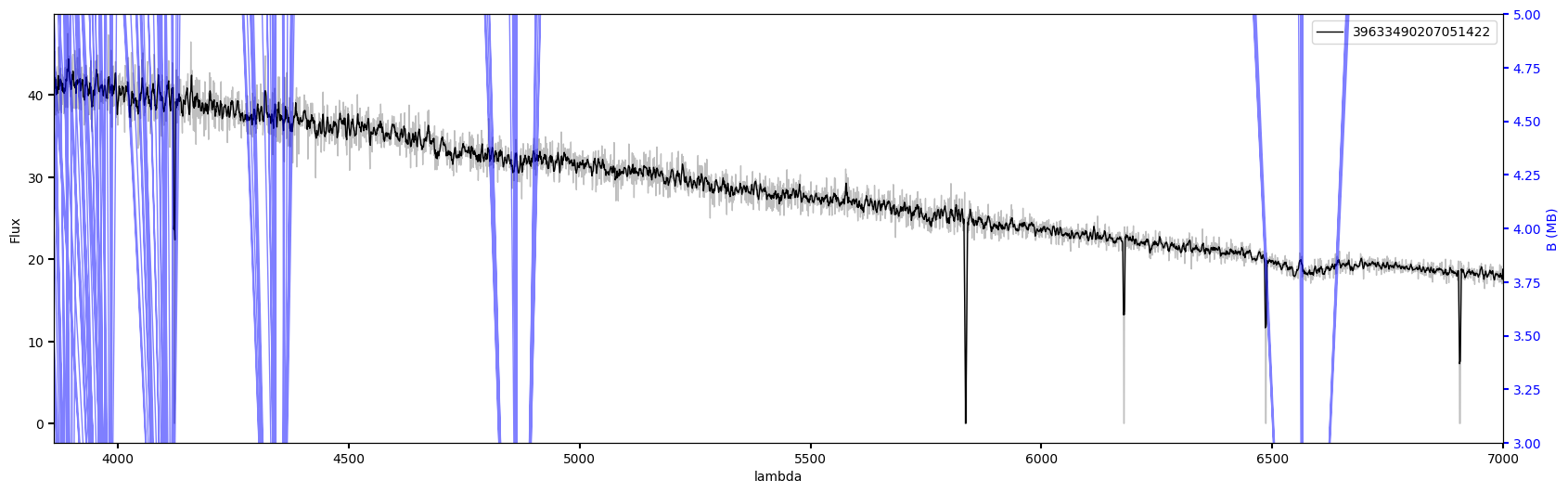}\\
\includegraphics[width=0.9\linewidth]{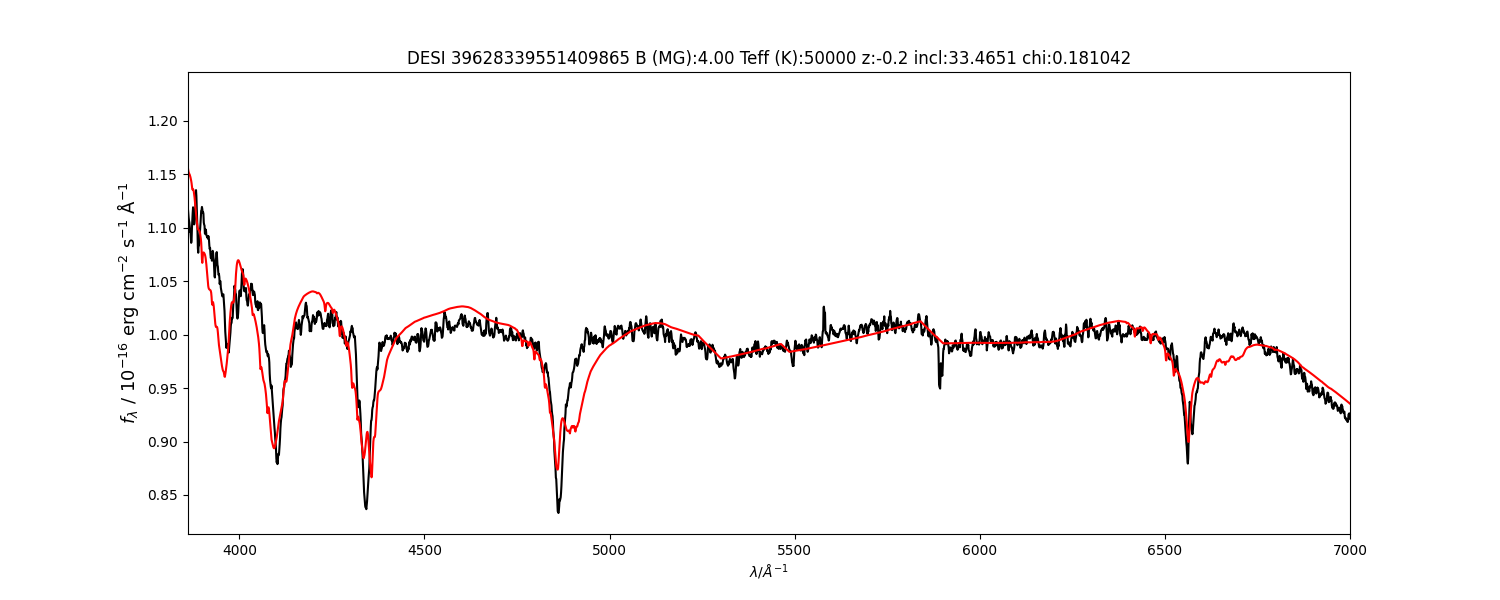}\\
\includegraphics[width=0.9\linewidth]{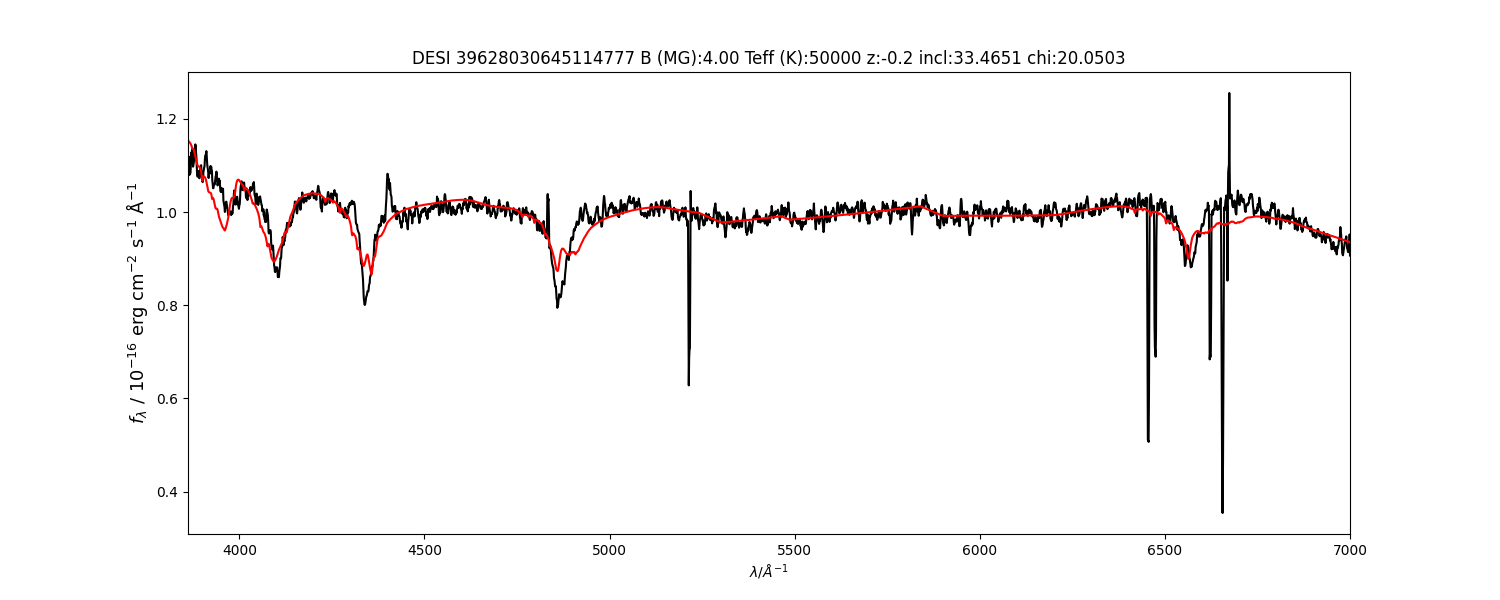}\\ 
\end{supertabular}
 \newpage \captionof{figure}{cont.}
\begin{supertabular}{c}
\includegraphics[width=0.9\linewidth]{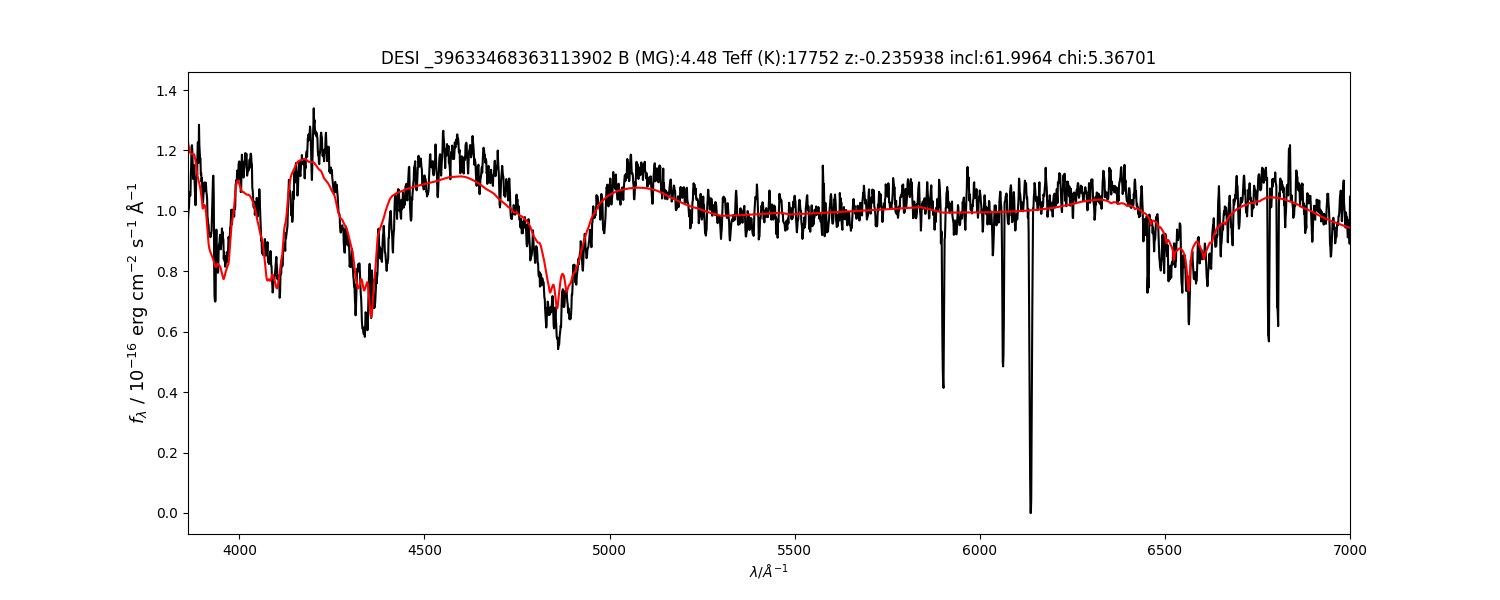}\\
\includegraphics[width=0.9\linewidth]{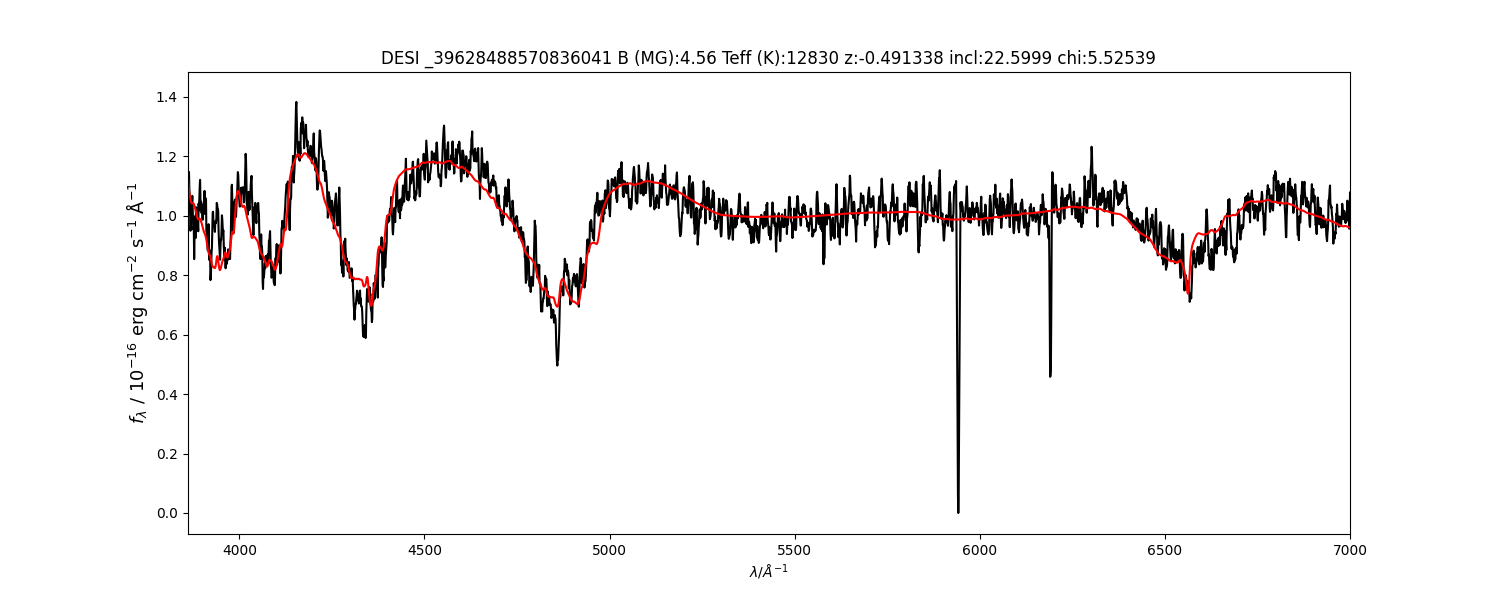}\\
\includegraphics[width=0.9\linewidth]{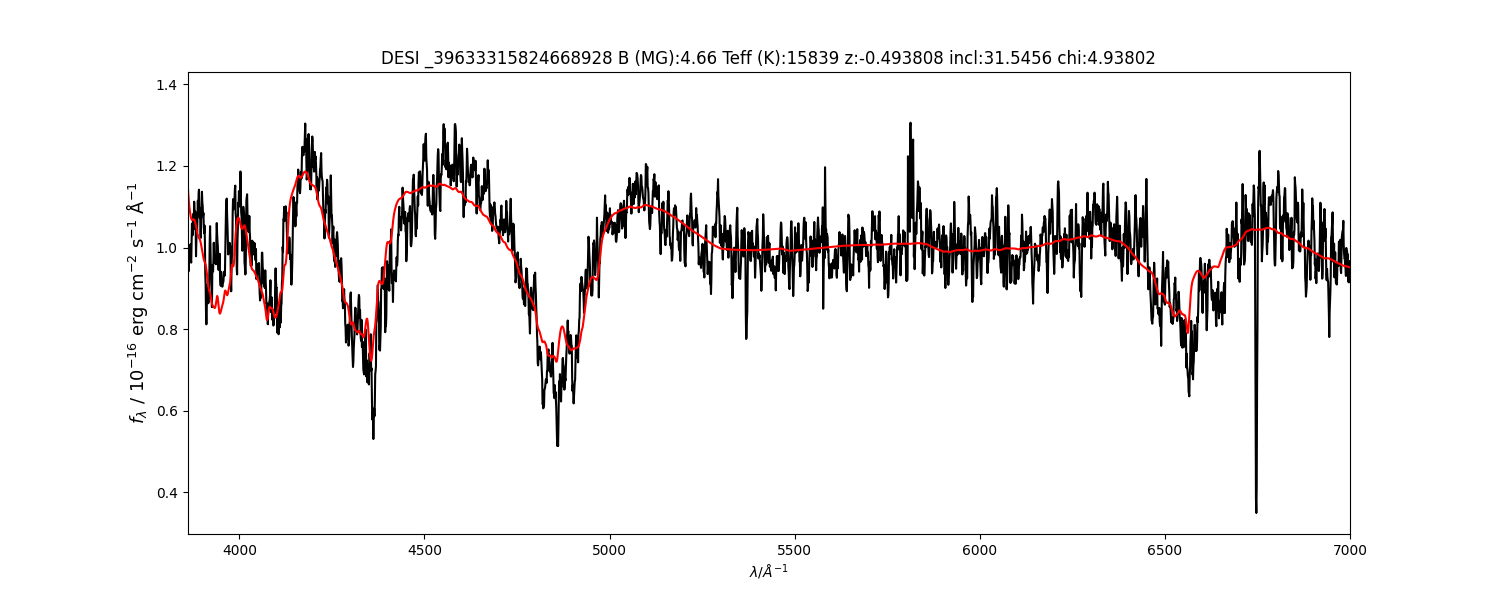}\\ 
\end{supertabular}
 \newpage \captionof{figure}{cont.}
\begin{supertabular}{c}
\includegraphics[width=0.9\linewidth]{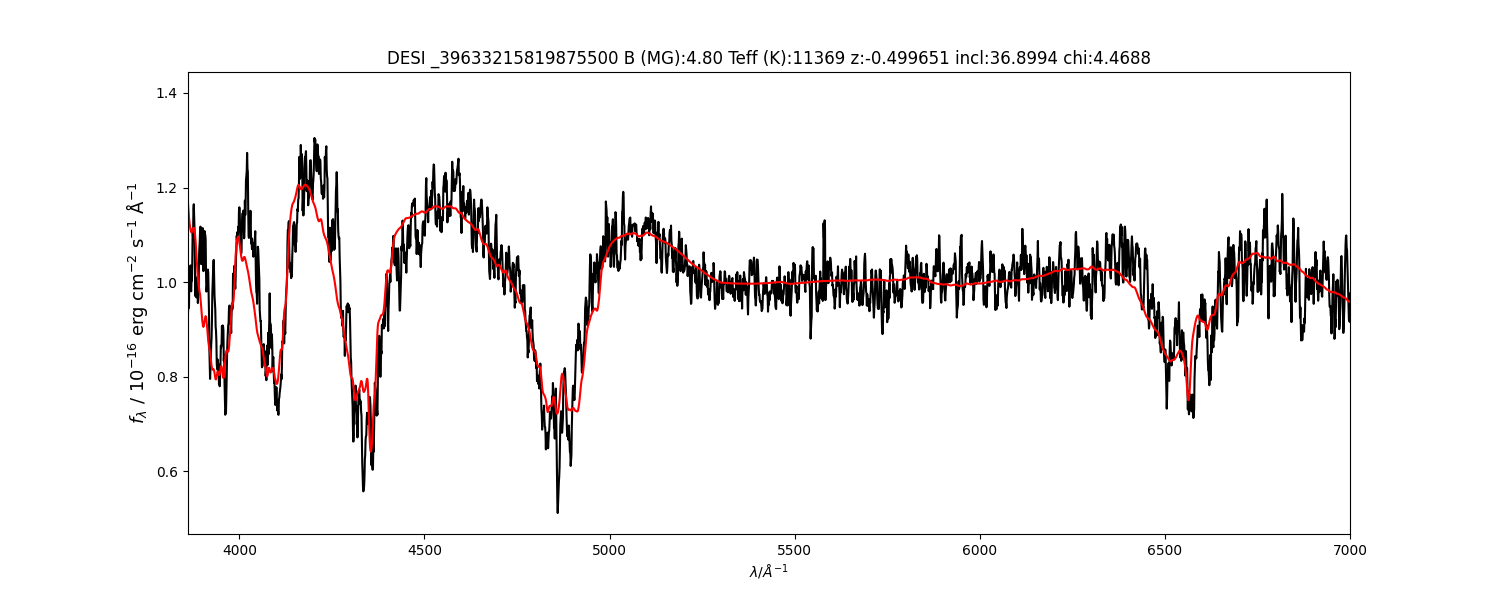}\\
\includegraphics[width=0.9\linewidth]{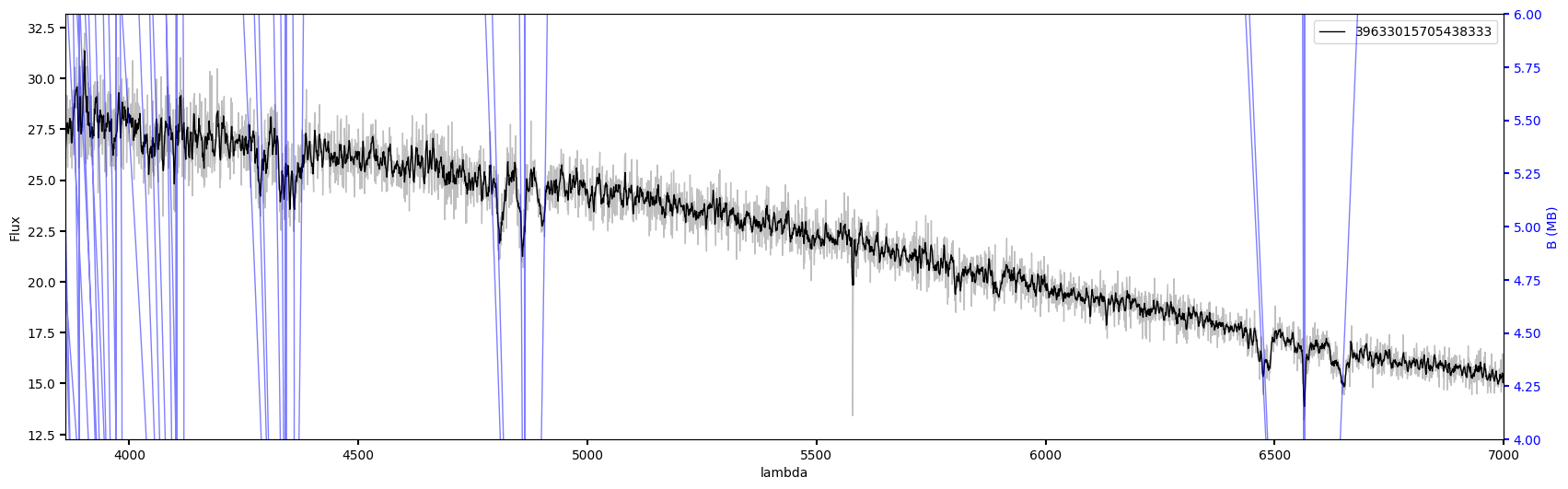}\\
\includegraphics[width=0.9\linewidth]{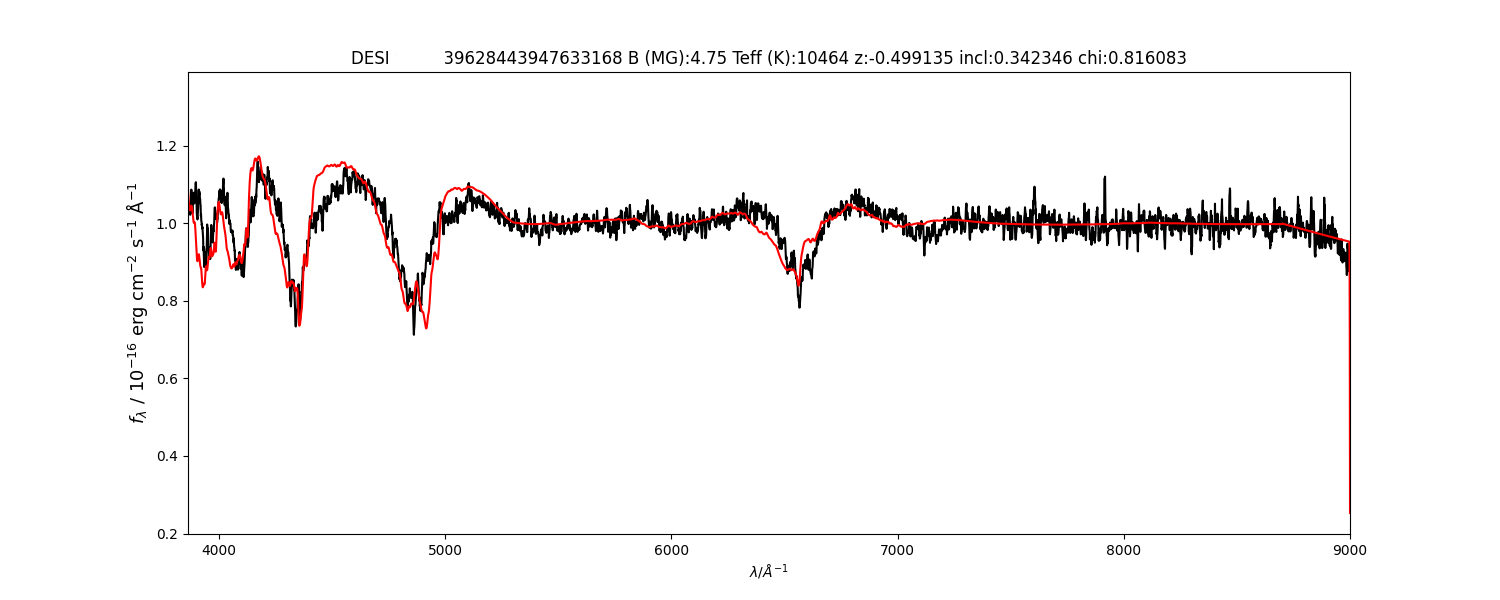}\\ 
\end{supertabular}
 \newpage \captionof{figure}{cont.}
\begin{supertabular}{c}
\includegraphics[width=0.9\linewidth]{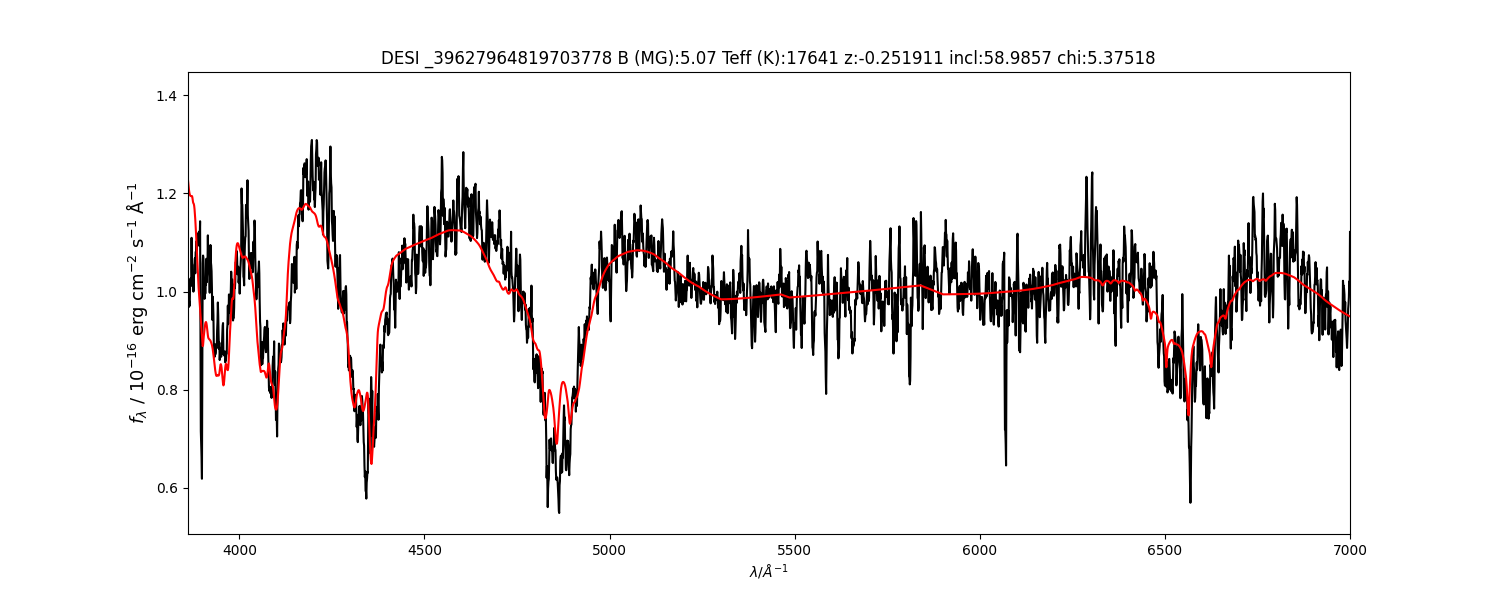}\\
\includegraphics[width=0.9\linewidth]{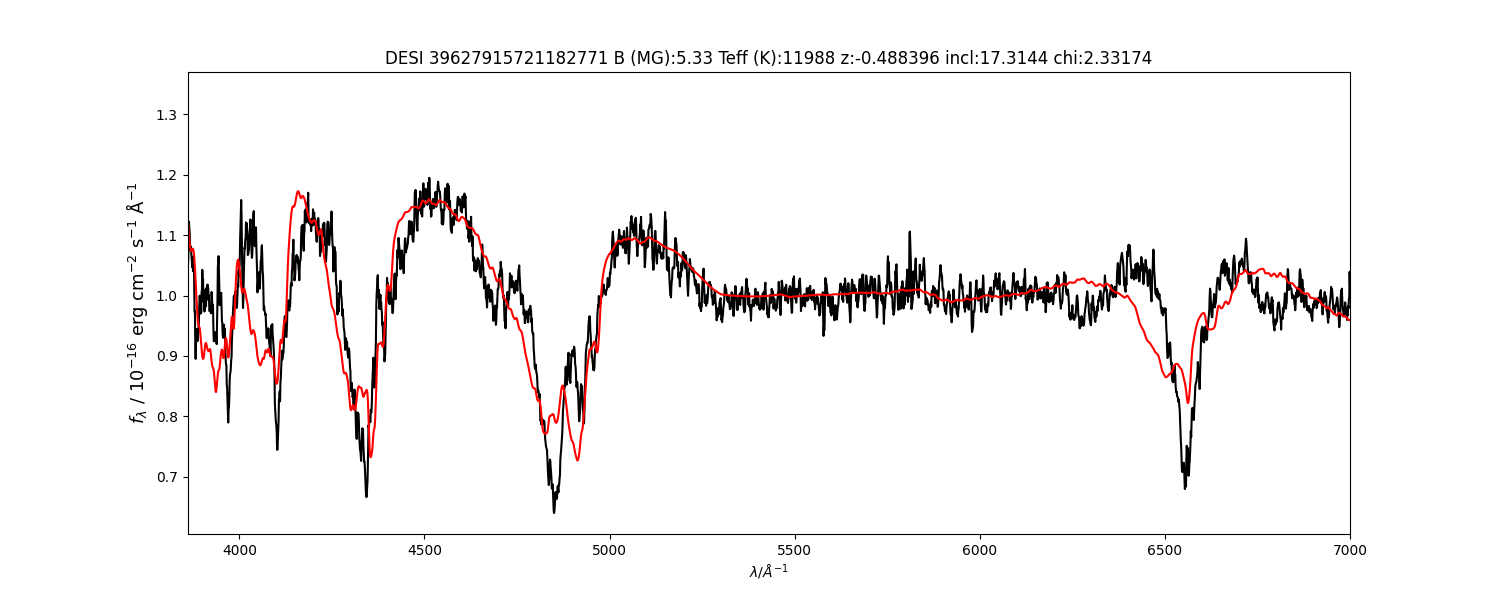}\\
\includegraphics[width=0.9\linewidth]{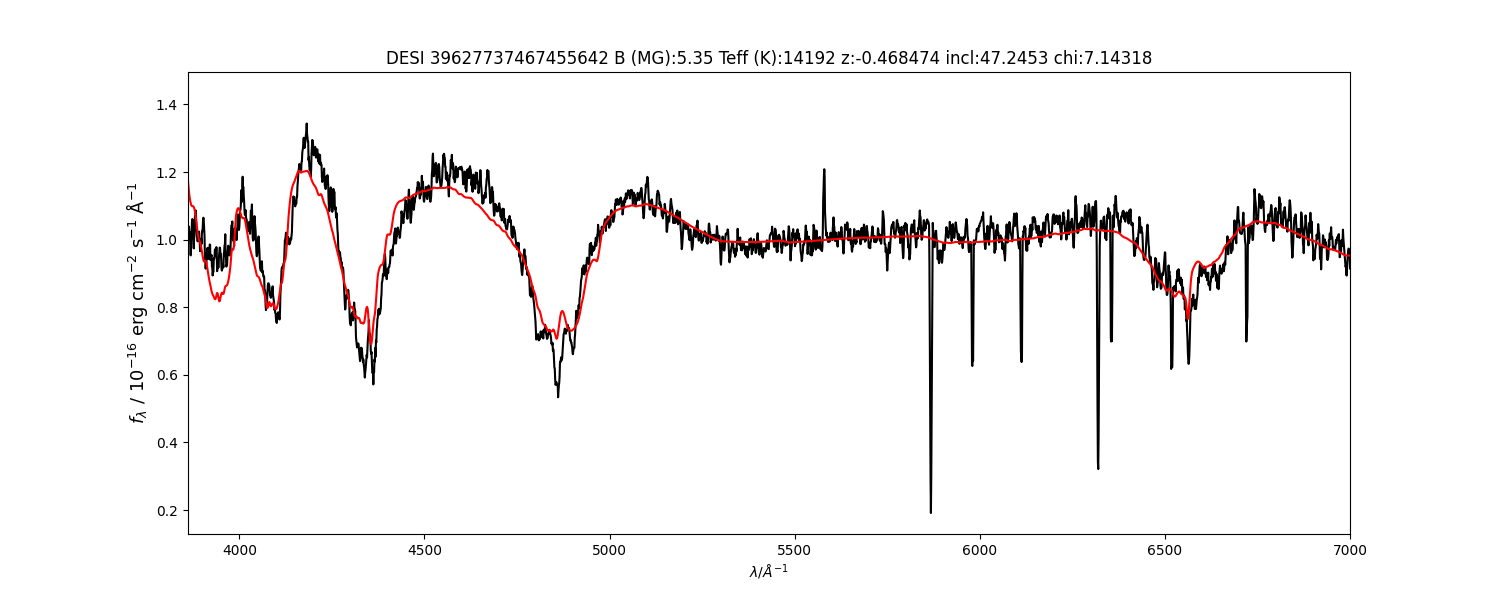}\\ 
\end{supertabular}
 \newpage \captionof{figure}{cont.}
\begin{supertabular}{c}
\includegraphics[width=0.9\linewidth]{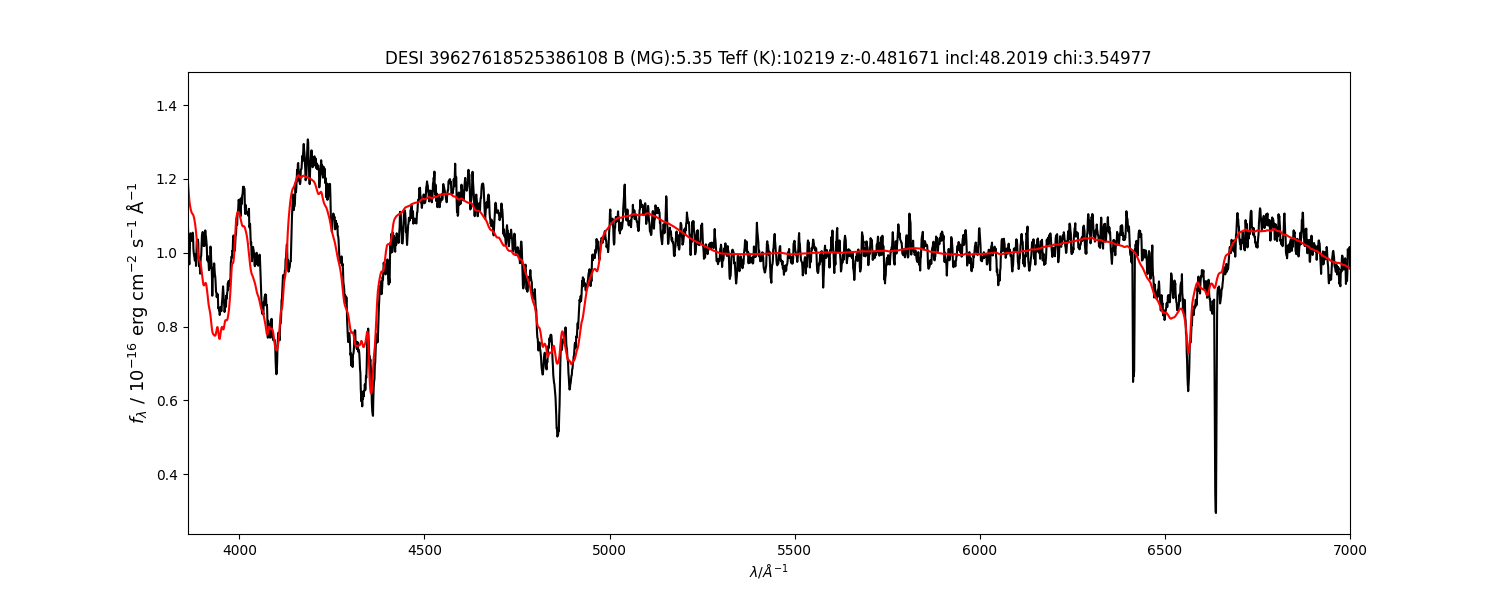}\\
\includegraphics[width=0.9\linewidth]{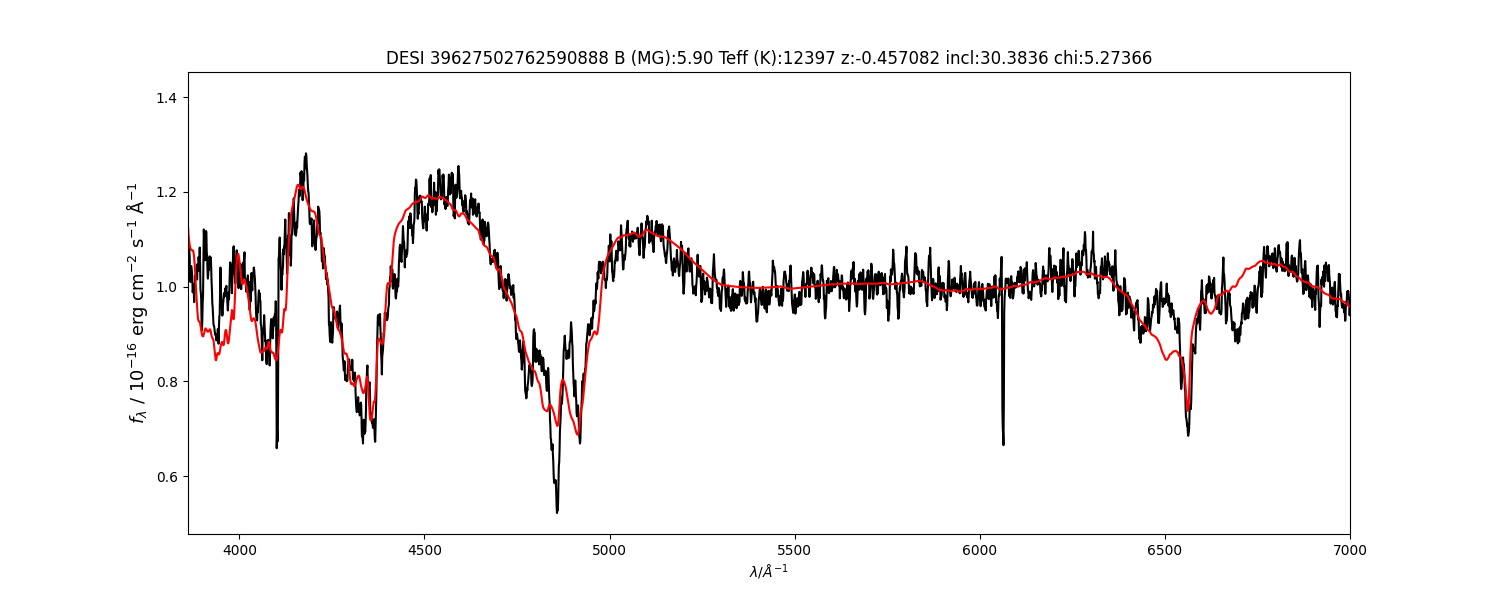}\\
\includegraphics[width=0.9\linewidth]{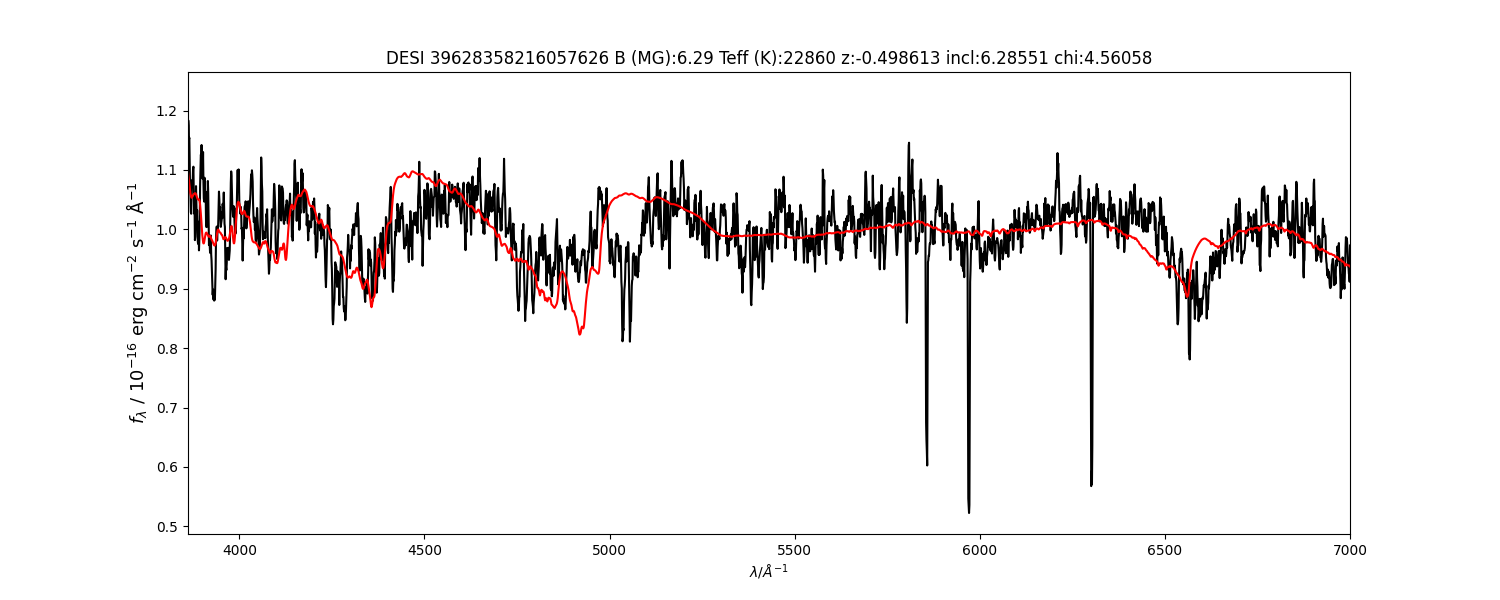}\\ 
\end{supertabular}
 \newpage \captionof{figure}{cont.}
\begin{supertabular}{c}
\includegraphics[width=0.9\linewidth]{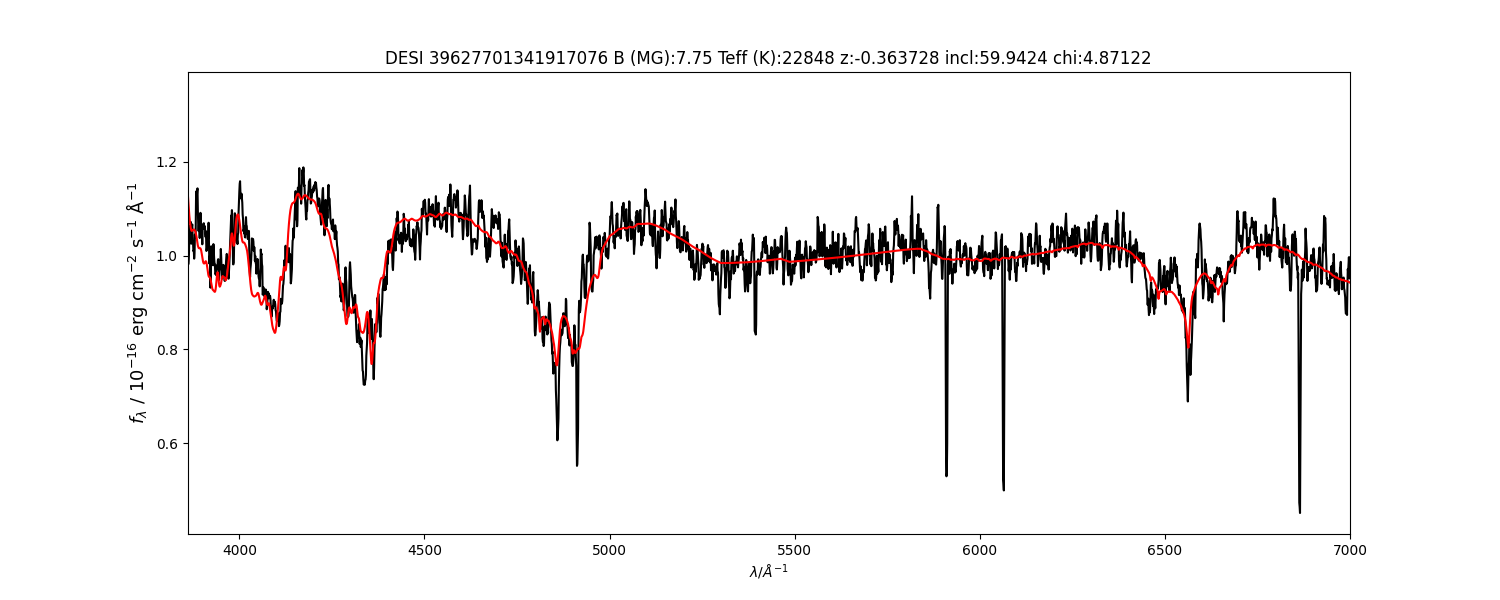}\\
\includegraphics[width=0.9\linewidth]{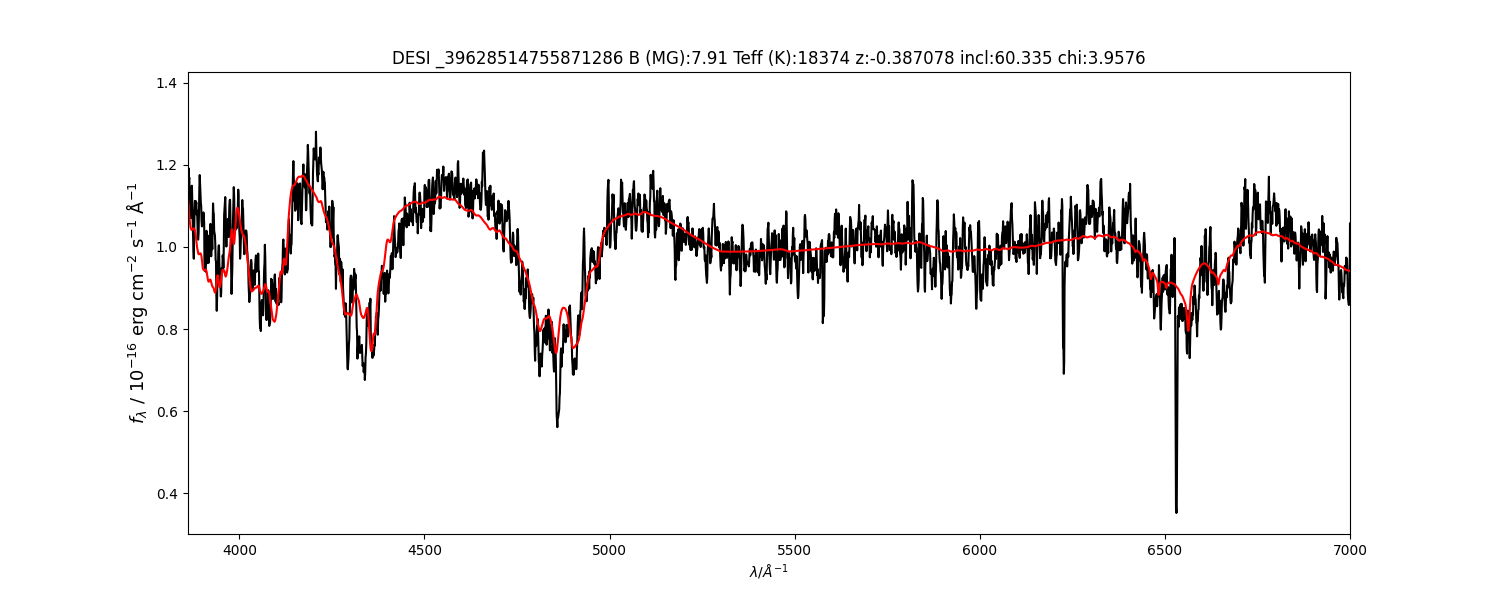}\\
\includegraphics[width=0.9\linewidth]{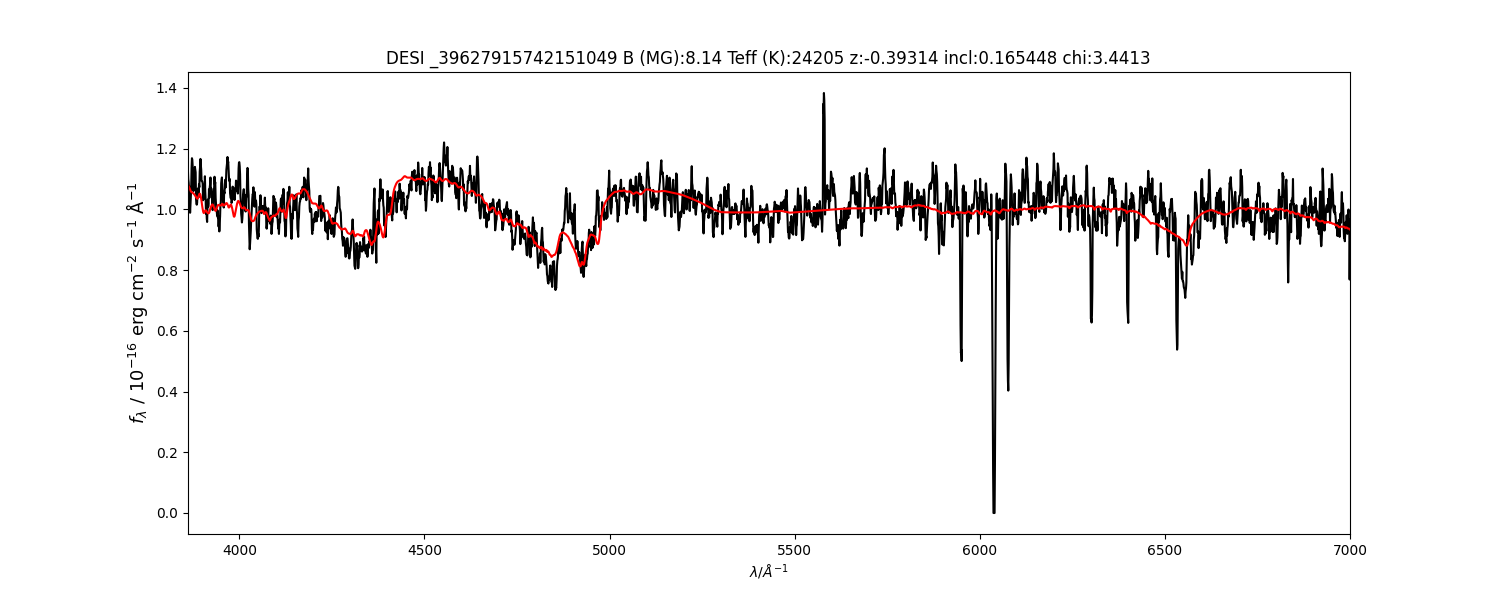}\\ 
\end{supertabular}
 \newpage \captionof{figure}{cont.}
\begin{supertabular}{c}
\includegraphics[width=0.9\linewidth]{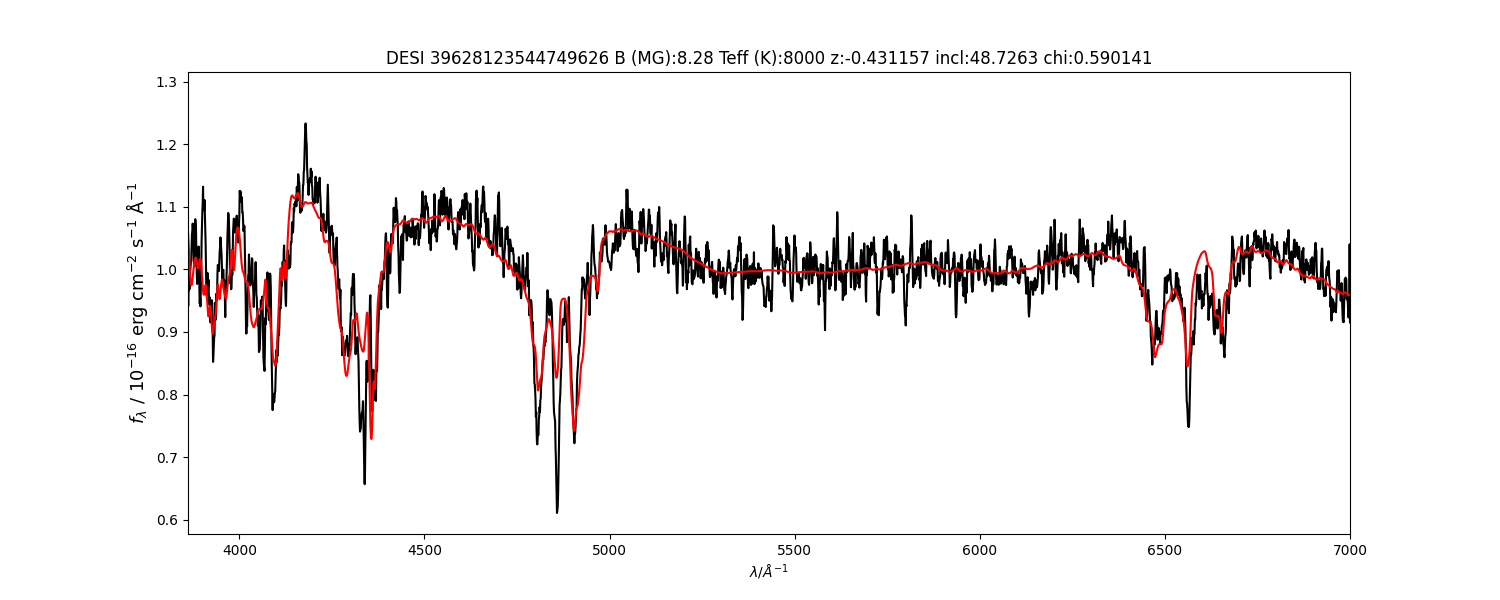}\\
\includegraphics[width=0.9\linewidth]{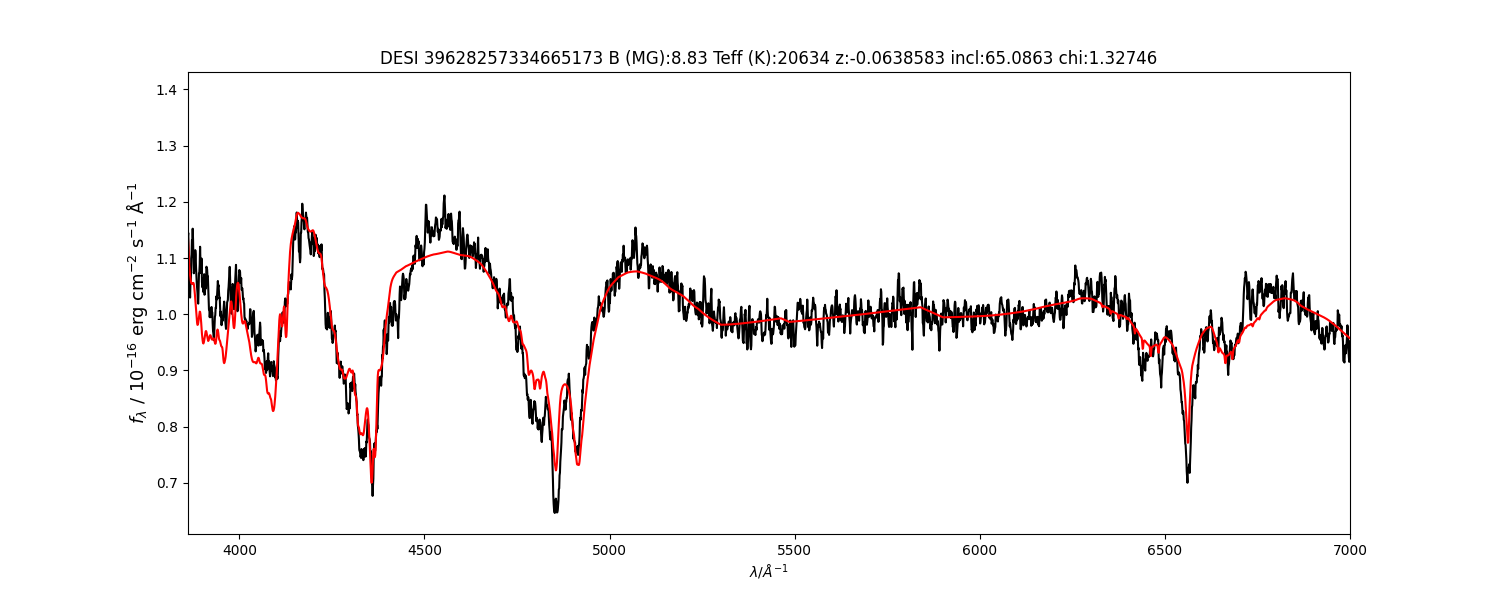}\\
\includegraphics[width=0.9\linewidth]{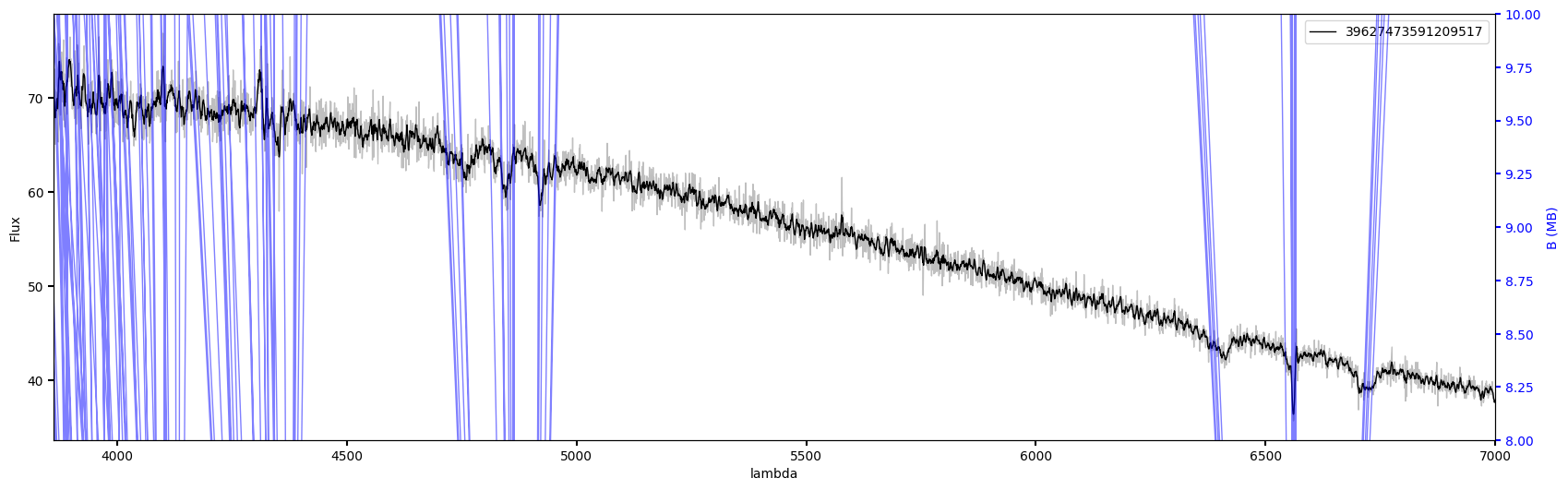}\\ 
\end{supertabular}
 \newpage \captionof{figure}{cont.}
\begin{supertabular}{c}
\includegraphics[width=0.9\linewidth]{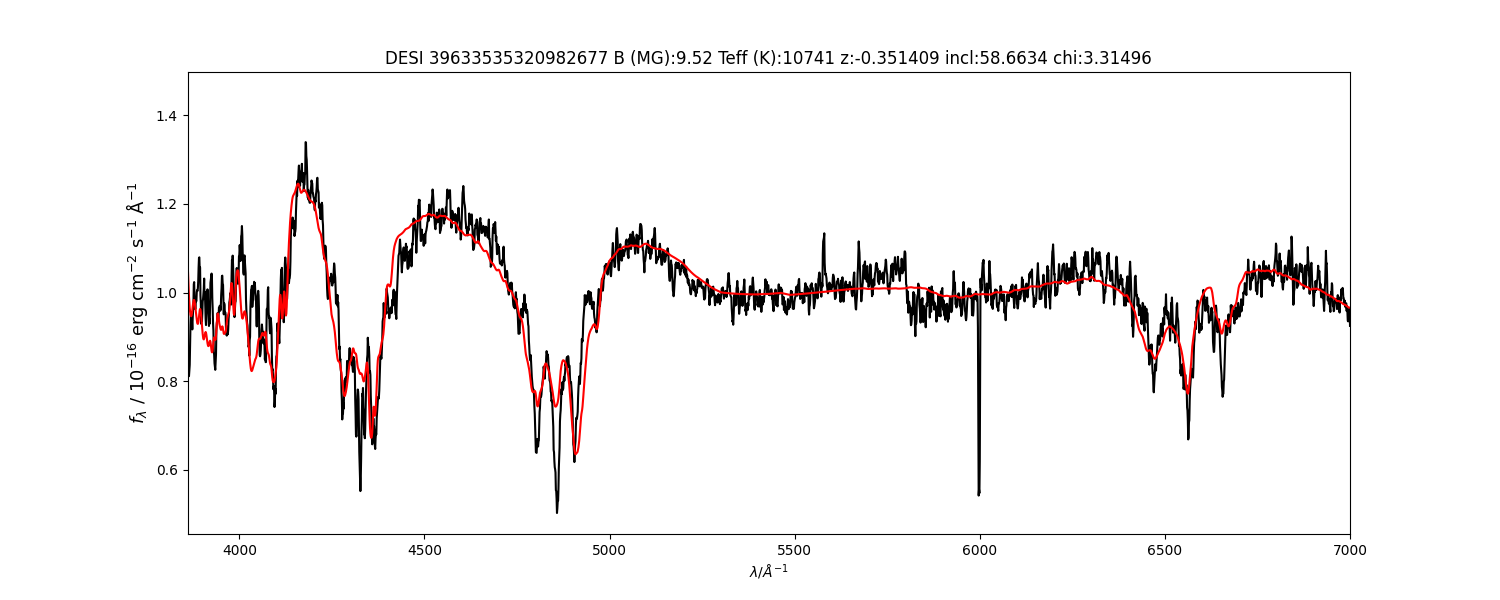}\\
\includegraphics[width=0.9\linewidth]{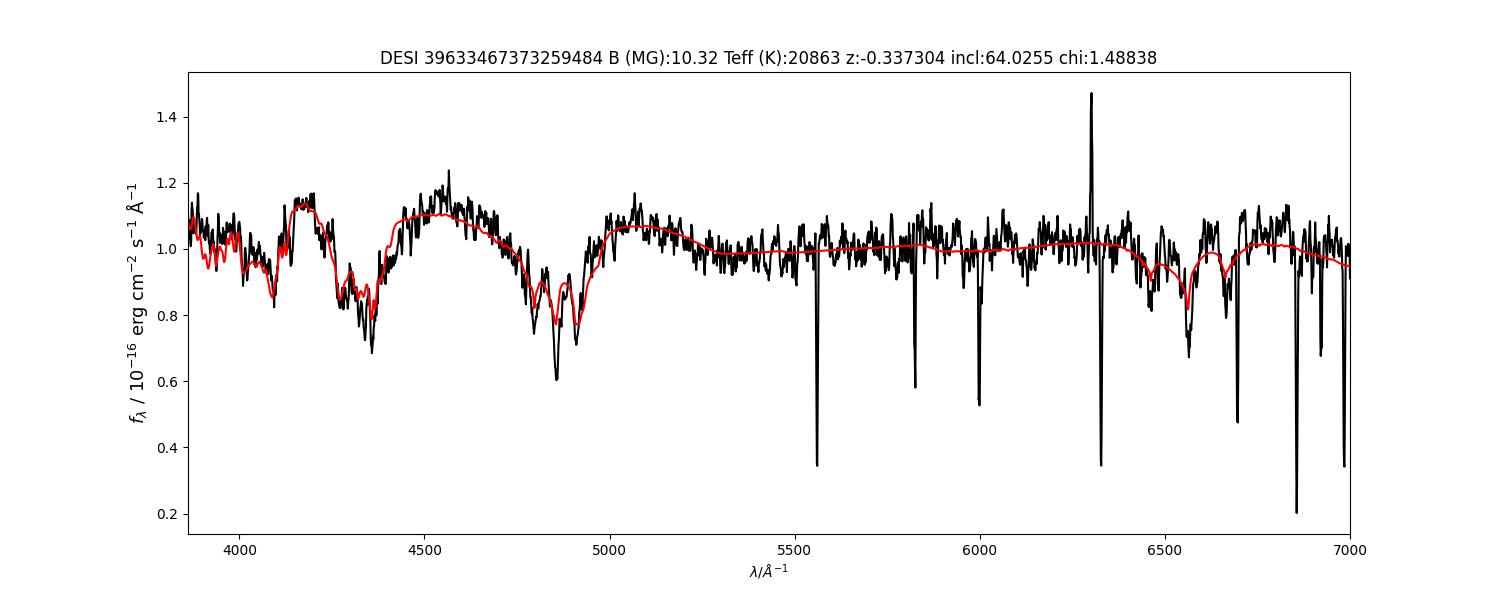}\\
\includegraphics[width=0.9\linewidth]{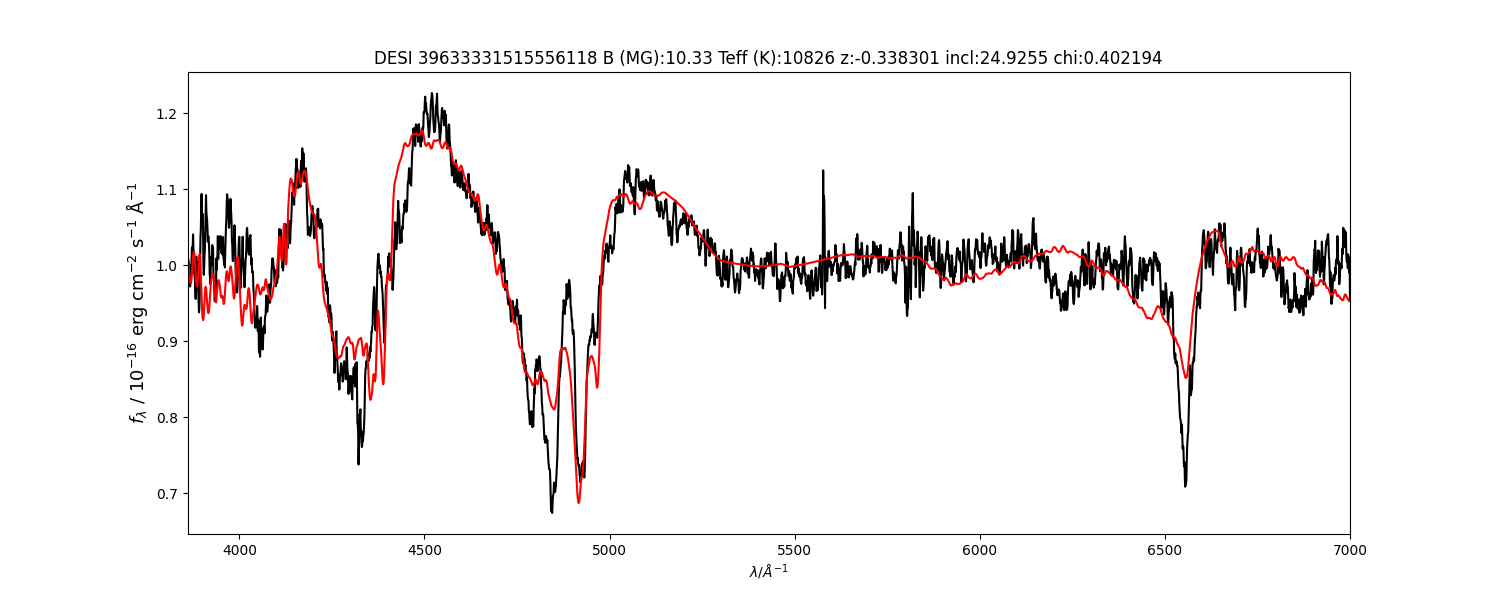}\\ 
\end{supertabular}
 \newpage \captionof{figure}{cont.}
\begin{supertabular}{c}
\includegraphics[width=0.9\linewidth]{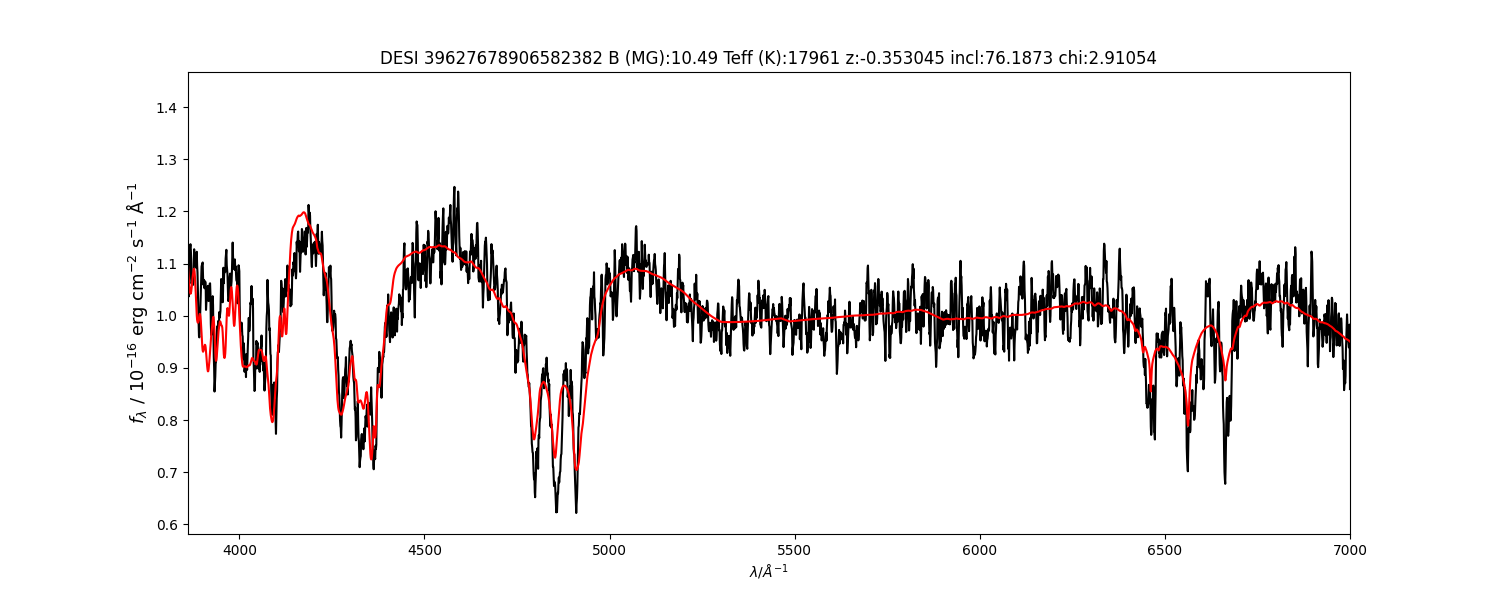}\\
\includegraphics[width=0.9\linewidth]{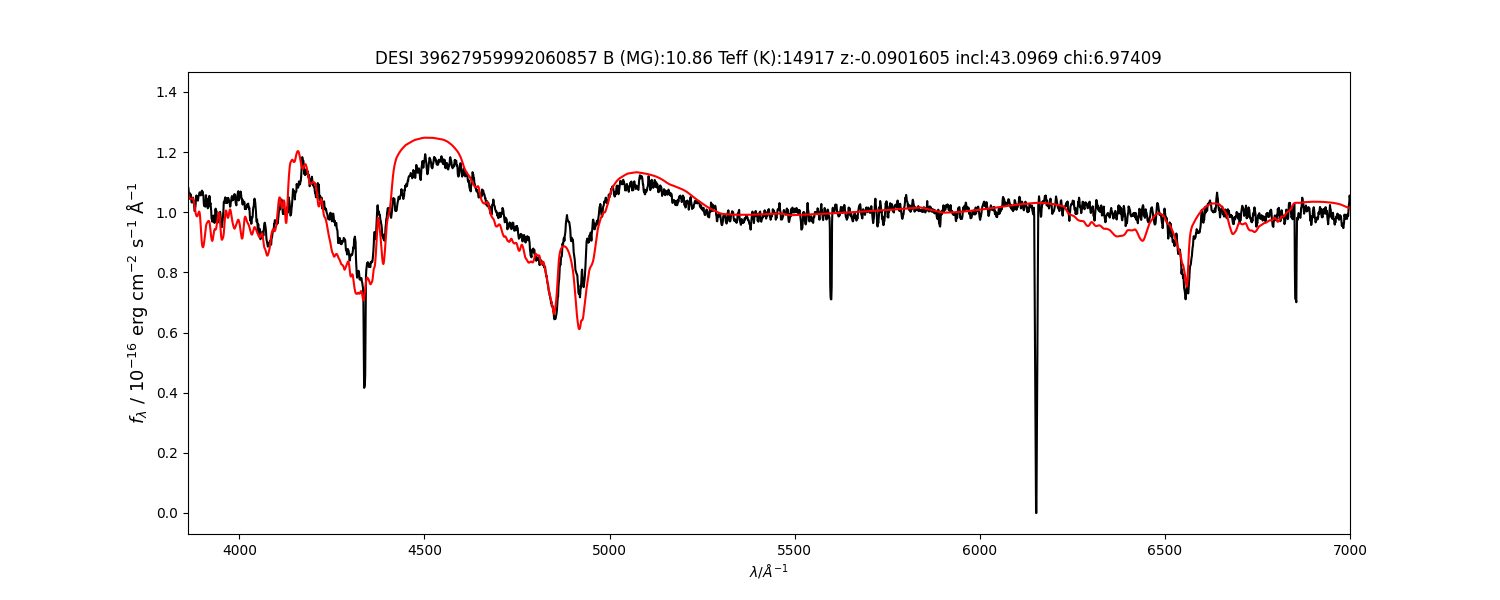}\\
\includegraphics[width=0.9\linewidth]{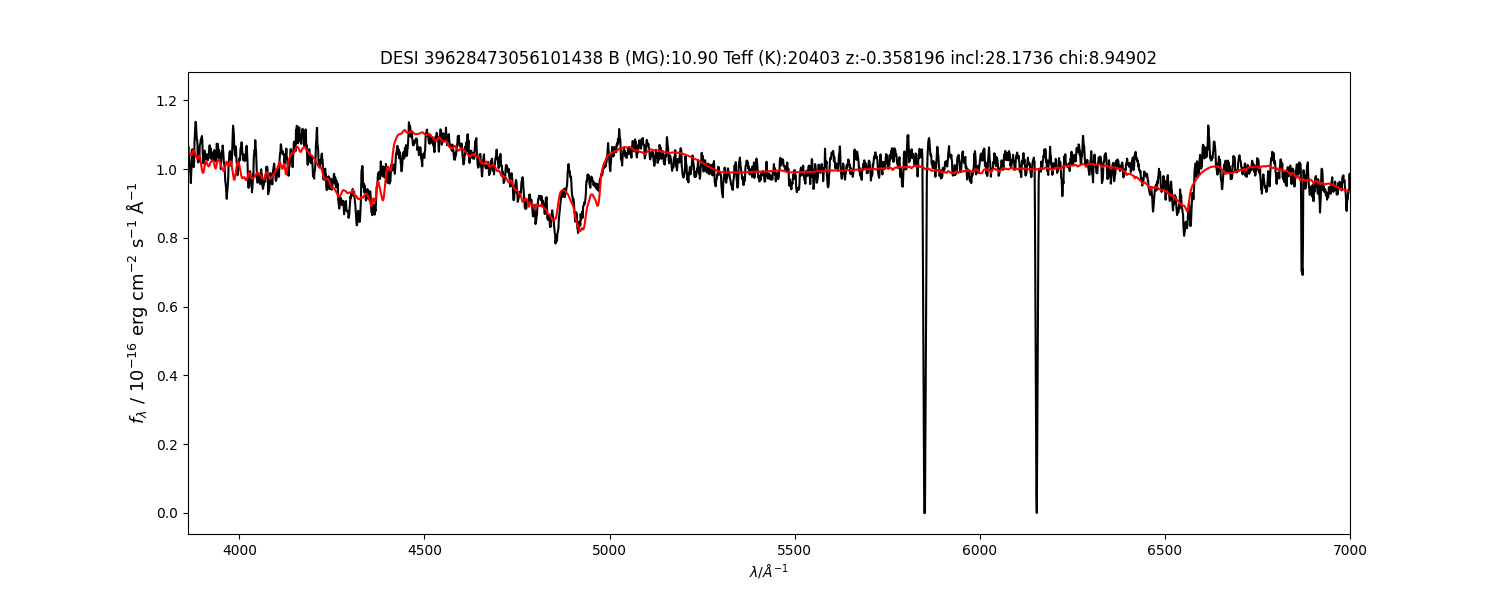}\\ 
\end{supertabular}
 \newpage \captionof{figure}{cont.}
\begin{supertabular}{c}
\includegraphics[width=0.9\linewidth]{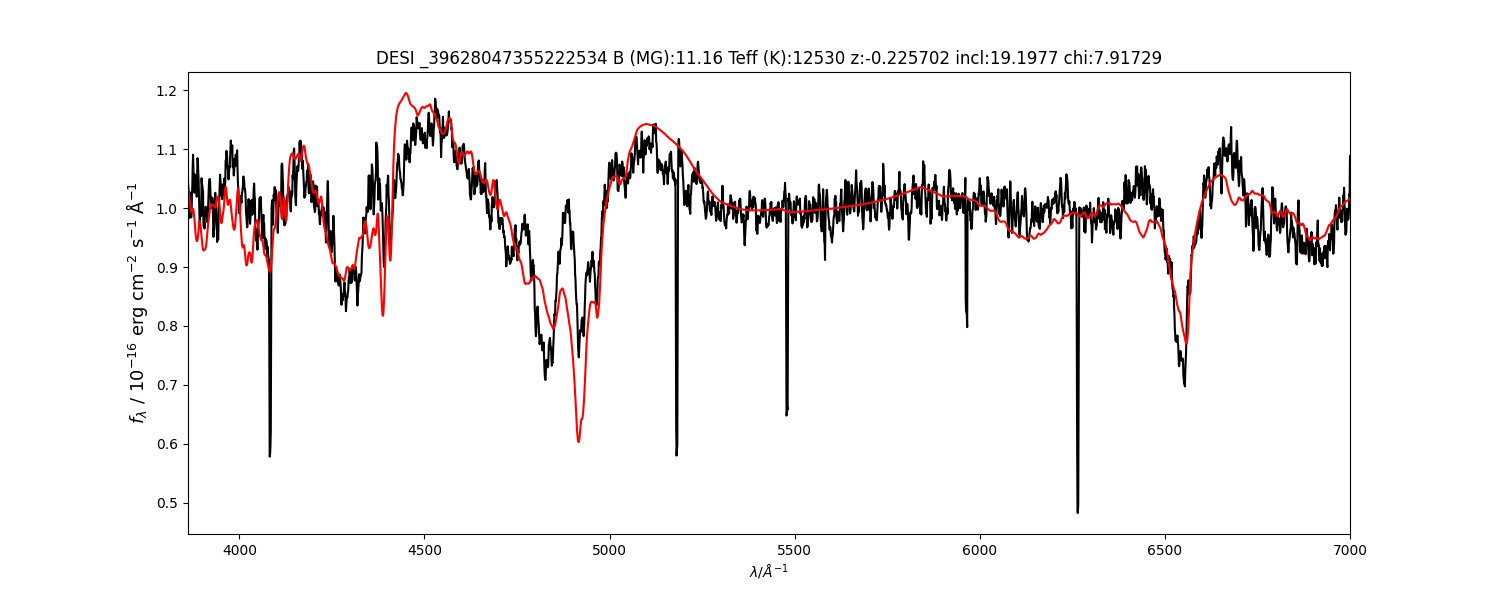}\\
\includegraphics[width=0.9\linewidth]{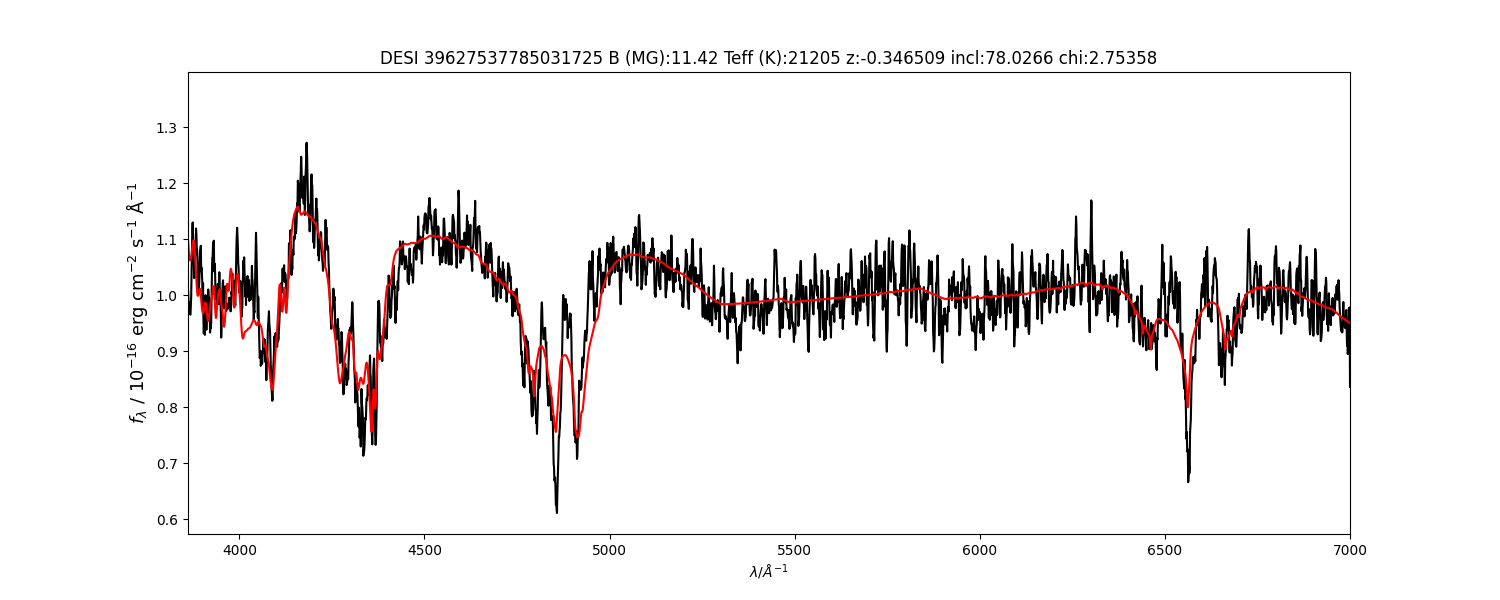}\\
\includegraphics[width=0.9\linewidth]{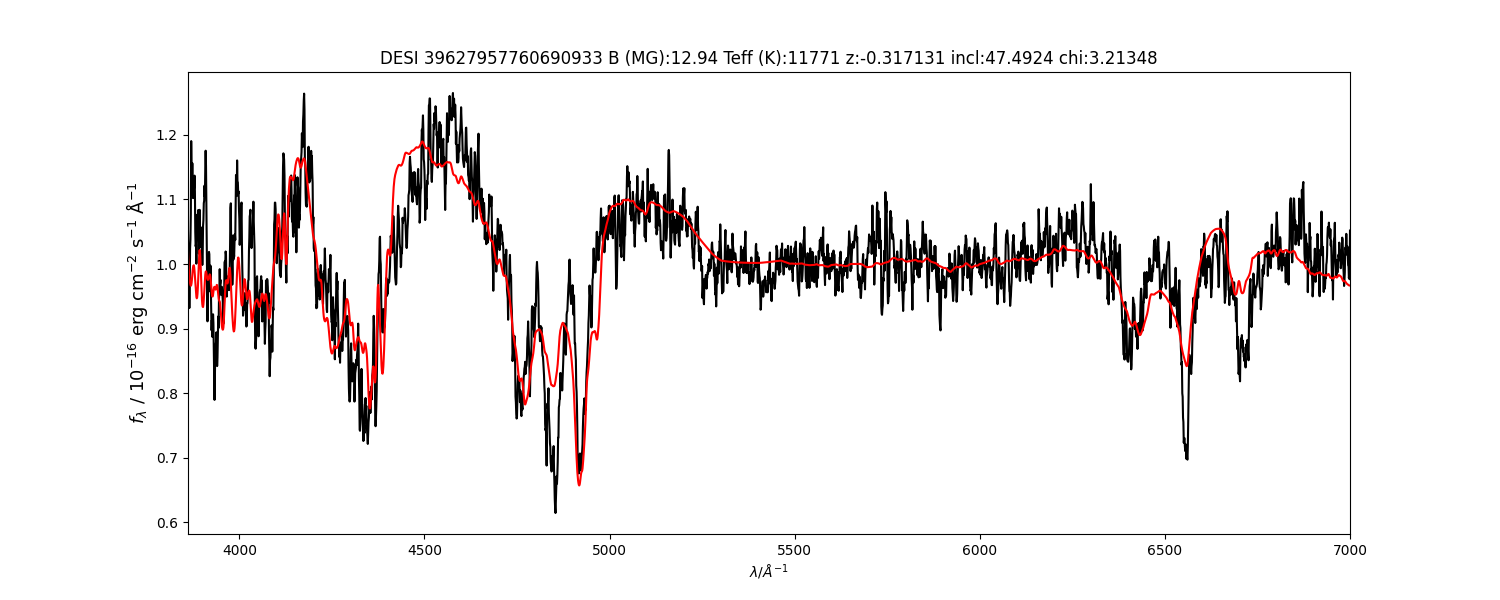}\\ 
\end{supertabular}
 \newpage \captionof{figure}{cont.}
\begin{supertabular}{c}
\includegraphics[width=0.9\linewidth]{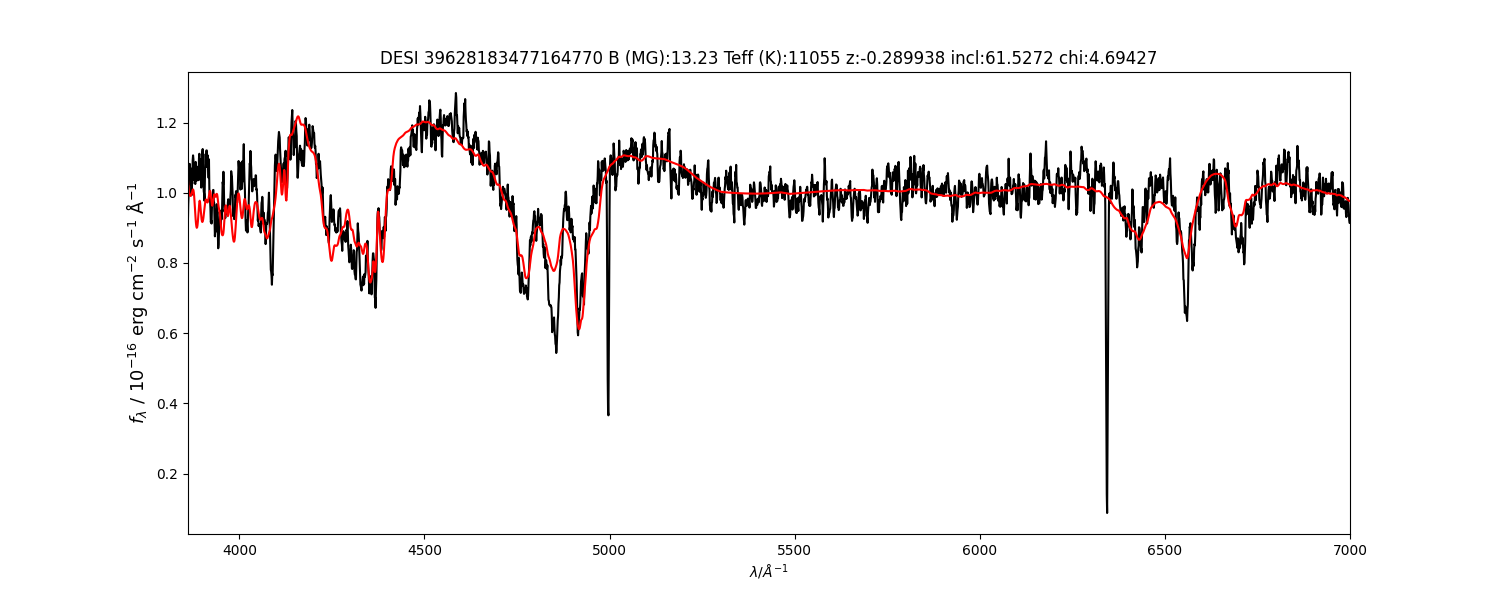}\\
\includegraphics[width=0.9\linewidth]{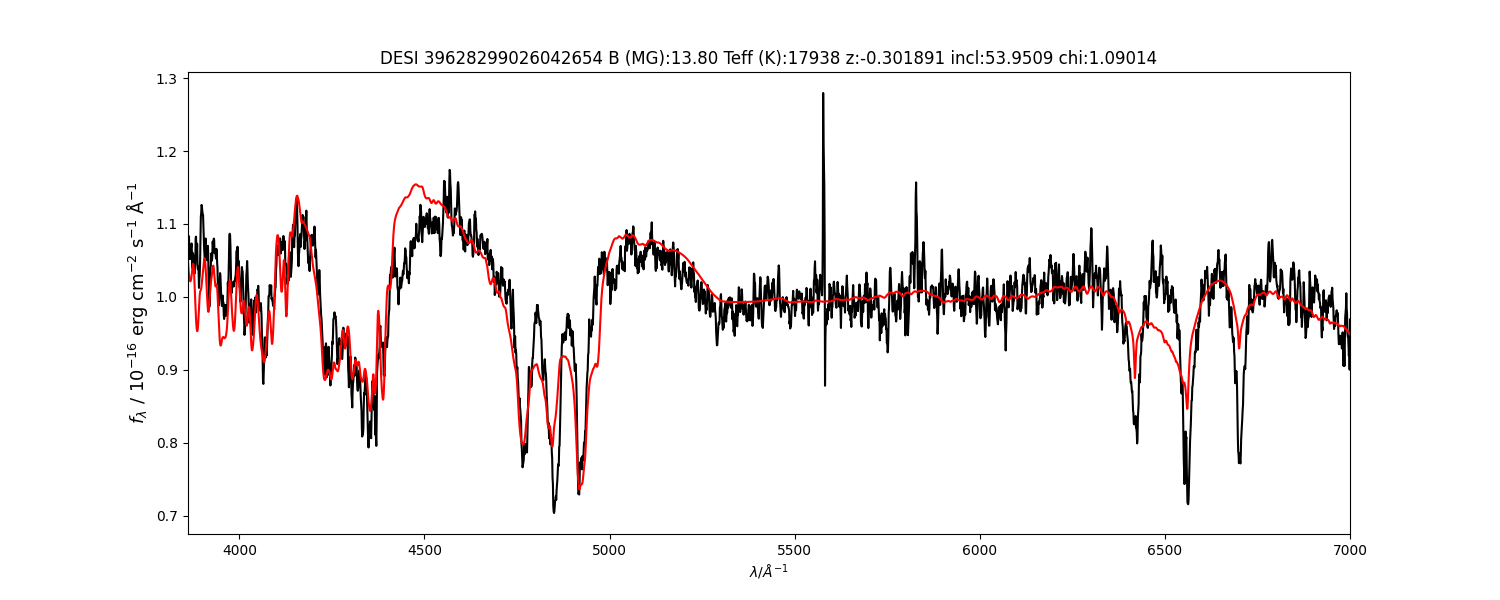}\\
\includegraphics[width=0.9\linewidth]{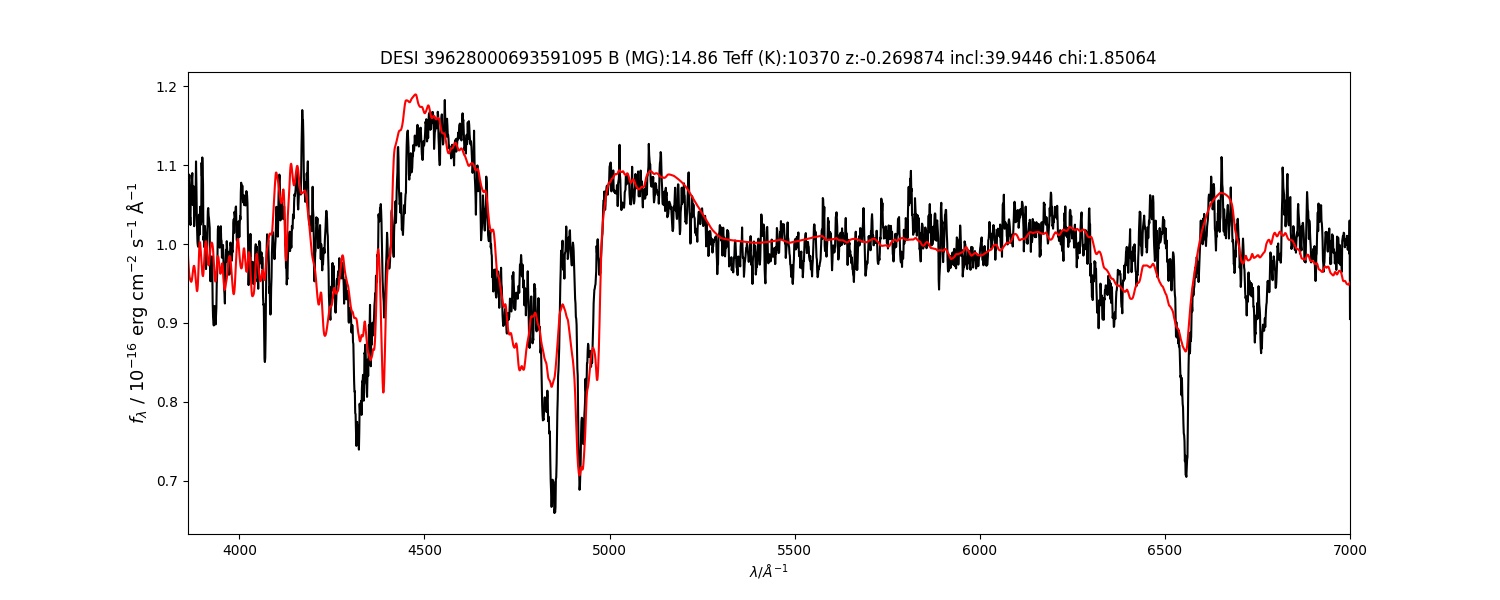}\\ 
\end{supertabular}
 \newpage \captionof{figure}{cont.}
\begin{supertabular}{c}
\includegraphics[width=0.9\linewidth]{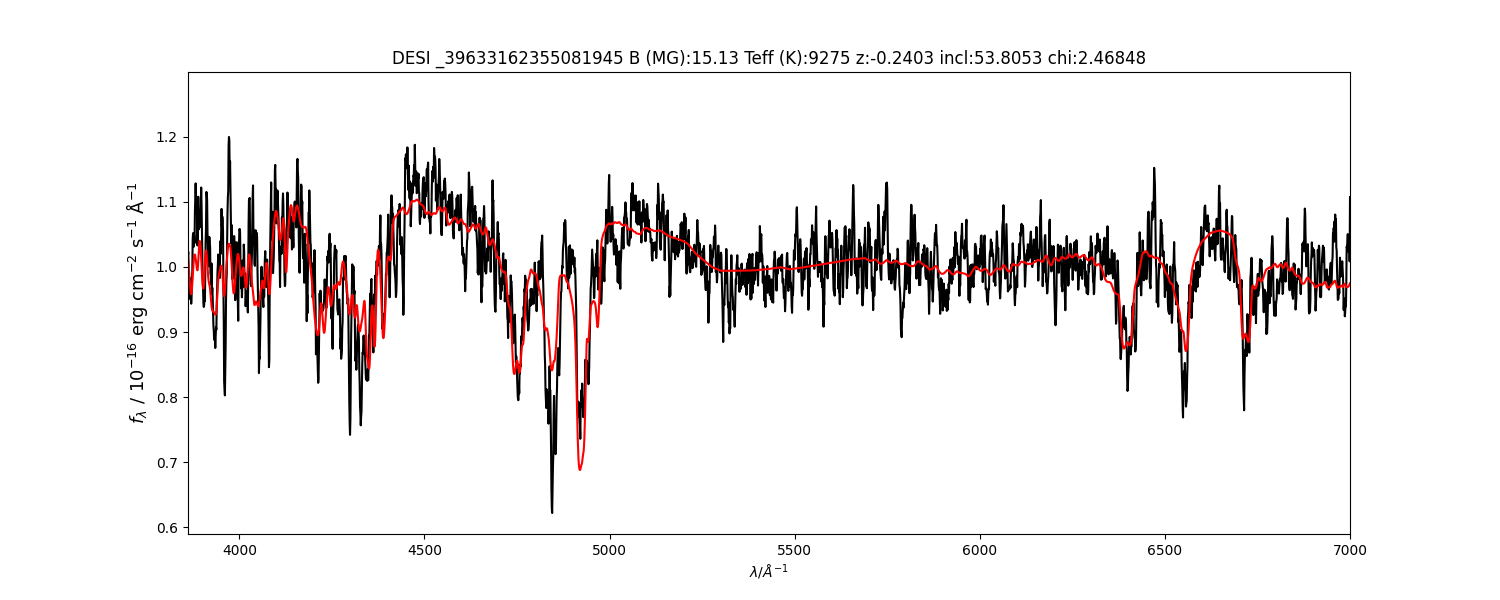}\\
\includegraphics[width=0.9\linewidth]{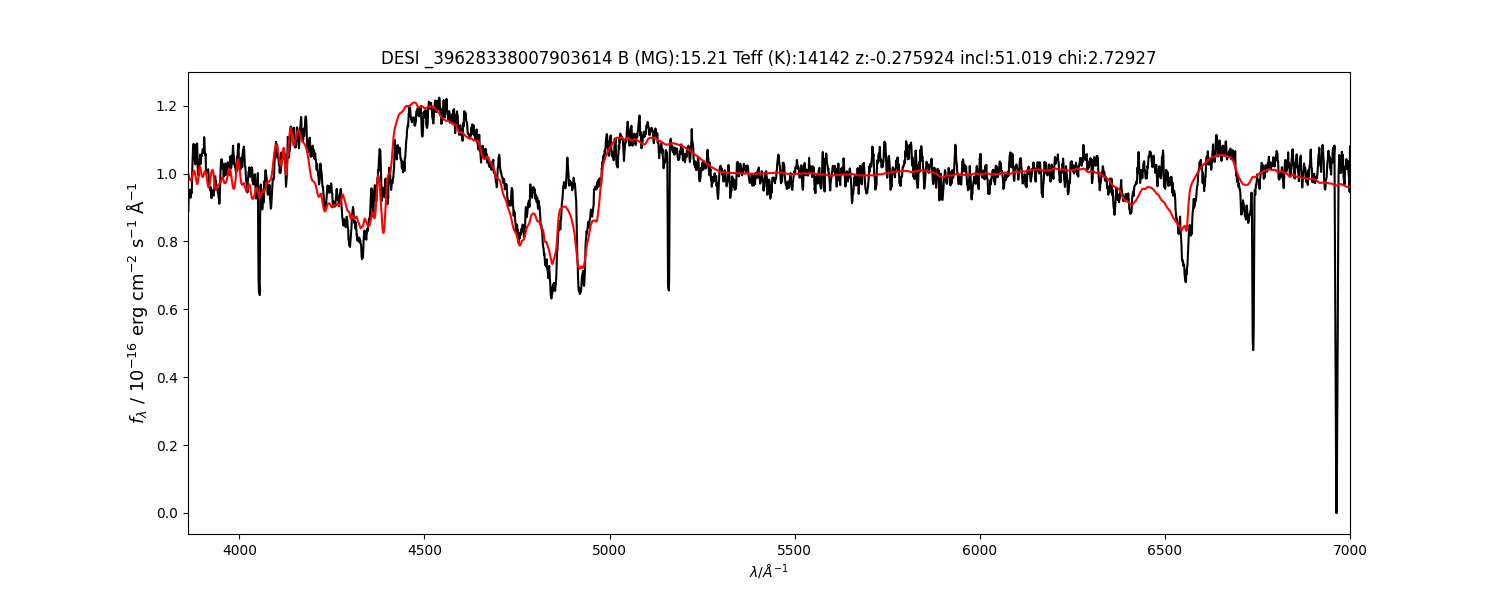}\\
\includegraphics[width=0.9\linewidth]{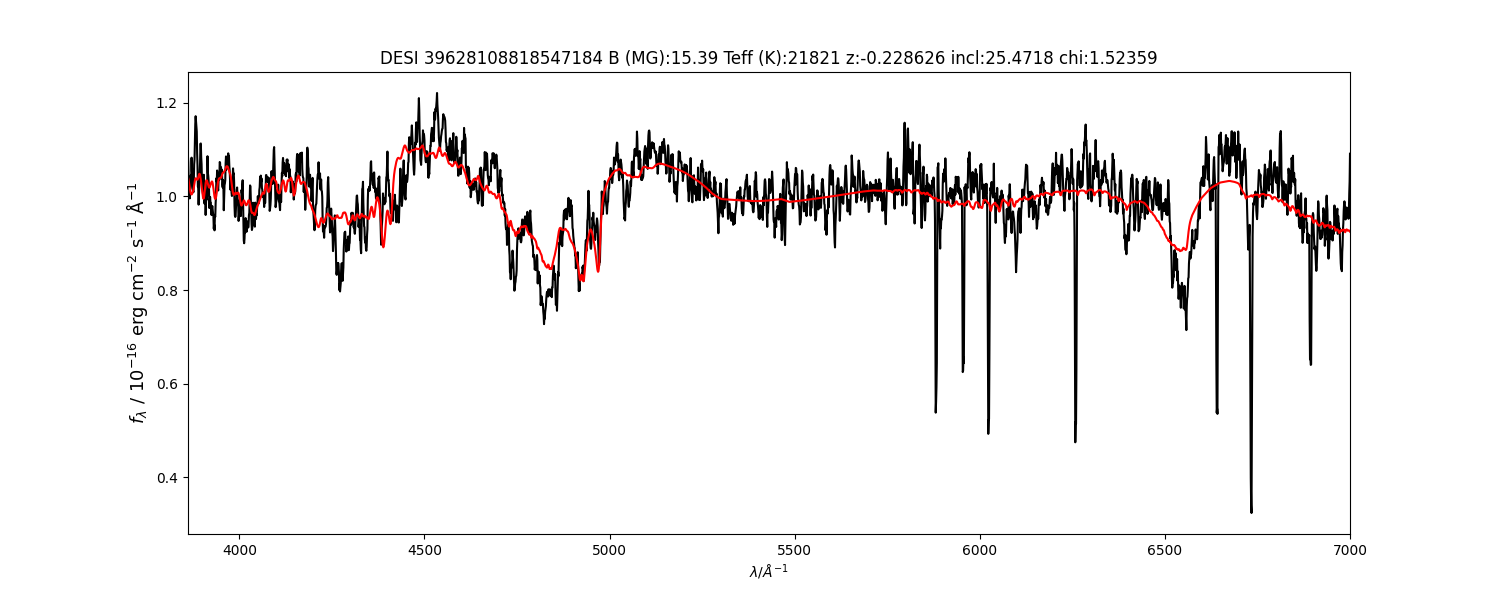}\\ 
\end{supertabular}
 \newpage \captionof{figure}{cont.}
\begin{supertabular}{c}
\includegraphics[width=0.9\linewidth]{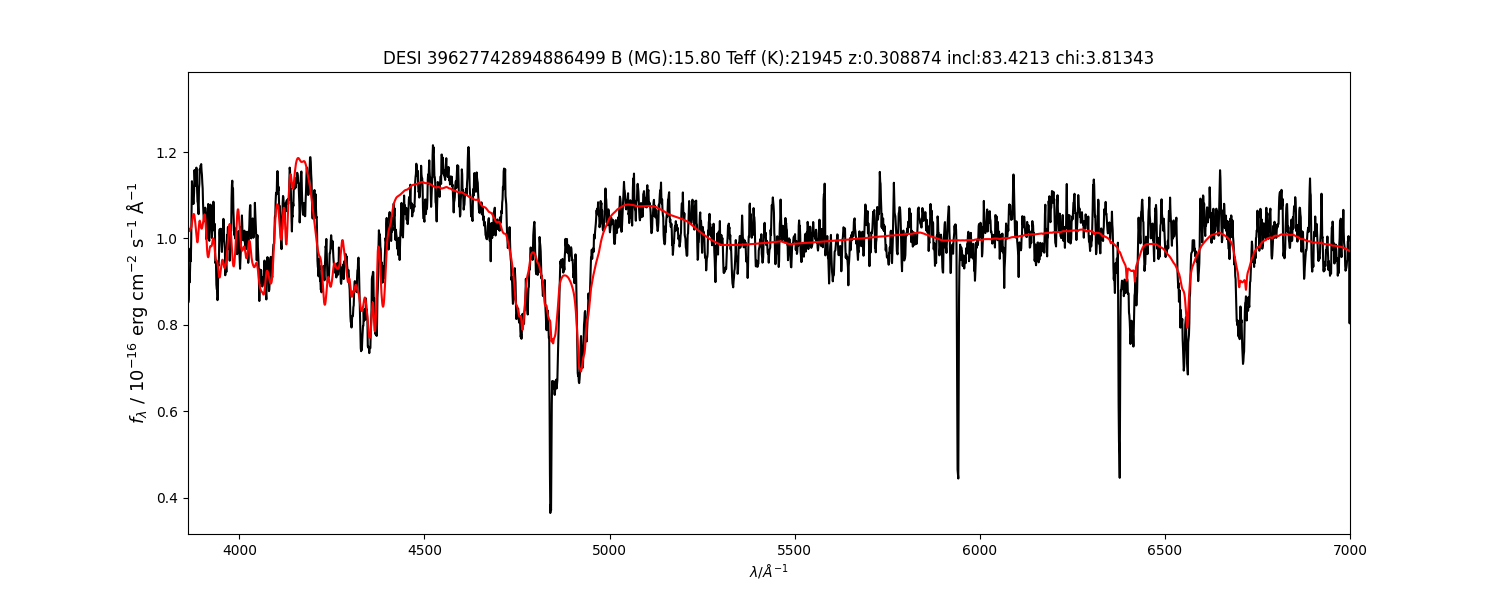}\\
\includegraphics[width=0.9\linewidth]{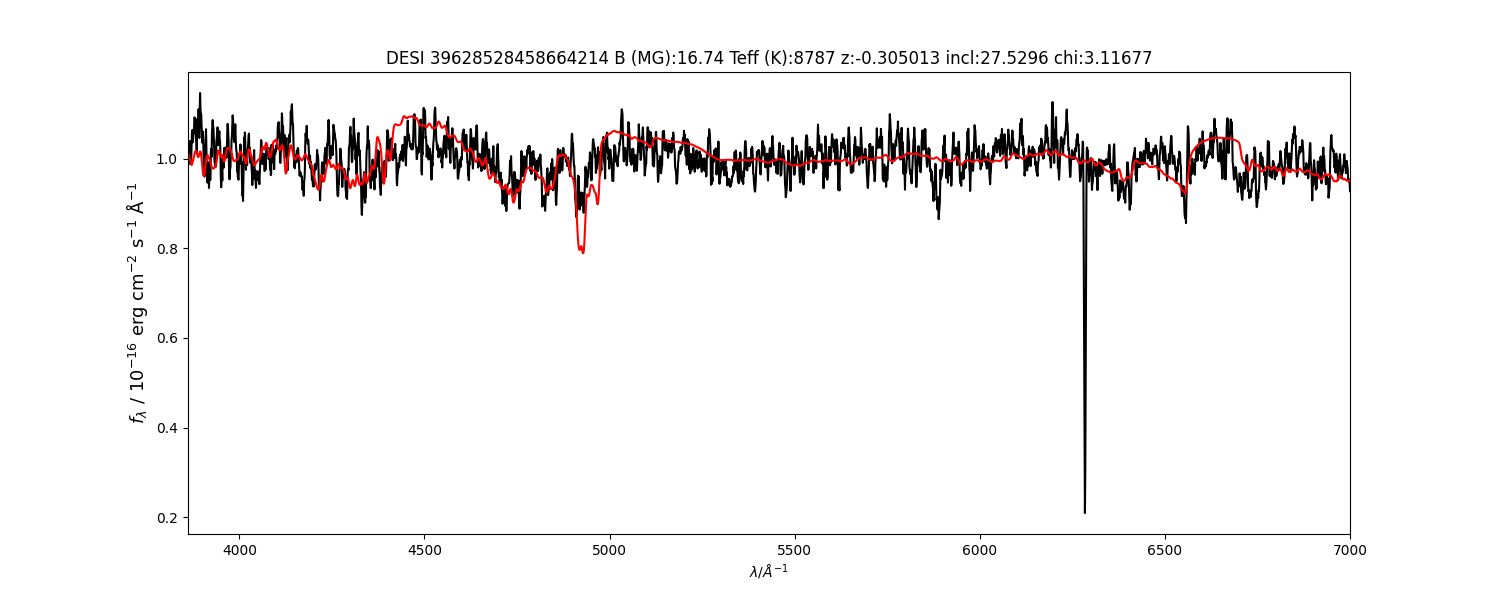}\\
\includegraphics[width=0.9\linewidth]{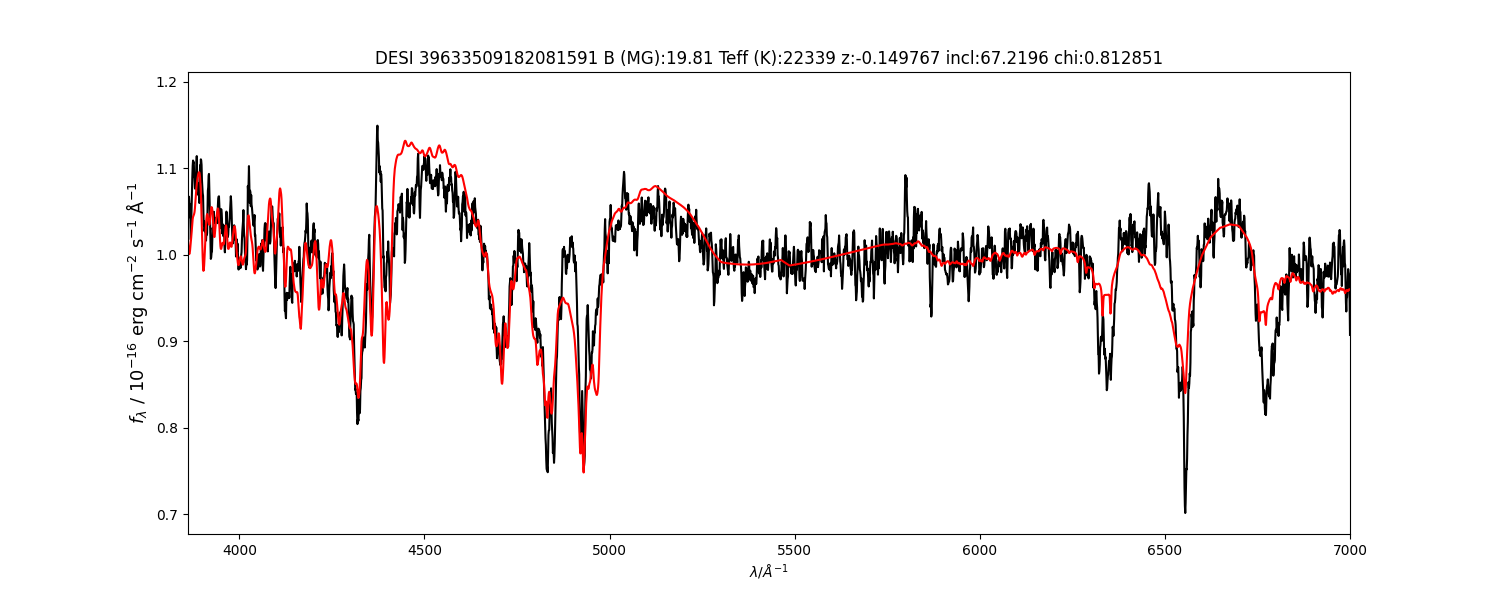}\\ 
\end{supertabular}
 \newpage \captionof{figure}{cont.}
\begin{supertabular}{c}
\includegraphics[width=0.9\linewidth]{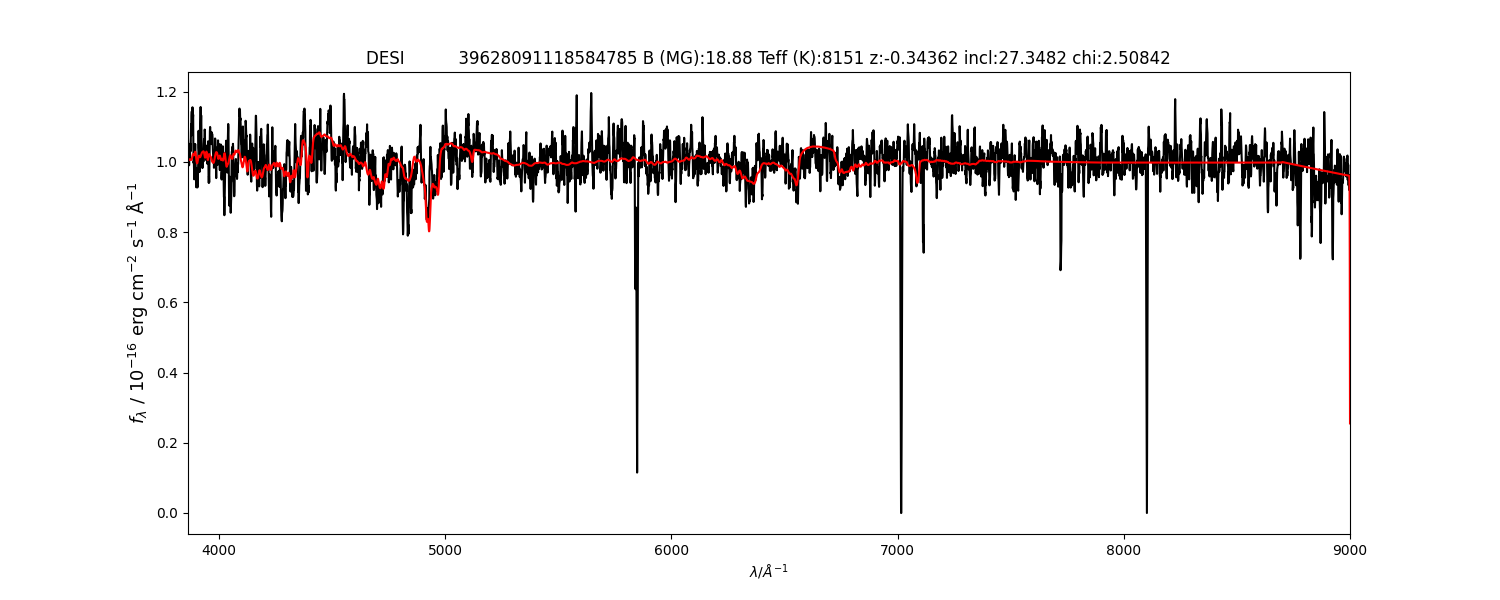}\\
\includegraphics[width=0.9\linewidth]{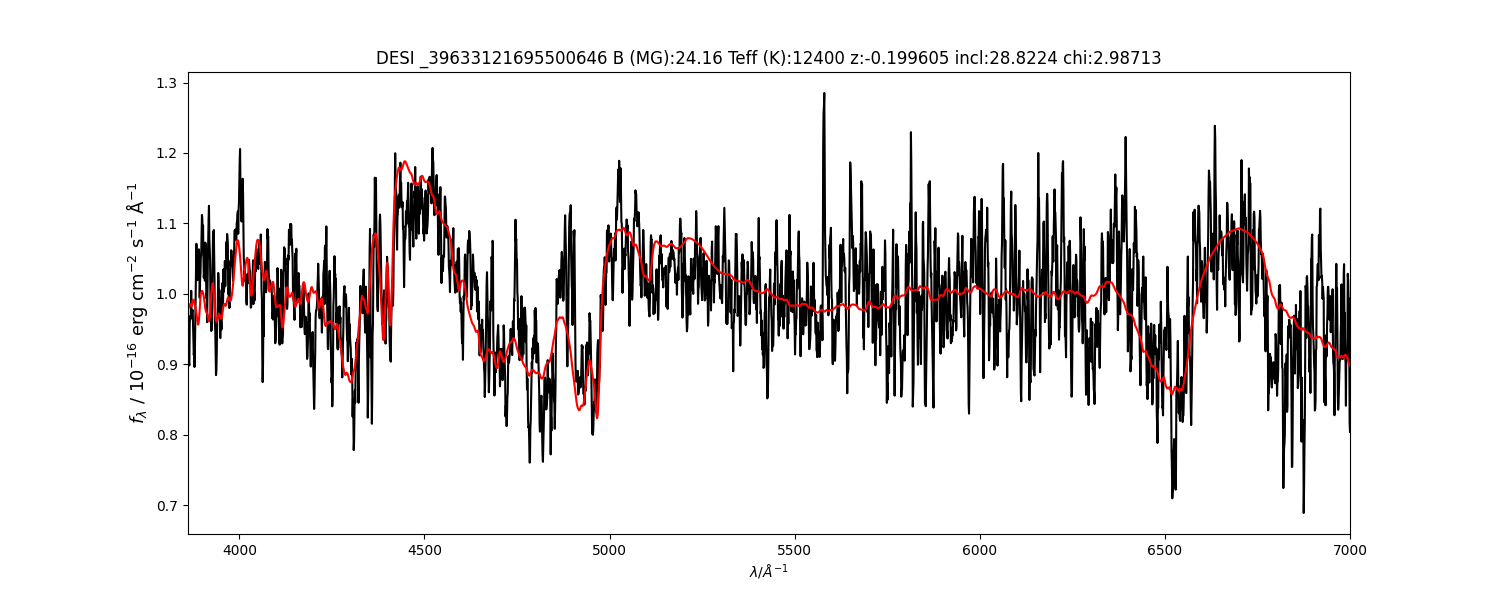}\\
\includegraphics[width=0.9\linewidth]{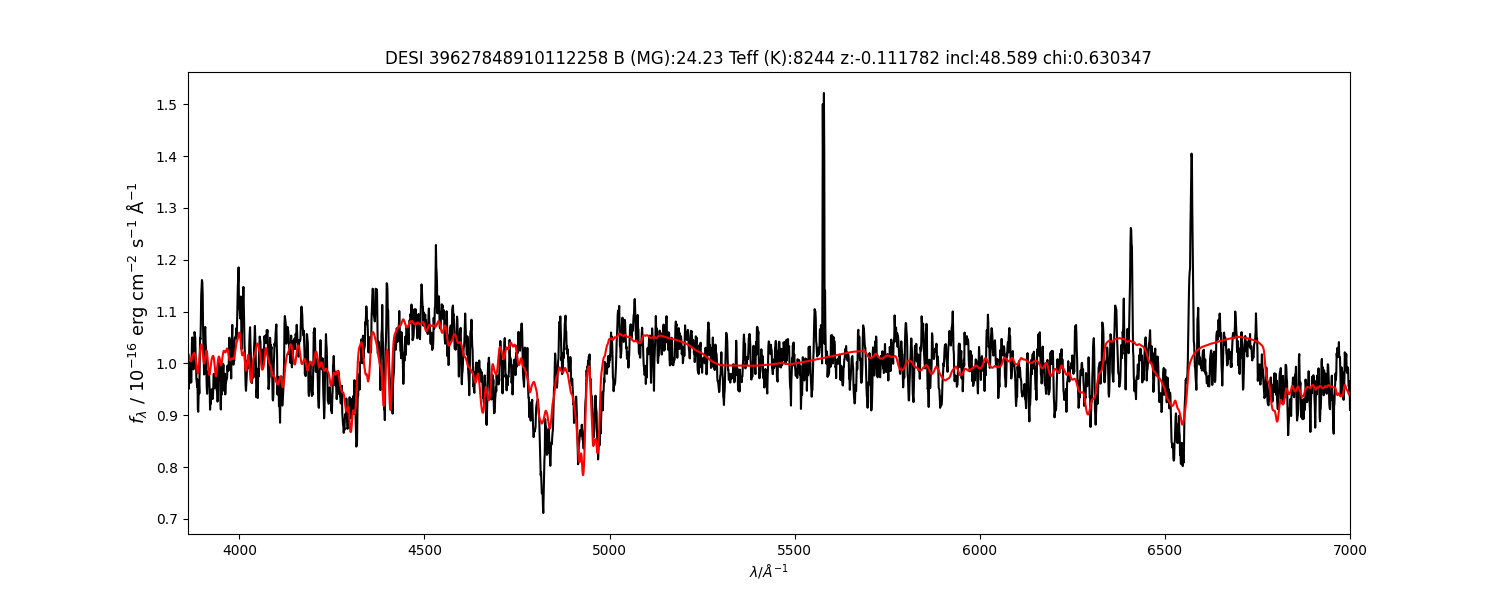}\\ 
\end{supertabular}
 \newpage \captionof{figure}{cont.}
\begin{supertabular}{c}
\includegraphics[width=0.9\linewidth]{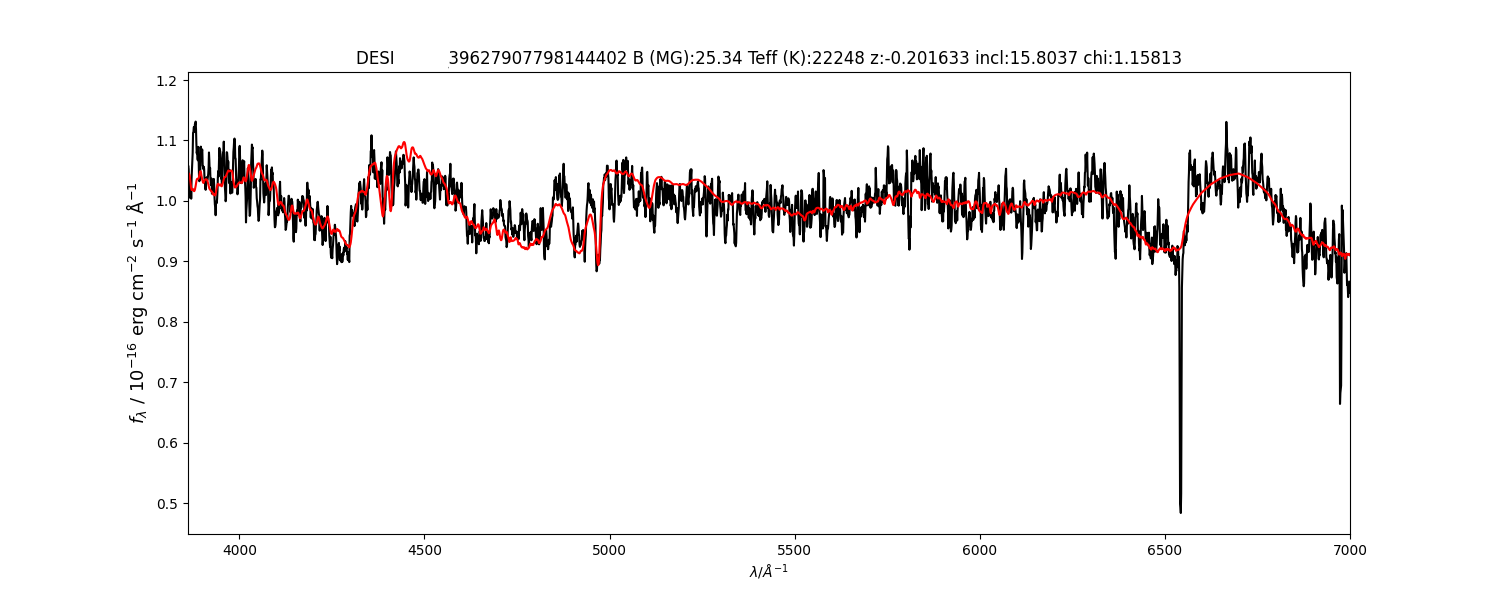}\\
\includegraphics[width=0.9\linewidth]{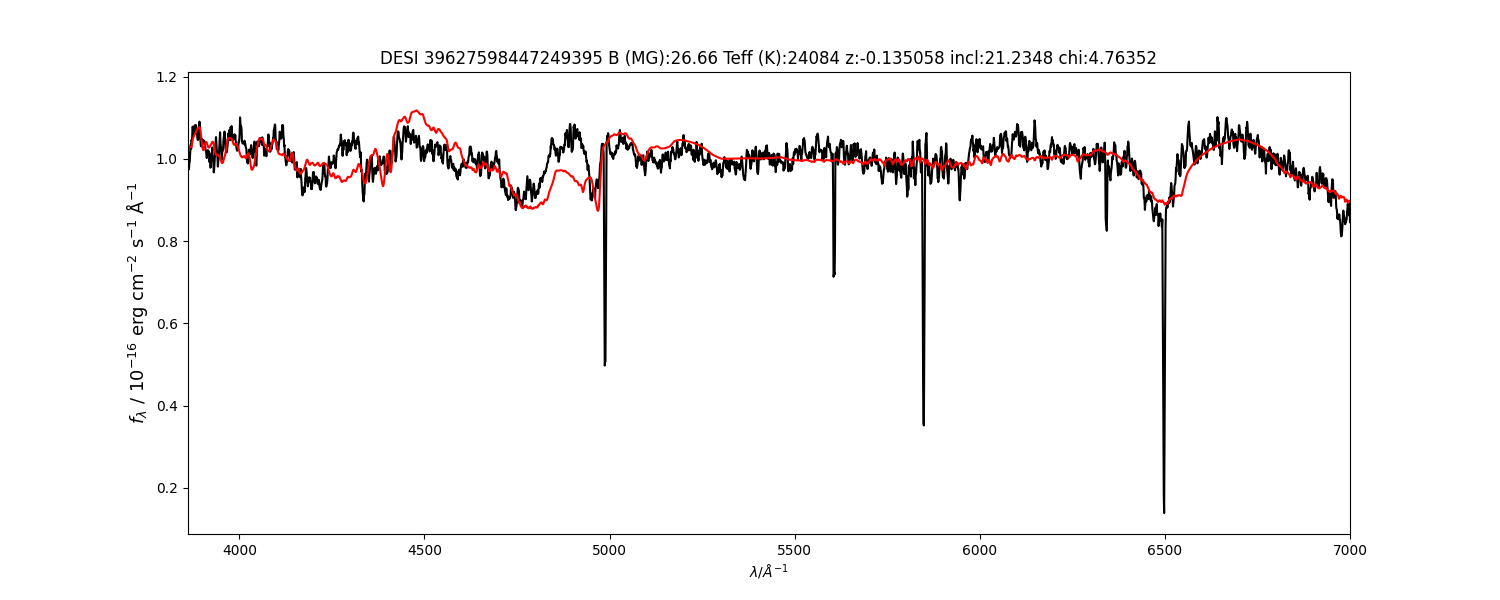}\\
\includegraphics[width=0.9\linewidth]{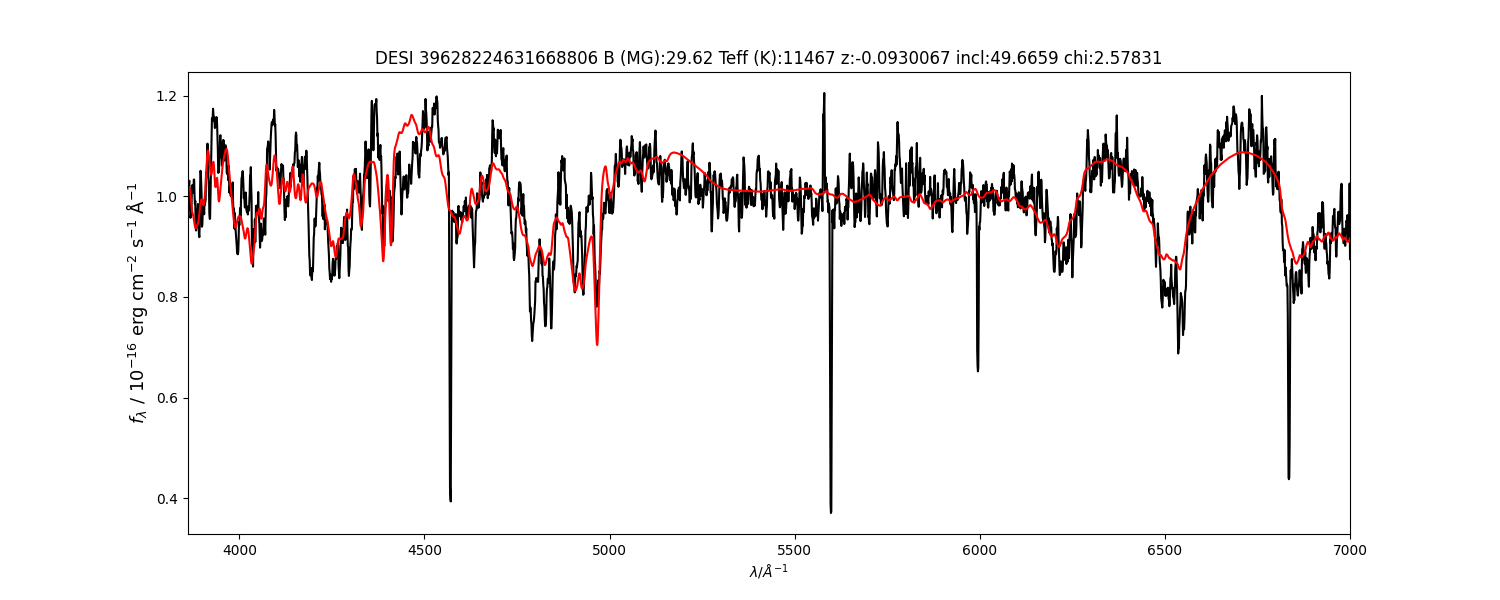}\\ 
\end{supertabular}
 \newpage \captionof{figure}{cont.}
\begin{supertabular}{c}
\includegraphics[width=0.9\linewidth]{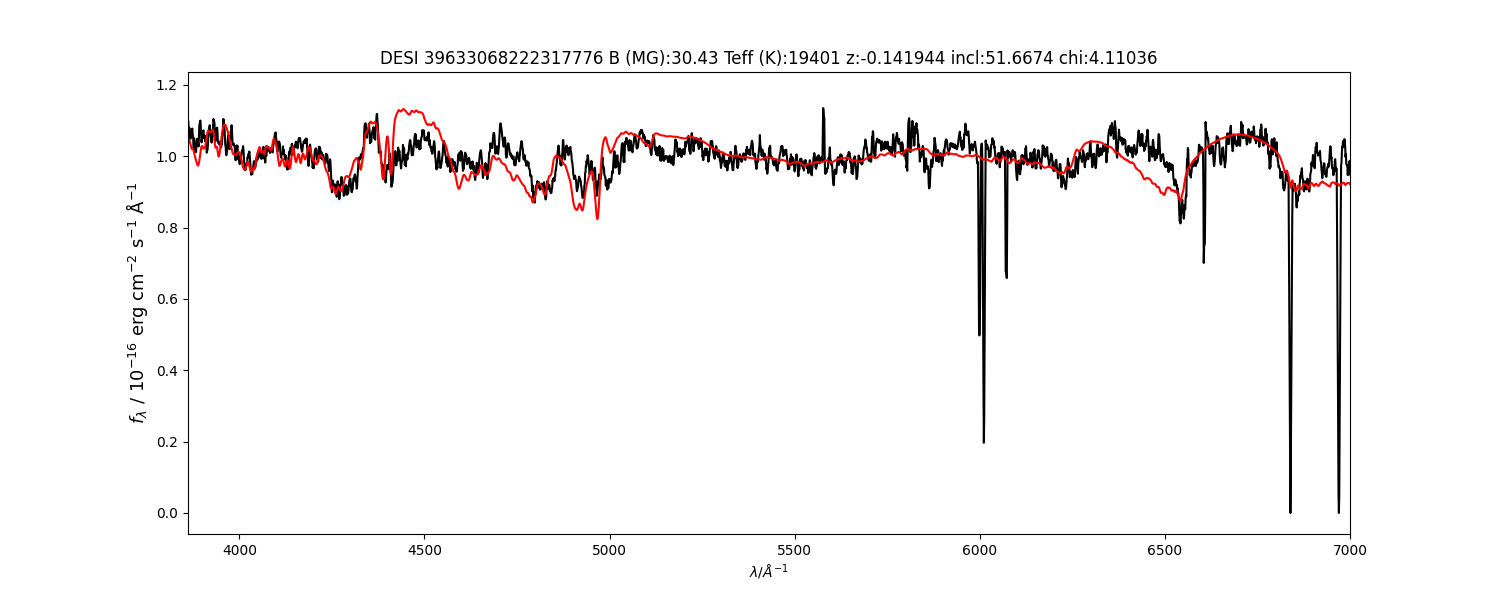}\\
\includegraphics[width=0.9\linewidth]{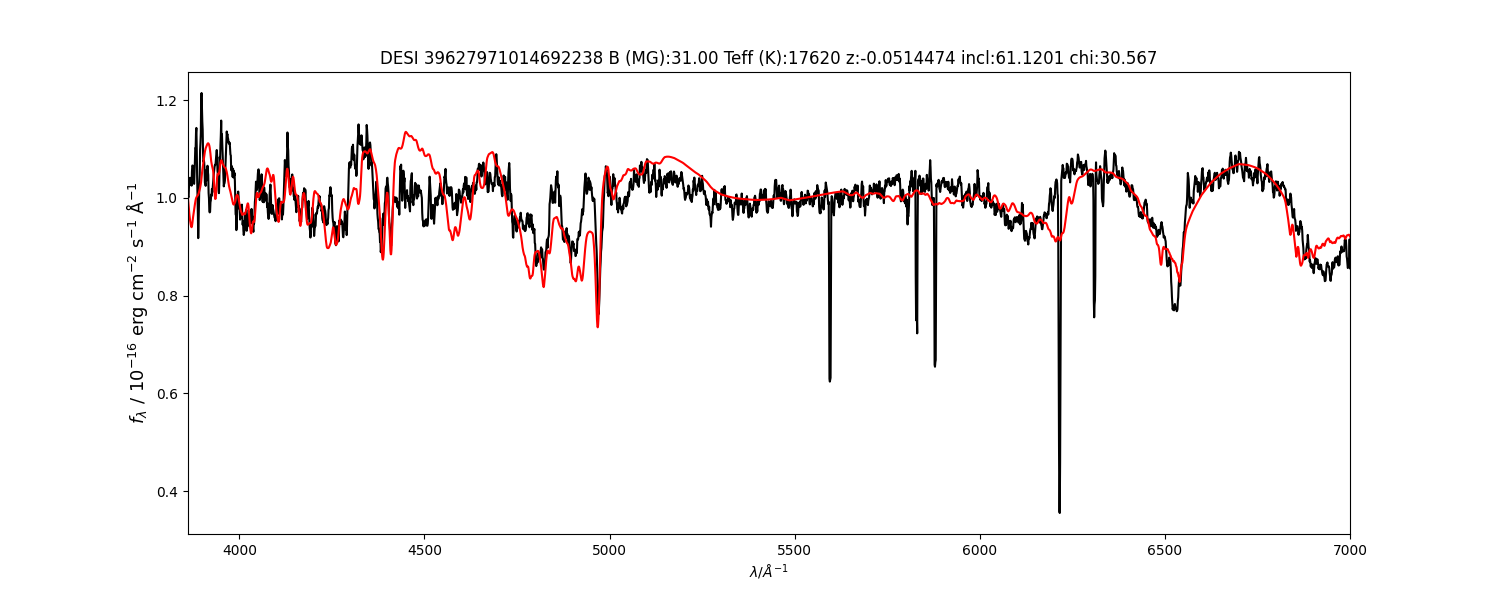}\\
\includegraphics[width=0.9\linewidth]{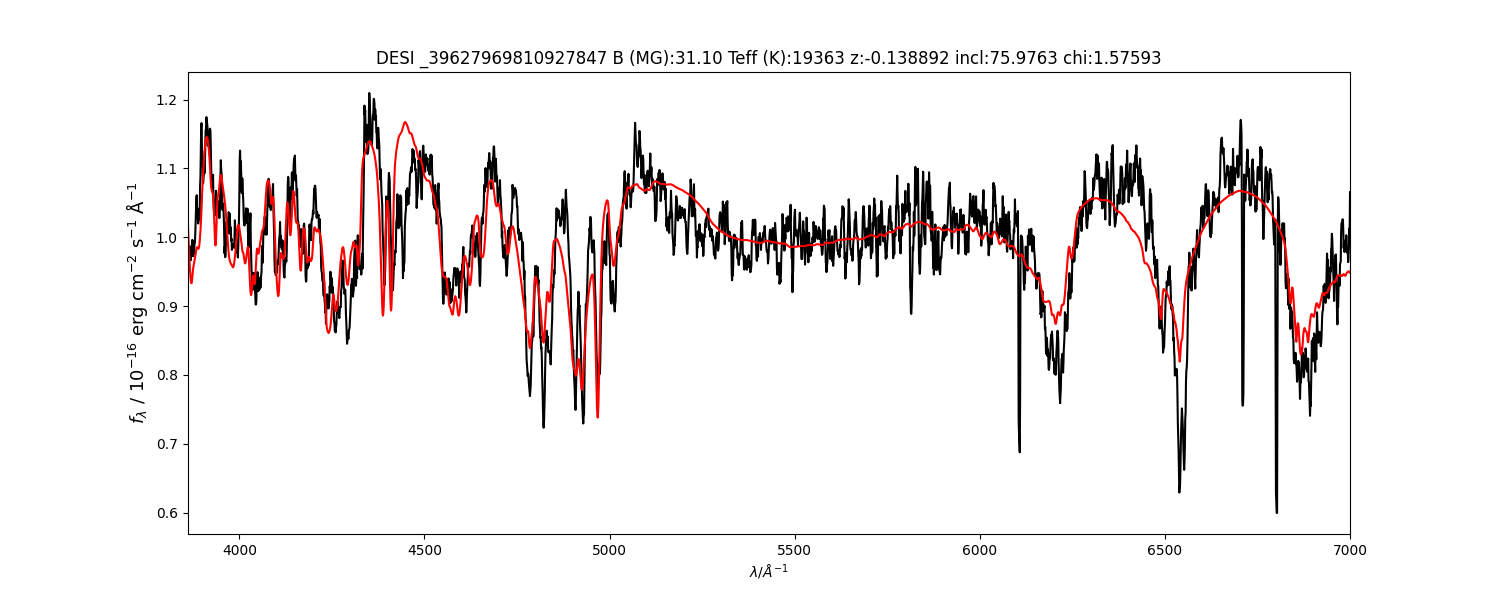}\\ 
\end{supertabular}
 \newpage \captionof{figure}{cont.}
\begin{supertabular}{c}
\includegraphics[width=0.9\linewidth]{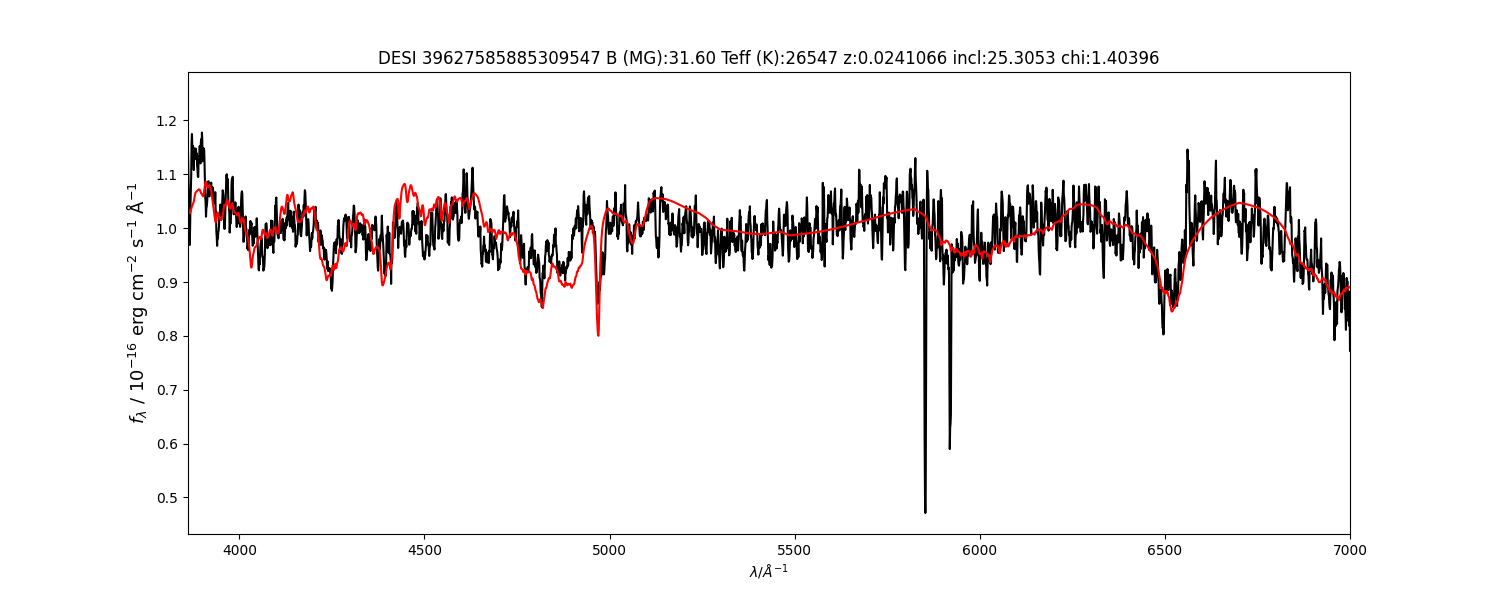}\\
\includegraphics[width=0.9\linewidth]{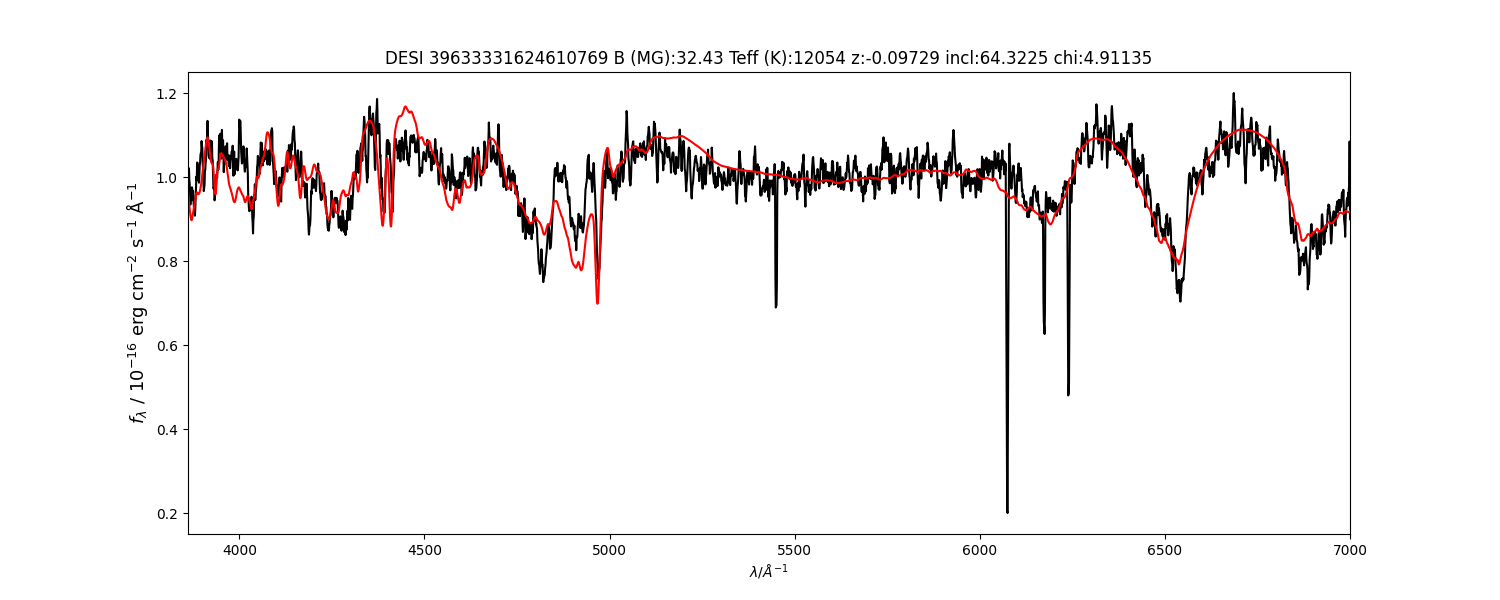}\\
\includegraphics[width=0.9\linewidth]{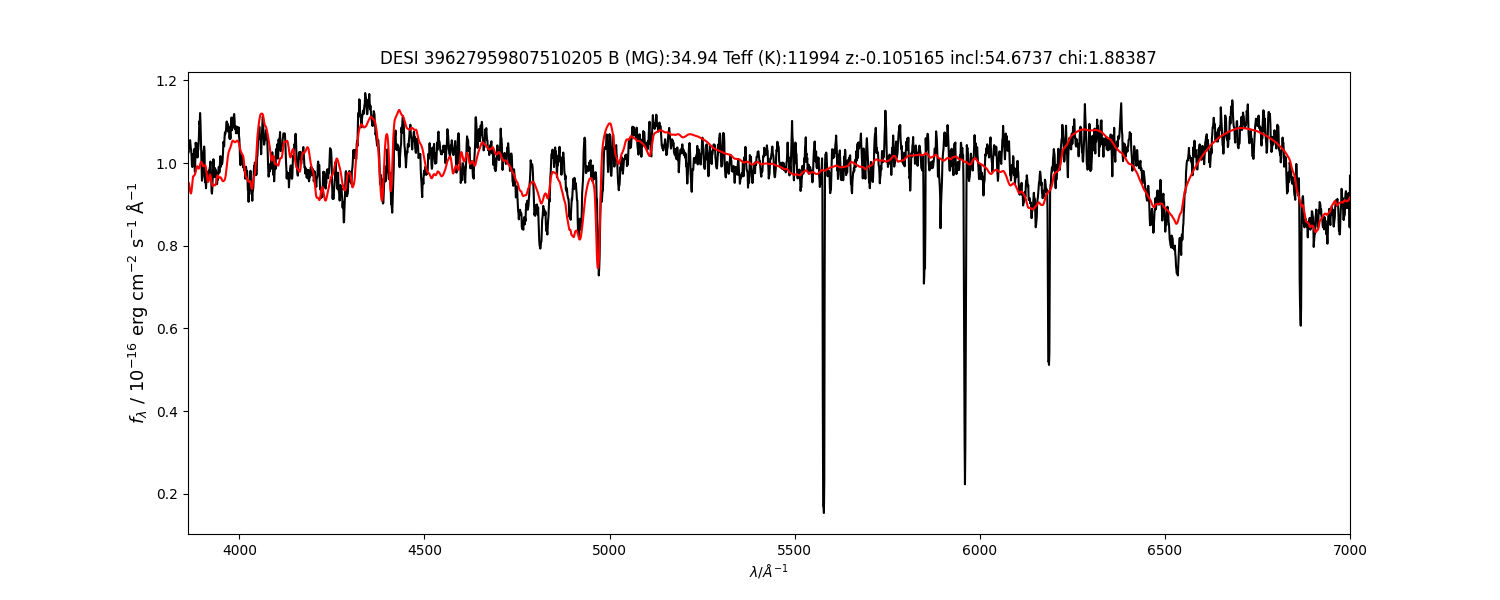}\\ 
\end{supertabular}
 \newpage \captionof{figure}{cont.}
\begin{supertabular}{c}
\includegraphics[width=0.9\linewidth]{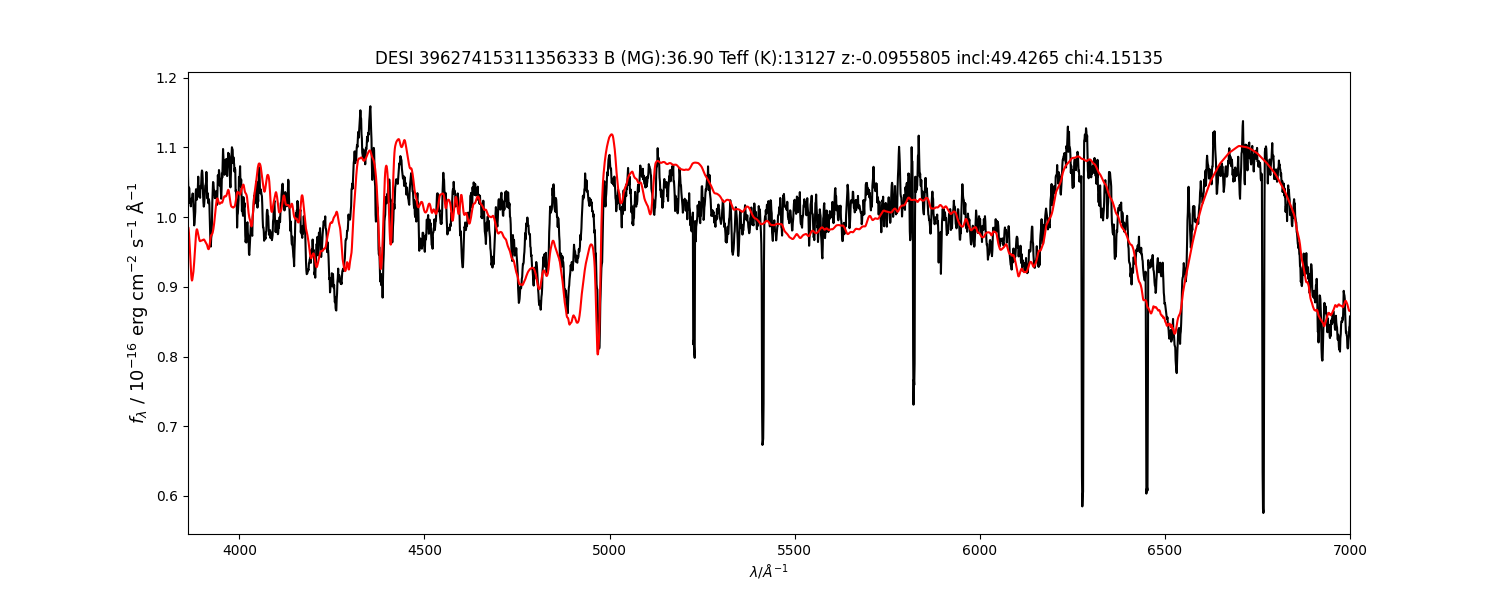}\\
\includegraphics[width=0.9\linewidth]{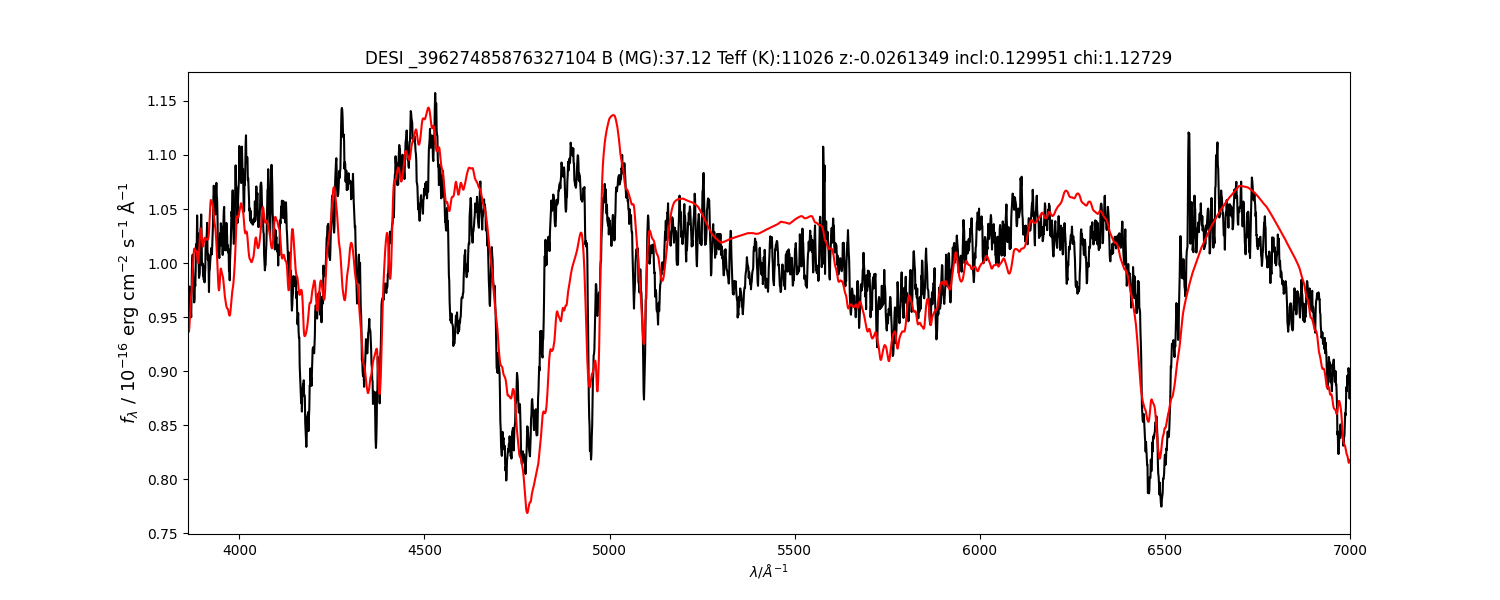}\\
\includegraphics[width=0.9\linewidth]{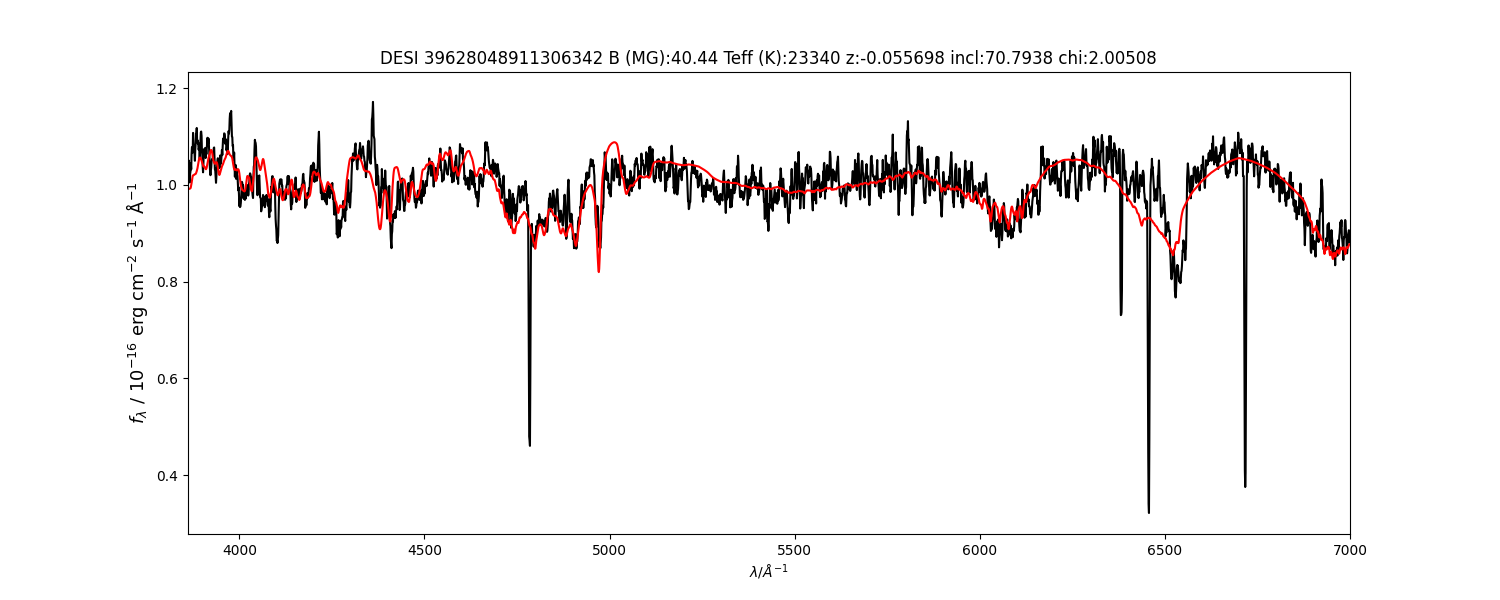}\\ 
\end{supertabular}
 \newpage \captionof{figure}{cont.}
\begin{supertabular}{c}
\includegraphics[width=0.9\linewidth]{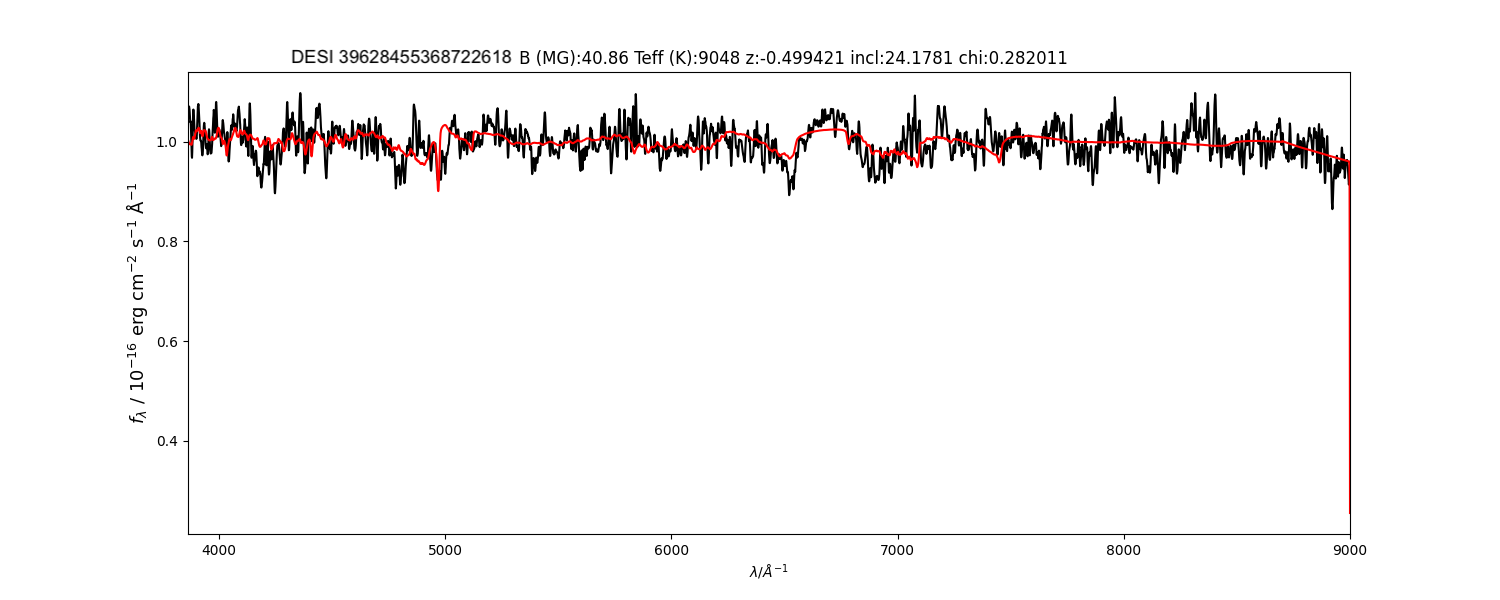}\\
\includegraphics[width=0.9\linewidth]{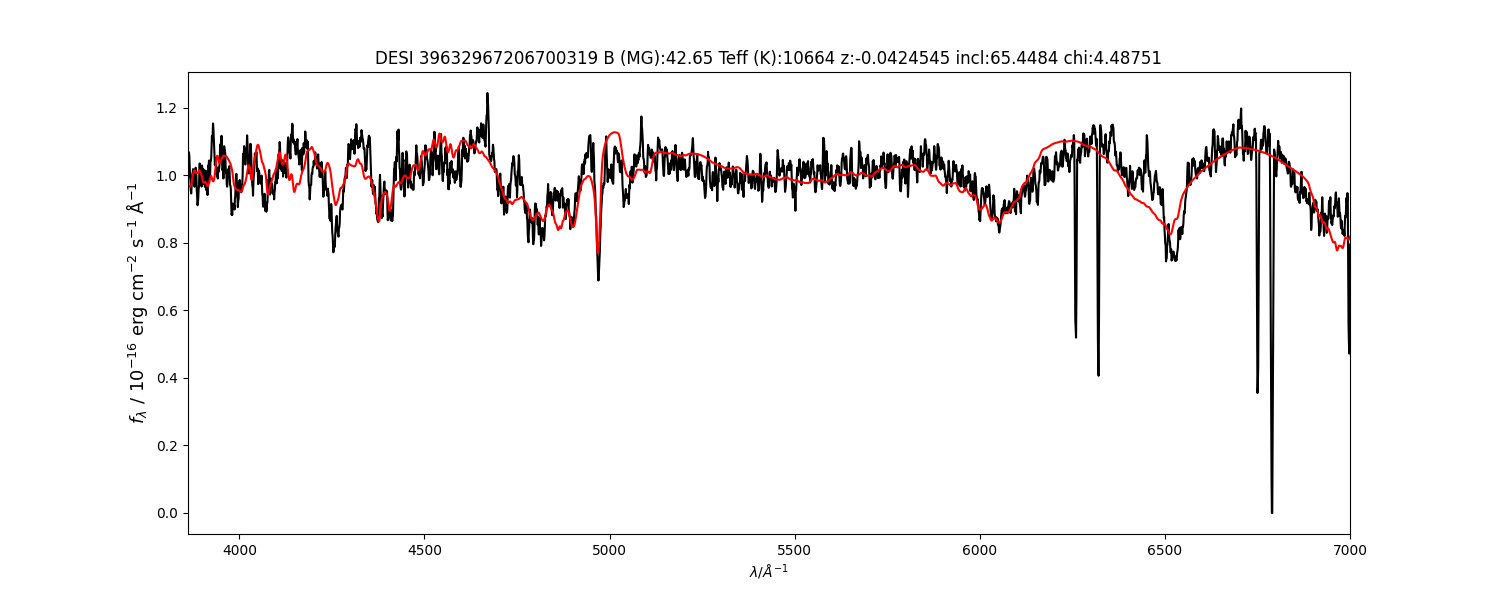}\\
\includegraphics[width=0.9\linewidth]{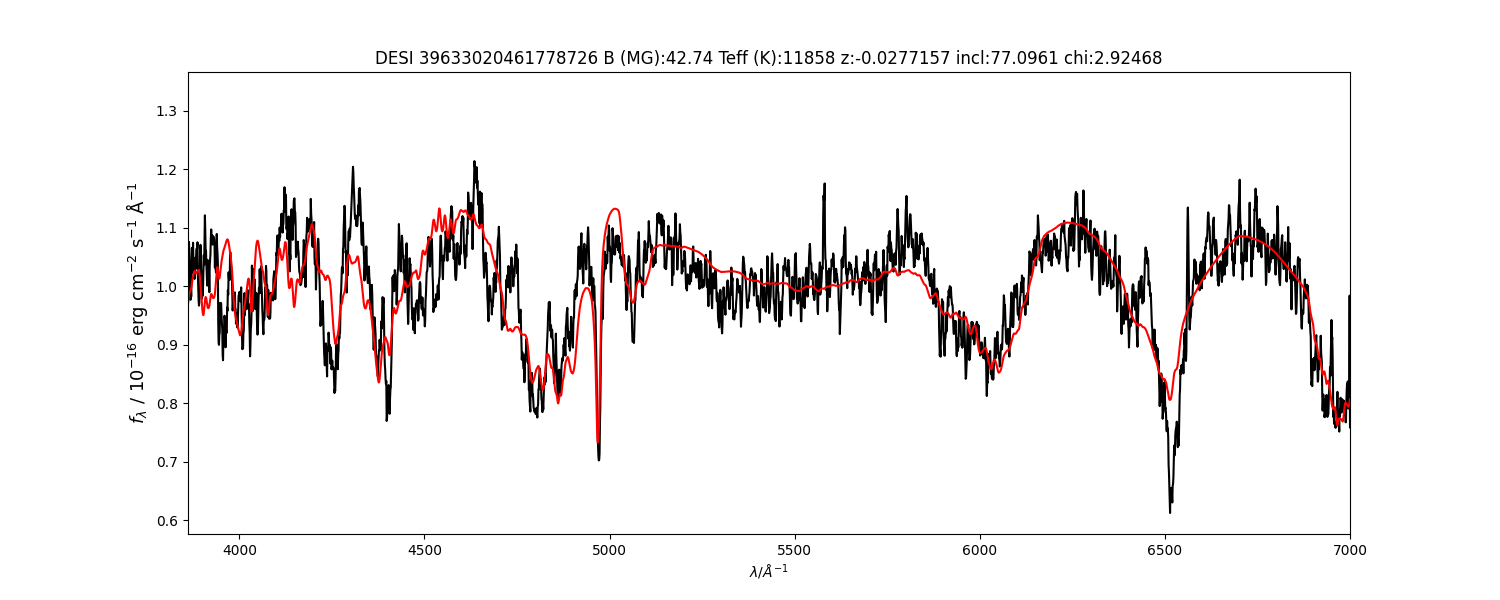}\\ 
\end{supertabular}
 \newpage \captionof{figure}{cont.}
\begin{supertabular}{c}
\includegraphics[width=0.9\linewidth]{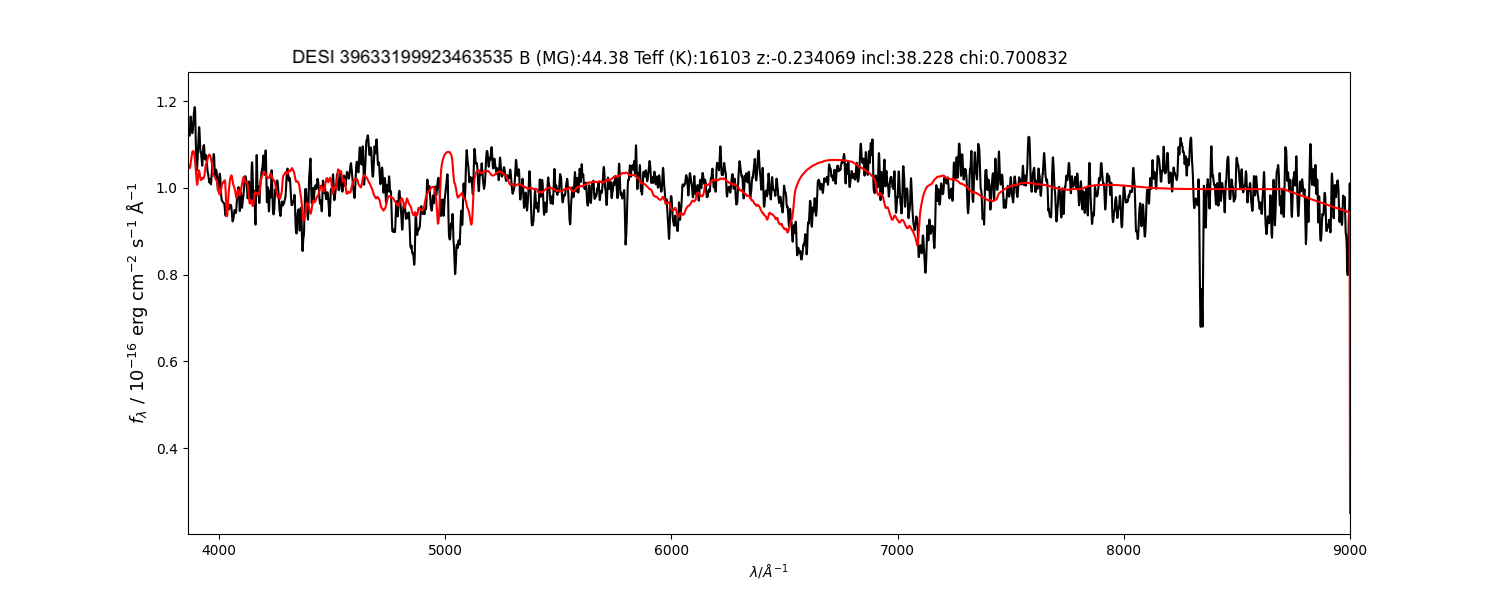}\\
\includegraphics[width=0.9\linewidth]{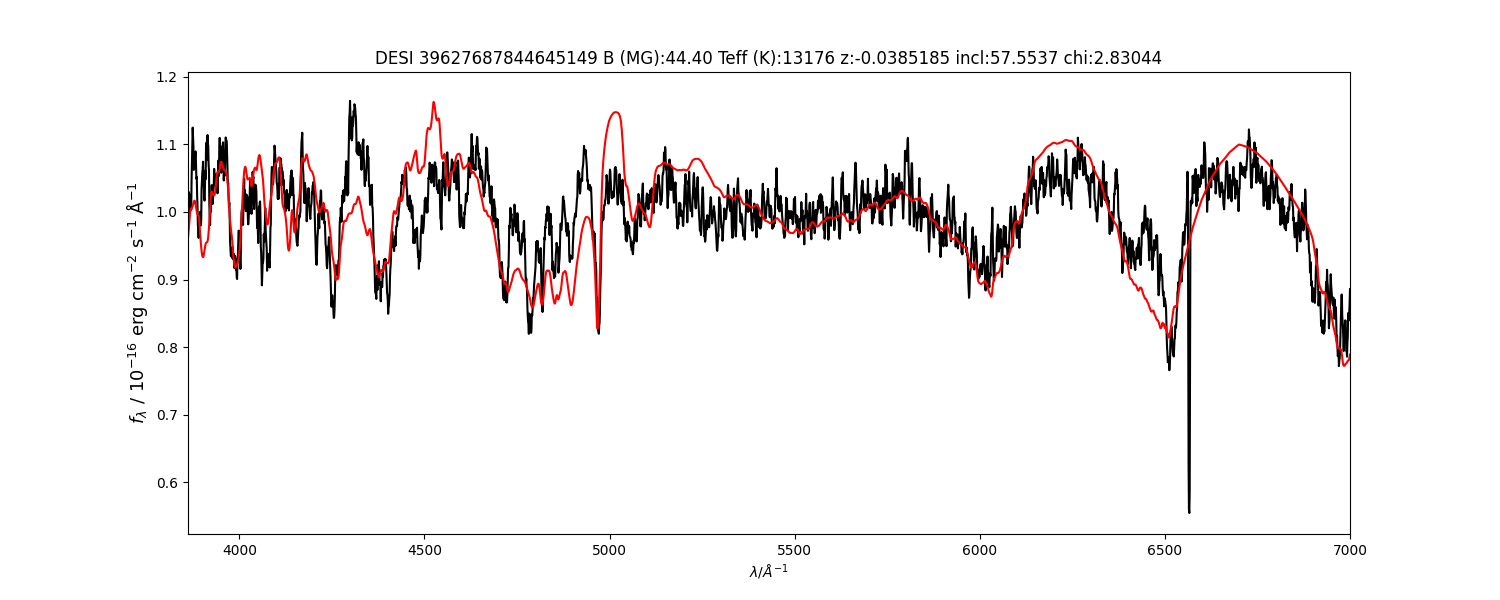}\\
\includegraphics[width=0.9\linewidth]{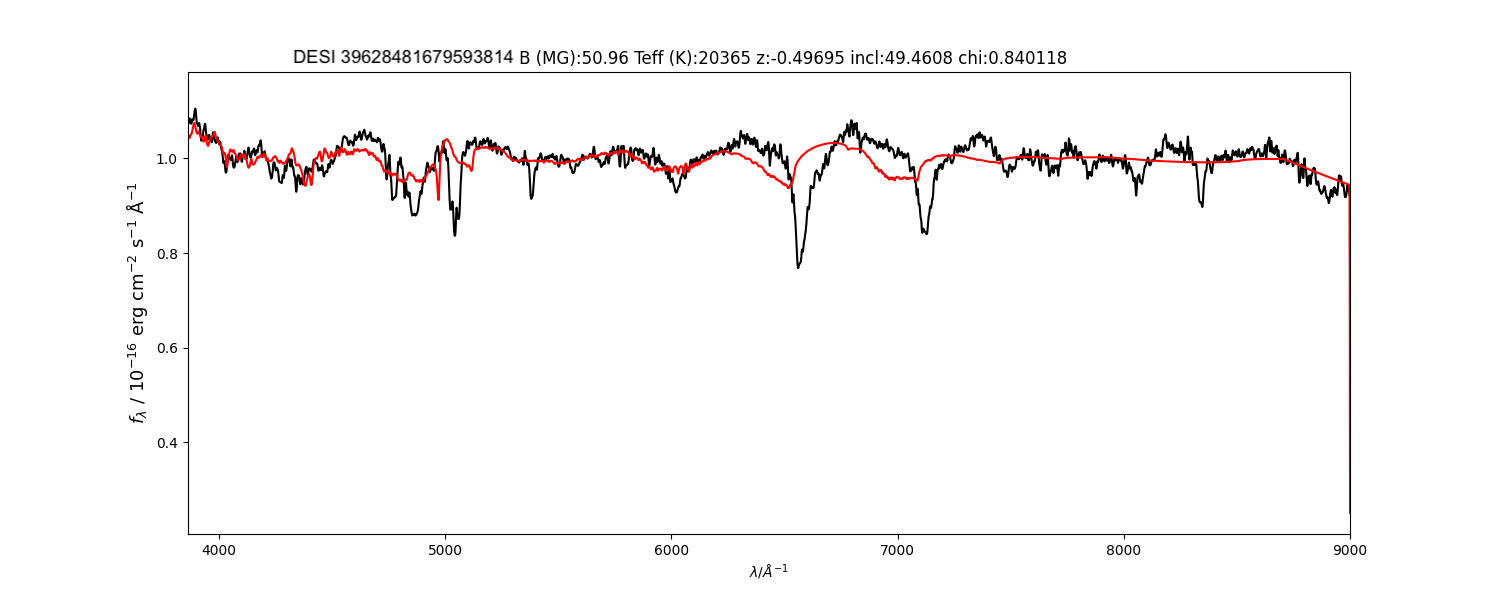}\\ 
\end{supertabular}
 \newpage \captionof{figure}{cont.}
\begin{supertabular}{c}
\includegraphics[width=0.9\linewidth]{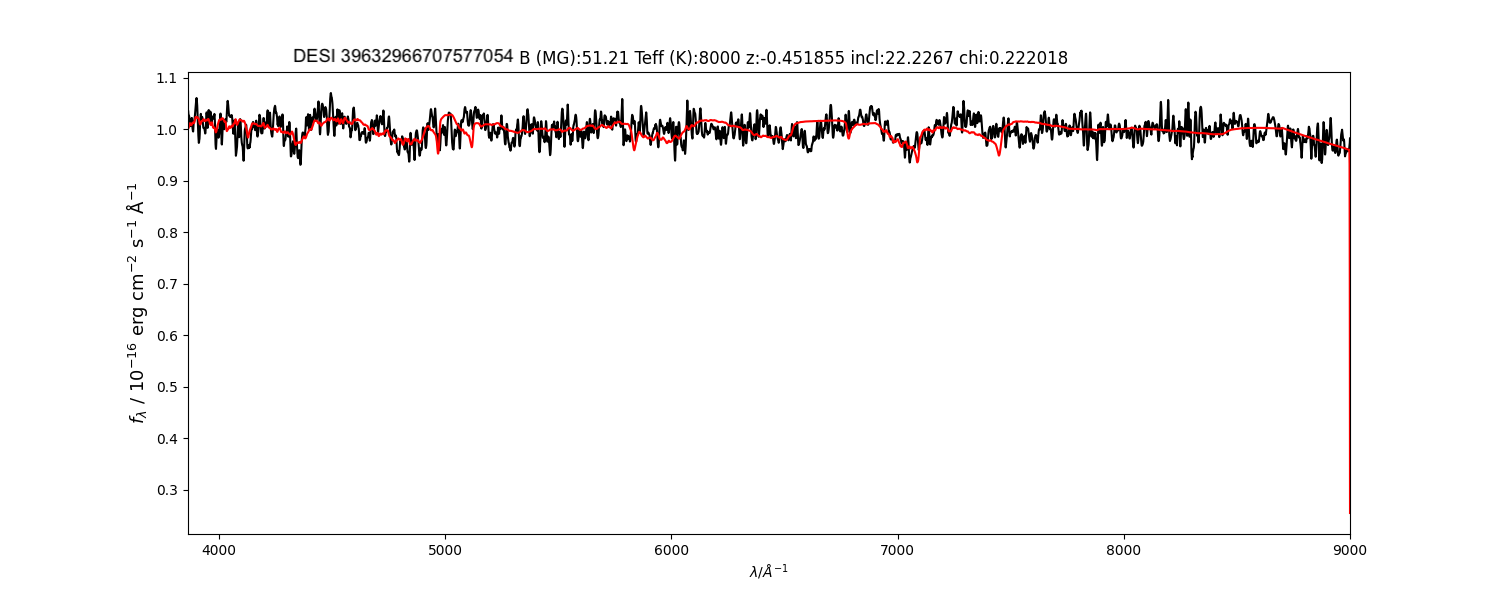}\\
\includegraphics[width=0.9\linewidth]{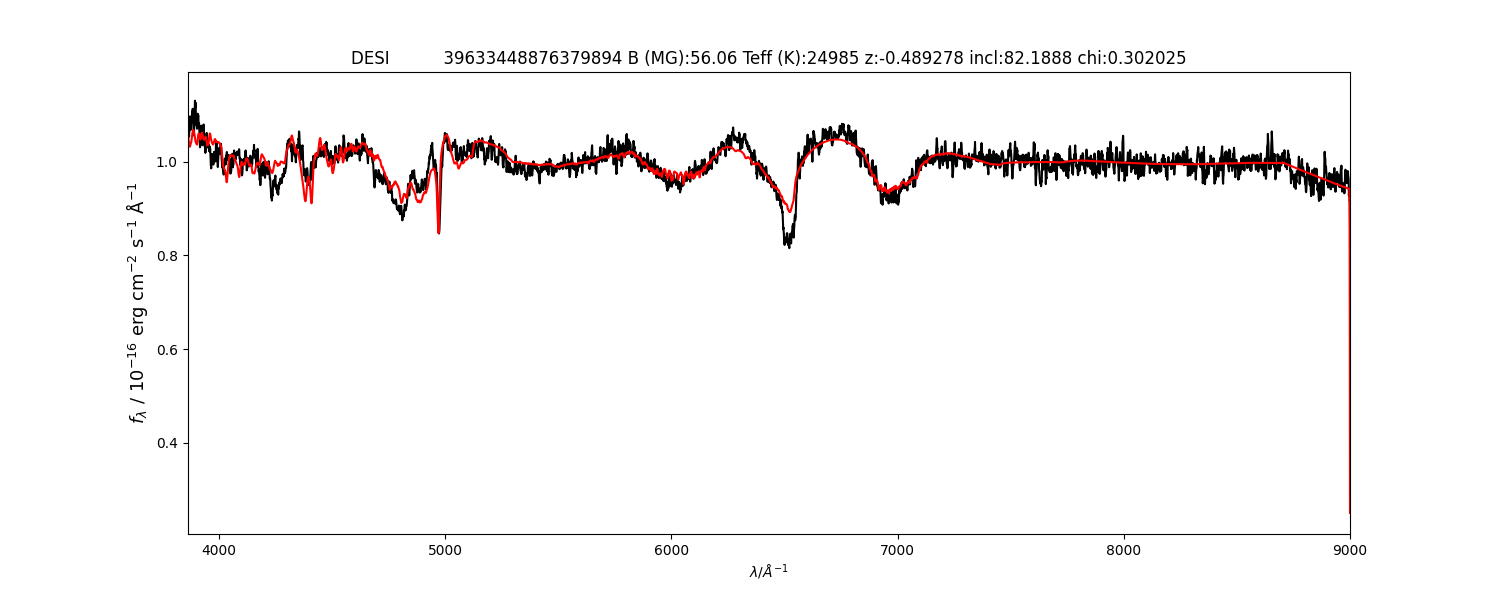}\\
\includegraphics[width=0.9\linewidth]{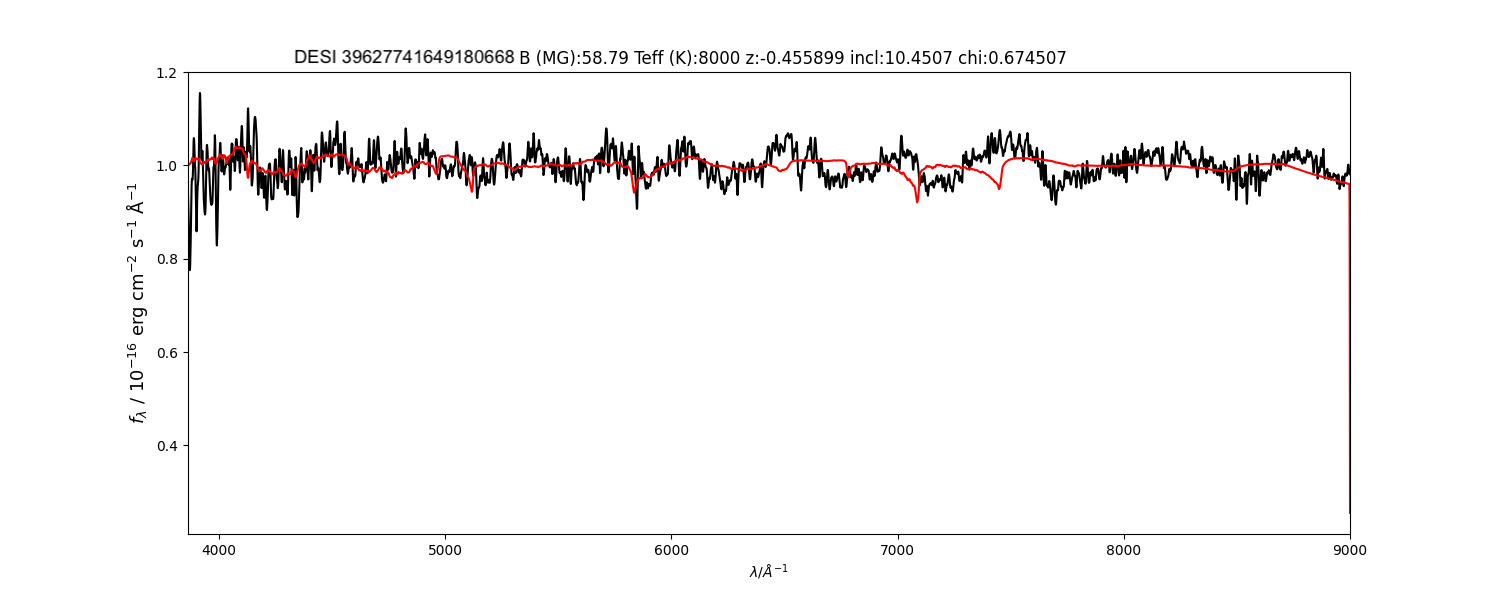}\\ 
\end{supertabular}
 \newpage \captionof{figure}{cont.}
\begin{supertabular}{c}
\includegraphics[width=0.9\linewidth]{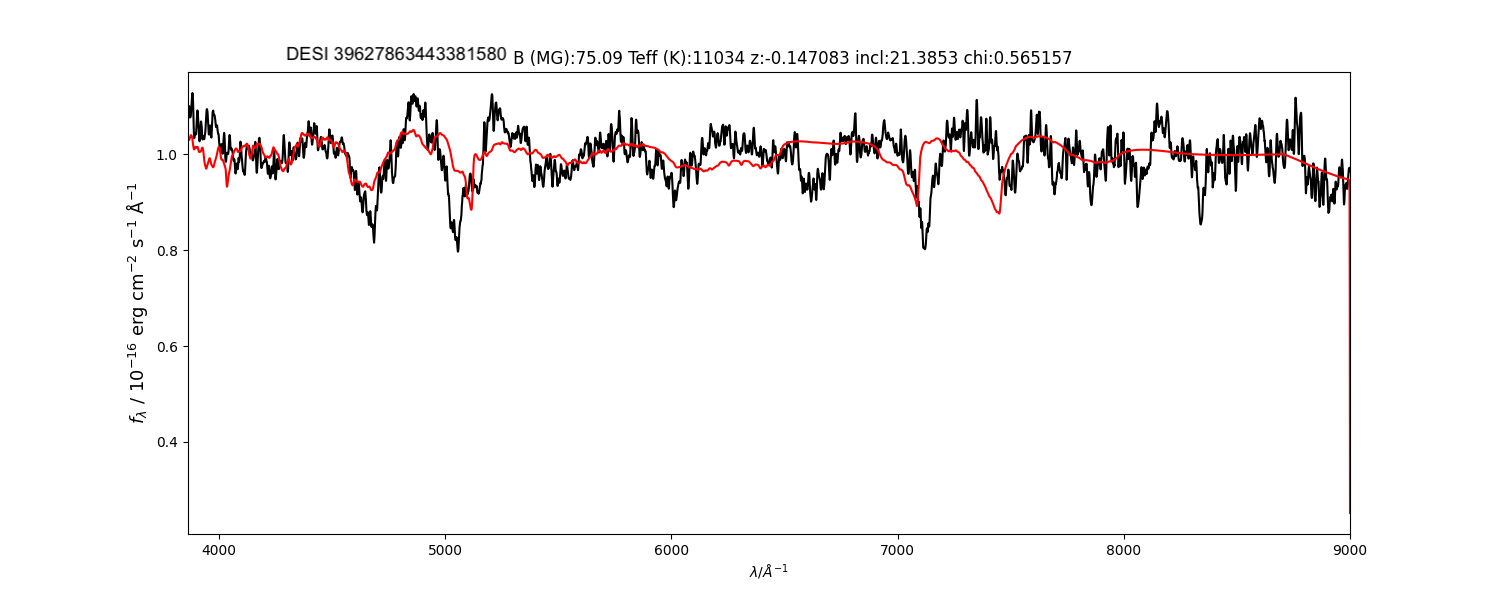}\\
\includegraphics[width=0.9\linewidth]{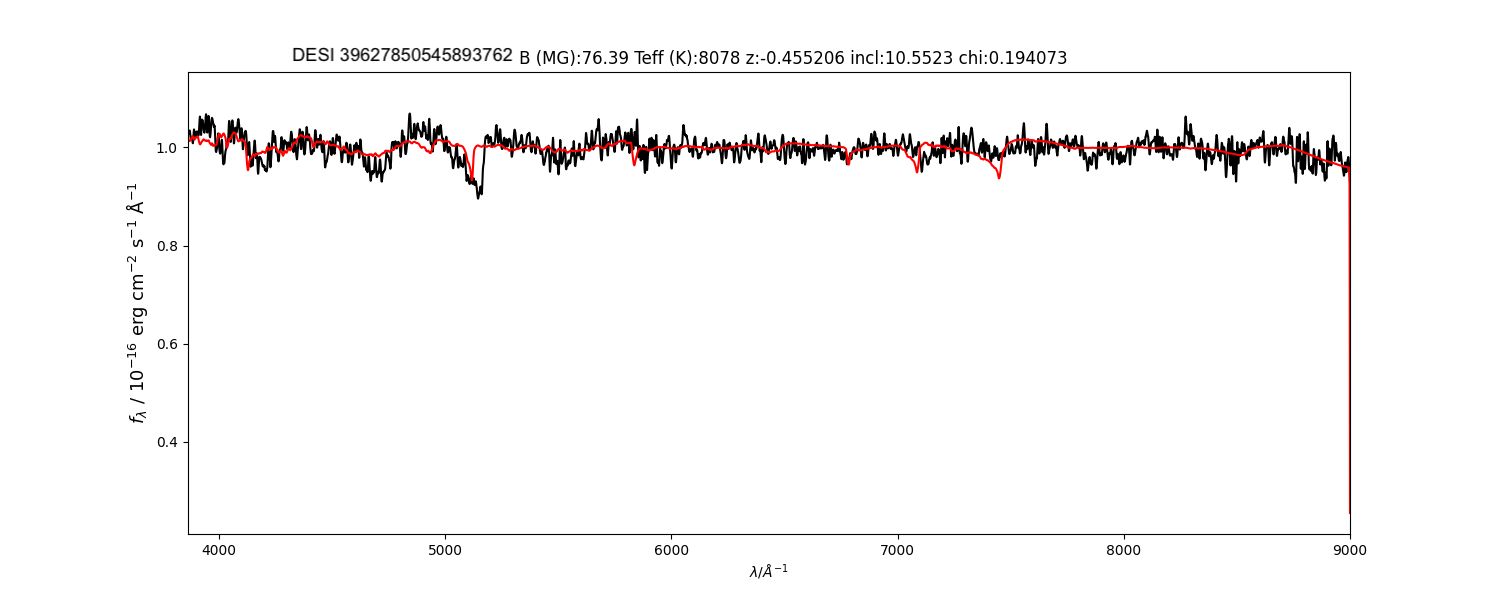}\\
\includegraphics[width=0.9\linewidth]{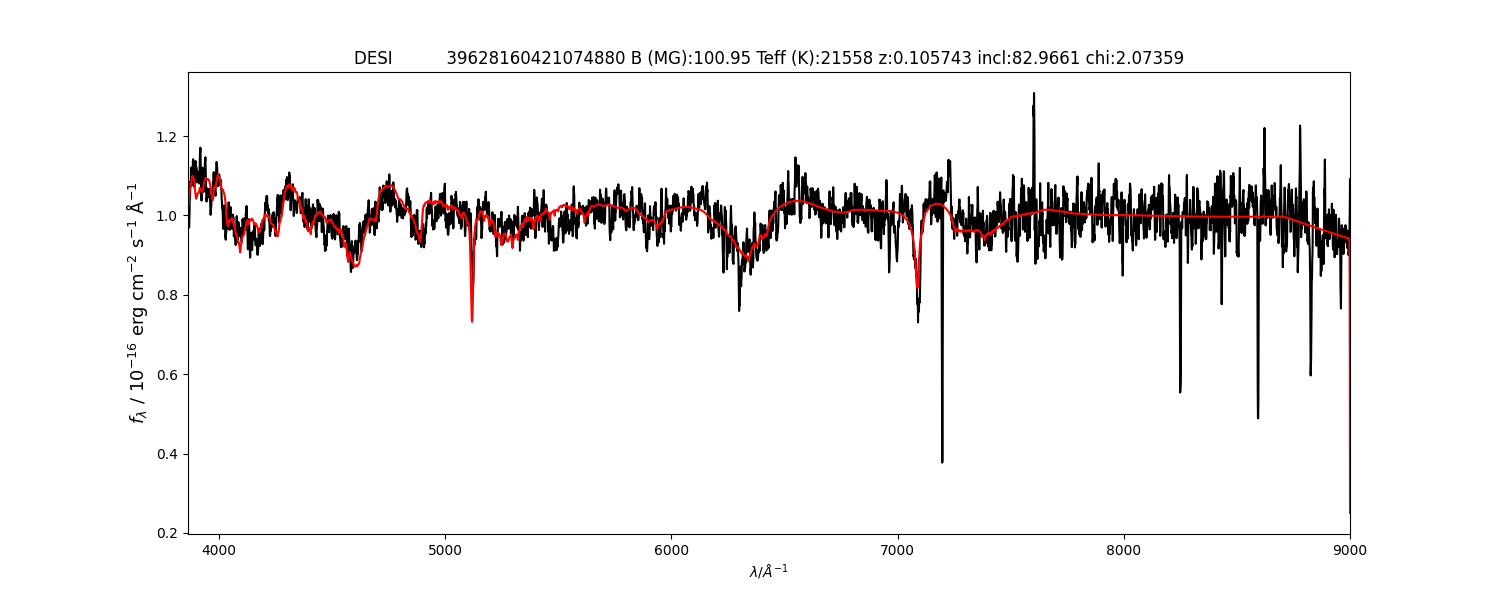}\\ 
\end{supertabular}
 \newpage \captionof{figure}{cont.}
\begin{supertabular}{c}
\includegraphics[width=0.9\linewidth]{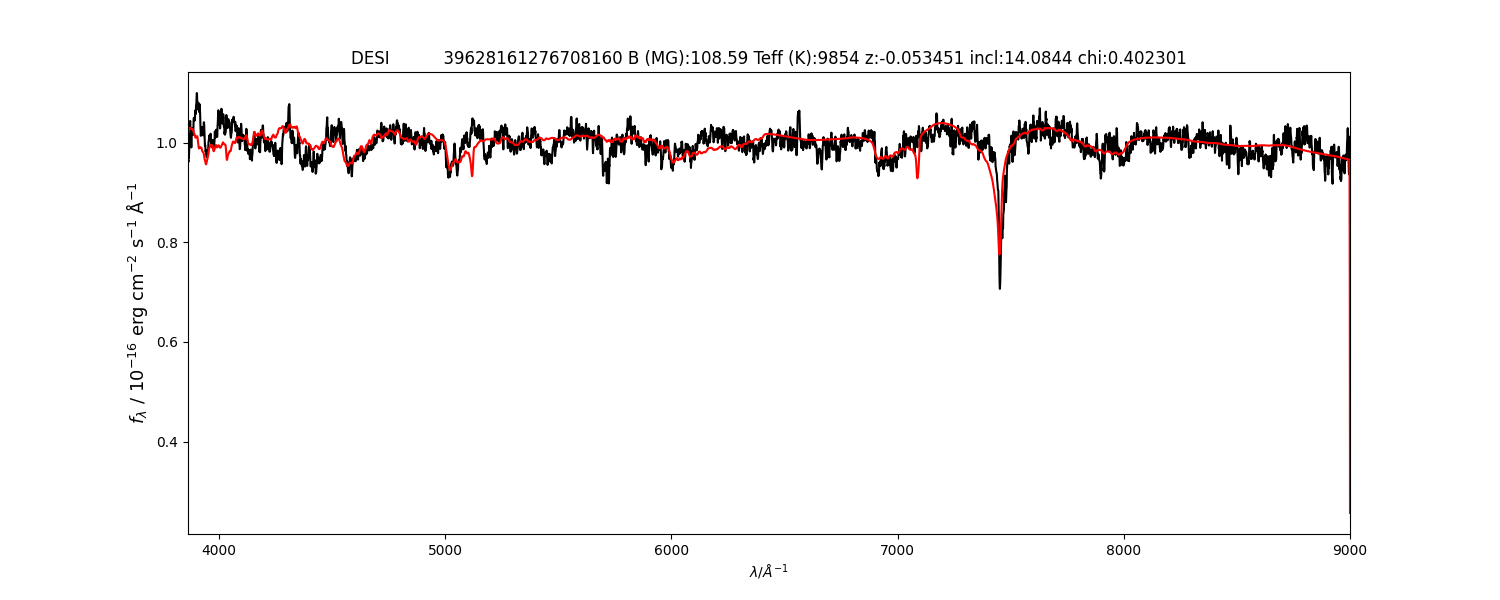}\\
\includegraphics[width=0.9\linewidth]{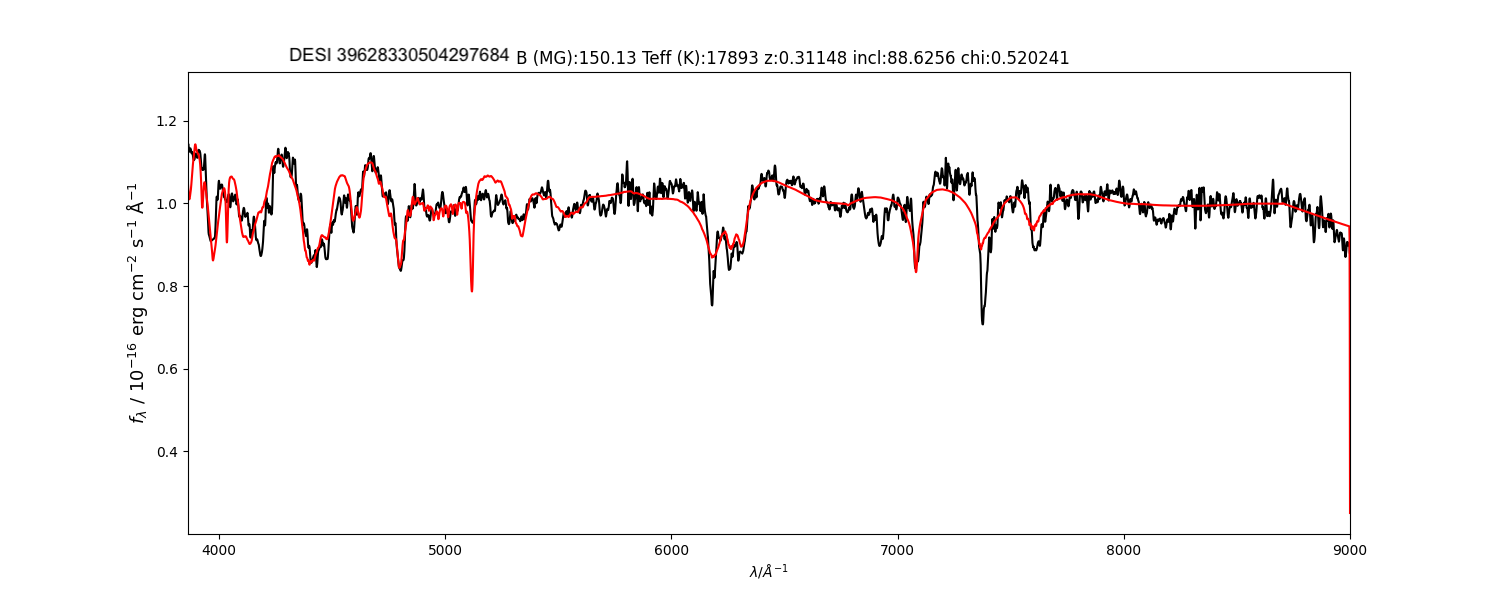}\\
\includegraphics[width=0.9\linewidth]{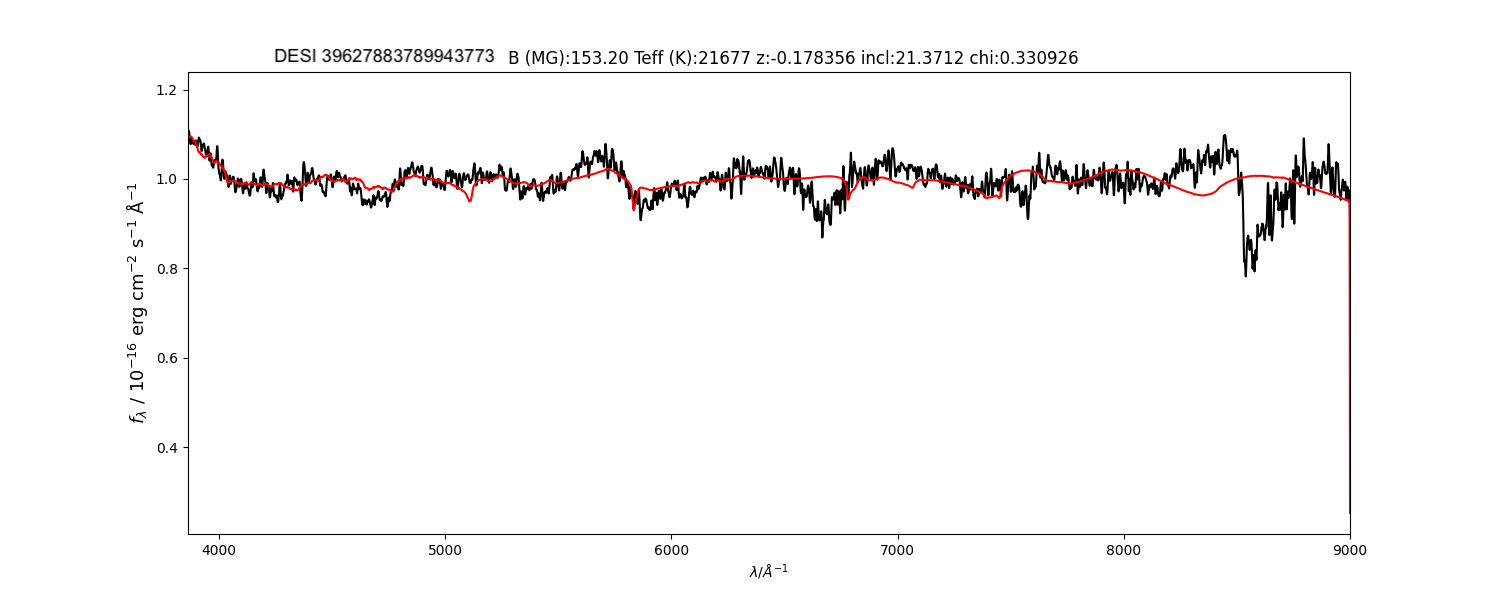}\\ 
\end{supertabular}
 \newpage \captionof{figure}{cont.}
\begin{supertabular}{c}
\includegraphics[width=0.9\linewidth]{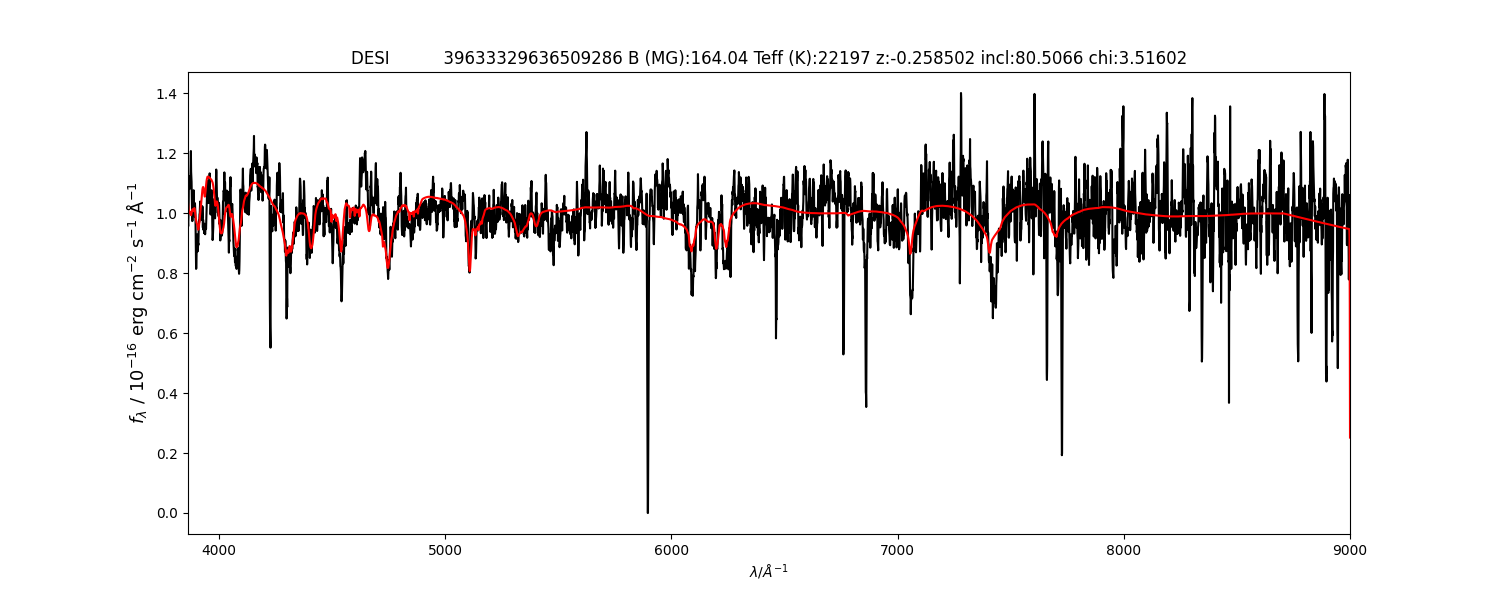}\\
\includegraphics[width=0.9\linewidth]{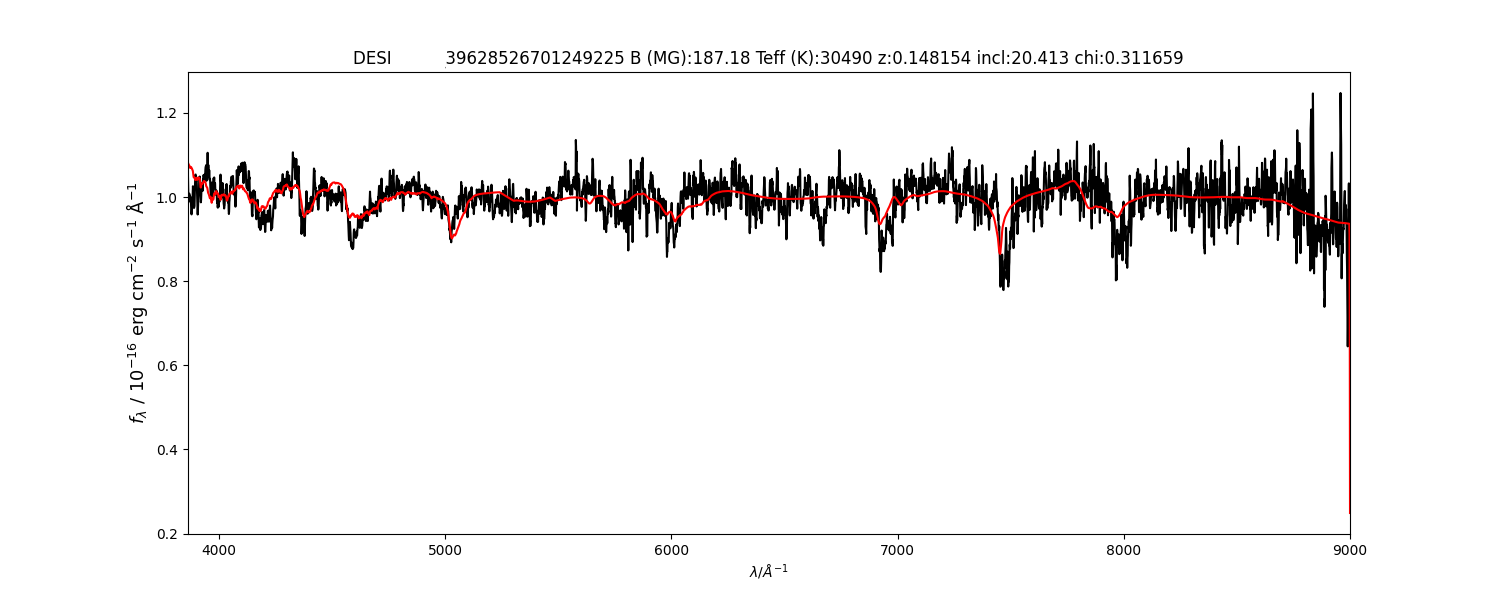}\\
\includegraphics[width=0.9\linewidth]{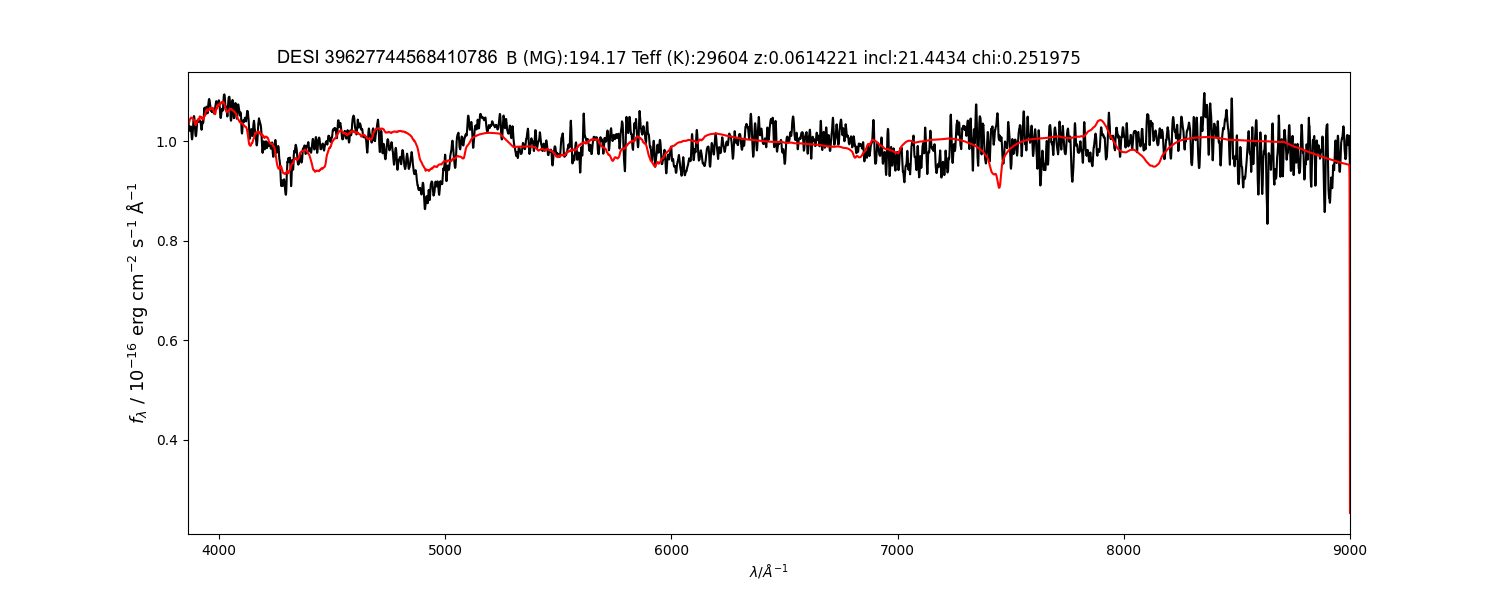}\\ 
\end{supertabular}
 \newpage \captionof{figure}{cont.}
\begin{supertabular}{c}
\includegraphics[width=0.9\linewidth]{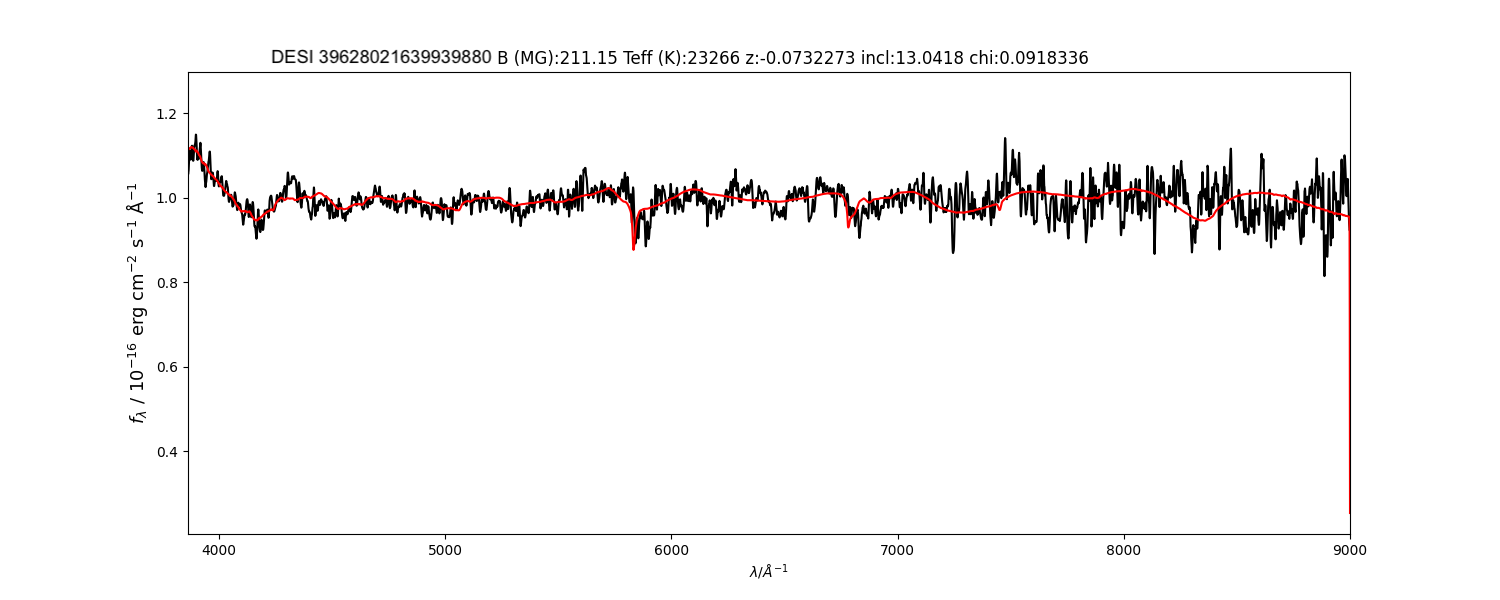}\\
\includegraphics[width=0.9\linewidth]{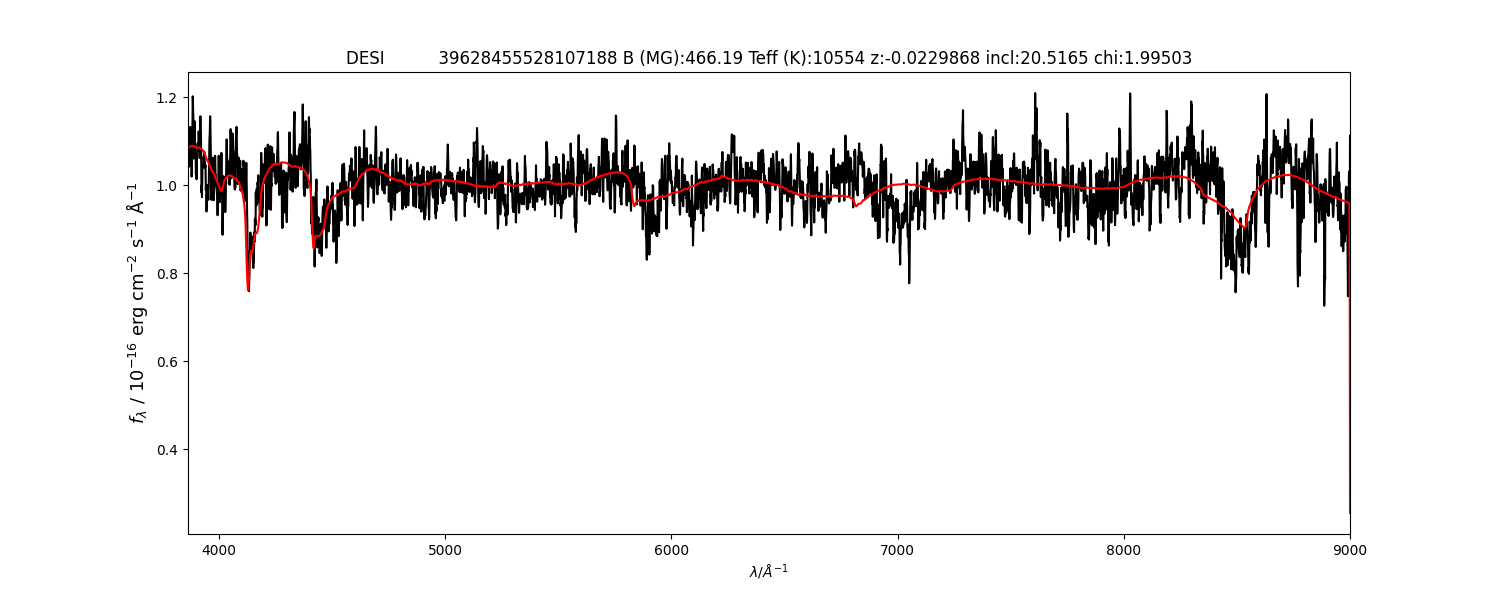}\\
\end{supertabular}

\end{document}